\documentclass[11pt]{article}
\pdfoutput=1
\usepackage{amsmath,amssymb,epsfig,amsfonts}
\usepackage{graphicx}
\usepackage{subfig}
\usepackage[usenames, dvipsnames]{color}
\usepackage{cite}
\usepackage{multirow}
\usepackage{hyperref}
\usepackage{verbatim} 
\usepackage{enumitem}
\usepackage{rotating}
\usepackage{xcolor}
\usepackage{multirow}
\usepackage{bm}
\usepackage[normalem]{ulem}
\usepackage{cleveref}
\usepackage{slashed}

\graphicspath{ {figs/} }

\makeatletter

\newcommand{\be}{\begin{equation}}
\newcommand{\ee}{\end{equation}}
\newcommand{\ba}{\begin{aligned}}
\newcommand{\ea}{\end{aligned}}

\newcommand{\bea}{\begin{eqnarray}}
\newcommand{\eea}{\end{eqnarray}}

\def\unit{{1\kern-.65ex {\rm l}}}
\def\1{{1\kern-.65ex {\rm l}}}

\newcount\hour \newcount\minute
\hour=\time \divide \hour by 60
\minute=\time
\def\now{%
\ifnum \hour<13
  \ifnum \hour=0 \advance \hour by 12 \number\hour:\else \number\hour:\fi%
     \ifnum \minute<10 0\fi%
     \number\minute%
\ A.M.%
\else \advance \hour by -12 \number\hour:%
  \ifnum \minute<10 0\fi%
  \number\minute%
  \ P.M.%
\fi%
}

\makeatother

\def\mb{\mathbb}
\def\mbf{\mathbf}
\def\mc{\mathcal}
\def\bp{\begin{pmatrix}}
\def\ep{\end{pmatrix}}

\begin{document}

\baselineskip=18pt  
\numberwithin{equation}{section}  
\allowdisplaybreaks  


%
%


\thispagestyle{empty}

\vspace*{0.8cm} 
\begin{center}
{\Huge {5d SCFTs from Decoupling and Gluing}
}

 \vspace*{1.5cm}
Fabio Apruzzi,  Sakura Sch\"afer-Nameki and Yi-Nan Wang\\

 \vspace*{1.0cm} 
{\it  Mathematical Institute, University of Oxford, \\
Andrew-Wiles Building,  Woodstock Road, Oxford, OX2 6GG, UK}\\

\vspace*{0.8cm}
\end{center}
\vspace*{.5cm}

\noindent
We systematically analyse 5d superconformal field theories (SCFTs) obtained by dimensional reduction from 6d $\mathcal{N}=(1,0)$ SCFTs. Such theories have a realization as M-theory on a singular Calabi-Yau threefold, from which we determine the so-called combined fiber diagrams (CFD) introduced in \cite{Apruzzi:2019vpe, Apruzzi:2019opn, Apruzzi:2019enx}. 
The CFDs are graphs that encode the superconformal flavor symmetry, BPS states, low energy descriptions, as well as descendants upon flavor matter decoupling. 
To obtain a 5d SCFT from 6d, there are two approaches: 
the first is to consider a circle-reduction combined with mass deformations. The second is to circle-reduce and decouple an entire gauge sector from the theory. 
The former is applicable e.g. for very Higgsable theories, whereas the latter is required to obtain a 5d SCFT from a non-very Higgsable 6d theory. 
In the M-theory realization the latter case corresponds to decompactification of a set of compact surfaces in the Calabi-Yau threefold. To exemplify this we consider the 5d SCFTs that descend from  non-Higgsable clusters and non-minimal conformal matter theories.
Finally, inspired by the quiver structure of 6d theories, we propose a gluing construction for 5d SCFTs from building blocks and their CFDs. 

\newpage

\tableofcontents




\section{Introduction}
5d superconformal field theories (SCFTs) are intrinsically non-perturbative. For instance, 5d gauge theories become
strongly coupled in the UV and they can only be low-energy effective descriptions of the putative superconformal field theories. In particular, their Coulomb branches can be used to effectively study the SCFTs at low-energies \cite{Seiberg:1996bd}. Generically 5d SCFTs show very interesting non-perturbative phenomena, such as enhancement of flavor symmetry at strong coupling, which characterize the spectrum of operators of the SCFT \cite{Seiberg:1996bd, Kim:2012gu, Tachikawa:2015mha, 
Bergman:2013aca,
Zafrir:2014ywa, Mitev:2014jza, Hwang:2014uwa, Gaiotto:2015una,
Yonekura:2015ksa, Zafrir:2015uaa, Bergman:2016avc, Ferlito:2017xdq, Cabrera:2018jxt, Apruzzi:2019vpe, Apruzzi:2019opn, Apruzzi:2019enx, Bourget:2019aer}.  

One of the recent successes of geometric approaches to string theory is the prediction for the existence of 6d and 5d  SCFTs. More specifically, the non-perturbative completions of string theory, i.e. F- and M-theory, not only predict the existence of these SCFTs by relying on the geometry of singular Calabi-Yau threefolds, but they also encode key physical features of the SCFTs.
 An F-theory geometric classification of 6d SCFTs with a tensor branch has been realized in \cite{Heckman:2013pva, DelZotto:2014hpa, Heckman:2015bfa}, whereas 5d SCFTs can be engineered from M-theory on a  non-compact singular Calabi-Yau threefold \cite{Morrison:1996xf, Intriligator:1997pq,DelZotto:2017pti,Xie:2017pfl, Apruzzi:2019vpe, Apruzzi:2019opn, Apruzzi:2019enx} and torically in \cite{Apruzzi:2018nre, Closset:2018bjz, Saxena:2019wuy}. 
Based on this approach, some partial classifications have been proposed in \cite{Jefferson:2017ahm,Jefferson:2018irk,Bhardwaj:2018yhy, Bhardwaj:2018vuu,  Apruzzi:2019vpe, Apruzzi:2019opn, Apruzzi:2019enx, Bhardwaj:2019jtr, Bhardwaj:2019fzv}. 
Complementing this, large classes of theories have been constructed via IIB brane webs \cite{Aharony:1997ju,Aharony:1997bh,DeWolfe:1999hj,Bergman:2015dpa, Zafrir:2015rga, Zafrir:2015ftn, Ohmori:2015tka, Hayashi:2017btw, Hayashi:2018bkd,Hayashi:2018lyv, Hayashi:2019yxj, Kim:2019dqn}.  

A natural question is whether 5d SCFTs are related to 6d SCFTs upon circle compactification, and even whether they can be classified by descending from 6d. {This approach was initiated in \cite{Ganor:1996pc}. 
Recently in} \cite{Apruzzi:2019vpe, Apruzzi:2019opn, Apruzzi:2019enx}, we utilized  the relation between 6d and 5d SCFTs by systematically studying the singularity resolutions of the singular  Calabi-Yau threefolds that underlie the construction of 6d theories in  F-theory. In M-theory, these geometries model the Coulomb branch of 5d gauge theories. 
The superconformal flavor symmetry of the 5d SCFTs is a key datum to characterize these theories, and in \cite{Apruzzi:2018nre, Apruzzi:2019vpe, Apruzzi:2019opn, Apruzzi:2019enx} we initiated a classification, which keeps manifestly track of the strongly coupled flavor symmetries. These can be computed geometrically, and more strikingly can be summarized in graphs, the {\it combined fiber diagrams (CFD)}. The CFDs are very powerful tools. They contain as marked subgraphs, the  Dynkin diagrams of the enhanced flavor symmetries as well as nontrivial information about the BPS states of the theory. Furthermore, they comprehensively encode all mass deformations that trigger RG-flows with new 5d UV fixed points. 

The classification strategy in \cite{Apruzzi:2019vpe, Apruzzi:2019opn, Apruzzi:2019enx} can be succinctly summarized as follows:
begin with a 6d SCFT, defined by a singular elliptic Calabi-Yau geometry in F-theory.  These geometries have so-called non-minimal singularities. Depending on the resolution of the singularities, we obtain different M-theory compactifications, which model 5d gauge theories on the Coulomb branch. From this geometry, we extract the CFD, which encodes the flavor symmetry at the origin of the Coulomb branch, as well as the mass deformations, which trigger RG-flows. 
The systematic exploration of 5d SCFTs hinges then on obtaining the CFDs for the marginal or KK-theories, from which all descendant 5d SCFTs can be obtained by simple graph operations on the CFDs.

The CFDs defined in \cite{Apruzzi:2019vpe} not only encode key non-perturbative information of the theory but also the trees of descendant 5d SCFTs with the same dimension of the Coulomb branch (rank). 
This approach is always applicable in the case of so-called {\it very Higgsable} 6d theories \cite{Ohmori:2015pia}, i.e. geometries where the base of the elliptic fibration is smooth.

In cases when the 6d theory is not very Higgsable, the approach requires substantial generalization. 
Specifically, the base of the elliptic fibration for a 6d SCFT is in general an orbifold $\mathbb C^2/\Gamma$, where $\Gamma$ is a finite subgroup of $U(2)$. 
These are called {\it non-very Higgsable theories}. Examples are the  non-Higgsable clusters (NHCs) and non-minimal conformal matter theories,  corresponding to $N>1$ M5-branes probing $\mathbb R \times \mathbb{C}^2/\Gamma_\text{ADE}$. 

Again the circle reduction of this class of 6d SCFTs lead to 5d KK-theories, which uplift back to 6d SCFTs in the UV, but differently from the very Higgsable case, they do have an IR description in terms of a marginal theory with flavor matter (as opposed to bifundamental matter).
In \cite{Ohmori:2015pia}, it was conjectured that the reduction to 5d of non-minimal conformal matter leads in the IR to a 5d quiver gauge theory coupled to 
an extra dynamical $SU(N)$ vector multiplet, and only the decoupling of the latter can result in a theory with a 5d UV fixed point. 
We prove this conjecture geometrically for the case of non-minimal conformal matter and some of the single node tensor branch theory with a gauge group, which contains the NHCs. For these theories the only possible way to get 5d SCFTs consist of the following two options:
either the theory in the IR allows decoupling of a bifundamental hypermultiplet, whereby the resulting 5d theory factorizes into two SCFTs -- this case is not of interest to us. 

The second option -- which is the main objective of this paper -- is the possibility of decoupling an entire gauge sector, which then allows for the existence of a UV fixed point. 
Geometrically, this means sending the volume of the compact divisors, that engineer this gauge sector, to infinity -- i.e. they are decompactified. 

We study the geometries for the single curve with gauge group as well as the non-minimal conformal matter theories. We also present the CFDs before and after decompactification, which match the expected flavor symmetry enhancements \cite{Ohmori:2015pia}. For the NHCs we always get the geometry and CFD corresponding in the IR to the 5d gauge theory analog to the 6d theory in the tensor branch. We also study the possible IR low-energy descriptions for non-minimal conformal matter theories as predicted by the CFDs. This can contain interesting strongly {coupled} matter, leading to the construction of 5d generalized quiver, which happens when non-perturbative flavor symmetries are gauged. 

Finally, we propose a gluing construction, motivated by the structure of the 6d tensor branch. 
{The insight that gauging is gluing was observed from the point of view of surfaces in \cite{Bhardwaj:2018vuu}.} 
{We propose gluing condition on the local Calabi-Yau geometry,} in order to realize higher rank theories, such as non-minimal conformal matter starting with lower rank building blocks, where we use the 6d tensor branch as a guide for this gluing procedure. In particular we implement this from the graph theoretic perspective of the CFDs, and describe some rules how to glue them. We verify this gluing from geometry, and from the perspective of the IR gauge theory description. The existence of strongly coupled matter as well as these gluing procedure resemble the punctured sphere and their combination for 4d $\mathcal N=2$ theories of class S, \cite{Benini:2009gi, Gaiotto:2009we, Chacaltana:2010ks}. 

The present paper is structured as follows: in section \ref{sec:back} we both revise some basics on 
5d SCFTs, gauge theories and M-theory geometry. We furthermore give an in depth analysis and derivation of the concept of CFDs as flop-invariants. In section \ref{sec:strategy}, we present the general strategy to get 5d SCFTs given a general 6d SCFTs, which include mass deformation and the decoupling of a subsector of the theory and complement it with a decompactification in the M-theory geometry. 
 In section \ref{sec:NHCs}, we list the CFDs for non-Higgsable clusters, before and after decompactification. In section \ref{sec:NMCM} we present the CFDs and geometries for non-minimal conformal matter, and discuss various, dual low-energy descriptions that are motivated by the CFDs. In section \ref{sec:gluing}, based on the geometry we propose gluing rules for CFDs. We conclude in section \ref{sec:conclusions}. 
Appendix \ref{sec:BB} has a summary of all building blocks, including the tensor branches in 6d, as well as the CFDs in 5d. The remaining appendices provide details of the geometric computations that underlie the main text, in particular the derivations of CFDs for NHCs and non-minimal conformal matter theories.

{\it Note added:} While we were completing this work, the paper \cite{Bhardwaj:2019xeg} appeared which proposes the decoupling idea from 6d to 5d as well. Our findings are consistent with the criteria described therein.

\section{5d SCFTs, CFDs and all that}
\label{sec:back}

In this section we review some basic and crucial aspects of 5d SCFTs and their construction from M-theory on non-compact Calabi-Yaus with a canonical singularity. We will then review how these geometries are captured into graphs called combined fiber diagrams (CFD), which encode the superconformal flavor symmetry, as well as information on the BPS states. 

\subsection{5d SCFTs and Gauge Theories}

5d SCFTs are always strongly coupled and do not have a Lagrangian description at all energy scales. However, they allow for effective descriptions at low energies. For instance, mass deformations of SCFTs with 
\begin{equation} \label{eq:gkterm}
\int d^5x \, g_{\mathcal O}\, \mathcal O(x) \  = \ \frac{1}{g^2_{\rm YM}} \int d^5x\, {\rm Tr}\left(F^{\mu \nu} F_{\mu \nu}\right) \,,
\end{equation}
lead to effective gauge theories at low energies. This implies that we can write an effective Lagrangian in the IR, \cite{Seiberg:1996bd,Intriligator:1997pq, Apruzzi:2019vpe, Jefferson:2017ahm, Apruzzi:2018nre, Closset:2018bjz}. The content of fields consists of vector multiplets and matter hypermultiplets. The vector multiplet is in the adjoint representation of the semi-simple gauge group 
\begin{equation}
  G_{\text{gauge}}= \prod_I G_I^{(r_I)}\,.
\end{equation} 
Note that this allows for quiver gauge theories. Here $r= \sum_i r_i$ is the rank with $r_i$ the ranks of the simple factor.  
Every vector multiplet corresponding to $G_I^{r_I}$ contains a real scalars that can take vev in the Cartan of the gauge group. This defines the Coulomb branch of the $G_{\text{gauge}}$ gauge theory 
\begin{equation}
  \mathcal{C} = \{ \phi_I \in \mathbb{R}^{r} \, | \, \langle \phi ,
  \alpha_j^{(I)}\rangle > 0
\text{ for all } j,I \} \,,
\end{equation}
where $\alpha_j^{(I)}$ are the positive simple roots of
$G_I^{(r_I)}$.

The dynamics of the  gauge theory on the Coulomb branch is parametrized by a real, one-loop exact prepotential
\begin{align}
\ba
\mathcal{F}= \mathcal{F}_\text{classical} + \mathcal{F}_\text{1-loop}=&\sum_I \left( \frac{1}{2g_{YM_I}^2}\, C^{(I)}_{ij} \phi_I^i \phi_I^j + \frac{ k_I}{6} \, d^{(I)}_{ij \ell} \phi_I^i \phi_I^j \phi_I^\ell \right) \cr 
& 
 +\frac{1}{12} \sum_I \left( \sum_{\alpha_i^{(I)} \in \Phi_{\mathfrak{g}}} |\alpha^{(I)} _i \, \phi_I^i|^3 - \sum_{{\bf R }^{(I)}_f} \sum_{ \lambda^{(I)} \in \mathbf W^{(I)} _{{\bf R}_f}} |\lambda^{(I)} _i \, \phi_I^i + m^{(I)} _f|^3 \right. \cr
 & \left. - \sum_J \sigma_{IJ} \sum_{\widetilde \lambda^{(I)} \in \mathbf{\widetilde W}^{(I)} _{{\bf R}_f}}   \sum_{ \widetilde \lambda^{(J)} \in \mathbf{\widetilde W}^{(J)} _{{\bf R}_f}}  |\widetilde \lambda^{(I)} _i \, \phi_I^i+\widetilde \lambda^{(J)} _i \, \phi_J^i + m^{(IJ)} _f|^3\right)  \, ,
\ea
\end{align} 
where {$d^{(I)}_{ij\ell}= \frac{1}{2} {\rm tr}_\text{fund}\left( T^{(I)}_i (T^{(I)}_j T^{(I)}_\ell +
T^{(I)}_\ell T_j^{(I)})\right)$}, $C^{(I)}_{ij}$ is the Cartan matrix and  $\mathbf W^{(I)}_{{\bf R}_f}$ are the weights of a representation ${\bf R}^{(I)}_f$ of the gauge group $G_I$. Moreover, $\sigma_{IJ}=1$ if there is a hypermultiplet connecting two gauge groups, it vanishes otherwise.

Evidence for the existence of a UV fixed point are provided if there exist a point in the physical Coulomb branch such that $\forall\, I$, $g_{YM}^I \rightarrow \infty$, where the physical Coulomb branch is defined by the subregions of $\mathcal C$ with positive definite metric and magnetic string tensions \cite{Jefferson:2017ahm}:
\be
\ba
     G^{(IJ)}_{ij} &= \frac{\partial^2\mathcal F}{\partial \phi^i_I \partial \phi^j_J} >0 \cr 
     T^{(I)}_i&= \frac{\partial\mathcal F}{\partial \phi^i_I} >0 \, .
     \ea\ee

A 5d SCFT can have many IR gauge theory descriptions, which are dual in the UV, meaning that they have the same UV fixed point. For this reason, the gauge redundancies apart from providing information about the Coulomb branch of the theory, they are not enough to specify the UV fixed points completely. At the SCFT point one can describe the theory in terms of operators, correlation functions and states, which are specified by quantum numbers. An important factor is, in fact, given by the flavor symmetry, which in the UV is different from the one observed in the IR. For instance, from a gauge theory perspective there can be a flavor symmetry rotating the hypermultiplets with an associated current $J_{\mu}^{\text{F}_{\text{cl}}}$, transforming in the adjoint representation of $G_{\text{F}_{\text{cl}}}$. In addition, there is a $U(1)$ topological current in 5d defined by,
\begin{align}
	J_T= \frac{1}{8\pi^2}  \star \text{Tr} (F \wedge F) \,.
\end{align}
From the gauge theory point of view there are massive non-perturbative states (like instanton particles), but they become massless at the UV fixed point. Moreover, when they are generically charged under the classical flavor current as well as the topological $U(1)_T$, it happens that these symmetries mix quantum mechanically, leading to enhanced flavor symmetry at strong coupling \cite{Tachikawa:2015mha, Yonekura:2015ksa}.

\subsection{5d SCFTs from M-theory}
\label{sec:SCFT-M-geometry}

M-theory on a non-compact Calabi-Yau threefold provides a geometric framework to model the Coulomb branch and UV fixed points of 5d theories. For instance, it allows to track the theory from the IR effective description in the Coulomb branch up to the fixed point in the UV. The SCFT corresponds to the singular point (canonical singularity), and the Coulomb branches is given by its crepant resolutions \cite{Intriligator:1997pq, Jefferson:2018irk, Apruzzi:2018nre, Closset:2018bjz, Apruzzi:2019opn, Apruzzi:2019enx},
The resolution introduces compact surfaces 
\begin{equation}
\mathcal S = \bigcup_{i=1}^r S_i \,, 
\end{equation}
which supply $(1,1)$ forms that model the Cartans of the gauge group. In particular $r$ is the rank of $G_{\text{gauge}}$. 
These surfaces intersect along curves $S_i \cdot S_j = C_{ij}$, and  they can also intersect with non-compact divisors $D_{\alpha} \cdot S_i=C_{i \alpha}$. 
Weakly coupled  gauge theory descriptions exist, if the reducible surface $\mathcal{S}$ admits a  ruling, i.e. admit a  $f_i= \mathbb{P}^1$ fibration over a collection of genus $g$ curves. 
In particular, if the surfaces intersect along section of these ruling such that they form a Dynkin diagram of $G_{\text{gauge}}$, they will define the Coulomb branch of the gauge theory in the following way
\begin{enumerate}
\item The Cartan of the gauge symmetry is given by expanding the 3-form  potential, {$C_3 = \sum_i \omega_i^{(1,1)}\wedge A^i$ where the $(1,1)$-form $\omega_i^{(1,1)}$ is} the Poincar\'{e} dual of $S_i$, and $A^i$ are $U(1)$ gauge potentials. 
\item The W-bosons are given by M2-branes wrapping the generic fiber of the ruling $f_i$, whose self intersection are $f_i\cdot_{S_i} f_i = 0$ and $S_i\cdot f_i= -2$.
\item The matter is given instead by M2-branes wrapping fibral $\mc{O}(-1)\oplus\mc{O}(-1)$ curves. 
\end{enumerate}
In this way $\mathcal S$ can then collapse to a curve of singularities of the type of $G_{\rm gauge}$ type, and we notice that in this limit all these gauge theory states become massless.

Another situation is given if the surfaces are still ruled, but some of them intersect at special fibers instead of sections. For example, we could have $S_i\cdot S_j\subseteq f_i,f_j$. In this case the geometry collapses to multiple intersecting curves of singularities, and realizes a 5d quiver gauge theories, where the matter transforming in two connected gauge groups $G_I \times G_J$ comes from M2-branes (antibranes) wrapping fibral $\mc{O}(-1)\oplus\mc{O}(-1)$ curves, which intersect the curve $S_i\cdot S_j$ between the two surfaces. If there is no consistent assignment of ruling and section curves that apply to all $S_i\cdot S_j$, the theory does not have a gauge description. In such instances, there can however still exist a low energy effective theory specified by some generalization of quivers, we will discuss this situation for example in section \ref{sec:CM-gauge}.

The geometric prepotential is computed from the triple intersection numbers $c_{ij\ell}=S_i \cdot S_j \cdot S_\ell$ in the Calabi-Yau  threefold
\begin{equation} \label{eq:geomprep}
\mathcal F_{\rm geo} = \frac{1}{6}c_{ij \ell}\phi^i \phi^j \phi^\ell  \,,
\end{equation} 
where $\phi^i$ are the K\"ahler parameter dual to the compact divisors $S_i$. 
This matches the {cubic terms in the Coulomb branch parameters of the }gauge theory prepotential, if the resolution realizes the ruling that corresponds to the  effective gauge theory description. In general the extended K\"ahler cone of the Calabi-Yau, including also the K\"ahler parameters of the  of the curves  $\mathcal S \cdot D_{\alpha}$ associated to the mass parameters $m_f$ of the gauge theory, corresponds to the extended Coulomb branch $\mathcal{K}(\phi^i,m_f)$. 
The slices at fixed $m_f$, $K(\phi^i, m_f)|_{{\rm fixed}\, m_f}$,  are identified with the Coulomb branch. 

\subsection{5d SCFTs from 6d and Flavor Symmetries}

An alternative, though closely related approach to studying 5d SCFTs, is to compactify 6d $(1,0)$ theories on a circle with  holonomies in the flavor symmetry turned on; these correspond to mass deformations in 5d. The 6d theories can be engineered from elliptic Calabi-Yau compactifications in F-theory,  and the associated 5d theories are constructed using M-theory/F-theory duality. 

Usually a standard circle compactification of a 6d theory leads to a KK-theory, which UV completes back in 6d. These KK-theory can sometimes have marginal gauge theory description in the IR \cite{Jefferson:2017ahm}, where marginal means that the metric of the Coulomb branch is positive semi-definite, see also \cite{Hayashi:2015zka, Hayashi:2015vhy}. Mass deformations of these marginal theories lead to a tree of {descendant} theories, which in the UV complete to 5d SCFTs. 
The mass deformations can be of two types:
\begin{enumerate}
\item \underline{Decoupling a matter hypermultiplet:}  \\
Field-theoretically this is giving  mass to a hypermultiplet that is charged under the flavor symmetry and sending the mass to infinity.
In the M-theory geometry, this corresponds to a flop of an $\mc{O}(-1)\oplus\mc{O}(-1)$ curve $C$, which is flopped out of the reducible surface $\mathcal S$.
In the singular limit, when vol$(\mathcal{S}) \rightarrow 0$, the state obtained by an M2-brane wrapping $C$ decouples. 
\item \underline{Decoupling of a gauge sector:} \\
One can also decouple an entire sector of the theory, such as a gauge vector multiplet. In the geometry, this corresponds to the decompactification, $\text{vol}(S_k) \rightarrow \infty$, of some of the compact surface components $S_k \subset \mathcal{S}$, thereby sending the associated gauge couplings to zero. 
\end{enumerate}
Both of these are key in the construction of 5d SCFTs from 6d, and will be important in the comprehensive study of all 5d theories obtained in this way. 
We will give a detailed discussion of these in section \ref{sec:strategy}. 

In 6d the flavor symmetries are encoded in the Kodaira singular fiber type over non-compact curves in the base $B_2$ of the elliptic Calabi-Yau threefold $Y_3$.  These flavor symmetry generators will be denoted by $D_i$ which are ruled surfaces, with fibers $\mathbb{P}^1_i$, intersect in the affine Dynkin diagram of the flavor symmetry group. 
{M-theory on the resolved Calabi-Yau threefold} results in 5d gauge theories on the Coulomb branch. 
The compact surfaces that arise in this resolution, $\mathcal{S} = \cup S_k$ correspond to the Cartans of the gauge group. 
{The} flavor curves that are contained in $\mathcal{S}$ (i.e. they are fibral curves $D_i \cdot \mathcal{S}$) determine the flavor symmetry at the UV fixed point \cite{Apruzzi:2018nre, Apruzzi:2019vpe}: in the limit $\text{vol}(\mathcal{S}) \rightarrow 0$  the states charged under the corresponding flavor symmetry become massless. 
In the next subsection we will review the geometric/graph theoretic tool, the CFD introduced in \cite{Apruzzi:2019vpe, Apruzzi:2019opn, Apruzzi:2019enx}, to track these flavor symmetries in the process of decoupling hypermultiplets. 

\subsection{CFDs and BG-CFDs}

We now summarize and extend the definition of CFDs and BG-CFDs introduced in \cite{Apruzzi:2019vpe, Apruzzi:2019opn, Apruzzi:2019enx}.
We associated to each 5d SCFT a graph, the combined fiber diagram (CFD), which encodes the flavor symmetry of the UV fixed point as well as the BPS states. Each CFD is an undirected, marked graph, where each vertex $C_\alpha$ has two integer labels $(n, g)$: the self-intersection number $C_\alpha^2=n$ and genus $g$. Two vertices $C_\alpha$ and $C_\beta$ are connected with $C_\alpha\cdot C_\beta=m_{\alpha\beta}$ edges.

Depending on the values of $n$ and $g$, the vertices can be divided into the following types:
\begin{enumerate}
\item[(V1)]{$(n,g)=(-2,0)$

The vertex is marked (in green), and it corresponds to a Cartan node of the non-Abelian part of superconformal flavor symmetry $G_F$.}
\item[(V2)]{$(n,g)=(-2p,-(p-1))$, $p>1$

This vertex can be thought as a combination of $p$ disconnected $(n,g)=(-2,0)$ vertices, which also  contributes to the non-Abelian part of $G_F$. It is hence a marked (green) vertex as well. This type of vertex only appears as the short root of a non-simply laced Lie algebra $H_F$, which is a subalgebra $H_F\subset G_F$.

}

\item[(V3)]{$(n,g)=(-1,0)$

This vertex is considered as the ``extremal vertex'' that generates a CFD transition, as we introduce it shortly after. It is unmarked and does not contribute to the non-Abelian part of $G_F$. }

\item[(V4)]{$(n,g)=(-p,-(p-1))$, $p>1$

This vertex is a combination of $p$ disconnected $(n,g)=(-1,0)$ vertices, which is unmarked. In the CFD transition, these $p$ vertices need to be removed together.

}

\item[(V5)]{All the other cases:

For other values of $(n,g)$, these vertices are unmarked and they do not contribute to the non-Abelian part of $G_F$. Nonetheless, they still could generate the Abelian part and $G_F$ and should be drawn in the figure. Note that certain combinations of $(n,g)$ are forbidden, such as the cases with $n<-2$, $g=0$. 

}

\end{enumerate}

In the M-theory geometry picture, an unmarked vertex $C_\alpha$ with $(n,g)=(-1,0)$ can be thought as a complex curve with normal bundle $\mc{O}(-1)\oplus\mc{O}(-1)$. The BPS state from M2-branes wrapping $C_\alpha$ is a 5d massive hypermultiplet in the Coulomb branch of IR gauge theory. Removing such a vertex will correspond to decoupling a hypermultiplet in the gauge theory. In the CFD language, this operation generates a ``CFD transition'' with the following modifications on the graph.

\vspace{0.2cm}

\noindent
\underline{\textit{CFD transition after removing a vertex $C_\alpha$ with $(n,g)=(-1,0)$}}

\begin{enumerate}
\item{$\forall C_\beta$ with $(n_\beta,g_\beta)$ and $m_{\alpha\beta}>0$, the $(n',g')$ of  $C_\beta'$ in the new CFD are:
\be
\ba
n'_\beta&=n_\beta+m_{\alpha\beta}^2\cr
g'_\beta&=g_\beta+\frac{m_{\alpha\beta}^2-m_{\alpha\beta}}{2}.
\ea
\ee
}
\item{$\forall C_\beta,C_\gamma$ $(\beta\neq\gamma)$ with $m_{\alpha\beta}>0$, $m_{\alpha\gamma}>0$, in the new CFD, the number of edges between $C_\beta'$ and $C_\gamma'$ is:
\be
m'_{\beta\gamma}=m_{\beta\gamma}+m_{\alpha\beta}m_{\alpha\gamma}.
\ee
}
\end{enumerate}

On the other hand, a vertex $C_\alpha$ with $(n,g)=(-p,-(p-1))$ $(p>1)$ can be thought as $p$ complex curves with normal bundle $\mc{O}(-1)\oplus\mc{O}(-1)$, which are homologous in the Calabi-Yau threefold and needs to be removed simultaneously. After the CFD transition generated by removing $C_\alpha$, the graph is modified as:

\vspace{0.2cm}

\noindent 
\underline{\textit{CFD transition after removing a vertex $C_\alpha$ with $(n,g)=(-p,-(p-1))$}}

\begin{enumerate}
\item{$\forall C_\beta$ with $(n_\beta,g_\beta)$ and $m_{\alpha\beta}=pm>0$, where $m\in\mb{Z}^+$, the $(n',g')$ of  $C_\beta'$ in the new CFD are:
\be
\ba
n'_\beta&=n_\beta+pm^2\cr
g'_\beta&=g_\beta+\frac{m^2-m}{2}.
\ea
\ee
}
\item{$\forall C_\beta$ with $(n_\beta,g_\beta)$ and $p\nmid m_{\alpha\beta}$, the $(n',g')$ of  $C_\beta'$ in the new CFD are:
\be
\ba
n'_\beta&=n_\beta+m_{\alpha\beta}^2\cr
g'_\beta&=g_\beta+\frac{m_{\alpha\beta}^2-m_{\alpha\beta}}{2}\,.
\ea
\ee
}
\item{$\forall C_\beta,C_\gamma$ $(\beta\neq\gamma)$ with $m_{\alpha\beta}>0$, $m_{\alpha\gamma}>0$, in the new CFD, the number of edges between $C_\beta'$ and $C_\gamma'$ is:
\be
m'_{\beta\gamma}=m_{\beta\gamma}+m_{\alpha\beta}m_{\alpha\gamma}\,.
\ee
}
\end{enumerate}

For the 5d KK-theory, obtained by circle-reduction of a given 6d (1,0) SCFT, 
we define an associated \textit{marginal CFD}, which often contains green vertices that form Dynkin diagrams of affine Lie algebras. For this marginal CFD, we can construct the CFD transitions in all possible ways, which generates all the 5d SCFT descendants from the KK theory. 

More generally, the non-Abelian part of $G_F$ can be clearly read off from the sub-diagram of marked (green) vertices of a CFD. The intersection matrix between the vertices $C_\alpha$ is exactly the symmetrized Cartan matrix
\be
m_{\alpha\beta}=-\frac{2\langle \alpha,\alpha\rangle_{\rm max}\langle \alpha_\alpha,\alpha_\beta\rangle}{\langle \alpha_\alpha,\alpha_\alpha\rangle \langle \alpha_\beta,\alpha_\beta\rangle}\,,
\ee
where the $\alpha$s are the roots of the Lie algebra. For non-simply laced Lie algebra factor $H_F$, the short roots correspond to vertices with $(n,g)=(-2p,-(p-1))$, where the integer $p$ is the ratio between the length of the long roots and the length of the short roots. 

For a given 5d SCFT, the CFD is not necessarily uniquely determined. There are two possible ways to get equivalent CFDs, {where equivalence here means, the CFD describes the same SCFT. Here we specify this as the same theories including the full set of BPS states:}

\begin{enumerate}
\item
Adding vertices into the CFD that are linear combinations of the existing ones: 
In the CFD, one can always add more copies $C_*$ of the existing vertex $C_\alpha$, which are connected with $m_{*,\alpha}=n_\alpha$ edges. The resulting CFD is trivially equivalent to the original one, but this procedure is useful in the gluing construction that we discuss in section \ref{sec:gluing}.

More generally, one can add linear combination of vertices 
\be\label{Cstardef}
C_*=\sum_{\alpha=1}^k a_\alpha C_\alpha\,, 
\ee
{where $a_\alpha$ are non-negative integer coefficients}, to the graph, which do not change the flavor symmetry or transitions.
The values of $(n_*,g_*)$ and number of edges with other vertices are
\be
\ba
\label{n-g-linearcomb}
n_*&=\left(\sum_{\alpha=1}^k a_\alpha C_\alpha\right)^2=\sum_{\alpha=1}^k a_\alpha^2 n_\alpha+2\sum_{\alpha=1}^k \sum_{\beta<\alpha} a_\alpha a_\beta m_{\alpha\beta} \cr
g_*&=1+\frac{\sum_{\alpha=1}^k a_\alpha(2g_\alpha-2-n_\alpha)+n_*}{2}\cr
m_{*,\alpha}&=a_\alpha n_\alpha+\sum_{\beta\neq\alpha}a_\beta m_{\alpha\beta}
\ea
\ee
The detailed criterion and derivation of these formula will be discussed in the next section. The main constraint is that the vertices of type (V1)-(V4) that can be added cannot change the flavor symmetry or transitions.

\item
Two CFDs with different marked subgraphs, but same $G_F$:\\
In this case, the sub-diagram of marked vertices are different in the two CFDs, but after inclusing of BPS states from the $(n,g)=(-1,0)$ and $(n,g)=(-p,-(p-1))$ vertices, one can combined them into non-trivial representations of a larger Lie algebra. Detailed examples will be presented in the CFD building block section of appendix \ref{sec:BB}. This geometrically means that we have chosen two different complex structures  of $\cup S_i$, resulting in two different looking CFDs. {Note that there can be multiple such distinct realizations, e.g. for the rank 1 theories there are CFDs with manifest $E_8$, $(E_7 \times SU(2))/\mathbb{Z}_2$, $(E_6 \times SU(3))/\mathbb{Z}_2$, $(SO(8) \times SO(8))/\mathbb{Z}_4$ flavor symmetries,} {\cite{Kim:2018lfo,Kim:2018bpg,Ohmori:2018ona}}.  These geometries correspond to different choices of complex structure on generalized del Pezzo singularities, which however collapse to the same SCFT.

\end{enumerate}

Finally, we introduce the way of reading off IR classical flavor symmetry $G_{F}^{\text{cl}}$ from the CFD, which leads to constraints on the IR gauge theory descriptions. We define a set of graphs, the \textit{box graph combined fiber diagram (BG-CFD)}, in table~\ref{tab:BGCFDs}. {These were introduced in order to characterize the IR descriptions using box graphs \cite{Hayashi:2014kca}, which succinctly characterize the Coulomb branch of 5d gauge theories. }
If $k$ of these BG-CFDs with $G_{\text{gauge},i}$ and $G_{F_{cl},i}$ $(i=1,\dots,k)$ can be embedded in the CFD without connected to each other, then the IR gauge theory can have gauge factors $\prod_{i=1}^k G_{\text{gauge},i}$ with classical flavor symmetry $\prod_{i=1}^k G_{F,i}^{\text{cl}}$. 

Note that the gauge groups are not fixed in this procedure, but they are constrained with the information of the total flavor rank and gauge rank, see \cite{Apruzzi:2019enx}.

\begin{table}
  \centering
  \begin{tabular}{|c|c|c|}
    \hline
$G_\text{gauge} \text{ with } N_{\bm{R}}\bm{R}$ &  $G_{F}^{\text{cl}}$& BG-CFD \cr\hline\hline
 $\begin{array}{c} 
    \cr 
      SU(N \geq 3) + N_{\bm{R}}\bm{F} \cr
      SU(N \geq 5) + N_{\bm{R}}\bm{AS} \cr
      SU(N \geq 3) + N_{\bm{R}}\bm{Sym} \cr
      E_6 + N_{\bm{R}}\bm{27} \cr 
      \cr 
    \end{array}$ 
    &
    $U(N_{R})$
    & 
    $\begin{gathered}
      \includegraphics[width= 4cm]{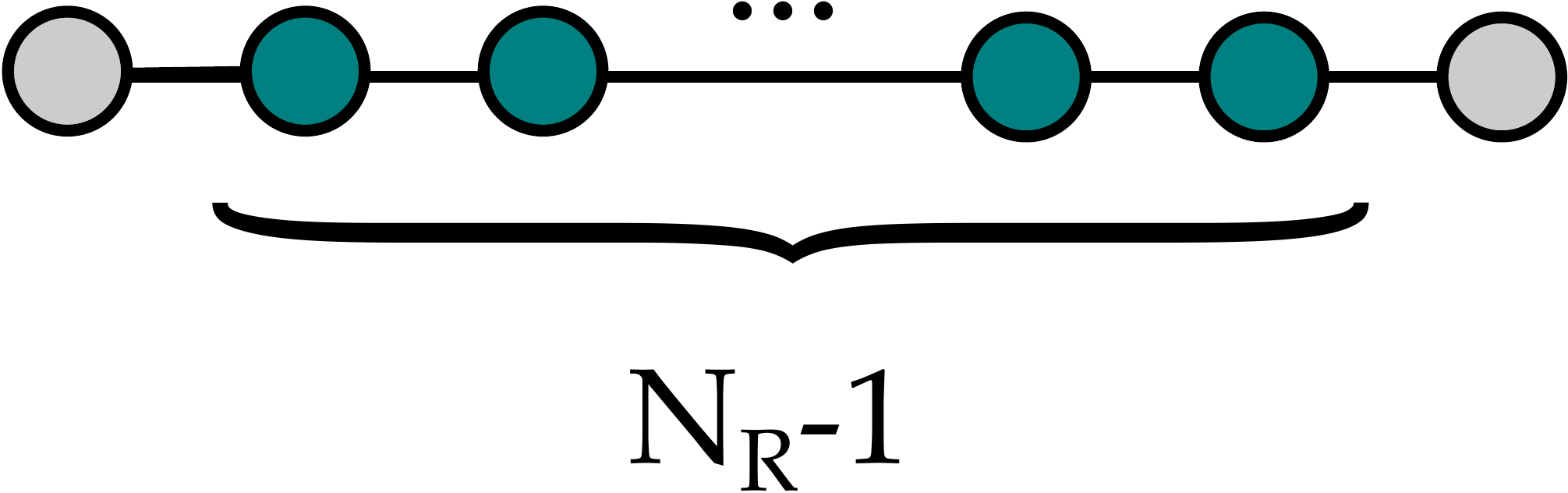}
    \end{gathered}$ \cr\hline
 $\begin{array}{c}
      Sp(N \geq 1) + N_{\bm{R}}\bm{F}\,, \cr
      E_7 + N_{\bm{R}}\bm{56} \,, N_{\bm{R}} \in \mathbb{N}
    \end{array}$ 
 
   &  $SO(2N_{\bm{R}})$& 
    $\ba\cr 
  N_{\bm{R}} = 1:&\qquad  \includegraphics[width=0.8cm]{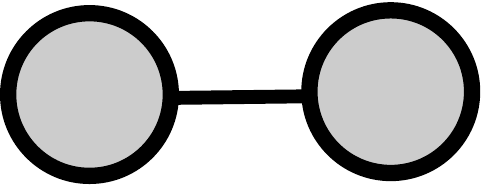}  \cr
     N_{\bm{R}} \in \mathbb{N}_{>1}&\qquad  \includegraphics[width= 4cm]{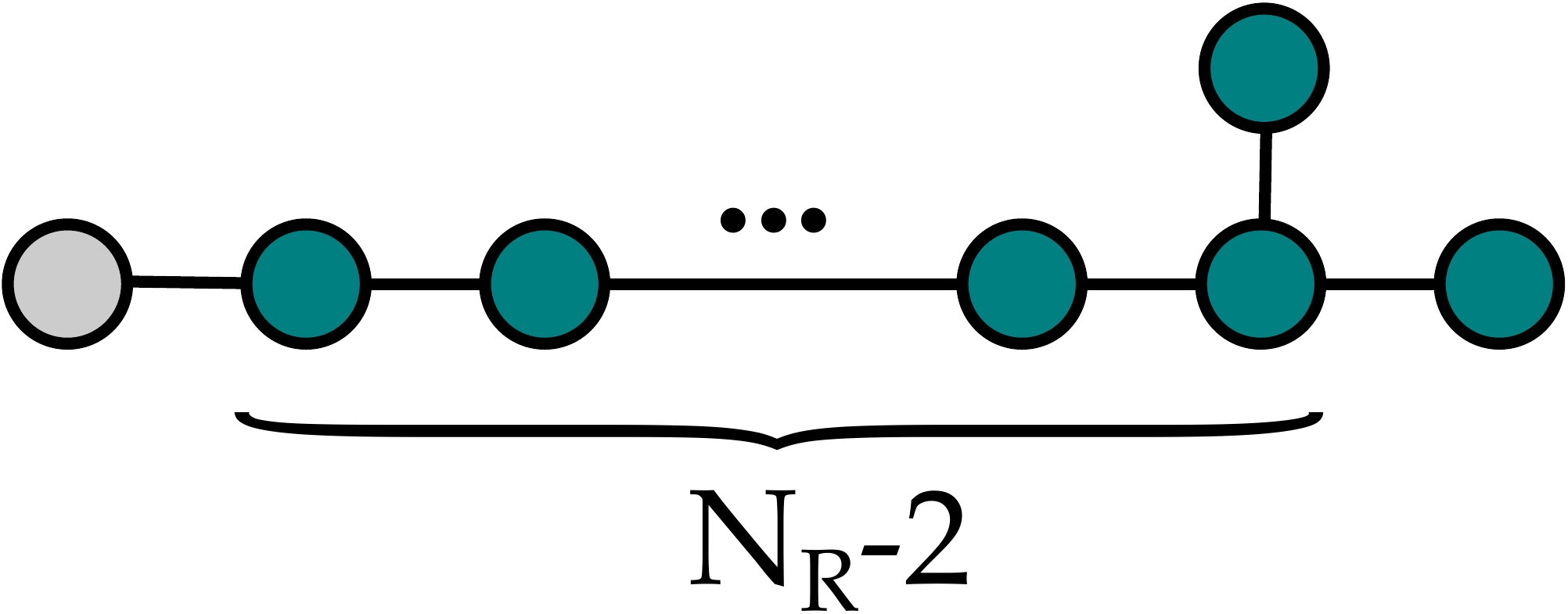} \cr 
    \ea$ \cr\hline
 $\begin{array}{c}
      E_7 + N_{\bm{R}}\bm{56} \,, \cr 
      N_{\bm{R}} \in \mathbb{N}+{1\over 2 }
    \end{array}$ 
    & 
    $SO(2N_{\bm{R}} +1)$
    & 
	$\ba\cr 
     \includegraphics[width= 4cm]{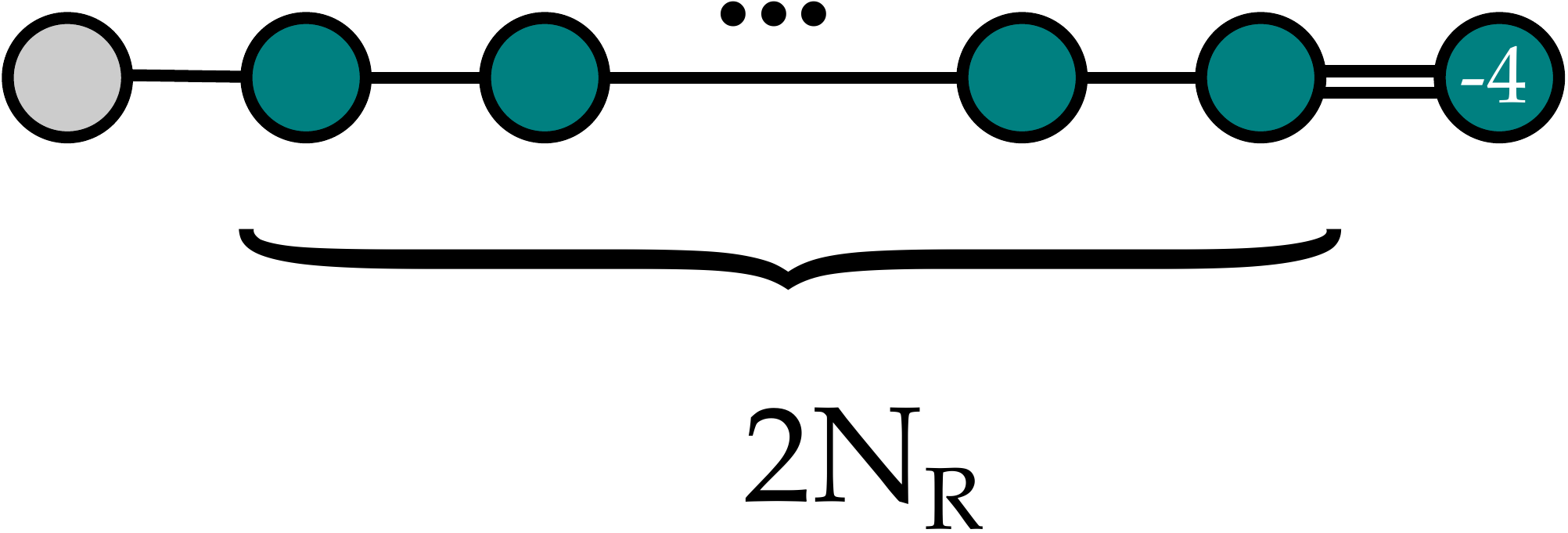}\cr 
     \ea $
  \cr\hline 
    $\begin{array}{c}\cr 
      Sp(N \geq 2) + N_{\bm{R}}\bm{AS} \cr
      SO(N \geq 5) + N_{\bm{R}}\bm{V} \cr
      SU(4) + N_{\bm{R}}\bm{AS} \cr
      G_2 + N_{\bm{R}}\bm{7} \cr
      F_4 + N_{\bm{R}}\bm{26} \cr 
      \cr 
    \end{array}$ & $Sp(N_{\bm{R}})$
    &
    $\ba
   \cr \ba
   &\includegraphics[scale=0.2]{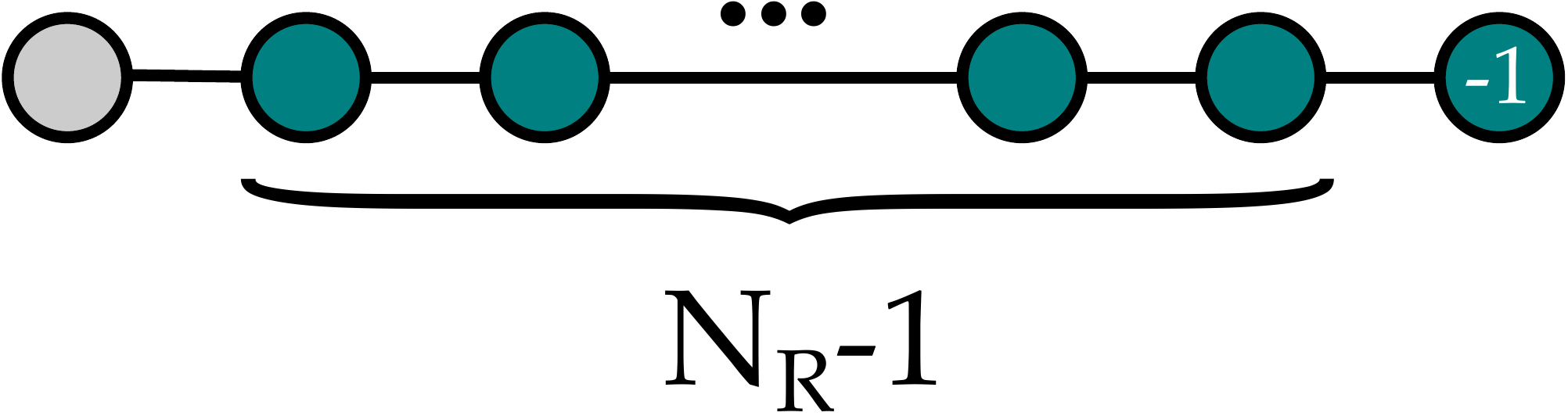} \cr 
   &\includegraphics[scale=0.25]{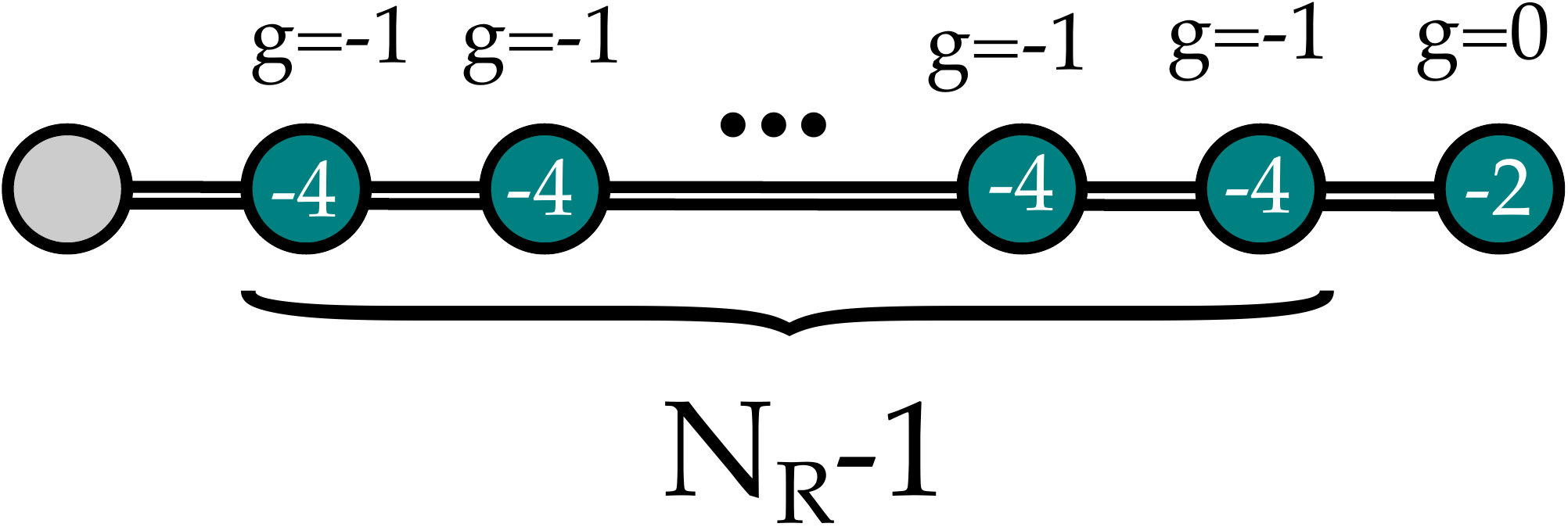}
\ea \ea$    \cr\hline
  \end{tabular}
  \caption{The list of BG-CFDs with the gauge theory descriptions and classical flavor symmetry $G_{F_{cl}}$. The grey vertices denote $(n,g)=(-1,0)$ vertices in the CFD that can be removed via a CFD transition. In the $Sp(N_R)$ case, the two BG-CFDs are equivalent in the sense that the intersection matrices are rescaled by a factor of two. Note that the grey node in the BG-CFD with $(n,g)=(-4,-1)$ vertices should be a type (V4) vertex with $(n,g)=(-2,-1)$.\label{tab:BGCFDs}}
\end{table}

\subsection{CFDs from Geometry as Flop-Invariants}
\label{sec:CFD-geo}


In this section, we present a systematic way of deriving the CFD of a 5d SCFT from the Calabi-Yau threefold geometry introduced in section~\ref{sec:SCFT-M-geometry}.

The complex curves on the surfaces include the intersection curves $S_i\cdot S_j$ $(i\neq j)$ between different compact surfaces and the intersection curves $C_{i\alpha}=S_i\cdot D_\alpha$ with a non-compact surface $D_\alpha$ in $Y_3$. In the CFD, each node $C_\alpha$ is essentially a linear combination of $C_{i\alpha}$ on each $S_i$, and we want to read off the labels $(n,g)$ of $C_\alpha$ directly from the geometry. In \cite{Apruzzi:2019opn,Apruzzi:2019enx}, the correspondence between resolution geometry and CFDs is developed only for compactifications of some classes of very Higgsable (VH) 6d theories, focusing on  minimal conformal matter, i.e. collisions of codimension one singularities at a smooth point in the base. 
We will generalize this in the present paper to include non-very Higgsable (NVH) theories, in particular non-minimal conformal matter theories, for which we now define CFDs more generally. This will be consistent with the previous cases, but generalizes it substantially. 

The idea is that the CFD should capture the geometric invariants under flop among different compact surface components $S_i\subset{\mathcal{S}}$.
This definition is motivated through the fact that such flops do not change the 5d SCFT fixed point, and only correspond to different gauge theory Coulomb branch phases, with the same UV fixed point.
Therefore,  CFDs, which are defined to capture the properties of the SCFTs, should be invariant under such geometric operations. 
For all curves $D_\alpha\cdot S_i$, we need to introduce a multiplicity factor $\xi_{i,\alpha}$, such that the normal bundle of the curve
\be
C_\alpha=\sum_{i=1}^r\xi_{i,\alpha}D_\alpha\cdot S_i
\ee
is invariant under such flops. Then the label $(n,g)$ of the corresponding vertex in the CFD is given by
\begin{eqnarray}
n_\alpha &=&\sum_{i=1}^r \xi_{i,\alpha} (D_\alpha)^2\cdot S_i  \label{CFD-node-n}
\\
g_\alpha&=&1+\frac{1}{2}\left[\sum_{i=1}^r \xi_{i,\alpha} (D_\alpha)^2\cdot S_i+D_\alpha\cdot \left(\sum_{i=1}^r \xi_{i,\alpha} S_i\right)^2\right]\label{CFD-node-g}
 \,.
\end{eqnarray}
With this multiplicity factor, the number of edges between two vertices in the CFD is 
\be
\label{CFD-edge-m}
m_{\alpha,\beta}=\sum_{i=1}^r \xi_{i,\alpha}D_\alpha\cdot D_\beta\cdot S_i \,.
\ee
In this formula, it is assumed that the two curves $D_\alpha\cdot S_i$ and $D_\beta\cdot S_i$ have the same multiplicity factor if they intersect each other on $S_i$. Note however that in certain cases there can be subtleties as we will discuss later in the rank 2 E-string example. 

Similarly, to properly define the linear combination of vertices $C_\alpha$ $(\alpha=1,\dots,k)$ in the CFD defined in (\ref{Cstardef})
\be
\ba
\label{linear-comb}
C_*&=\sum_{\alpha=1}^k a_\alpha C_\alpha\cr
&=\sum_{\alpha=1}^k\sum_{i=1}^r \xi_{i,\alpha}a_\alpha D_\alpha\cdot S_i\,,
\ea
\ee
one of the following two conditions need to be satisfied for a particular flopped geometry, and the formula (\ref{n-g-linearcomb}) hold:
\begin{enumerate}
\item
All the multiplicity factors $\xi_{i,\alpha}$ are the same if $a_\alpha S_i\cdot D_\alpha$ is non-zero in the second line of (\ref{linear-comb}). 
If this is the case, then 
\be
C_*=\left(\sum_{\alpha=1}^k \xi_{i,\alpha}a_\alpha D_\alpha\right)\cdot \left(\sum_{i=1}^r  S_i\right)\,,
\ee
 which is in the form of a complete intersection curve.

\item
All the curve components $D_\alpha\cdot S_i$ in (\ref{linear-comb}) lie on the same surface $S_j$. 
If this happens, then 
\be
C_*=\left(\sum_{\alpha=1}^k \xi_{j,\alpha}a_\alpha D_\alpha\right)\cdot S_j\,,
\ee
which is also in the form of a complete intersection curve.
\end{enumerate}

\begin{figure}
\centering
\includegraphics[width=11cm]{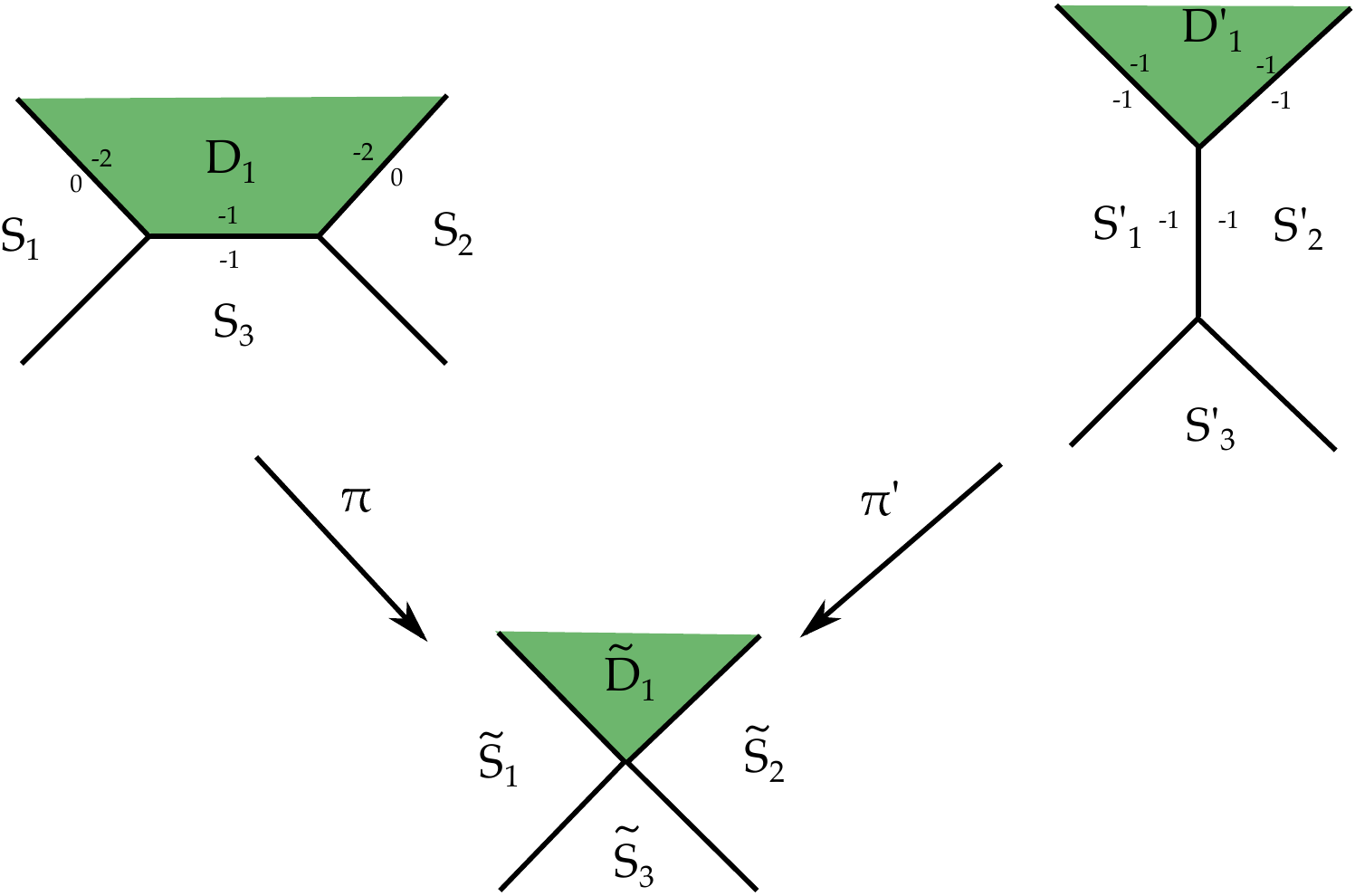}
\caption{The flop operation on an example with $D_1^2\cdot S_3=-1$, $D_1\cdot S_1\cdot S_3=1$ and $D_1\cdot S_2\cdot S_3=1$. In the picture, each line segment denotes an intersection curve $S\cdot S'$, and the integer label on the side of $S$ (or $S'$) is the triple intersection number $S\cdot (S')^2$ (or $S'\cdot S^2$). After shrinking the curve $D_1\cdot S_3$, the surface geometry of $\tilde{D}_1$ and $\tilde{S}_i$ has a conifold singularity. Then after blowing up $\tilde{S}_1$ and $\tilde{S}_2$, the geometry will become the $D_1'$ and $S_i'$, which is the flopped geometry of the original one.}
\label{f:flop1}
\end{figure}

To define the multiplicity factors, we first study the curve components $D_\alpha\cdot S_i$ with normal bundle $\mc{O}(-1)\oplus\mc{O}(-1)$, which intersects other $S_j$s at one or more points. As an example, consider the geometric configuration in the figure~\ref{f:flop1}, where the curve $D_1\cdot S_3$ intersects both $S_1$ and $S_2$. After the curve $D_1\cdot S_3$ is shrunk, both $S_1$ and $S_2$ are blown up at one point, and the curves $D_1\cdot S_1'$ and $D_1\cdot S_2'$ will have normal bundle $\mc{O}(-1)\oplus\mc{O}(-1)$. As we can see, if we define the multiplicity factors $\xi_{3,1}=2$, $\xi_{1,1}'=\xi_{2,1}'=1$, where the notation with ``$\prime$'' denotes the quantities after the flop, then the linear combination (\ref{CFD-node-n}) is invariant.

For more general cases, in principle one needs to perform a sequence of flops $\bigcup_{i=1}^r S_i\rightarrow \bigcup_{i=1}^r S_i'$ among the compact surface components, such that all the curve components  $D_\alpha\cdot S_i'$ for a fixed $\alpha$ only intersect other $S_j'$ at one or zero points. From the perspective of the non-compact surface $D_\alpha$, this corresponds to a blow down sequence of $D_\alpha$ which terminates when $D_\alpha$ cannot be blown down, or all the $\mc{O}(-1)\oplus\mc{O}(-1)$ curves $D_\alpha\cdot S_i'$ only intersect another $S_j'$ at one point.

In this ``terminated'' geometry, all the multiplicity factors $\xi_{i,\alpha}'$ for $S_i'\cdot D_\alpha$ would be trivially one. Then the multiplicity factor $\xi_{i,\alpha}$ of the original curve $D_\alpha\cdot S_i$ can be counted as:
\be
\label{multiplicity-flop}
\xi_{i,\alpha}=\sum_{j\neq i} [-(D_\alpha)^2\cdot S_i'+(D_\alpha)^2\cdot S_i].
\ee
An equivalent simple way to read off the multiplicity factors is to consider the collapse of the following sequence of curves:
\be
\label{fm-shrinking}
\begin{array}{c}
\overset{\xi_{p_1,\alpha}}{(D_\alpha\cdot S_{p_1}^2)}-\overset{\xi_{p_2,\alpha}}{(D_\alpha\cdot S_{p_2}^2)}-\dots-\overset{\xi_{p_q,\alpha}}{(D_\alpha\cdot S_{p_q}^2)}\\
\downarrow\\
\vdots\\
\downarrow\\
\overset{1}{(D_\alpha\cdot S_{p_1'}'^2)}-\overset{1}{(D_\alpha\cdot S_{p_2'}^2)}-\dots-\overset{1}{(D_\alpha\cdot S_{p_{q'}'}^2)} 
\end{array}
\ee
We put the self-intersection number of curves in the brackets and the multiplicity factors over them. In the terminated geometry on the bottom line of (\ref{fm-shrinking}), we assign trivial multiplicity factor one to all of the curves. {We conjecture that such a terminated geometry always exists and it would be interesting to develop the algebraic geometry associated to this problem. We have shown the existence in all the theories that have been studied in this paper.}
Then when we go up (blow up $D_\alpha$), the multiplicity of the old curves remain the same, while the multiplicity of the new exceptional $(-1)$-curve is given the sum of the two multiplicity factors on each side. After repeating the procedure, we can get all the multiplicity factors in the original geometry. The procedure can be easily generalized to non-toric $D_\alpha$ as well, where the multiplicity of the new exceptional $(-1)$-curve is given the sum of the multiplicity factors on all the old curves that intersect the new $(-1)$-curve.

In fact, it can be shown that the quantities (\ref{CFD-node-n}) and (\ref{CFD-node-g}) are invariant under the operation (\ref{fm-shrinking}).
For the example in figure~\ref{f:flop1}, this sequence is
\be
\label{flop1-shrinking}
\begin{array}{c}
\overset{1}{(-2)}-\overset{2}{(-1)}-\overset{1}{(-2)}\\
\downarrow\\
\overset{1}{(-1)}-\overset{1}{(-1)}
\end{array}
\ee
We can see that the multiplicity of the middle curve $D_1\cdot S_3$ on $S_3$ is indeed two.

If initially we already have
\be
\sum_{j\neq i}D_\alpha\cdot S_i\cdot S_j=1 \,,
\ee
then we can still construct such a flop, where $D_\alpha\cdot S_i$ is flopped into another $S_j'$. Then from the formula (\ref{multiplicity-flop}), we can compute $\xi_{i,\alpha}=1$, which corresponds to the trivial case.

From the procedure (\ref{fm-shrinking}), actually we have determined the multiplicity factors of some $\mc{O}\oplus\mc{O}(-2)$ curves as well. For the other curve $D_\alpha\cdot S_i$ with normal bundle that is not $\mc{O}(-1)\oplus\mc{O}(-1)$, if it intersects another curve $D_\beta\cdot S_i$ with a well defined multiplicity factor $\xi_{i,\beta}$ on $S_i$ , then we define $\xi_{i,\alpha}=\xi_{i,\beta}$. This is called ``neighbor principle'' in the later references. If this does not happen, then we simply take $\xi_{i,\alpha}=1$.

Finally, for a curve $D_\alpha\cdot S_i$ with normal bundle $\mc{O}(-1)\oplus\mc{O}(-1)$ that does not intersect another $S_j$, one can directly flop it out of $\bigcup_{i=1}^r S_i$ and that corresponds to a CFD transition. Its multiplicity $\xi_{i,\alpha}$ equals to a previously defined $\xi_{i,\beta}$ if $S_i\cdot D_\alpha\cdot D_\beta>0$, unless there are two different $\xi_{i,\beta}\neq \xi_{i,\gamma}$, where $S_i\cdot D_\alpha\cdot D_\beta,S_i\cdot D_\alpha\cdot D_\gamma>0$. If the latter situation happens, then $\xi_{i,\alpha}$ is not uniquely defined. We will discuss this situation in the rank-two E-string example latter.

To illustrate this general framework, we consider two, somewhat more complicated examples: 

\subsubsection*{Example 1: if $\forall j\neq i$, $D_\alpha\cdot S_i\cdot S_j\leq 1$.}

\begin{figure}
\centering
\includegraphics[width=15cm]{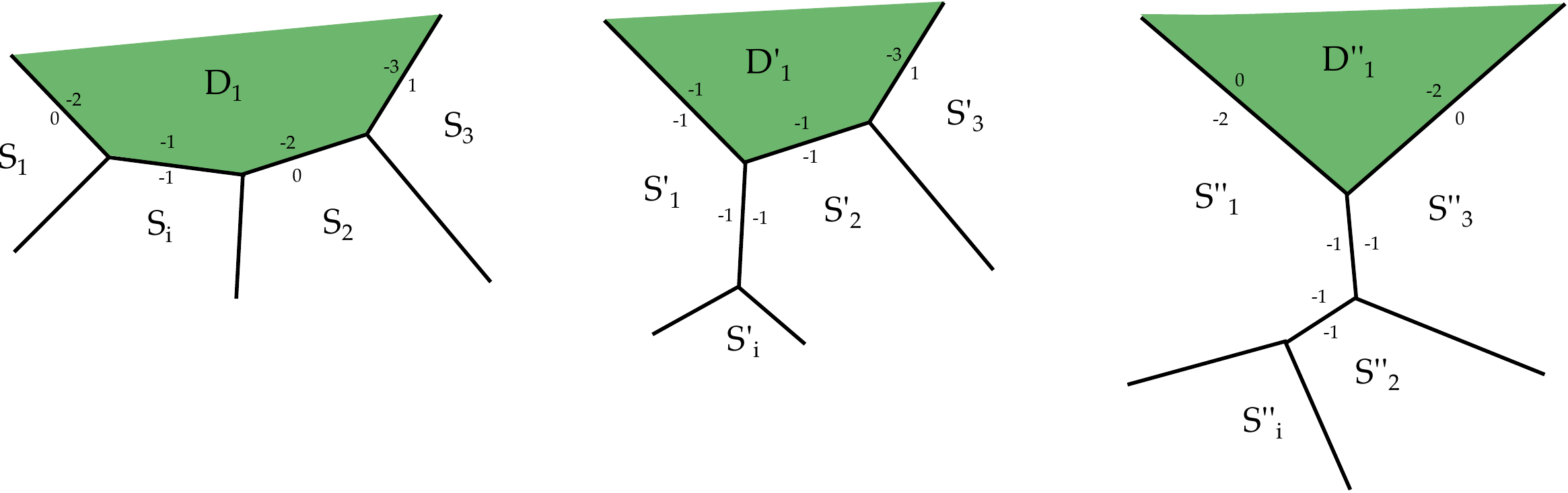}
\caption{\label{f:flop2} The necessary flop operations to determine the multiplicity factor $\xi_{i,1}$ of the non-compact surface $D_1$ on $S_i$. The procedure includes two conifold transitions, where we do not draw the singular geometry explicitly.}
\end{figure}
In figure~\ref{f:flop2}, we show a non-compact surface $D_1$ with four compact surfaces $S_i$, $S_1$, $S_2$ and $S_3$. In the first flop, we shrink the curve $D_1\cdot S_i$ and blow up the surface $S_1$, $S_2$. After this flop, $D_1\cdot S_2'$ still intersects $S_1'$ and $S_3'$, hence we need to further shrink the curve $D_1\cdot S_2'$ and blow up the surface $S_1'$, $S_3'$. Finally, on the surface components $S_i''$, all the multiplicity factors associated to $D_1$ equal to one, and we can see that $D_1$ should correspond to a vertex with $n(D_1)=-2$. From (\ref{multiplicity-flop}) applied with $S_i''$, we can see that the correct multiplicity factor of $D_1\cdot S_i$ on the original surface $S_i$ is $\xi_{i,1}=3$, and the multiplicity factor of $D_1\cdot S_2'$ on $S_2'$ is $\xi_{2,1}'=2$. 

We can also apply the procedure (\ref{fm-shrinking}), which explicitly generates the correct multiplicity factors:
\be
\label{flop2-shrinking}
\begin{array}{c}
\overset{1}{(-2)}-\overset{3}{(-1)}-\overset{2}{(-2)}-\overset{1}{(-3)}\\
\downarrow\\
\overset{1}{(-1)}-\overset{2}{(-1)}-\overset{1}{(-3)}\\
\downarrow\\
\overset{1}{(0)}-\overset{1}{(-2)}
\end{array}
\ee

\subsubsection*{Example 2: if $\exists j\neq i$, $(D_\alpha)^2\cdot S_i=D_\alpha\cdot (S_i)^2=-1$ and $D_\alpha\cdot S_i\cdot S_j> 1$.}

In this case, the shrinking of $D_\alpha\cdot S_i$ will change the genus of the compact intersection curve $S_i\cdot S_j$. In \cite{Jefferson:2018irk}, it was shown that such genus changing transition involves changing the complex structure moduli of the surfaces. For example, if one wants to transform a genus-one curve $S_i\cdot S_j$ into a genus-zero curve, then one needs to first take the singular limit of $S_i\cdot S_j$ where the torus is pinched at a point and then blow up that double point singularity. Nonetheless, it is still possible to define invariant quantities $n_\alpha$ under this kind of geometric transition.

\begin{figure}
\centering
\includegraphics[width=14cm]{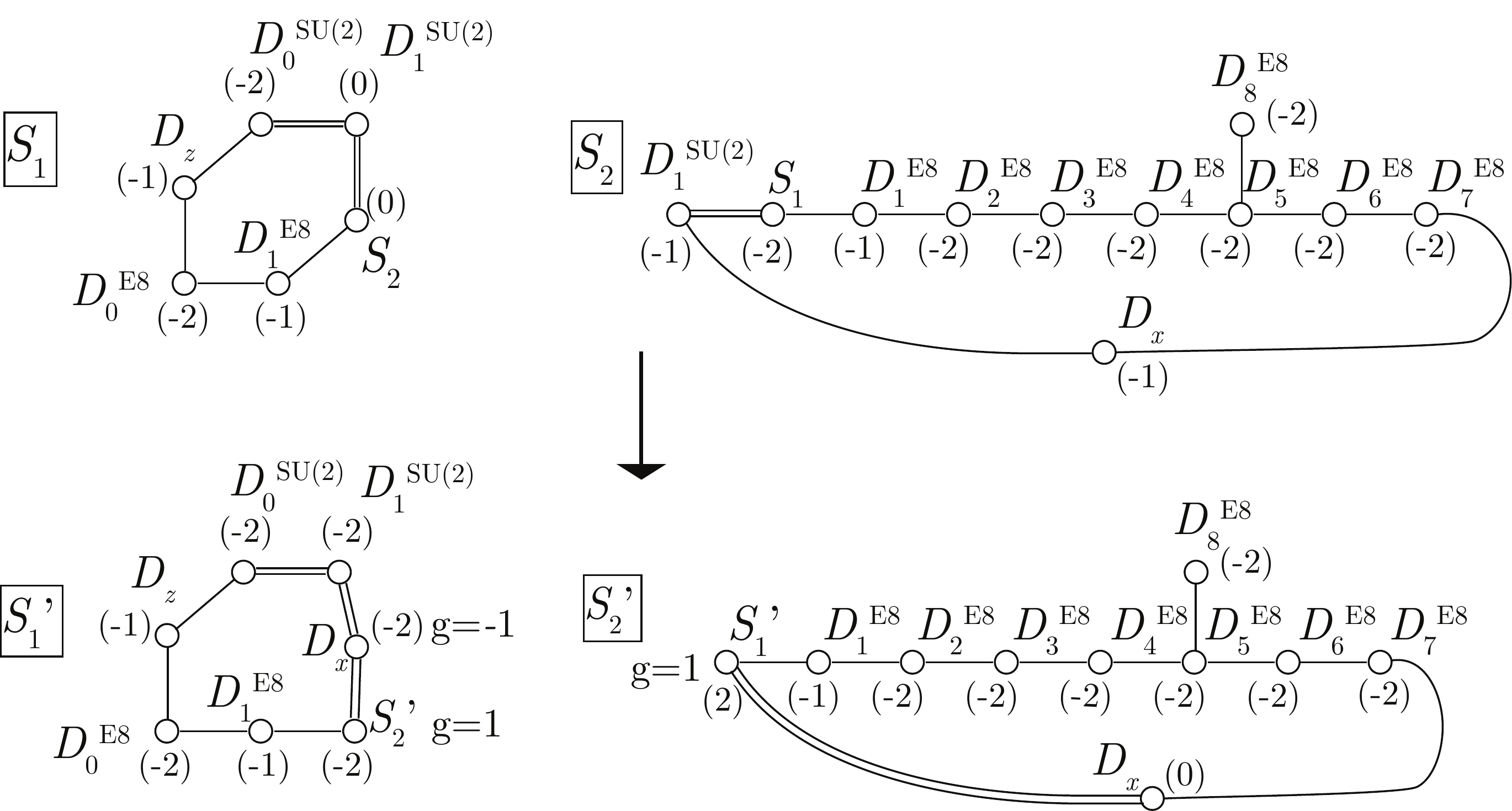}
\caption{The flop operation on the resolution geometry of $(E_8,SU(2))$ conformal matter (rank-two E-string). On each surface component $S_i$ (labeled by the letter in the box), each node $D_\alpha$ corresponds to the intersection curve $S_i\cdot D_\alpha$. The number besides the node is the intersection number $(D_\alpha)^2\cdot S_i$. The genus of such a curve is by default zero unless otherwise labeled. In this geometry, the intersection curve $S_1\cdot S_2$ has genus-zero. After the flop, the intersection curve $S_1'\cdot S_2'$ has genus-one.}
\label{f:rank-2-Estring-flop}
\end{figure}

For example, see the resolution geometries of $(E_8,SU(2))$ conformal matter (rank-two E-string) in figure~\ref{f:rank-2-Estring-flop}, which was discussed in \cite{Apruzzi:2019opn}. The non-compact divisors $D_i^{E_8}$ and $D_i^{SU(2)}$ correspond to the Cartan divisors of the affine $E_8$ and affine $SU(2)$ respectively. On $S_2$, the $\mc{O}(-1)\oplus\mc{O}(-1)$ curve $D_1^{SU(2)}\cdot S_2$ intersects $S_1$ at two points. After this curve is shrunk, the geometry is flopped such that the intersection curve $S_1'\cdot S_2'$ has genus one instead of zero. On the new geometry, we have $S_1'\cdot (D_1^{SU(2)})^2=-2$. Then from the formula (\ref{multiplicity-flop}), we can see that the multiplicity factor of $D_1^{SU(2)}\cdot S_2$ on the original geometry equals to two. 

In this case, we can also apply the procedure (\ref{fm-shrinking}), keeping in mind that $D_1^{SU(2)}\cdot S_1$ is actually a combination of two disjoint $\mc{O}\oplus\mc{O}(-2)$ curves. The sequence
\be
\label{rank-2-Estring-shrinking}
\begin{array}{c}
\overset{1}{(-2)}-\overset{2}{(-1)}-\overset{1}{(-2)}\\
\downarrow\\
\overset{1}{(-1)}-\overset{1}{(-1)}
\end{array}
\ee
is exactly the same as (\ref{flop1-shrinking}), but the $D_1^{SU(2)}\cdot S_1$ corresponds to the two $(-2)$-curves on the first line. On the bottom line, the combination of the two $(-1)$-curves correspond to $D_1^{SU(2)}\cdot S_1'$, which is also $\mc{O}\oplus\mc{O}(-2)$ curve with $(D_1^{SU(2)})^2\cdot S_1'=-2$, $D_1^{SU(2)}\cdot (S_1')^2=0$.

Then we can determine the multiplicity factors of the other curves with normal bundle $\mc{O}(-1)\oplus \mc{O}(-1)$, that intersects another $S_j$:
\be
\xi_{2,1^{E_8}}=1\ ,\ \xi_{1,x}'=1.
\ee
Note that the curve $D_x\cdot S_1'$ is a combination of two disjoint $(-1)$-curves. Each of the $(-1)$-curve only intersects $D_1^{SU(2)}$ and $S_2'$ at a single point.

For the other curves, the multiplicities can be read off by the neighbor principle, which all equal to one except for $D_x\cdot S_2$ on $S_2$. $D_x\cdot S_2$ is a $\mc{O}(-1)\oplus\mc{O}(-1)$ curve connected to a curve $D_1^{SU(2)}\cdot S_2$ with multiplicity two and another curve $D_7^{E_8}\cdot S_2$ with multiplicity one. 
Hence the multiplicity of such ``interpolating curve'' is not uniquely defined. Despite of this subtlety, there are two equivalent ways to present it in the CFD:

\begin{enumerate}
\item
Draw $D_1^{SU(2)}$ as a $(n,g)=(-2,0)$-node, and the node $D_x$ is drawn as a node with $(n,g)=(-2,-2)$, or two $(-1)$-nodes in the same circle (as in the $(E_7,SO(7))$ case in \cite{Apruzzi:2019enx}). In the edge multiplicity formula (\ref{CFD-edge-m}), the multiplicity factor $\xi_{i,\alpha}$ is always taken as that of $D_1^{SU(2)}$ and $D_7^{E_8}$. Hence the nodes $D_x$ and $D_1^{SU(2)}$ are connected with two edges, while the nodes $D_x$ and $D_7^{E_8}$ are connected with one edge. After the CFD transition of removing $D_x$, the node $D_1^{SU(2)}$ becomes a node with $(n,g)=(0,0)$, but the node $D_7^{E_8}$ will become a node with $(n,g)=(-1,0)$.

\item
Draw $D_1^{SU(2)}$ as a marked ``green $(-1)$-vertex' with $(n,g)=(-1,0)$, but it still contributes to the non-Abelian flavor symmetry. The node $D_x$ is simply drawn as an unmarked $(n,g)=(-1,0)$ node, which connects both $D_1^{SU(2)}$ and $D_7^{E_8}$ with one edge. Then after the CFD transition of removing $D_x$, the node $D_1^{SU(2)}$ becomes a node with $(n,g)=(0,0)$, and the node $D_7^{E_8}$ becomes a node with $(n,g)=(-1,0)$.
\end{enumerate}
The descendant CFDs are exactly the same, no matter which convention is used to compute them.

\section{5d SCFTs from Decoupling in 6d \label{sec:strategy}}

Our goal is to understand 5d SCFTs that descend from $S^1$-compactifications of a general 6d SCFTs. A trivial compactification does not lead to a 5d SCFT, but rather to a KK-theory \cite{Jefferson:2017ahm}, which can have many IR descriptions in terms of a marginal gauge theories. In order to obtain a genuine 5d SCFT in the UV, one usually needs to mass deform the KK-theory, which corresponds to turning on Wilson lines for the flavor symmetry. For some 6d SCFT, different choices of mass deformation lead to different 5d SCFT. This, however, is not always the case and it highly depend on the 6d theory we start with. 

A general 6d SCFT can be characterized by the tensor branch, which can be geometrically classified, and is comprised of smaller building blocks -- in the geometry these are curves in the base of the elliptic Calabi-Yau threefold, which intersect in a quiver, that obeys certain rules \cite{Heckman:2013pva}.
Field-theoretically, this is modeled by  constructing higher rank tensor branches by consistently gauging and adding tensor multiplets. 

The same logic can be implemented for the 5d SCFTs obtained by circle compactification and deformations. In particular, we start by defining some fundamental building blocks, which are reduction of 6d SCFTs on $S^1$. 
We then develop rules how these building blocks are consistently glued together. 
We implement this both from the (gauge) effective field theory prospective as well as using the geometry and 
 CFDs intoduced in \cite{Apruzzi:2019opn, Apruzzi:2019vpe}. 
 We note that the theories discussed in \cite{Apruzzi:2019opn, Apruzzi:2019vpe} form one class of building blocks in 5d, which descend from 6d conformal matter type theories. In the present paper, we develop the methodology how to generalize this to an arbitrary 6d theory as a starting point.

\subsection{Mass Deformations vs. Decoupling}
It is relevant for our purpose to divide 6d SCFTs in two classes \cite{Ohmori:2015pua, Ohmori:2015pia, Mekareeya:2017sqh}. 
We give in each case the field theoretic description, the Calabi-Yau threefold geometry in F-theory, as well as the tensor branch structure. The latter is characterized by a collection of intersection rational curves, with self-intersection numbers $\Sigma^2 = (-n)$. Blowing down $(-1)$ curves allows moving to the origin of the tensor branch, which transforms their self-intersection numbers as follows
\be
(-n)-(-1)-(-m) \quad \rightarrow \quad (-n+1)-(-m+1) \,.
\ee
We denote the endpoint of the tensor branch by $B_{\text{end}}$. The two types of theories in 6d are distinguished as follows:

\begin{itemize}
\item \textbf{Very Higgsable Theories (VH Theories):} \\
These are 6d theories which can be Higgsed completely to free hypermultiplets. 
Geometrically, this means that the non-minimal singularity of the F-theory model occurs at a smooth point in the base. 
In terms of the 
the resolved tensor branch geometry (which is a collection of rational curves in the base of the elliptic Calabi-Yau threefold) we get the endpoint configuration, which for very Higgsable thoeries is
\begin{equation}
B_{\text{end}} = \emptyset \,.
\end{equation}
\item \textbf{non-very Higgsable Theories (NVH Theories):} \\
These are 6d theories which cannot be Higgsed completely, but always have residual non-trivial 6d SCFTs in the Higgs branch. In F-theory geometry these correspond to singular elliptic fibrations over an orbifold base $\mathbb C ^2/ \Gamma$, with $\Gamma \subset U(2)$ \cite{Heckman:2013pva}, where the endpoint configuration is
\begin{equation}
B_{\text{end}} \neq \emptyset \,.
\end{equation}
\end{itemize}

\subsection{5d SCFTs from very Higgsable Theories}

A large class of 5d SCFTs arise from the dimensional reduction of VH theories, and mass deformations. 
Rank one and two theories are of this type \cite{Jefferson:2018irk, Apruzzi:2019opn, Apruzzi:2019vpe,Apruzzi:2019enx}, and more generally minimal conformal matter theories, whose descendants and flavor symmetry enhancements were systematically studied in \cite{Apruzzi:2019opn, Apruzzi:2019vpe,Apruzzi:2019enx}. 

This approach  generates a tree of 5d SCFTs connected by RG-flows triggered by mass deformations, where the tree originates from the marginal theory,  i.e. the 6d SCFT on $S^1$ without Wilson-lines. Most of these SCFTs have at least one IR effective gauge theory description and the mass deformation corresponds to decoupling an hypermultiplet at a time by sending their mass, $m_f \rightarrow \pm \infty$. 

From the point of view of M-theory geometry, a 5d SCFT is defined by M-theory on a Calabi-Yau threefold with a canonical singularity. This implies that the resolution is given by a collection of intersecting compact surfaces, which 
collapse to a point at the UV fixed point. 
Starting with the marginal theory, on an $S^1$ results in a 5d theory with an additional KK-$U(1)$. To get a theory that UV completes in 5d, we first need to mass deform the $U(1)$-KK. In the geometry this means we need to flop the $(-1)$-curve that corresponds to the states charged under the affine node of the 6d flavor symmetry. {Once the curve is flopped one needs to decouple the states associated to the wrapped M2-branes. This is done by sending its volume to infinity, which means that in the geometry the $T^2$-fiber has now infinite volume. }

\subsection{5d SCFTs from non-very Higgsable Theories}
\label{sec:nvht}

If the starting point is a NVH 6d SCFT, one needs to do something more drastic in order to actually get a 5d SCFT. In fact, the circle reduction in the Higgs branch gives a 5d SCFT coupled to an extra sector, which is usually an extra gauge vector multiplet \cite{Ohmori:2015pia}. Since, the Higgs branch moduli space does not mix with the Coulomb branch in 5d, one can turn off the Higgs branch vevs, without decoupling the gauge theory. At the origin of the the Higgs and Coulomb branch the resultant KK-theory will be a 5d SCFT non-trivially coupled to a gauge theory \cite{Ohmori:2015pia}. In this cases we will encounter the following situation
\begin{equation}
\mathcal T^{\text{6d}}_{S^1}= \mathcal S^{\text{5d}}(G)/G  \,,
\end{equation}
where the 5d SCFT $\mathcal S^{\text{5d}}$ whose flavor symmetry is (or contains) $G$, is modded out by gauge group $G$ redundancies, or in other words, part of its flavor symmetry is gauged. Let us assume that the effective gauge theory of $\mathcal S^{\text{5d}}$ has a quiver gauge theory effective description at low energies, which is indeed usually given by the 6d quiver theory in the tensor branch $\mathcal S^{\text{5d}}=G^{(6d)}_1 \times \ldots \times G^{(6d)}_i \times \ldots \times  G^{(6d)}_{\text{rank}(G)}$ (where $G^{(6d)}_i$ can also be trivial). $\mathcal S^{\text{5d}}$ couples to the extra gauge theory with gauge group $G$, and we can explicitly illustrate this coupling in terms of an effective Lagrangian  
\begin{equation}
\mathcal L_{\rm eff} \supset \Omega_{ij} \left( \frac{1}{4} \Phi^i {\rm Tr} ( F^j \wedge \ast F^j) +  \frac{1}{4} A^i {\rm Tr} ( F^j \wedge  F^j)\right) \,,
\end{equation}
where $(\Phi^i=2\pi R \varphi_{\text{6d}}^i, A^i=2\pi R a^i)$ are the Coulomb branch parameters and $U(1)$ Cartan gauge vector fields for the gauge theory with gauge group $G$, and $R$ is the radius of $S^1$. In particular, $\varphi_{\text{6d}}^i$ correspond to the tensor branch scalars of the 6d theory, and $a_i = \int_{S^1} B^i$ where $B^i$ are the two-form fields of the 6d tensor multiplets. Finally the pairing $\Omega_{ij} \in \mathbb{Z}$ is the Dirac pairing on the string charges lattice and the (anti) self-dual tensors lattice of the 6d theory \cite{Heckman:2018jxk}. We can notice that this extra gauge theory couples to the kinetic terms of the quiver gauge theory, as well as non-perturbatively to their $U(1)_T \left( G^{(6d)}_i \right)$ currents. Moreover, the gauge coupling is
\begin{equation}
\frac{8\pi^2}{g_G^2}=\frac{1}{R}\,.
\end{equation}
The couplings of the quiver theory and the extra gauge theory have different dependence in terms of the $S^1$ radius, $g_G^2 \sim R$ and $g_i^2 \sim R^{-1}$.  The 5d limit consists of sending $R\rightarrow 0$, and in this limit we conclude that the extra $G$ gauge theory and the coupled quiver cannot have a common strongly coupled regime. This implies that, in order to obtain a 5d SCFT we need to isolate $\mathcal S^{\text{5d}}$ and decouple the extra gauge theory with gauge group $G$. 

NVH 6d SCFTs are geometrically constructed from F-theory on a non-compact singular elliptically fibered Calabi-Yau threefold, where the base is itself an orbifold singularity. In order to get 5d SCFTs from the circle compactification of these 6d theories we have two possible geometric transitions:
\begin{itemize}
\item The only situation we encounter in where we can flop out a curve is when two compact surfaces intersect in a curve with $\mathcal{O}(1)\oplus \mathcal{O}(-1)$ normal bundle, i.e.
\begin{equation}
S_1 \cdot S_2 = C,\qquad C^2|_{S_1}=C^2|_{S_2} =-1.
\end{equation}
In this case the resulting geometric transition and decompactification of that curve leads to two disconnected, reducible surface components, and thus a reducible SCFT. 
 Field theoretically this  procedure corresponds to a mass deformation, which from 5d SCFT or KK-theory leads to multiple factorized 5d SCFTs. In terms of effective gauge theory, it correspond to a bifundamental hypermultiplet getting decoupled. This case is in fact excluded on purpose in the description of CFDs, since we do not allow the factorization into lower rank 5d SCFTs after a CFD transition as these are expected not to result in new lower rank SCFTs.

\item The second possibility corresponds to decoupling the extra gauge theory, and this is achieved by a decompactification limit of the compact surfaces 
\be
S_{i}, \qquad i\in s_G\,,\qquad |s_G|=\text{rank}(G)\,,
\ee 
which are dual to the Cartans of the extra gauge group $G$ that needs to be decoupled. 
In particular this decompactification retains a compact part of the theory, in particular taking $\text{vol}(S_{i\in s_G})\rightarrow \infty$ whilst keeping the volume of all other compact surfaces in $\mathcal{S}$ finite, a necessary condition is that the curves 
$S_{i} \cdot S_j$ for $i\in s_G$ and any $j\notin s_G$
remain at finite volume. This in particular requires potentially flopping curves before decoupling the surfaces. In the cases we analyze, the limit $\text{vol}(S_{i\in s_G})\rightarrow \infty$ in M-theory corresponds to sending the $T^2$ fiber to infinite volume, \cite{DelZotto:2017pti, DelZotto:2018tcj}. This can be seen via M/F-theory duality (circle reduction and T-duality), where the radius of the compactification to 5d, $R$, is mapped to the inverse radius of one of the two circles of the $T^2$-fiber in M-theory. Therefore the honest 5d limit is when the $T^2$-fiber has infinite volume. 
\end{itemize}
An example of these theories are Non-Higgsable Clusters (NHCs), and these two possible geometric operations in order to get 5d SCFTs were discussed in \cite{DelZotto:2017pti}. More generally, we also discuss non-minimal conformal matter theories in detail in this paper. 

This decoupling/decompactification process will be one of the main foci of this paper, and before describing these geometric operations in many examples, we briefly illustrate what happens in terms of the effective IR field theories for cases, where the theory is very Higgsable to 6d $(2,0)$ SCFTs.

\subsection{Non-Minimal Conformal Matter}
\label{sec:NMCMGauge}

An illustrative class of theories Higgsable to 6d $(2,0)$ SCFTs is provided by non-minimal 6d conformal matter theories. They are defined as $N$ M5 branes probing an ADE singularity $\mathbb{C}^2/\Gamma_{\text{G}}$. Their circle compactification leads to KK-theories which UV complete into the 6d SCFT they originate from. At low-energy they admit an effective gauge theory in terms of the quivers in table~\ref{tab:affine-quivers}.

\begin{table}
\begin{center}
\begin{tabular}{c||c}

$G$ & Quiver \cr \hline \hline 
$A_{n-1}$ &
$
//\underset{n}{\underbrace{SU(N)-...-SU(N)-...-SU(N)}}//
$
\cr \hline 
$D_{n>3}$ 
&
$
SU(N)-\underset{n-3}{\underbrace{\overset{%
\begin{array}
[c]{c}%
SU(N)\\
|
\end{array}
}{SU(2N)}-SU(2N)-...-SU(2N)-\overset{%
\begin{array}
[c]{c}%
SU(N)\\
|
\end{array}
}{SU(2N)}}}-SU(N)
$
\cr \hline 
$E_6$
&
$SU(N)-SU(2N)-\overset{%
\begin{array}
[c]{c}%
SU(N)\\
|\\
SU(2N)\\
|
\end{array}
}{SU(3N)}-SU(2N)-SU(N)
$\cr \hline
 $E_7$
 &
$
 SU(N)-SU(2N)-SU(3N)-\overset{%
\begin{array}
[c]{c}%
SU(2N)\\
|
\end{array}
}{SU(4N)}-SU(3N)-SU(2N)-SU(N) 
$ \cr \hline
 $E_8$
 &
{\footnotesize $
SU(N)-SU(2N)-SU(3N)-SU(4N)-SU(5N)-\overset{%
\begin{array}
[c]{c}%
SU(3N)\\
|
\end{array}
}{SU(6N)}-SU(4N)-SU(2N) 
$ } \cr \hline 
\end{tabular}
\caption{The 5d affine quiver gauge theory descriptions of the KK reductions of  
the non-minimal conformal matter theories of $N$ M5-branes probing $\mathbb{C}^2/\Gamma_G$.
}
\label{tab:affine-quivers}
\end{center}
\end{table}

Note that for $A_{n-1}$, the first and last $SU(N)$ are {connected via a hypermultiplet, resulting in a circular quiver with $n$ nodes}.
The rank of the classical flavor symmetries, which is obtained by counting the baryonic and topological $U(1)$s, matches the dimension of the following 6d flavor groups
\be \label{eq:flavsymm6d}
\ba
A_{n-1}:& \quad G^{(6d)}_F= S(U(n) \times U(n))\cr 
D_n: &\quad G^{(6d)}_F= SO(2n) \times SO(2n)\cr 
E_n: & \quad G^{(6d)}_F= E_n \times E_n, \quad n=6,7,8 \,, 
\ea
\ee
where the dimension of the flavor groups is given by the total number of nodes of the Dynkin diagrams respectively, plus an extra node which can be interpreted as a shared affine extension of the flavor symmetry algebras.  
This is consistent with the fact that they uplift to 6d in the UV. Moreover, the group structure of the flavor symmetries \eqref{eq:flavsymm6d} is actually given by modding out a common diagonal center symmetry, \cite{Kim:2018lfo,Kim:2018bpg,Ohmori:2018ona}. This can be seen in 5d from the spectrum of BPS states, in particular, by analyzing the representation content with respect to the flavor symmetry corresponding to \eqref{eq:flavsymm6d}. As already anticipated, a useful tool to study the BPS states can be provided by the CFDs, which encode the flavor symmetries of the SCFTs. The CFDs for these KK-theories are given in tables \ref{tab:nm-CM-CFD} and \ref{tab:nm-E-CM-CFD} (the same applies for minimal conformal matter theories \cite{Apruzzi:2019enx}). These diagrams have some (discrete) symmetries. We argue that these are redundancies, and therefore they must be modded out also when studying the BPS state. In fact the action of these symmetries has been already modded out in order to understand some other physical properties such as the trees of descendant theories after mass deformations \cite{Apruzzi:2019vpe, Apruzzi:2019enx}. This reflects the global structures of the flavor symmetry groups for conformal matter predicted in \cite{Kim:2018lfo,Kim:2018bpg,Ohmori:2018ona,Dierigl:2020myk}.

We can first notice that these theories have only bifundamental matter charged under the gauge groups, and they do not have any flavor matter. As already anticipated, by giving mass to these the quiver will factorize into subquivers, which might lead to fixed points.

The second,  more interesting prospect is to decouple the extra gauge theory, as first proposed in \cite{Ohmori:2015pia}. These KK-theories consist of 5d SCFTs coupled to a 5d $\mathcal{N}=1$ $SU(N)$ gauge theory. From the point of view of the classical gauge theories mentioned above, the difference now is that the $SU(N)$ gauge node corresponding to the affine becomes a flavor group. The theories are summarized in table \ref{tab:decoupquivers}.

\begin{sidewaystable}
\begin{tabular}{|c|c|c|}\hline
$\mathbb{C}^2/\Gamma_{G}$& Decoupled Quiver & $G_{\text{F}}^{(5d)}$\cr \hline \hline
&&\cr 
$A_{n-1}$ 
&
$[SU(N)]-\underset{n-1}{\underbrace{SU(N)-...-SU(N)-...-SU(N)}}-[SU(N)]$
& 
$S(U(n) \times U(n)) \times S(U(N)\times U(N))$
\cr \hline
$D_{n>3}$& 
$[SU(N)]-\underset{n-3}{\underbrace{\overset{%
\begin{array}
[c]{c}%
SU(N)\\
|
\end{array}
}{SU(2N)}-SU(2N)-...-SU(2N)-\overset{%
\begin{array}
[c]{c}%
SU(N)\\
|
\end{array}
}{SU(2N)}}}-SU(N),
$
& $SO(2n) \times SO(2n)\times SU(N)$
\cr \hline 
$E_6$ & 
$
[SU(N)]-SU(2N)-\overset{%
\begin{array}
[c]{c}%
SU(N)\\
|\\
SU(2N)\\
|
\end{array}
}{SU(3N)}-SU(2N)-SU(N).
$
& 
$E_6 \times E_6 \times SU(N)$
\cr\hline 
$E_7$
&
$ [SU(N)]-SU(2N)-SU(3N)-\overset{%
\begin{array}
[c]{c}%
SU(2N)\\
|
\end{array}
}{SU(4N)}-SU(3N)-SU(2N)-SU(N) 
$
& $E_7 \times E_7 \times SU(N)$
\cr\hline
$E_8$
&
$
[SU(N)]-SU(2N)-SU(3N)-SU(4N)-SU(5N)-\overset{%
\begin{array}
[c]{c}%
SU(3N)\\
|
\end{array}
}{SU(6N)}-SU(4N)-SU(2N) 
$ 
& $E_8 \times E_8 \times SU(N)$\cr \hline
\end{tabular}
\caption{Conformal Matter for $N$ M5-branes at $\mathbb{C}^2/\Gamma_{G}$ singularity. We tabulate the quivers that are obtained after decoupling, as well as the superconformal flavor symmetry in 5d.  \label{tab:decoupquivers} }
\end{sidewaystable}

The dimension of the flavor symmetries is given by the number of nodes of the Dynkin diagrams, consistently with the fact that these theory leads to 5d SCFTs in the UV. 

Once we have decoupled the extra $SU(N)$ gauge theory, a low energy alternative description of these theories is the given by the 5d analog of the partial tensor branch quivers in 6d,
\begin{equation}
[G] \overset{\rm cm}{-} G \overset{\rm cm}{-} G \overset{\rm cm}{-} \ldots \overset{\rm cm}{-} G \overset{\rm cm}{-}  [G] \,,
\end{equation}
where $G$ is of ADE type, and $\text{cm}$ stands for the circle compactification of the conformal matter, where a $G$ or a $G\times G$ subgroup of the superconformal flavor symmetry has been gauged. In particular, for the $G\times G$ gauging, if we assume that the matter has a weakly coupled gauge theory description, we have a contradiction. That is, if there exists a gauge theory, the compact surfaces describing this generalized matter are ruled, but not all the $G\times G$ generator curves are fibers of this ruling. In particular, one of them corresponds to the topological $U(1)$, and it is a section. Field theoretically, it means that the non-perturbative symmetry $U(1)_T$ is gauged, which implies that there is no weakly coupled matter charged under the hypothetical gauge groups. However, this is very analogous to what happens between gluing by tubes of sphere with punctures of 4d $\mathcal N=2$ Gaiotto theories \cite{Gaiotto:2009we}.

We have seen that the decoupling of the $SU(N)$ vector leads to a 5d SCFT with at least two effective descriptions, which might not be always weakly coupled. Geometrically, this is realized by decompactification of $N-1$ divisors of the KK-geometry. A very important point is that if we have a resolution geometry with a ruling of the affine quiver theory in table~\ref{tab:affine-quivers}, we can immediately identify the $N-1$ surfaces responsible for the $SU(N)$ gauge enhancement, i.e. the affine $SU(N)$ node.  However, we also need to make sure that the $U(1)_{T_{SU(N)}}$ associated to the $SU(N)$ gauge theory is decoupled from the 5d SCFT. Geometrically, 
this can require {to flop  $\mc{O}(-1)\oplus\mc{O}(-1)$ curves} before decompactification of the $N-1$ surfaces.
Indeed, $U(1)_{T_{SU(N)}}$ is related to the affine node in the elliptic fibration, and the additional {flops  make} sure that $U(1)_T$ decouples in the decompactification process. This breaks the affine structure of the flavor symmetry, and the resulting theory no longer UV-completes to a 6d SCFT.

Keeping track of these operations is very important in order to define a good geometry where all the surfaces are shrinkable to a point. Moreover it will allow us to write a CFD for the decompactified geometry, and study its descendant automatically. 


\section{CFDs for NHCs \label{sec:NHCs}}

Non-Higgsable clusters (NHCs) are an example of NVH theories and they are key building blocks for 6d SCFTs. We now discuss their counterpart in the reduction to 5d and determine the associated CFDs, implementing the decoupling philosophy. 

NHCs are characterized by a single $(-n)$ self-intersection curve with the following gauge algebra
\be
\begin{tabular}{c||c|c|c|c|c|c|c|c}
$\Sigma^2=(-n)$ & $-3$& $-4$& $-5$& $-6$& $-7$& $-8$ &$-12$\cr \hline
$\mathfrak{g}$ & $\mathfrak{su}(3)$ &$\mathfrak{so}(8)$ &$\mathfrak{f}_4$ &$\mathfrak{e}_6$ &$\mathfrak{e}_7+ {1\over 2} \bm{56}$ &$\mathfrak{e}_7$ &$\mathfrak{e}_8$ 
\end{tabular}
\ee
To determine the CFDs we first need to compute the resolution geometries. The surface components in the marginal resolution geometry was presented in \cite{Haghighat:2014vxa, DelZotto:2017pti,Bhardwaj:2018vuu}. In table \ref{t:CFD-NHC}, we summarize the CFDs read off from the geometry. In this section, we will only discuss the case of a single $(-3)$-curve in detail, in order to show an explicit example of the decoupling action. The geometry of the surface components and curves in the other cases are presented in appendix \ref{app:NHCGeo}. 

\begin{table}
\centering
\begin{tabular}{|c|c|c|c|}\hline
$\Sigma^2=(-n)$ & $\mathfrak{g}$ & Pre-decoupling CFD & CFD \cr \hline \hline
-3 & $\mathfrak{su}(3)$ &\includegraphics[height=2cm]{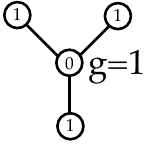}&\includegraphics[height=2cm]{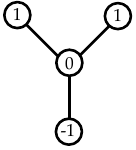} \cr \hline
-4& $\mathfrak{so}(8)$ &\includegraphics[height=2.2cm]{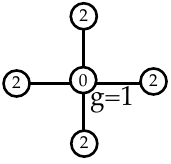}&\includegraphics[height=2.2cm]{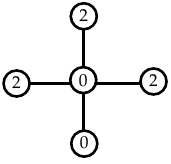} \cr \hline
-5&$\mathfrak{f}_4$ &\includegraphics[height=2cm]{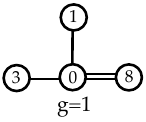}&\includegraphics[height=2cm]{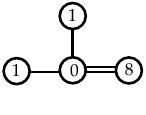} \cr \hline
-6&$\mathfrak{e}_6$&\includegraphics[height=2cm]{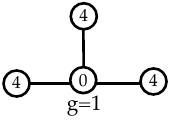}&\includegraphics[height=2cm]{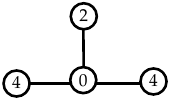} \cr \hline
-7&$\mathfrak{e}_7+ {1\over 2} \bm{56}$ &\includegraphics[height=2cm]{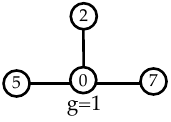}&\includegraphics[height=2cm]{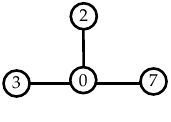} \cr \hline
-8&$\mathfrak{e}_7$ &\includegraphics[height=2cm]{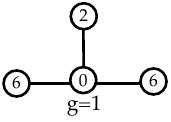}&\includegraphics[height=2cm]{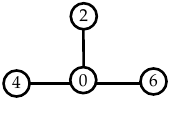} \cr \hline
-12&$\mathfrak{e}_8$ &\includegraphics[height=2cm]{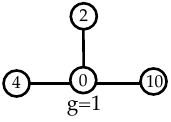}&\includegraphics[height=2cm]{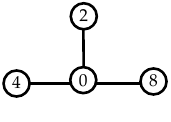} \cr \hline
\end{tabular}

\caption{CFDs for NHCs: the first column denotes the self-intersection number of the rational curve in the base of the elliptic fibration, $\mathfrak{g}$ is the non-Higgsable gauge group. The last two columns show the CFDs before and after the decoupling of the gauge sector.}\label{t:CFD-NHC}
\end{table}

For a single $(-3)$ curve, the 6d non-Higgsable gauge group is $SU(3)$, which is realized by a type $IV$ split Kodaira fiber. In the marginal geometry, there are three Hirzebruch surfaces $\mb{F}_1$ sharing a common $\mc{O}(-1)\oplus\mc{O}(-1)$ curve in the middle, which correspond to the Cartan divisors of the affine $SU(3)$. They are denoted by $U$, $u_1$ and $u_2$, where $U$ corresponds to the affine node.

We plot the curve configurations on the surfaces as follows:
\be
\label{3-topresol}
\includegraphics[width=7cm]{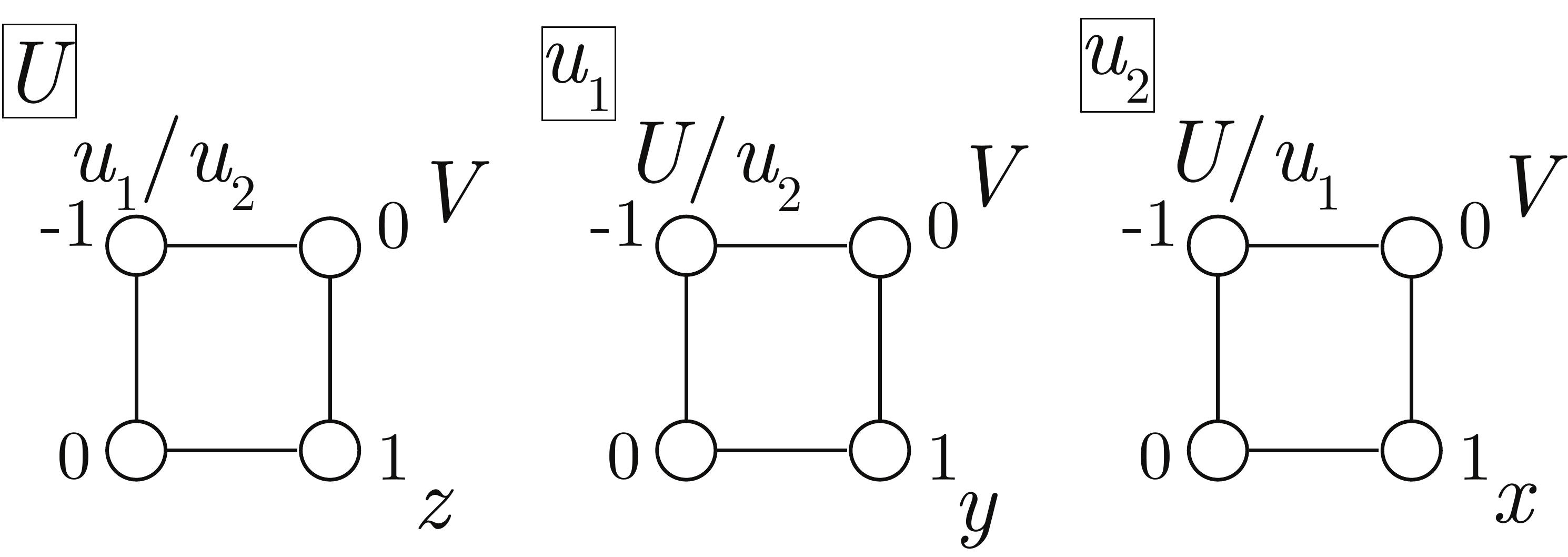} \,.
\ee
The letter in the box labels the compact surface component, and each node on each surface component denotes a complete  intersection curve between two surfaces. In this case, the letter $V$, $z$, $x$ and $y$ correspond to non-compact surfaces. The number next to a node is the self-intersection number of such complete intersection curve. By default, these curves are rational (with genus-0). If this is not the case, we will label it out with $g=$ the genus. When $g<0$, it describes a reducible curve with multiple (rational) components. For example, here on the surface component $U$, the complete intersection curve $V\cdot U$ is a rational 0-curve on $U$, and the curve $z\cdot U$ is a rational 1-curve on $U$. The complete intersection curve $U\cdot u_1$, $U\cdot u_2$ and $u_1\cdot u_2$ coincides, which are all $(-1)$-curves on $U$, $u_1$ and $u_2$. One can check that the adjunction formula
\be
D_1^2 D_2+D_2^2 D_1=2g(D_1\cdot D_2)-2\label{adjunction}
\ee
is always satisfied, where $g(D_1\cdot D_2)$ is the genus of the complete intersection curve $D_1\cdot D_2$.

In the pre-decoupled CFD, there are four nodes that correspond to the non-compact surfaces $V,x,y,z$. The self-intersection number $n$ and genus $g$ of each node are read off by:
\be
\ba
&V^2(U+u_1+u_2)=0\ ,\ V(U+u_1+u_2)^2=0,\cr
&x^2(U+u_1+u_2)=y^2(U+u_1+u_2)=z^2(U+u_1+u_2)=1,\cr  &x(U+u_1+u_2)^2=y(U+u_1+u_2)^2=z(U+u_1+u_2)^2=-3
\ea
\ee
and the adjunction formula (\ref{adjunction}). Hence $V$ is a node with $(n,g)=(0,1)$, and $x$, $y$, $z$ all corresponds to nodes with $(n,g)=(1,0)$. The number of edges between each nodes are read off by
\be
V\cdot x\cdot (U+u_1+u_2)=1\ ,\ V\cdot y\cdot (U+u_1+u_2)=1\ ,\ V\cdot z\cdot (U+u_1+u_2)=1\,.
\ee
We hence get the CFD before decoupling
\be
\includegraphics[height=2cm]{NHC-CFD-3-pre.pdf}\,.
\ee
To get a 5d SCFT, we decompactify the surface component $U$, see also \cite{DelZotto:2017pti}. Then the remaining compact surfaces are two $\mb{F}_1$ intersecting along an $\mc{O}(-1)\oplus\mc{O}(-1)$ curve, which corresponds to the rank-2 5d SCFT with gauge theory description $SU(3)_0$~\cite{Jefferson:2018irk}. In \cite{Apruzzi:2019opn}, it was shown to have the following CFD:
\be
\includegraphics[width=5cm]{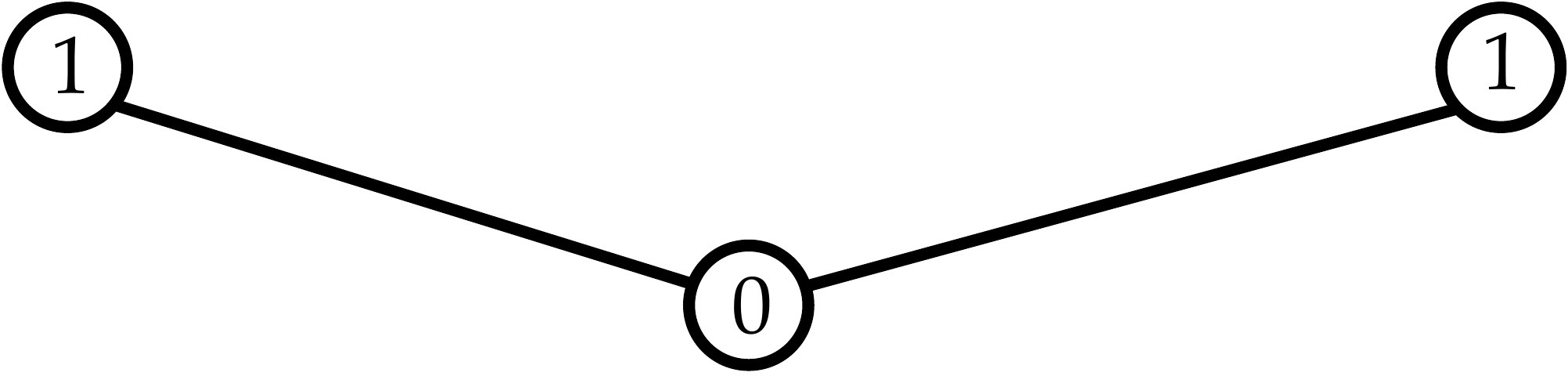}\,.
\ee

However, in the geometry (\ref{3-topresol}), the intersection curve $u_1\cdot u_2$ can also be interpreted as complete intersection curve between the non-compact surface $U$ after the decoupling. Indeed, if this curve is shrunk, the two remaining compact surface components become two disconnected $\mb{P}^2$, which leads to two decoupled copies of rank-one 5d SCFTs. In \cite{Apruzzi:2019opn}, such flop transition is not allowed as the CFD tree only includes irreducible rank-two theories. However, for the purpose of gluing, it is convenient to attach an additional $(-1)$-node corresponding to $U$, which leads to the final CFD
\be
\includegraphics[height=2.5cm]{NHC-CFD-3.pdf}\,.
\ee
As the 0-node is actually a combination of two 0-curves, after flopping this $(-1)$-node, the CFD will become two disconnected CFD of the rank-one with $(+1)$-nodes.

Note that for all the single curve NHC geometries, the decoupled surface $U$ is always a Hirzebruch surface $\mb{F}_m$, and the intersection curve $U\cdot u_1$ with another compact surface is the section curve $\Sigma$ with self-intersection $\Sigma^2=-m$. Then we can always decompactify the $\mb{P}^1$ fiber of $\mb{F}_m$, 
which is consistent with the decoupling criterion,  \cite{Bhardwaj:2019xeg} and section \ref{sec:nvht}.
The remaining cases are discussed in the appendix and are summarized in table \ref{t:CFD-NHC}.

\section{Non-Minimal Conformal Matter \label{sec:NMCM}}

As we already discussed in section \ref{sec:NMCMGauge} the non-minimal conformal matter theories in 6d are examples of NVH theories, and they can be Higgsed to 6d $\mathcal N=(2,0)$ theories. This implies that upon circle reduction, the KK-theory is described by a 5d SCFT coupled to an $SU(N)$ vector multiplet. The 5d SCFT are isolated by decoupling the extra sector via decompactification of the M-theory geometry. In this section we derive the 5d CFDs before and after decoupling for these models of type $(G,G)$ from which all descendants can be obtained by the usual CFD transition rules \cite{Apruzzi:2019vpe, Apruzzi:2019opn, Apruzzi:2019enx}. The geometric derivation of the CFDs starts with the tensor branch geometries in 6d. 

\subsection{Tensor Branch Geometries}

The non-minimal $(G, G)$ conformal matter theories of rank $N$ correspond to the 6d theory of $N$ M5-branes probing a $\mathbb{C}^2/\Gamma_{G}$ singularity, which have flavor symmetry is $G\times G$. We will first summarize the tensor branch geometries in 6d.

The tensor branch for the $(A_{n-1}, A_{n-1})$ non-minimal conformal matter theory is a quiver with nodes $\mathfrak{su}(n)$ on $(-2)$, i.e. 
denote by $N-1$ the number of $(-2)$-curves. In the standard 6d notation\footnote{In particular, in 6d the standard notation is to write $-n$ instead of $n= \Sigma^2$.}, the tensor branch is 
\be\label{sun2}
[SU(n)]-\stackrel{\mathfrak{su}(n)}{2}- \cdots- \stackrel{\mathfrak{su}(n)}{2}-[SU(n)]\,.
\ee
The non-minimal $(D_n,D_n)$ conformal matter is contructed by the following base geometry in 6d F-theory:
\be
\label{DkDk-base}
[SO(2n)]-\overset{\mathfrak{so}(2n)}{2}-...-\overset{\mathfrak{so}(2n)}{2}-[SO(2n)],
\ee
where there are $N-1$ $(-2)$ curves in the middle.
In the full tensor branch, the base geometry becomes
\be
\label{DkDk-tensorbranch}
[SO(2k)]-\overset{\mathfrak{sp}(k-4)}{1}-\overset{\mathfrak{so}(2k)}{4}-\overset{\mathfrak{sp}(k-4)}{1}-...-\overset{\mathfrak{so}(2k)}{4}-\overset{\mathfrak{sp}(k-4)}{1}-[SO(2k)].
\ee
There are $N$ $(-1)$ curves and $N-1$ $(-4)$ curves in the middle.

The non-minimal $(E_6,E_6)$ conformal matter is contructed by the following base geometry in 6d F-theory:
\be
\label{E6E6-base}
[E_6]-\overset{\mathfrak{e}_6}{2}-...-\overset{\mathfrak{e}_6}{2}-[E_6],
\ee
where there are $N-1$ $(-2)$ curves in the middle.

In the full tensor branch, the base geometry beecomes
\be
\label{E6E6-tensorbranch}
[E_6]-1-\overset{\mathfrak{su}(3)}{3}-1-\overset{\mathfrak{e}_6}{6}-...-\overset{\mathfrak{e}_6}{6}-1-\overset{\mathfrak{su}(3)}{3}-1-[E_6].
\ee
There are $N$ $(-3)$-curves, $2N$ $(-1)$-curves and $N-1$ $(-6)$-curves in the middle. 

The non-minimal $(E_7,E_7)$ conformal matter is contructed by the following base geometry in 6d F-theory:
\be
\label{E7E7-base}
[E_7]-\overset{\mathfrak{e}_7}{2}-...-\overset{\mathfrak{e}_7}{2}-[E_7],
\ee
where there are $N-1$ $(-2)$ curves in the middle.

In the full tensor branch, the base geometry becomes
\be
\label{E7E7-tensorbranch}
[E_7]-1-\overset{\mathfrak{su}(2)}{2}-\overset{\mathfrak{so}(7)}{3}-\overset{\mathfrak{su}(2)}{2}-1-\overset{\mathfrak{e}_7}{8}-...-\overset{\mathfrak{e}_7}{8}-1-\overset{\mathfrak{su}(2)}{2}-\overset{\mathfrak{so}(7)}{3}-\overset{\mathfrak{su}(2)}{2}-1-[E_7].
\ee
There are $N$ $(-3)$-curves, $2N$ $(-2)$-curves, $2N$ $(-1)$-curves and $N-1$ $(-8)$-curves in the middle.

Similarly, the non-minimal $(E_8,E_8)$ conformal matter is given by:
\be
\label{E8E8-base}
[E_8]-\overset{\mathfrak{e}_8}{2}-...-\overset{\mathfrak{e}_8}{2}-[E_8],
\ee
where there are $N-1$ $(-2)$ curves in the middle.

In the full tensor branch, the base geometry is
\be
\label{E8E8-tensorbranch}
[E_8]-1-2-\overset{\mathfrak{su}(2)}{2}-\overset{\mathfrak{g}_2}{3}-1-\overset{\mathfrak{f}_4}{5}-1-\overset{\mathfrak{g}_2}{3}-\overset{\mathfrak{su}(2)}{2}-2-1-\overset{\mathfrak{e}_8}{12}-1-...-1-[E_8].
\ee

There are in total $N$ $(-5)$-curves, $2N$ $(-3)$-curves, $4N$ $(-2)$-curves, $4N$ $(-1)$-curves and $(N-1)$ $(-12)$-curves in the middle.

\subsection{Example Geometry: $(SO(8), SO(8))$ non-minimal CM}

We now determine the CFDs for the decoupled theories from the tensor branch geometry. 
We exemplify this for one Calabi-Yau threefold geometry, the non-minimal $(SO(8),SO(8))$ conformal matter. 
We will discuss the decoupling procedure, CFD and the IR gauge theory descriptions from the geometric perspective. The remaining cases are discussed in the appendix \ref{app:CMN} and are summarized in tables \ref{tab:nm-CM-CFD} and \ref{tab:nm-E-CM-CFD}.

The minimal $(SO(8), SO(8))$ conformal matter theory is equivalent to the rank-one E-string theory with $E_8$ flavor symmetry. In the resolution geometry, there is a generalized $dP_9$ (rational elliptic surface) over the $(-1)$-curve on the base, which has the following set of genus-zero curves\footnote{This is one of the semi-toric surfaces (C.13) in \cite{Martini:2014iza}.}:
\be
\label{CFD:D4-D4}
\includegraphics[width=3cm]{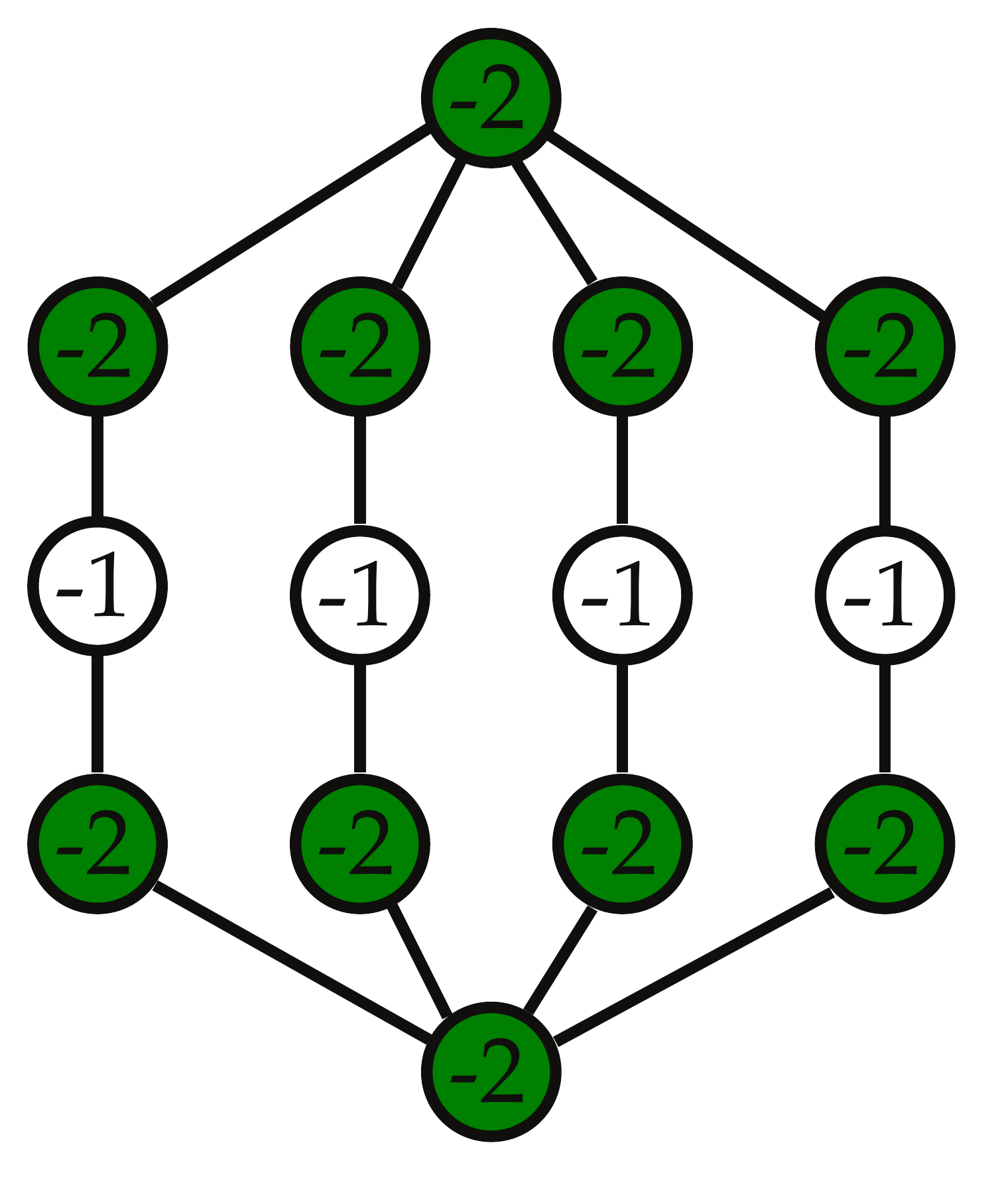}.
\ee
This figure is exactly the $(SO(8), SO(8))$ marginal CFD in table~\ref{tab:Rank1CFDs}.

With this rational elliptic surface as building blocks, we study the non-minimal $(SO(8),SO(8))$ conformal matter with $N=2$, which has the following tensor branch:
\be
\label{D4D4:tensorbranch}
[SO(8)]-1-\overset{\mathfrak{so}(8)}{4}-1-[SO(8)].
\ee
For each $SO(8)$, there are five complex surfaces connected in form of an affine $SO(8)$ Dynkin diagram. They  are denoted by $(V,v_1,\dots,v_4)$, $(U,u_1,\dots,u_4)$, $(W,w_1,\dots,w_4)$ from left to the right. The surfaces $(V,v_1,\dots,v_4)$ and $(W,w_1,\dots,w_4)$ are non-compact, while $(U,u_1,\dots,u_4)$ are compact. Here $U,V,W$ corresponds to the central node of the affine $SO(8)$, and $u_4,v_4,w_4$ corresponds to the affine node of $SO(8)$ which intersect the zero section $z$ of the resolved elliptic CY3. Finally, the rational elliptic surfaces over the two compact $(-1)$-curves in (\ref{D4D4:tensorbranch}) are denoted by $S_1$ and $S_2$, from left to the right.

\begin{figure}
\centering
\includegraphics[width=10cm]{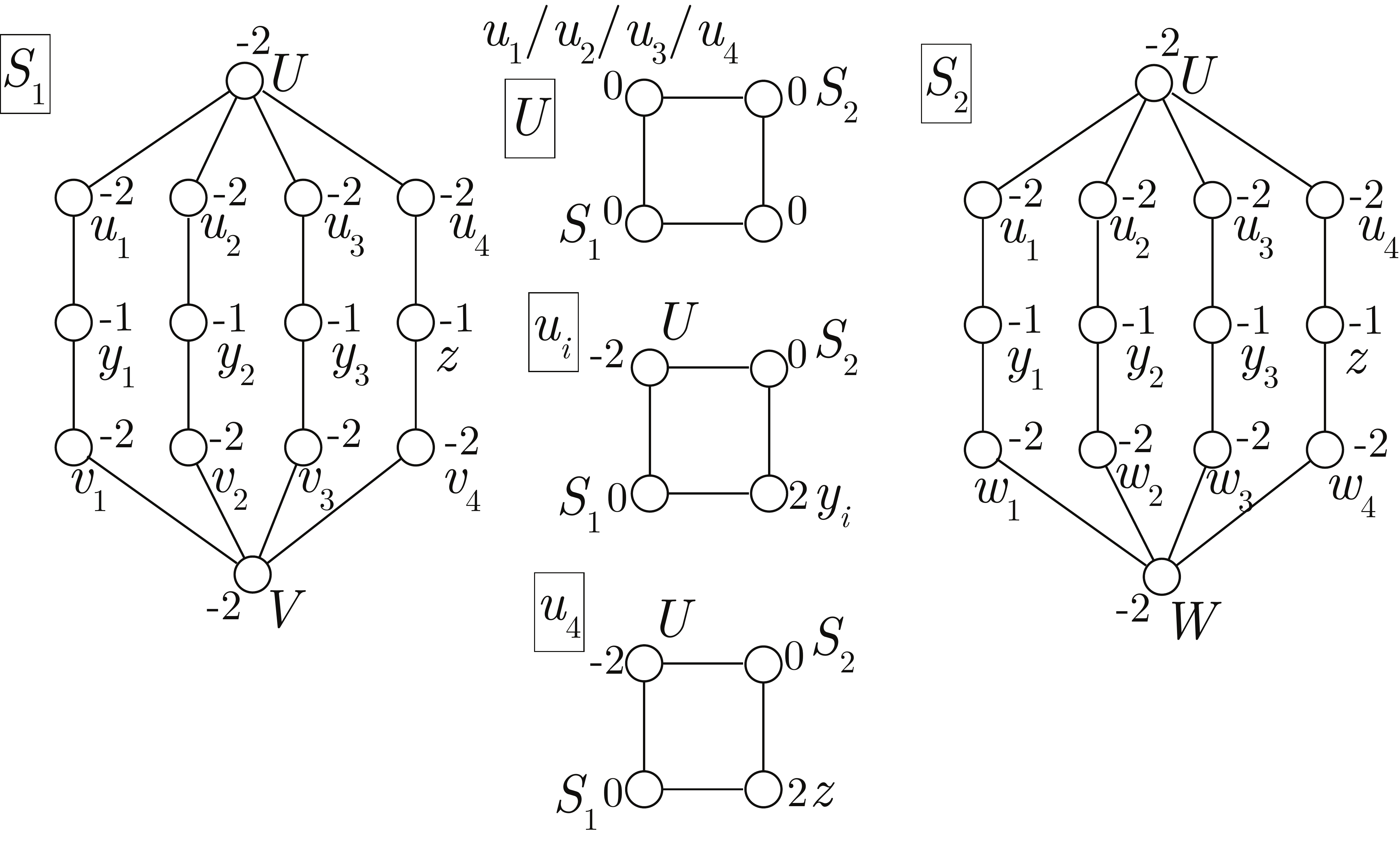}
\caption{The configuration of curves on the seven compact surfaces ($U,u_1,u_2,u_3,u_4,S_1,S_2$) in the resolution geometry of $(SO(8),SO(8))$ $N=2$ non-minimal conformal matter theory. Here $u_i$ denotes $u_1$, $u_2$ and $u_3$, which has the topology of Hirzebruch surface $\mb{F}_2$.}\label{f:D4D4N=2-topresol}.
\end{figure}

We plot the configuration of curves on the seven compact surfaces ($U,u_1,u_2,u_3,u_4,S_1,S_2$) in figure~\ref{f:D4D4N=2-topresol}. The two surfaces $S_1$ and $S_2$ have exactly the same curve configurations as (\ref{CFD:D4-D4}). The surface $u_1,u_2,u_3,u_4$ are all Hirzebruch surface $\mb{F}_2$ and the surface $U$ is the Hirzebruch surface $\mb{F}_0$, as expected in \cite{DelZotto:2017pti}.

From this geometry, the CFD vertex corresponding to non-compact surface $z$ is given by the following combination of curves
\be
C_z=z\cdot (S_1+u_4+S_2)\,,
\ee
where all the multiplicity factors equal to one. From (\ref{CFD-node-n}), (\ref{CFD-node-g}), it corresponds to a node $C_z$ with $(n,g)=(0,0)$. Similarly, the nodes $C_{y_i}$ corresponding to $y_i(i=1,\dots,3)$ have $(n,g)=(0,0)$ as well. Along with the number of edges computed with (\ref{CFD-edge-m}), the expected marginal CFD is
\be
\label{CFD:D4-D4N=2marginal}
\includegraphics[width=4cm]{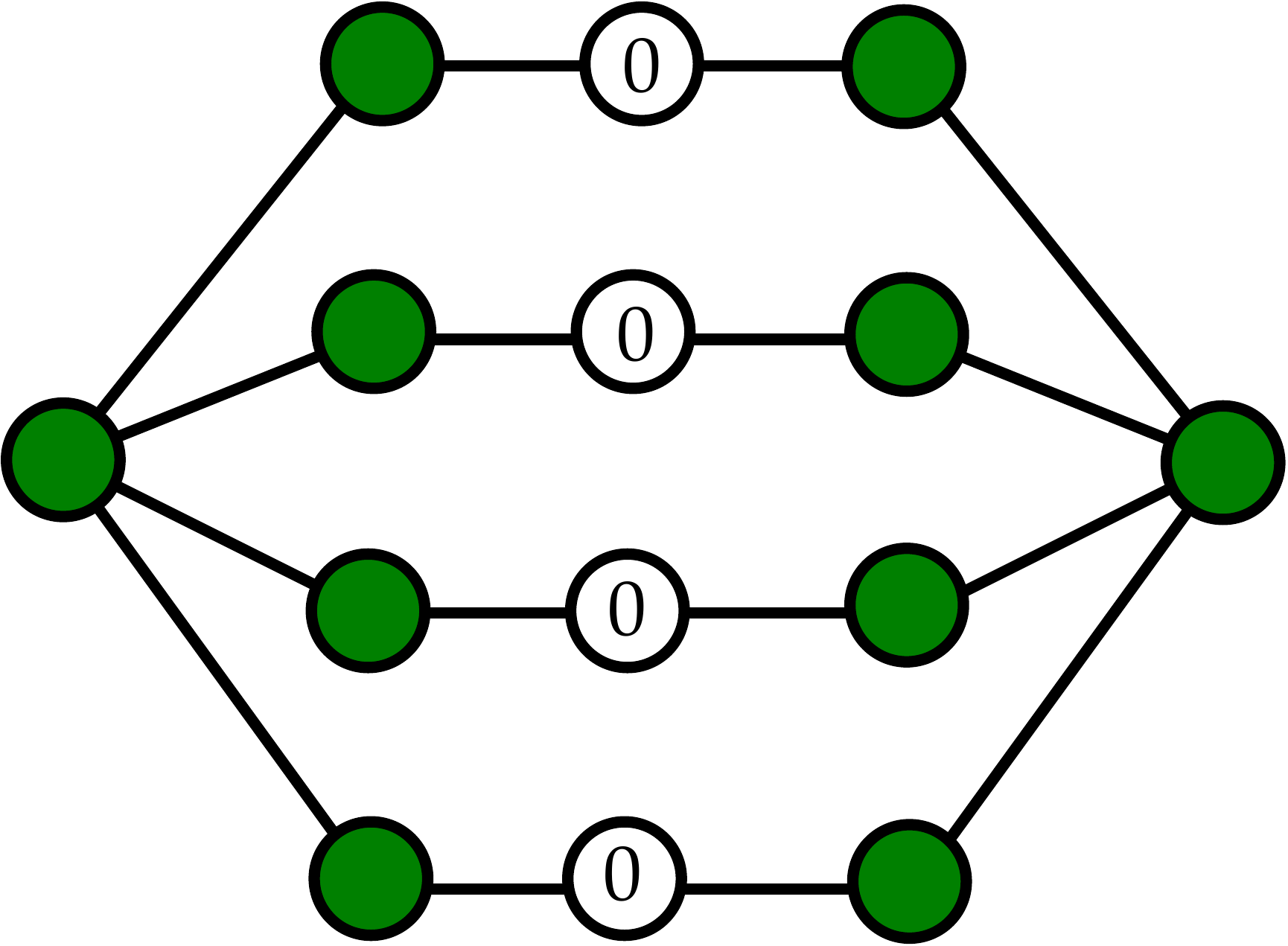}\,.
\ee
This marginal CFD corresponds to the 5d KK theory of the 6d (1,0) SCFT with the tensor branch (\ref{D4D4:tensorbranch}), which is a 5d SCFT coupled to a 5d $\mc{N}=1$ $SU(2)$ gauge theory. In this case, one cannot directly generate descendant 5d SCFTs via CFD transitions, because of the absence of extremal $(n,g)=(-1,0)$ vertices.

To get a 5d SCFT with descendants, we need to decouple the extra $SU(2)$ gauge theory by decompactifying the surface $u_4$ in figure~\ref{f:D4D4N=2-topresol}. The surface $u_4$ will give rise to a new vertex in the CFD after this operation. However, from this geometry we will naively get
\be
n(u_4)=u_4^2\cdot (S_1+S_2+U)=-4\,,
\ee
and $g(u_4)=0$, which is not allowed in a valid CFD. Moreover, if we want to keep the curves $U\cdot u_4$, $S_1\cdot u_4$ and $S_2\cdot u_4$ compact, since they are parts of the remaining compact surfaces, then all the curves on $u_4$ are compact. This is because the 0-curves and the (-2)-curve on $u_4$ generate the Mori cone of $u_4$, and the decompactification of $u_4$ will not be allowed~\cite{Bhardwaj:2019xeg}. To resolve this issue, we need to flop the curves $z\cdot S_1$ and $z\cdot S_2$ on $S_1$ and $S_2$ into $u_4$, which results in the following geometry
\be
\label{D4D4N=2-subtopresol}
\includegraphics[width=10cm]{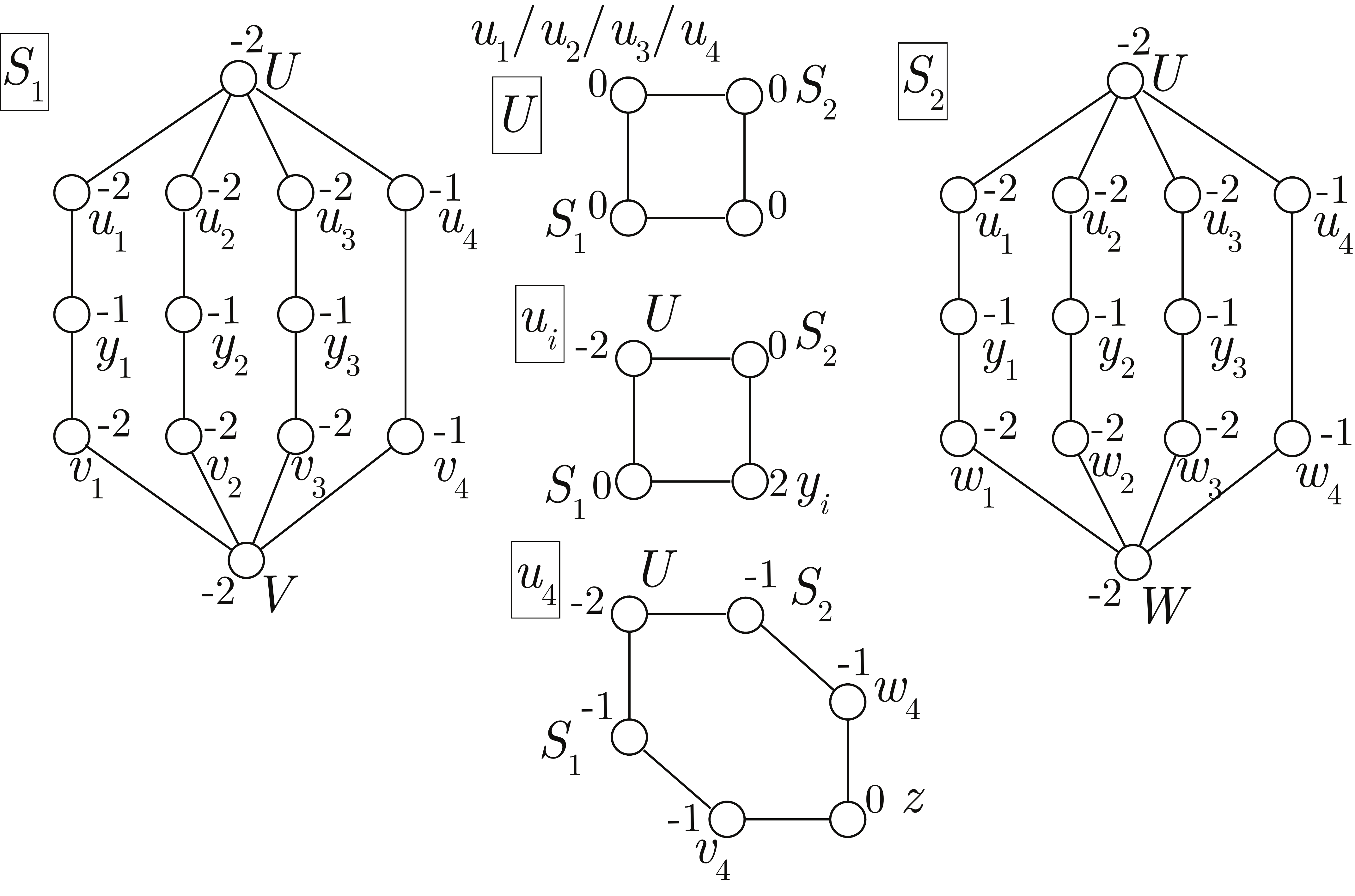}\,.
\ee
Now the surface $u_4$ has two more Mori cone generators $u_4\cdot v_4$ and $u_4\cdot w_4$, which can be made non-compact. Then there is no issue in decompactifying $u_4$. In the general case of non-minimal conformal matter, this flop should always happen before the decompactification, as expected in the field theory analysis in section~\ref{sec:strategy}.

Since all the multiplicity factors are trivially one, and we can read off the corresponding CFD (the letters label the corresponding non-compact surfaces)
\be
\label{CFD:D4-D4N=2subtop}
\includegraphics[width=4.5cm]{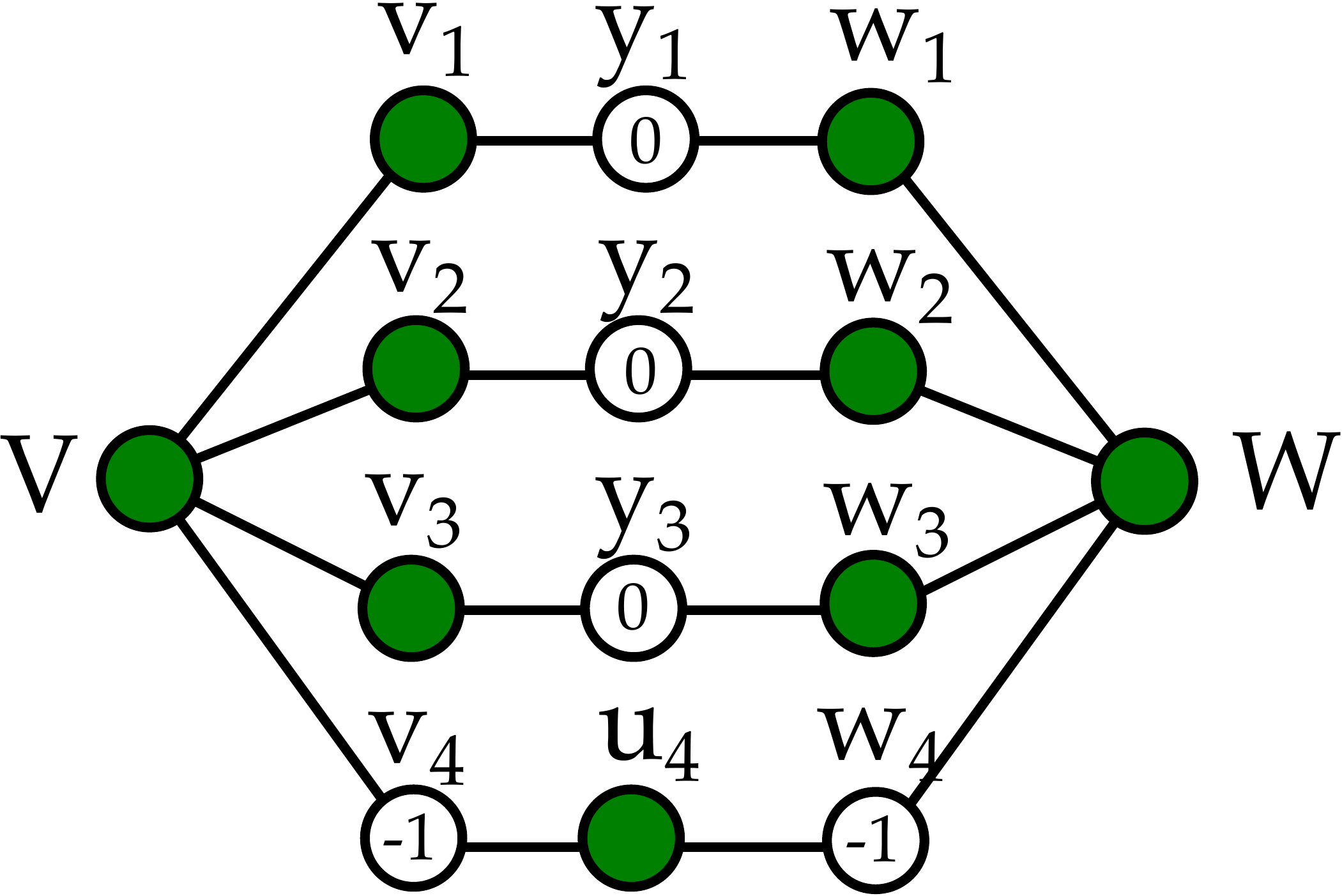}\,.
\ee
In the geometry (\ref{D4D4N=2-subtopresol}), we can assign the following $\mb{P}^1$ rulings on each surface component:
\be
\ba
f(S_1)&=(U+u_1+u_2+u_3+u_4+v_4)\cdot S_1\cr
f(S_2)&=(U+u_1+u_2+u_3+u_4+w_4)\cdot S_2\cr
f(U)&=S_1\cdot U=S_2\cdot U\cr
f(u_i)&=S_1\cdot U=S_2\cdot U\ (i=1,\dots,3)\,.
\ea
\ee
With this assignment, the surfaces $U,u_1,u_2,u_3$ will form the Cartans of an $SO(8)$ gauge group after the above ruling curves are shrunk to zero size, while $S_1$ and $S_2$ gives rise to two $SU(2)$s.

We hence have the following quiver gauge theory description:
\be
\label{D4D4N=2:quiver2}
3\bm{F}-SU(2)-SO(8)-SU(2)-3\bm{F}\,.
\ee
Although the geometry (\ref{CFD:D4-D4N=2subtop}) does not apparently have an $SU(4)\times SU(2)^3$ quiver gauge theory description, we can do a few flops to get it. We shrink the curves $y_i\cdot S_1$ $(i=1,2,3)$ on $S_1$ and $y_i\cdot S_2$ $(i=1,2,3)$ on $S_2$, and consequently blow up the compact surfaces $u_i$ $(i=1,2,3)$ two times for each. After the six flops, the curve configurations are
\be
\label{D4D4N=2-subtopflop}
\includegraphics[width=11cm]{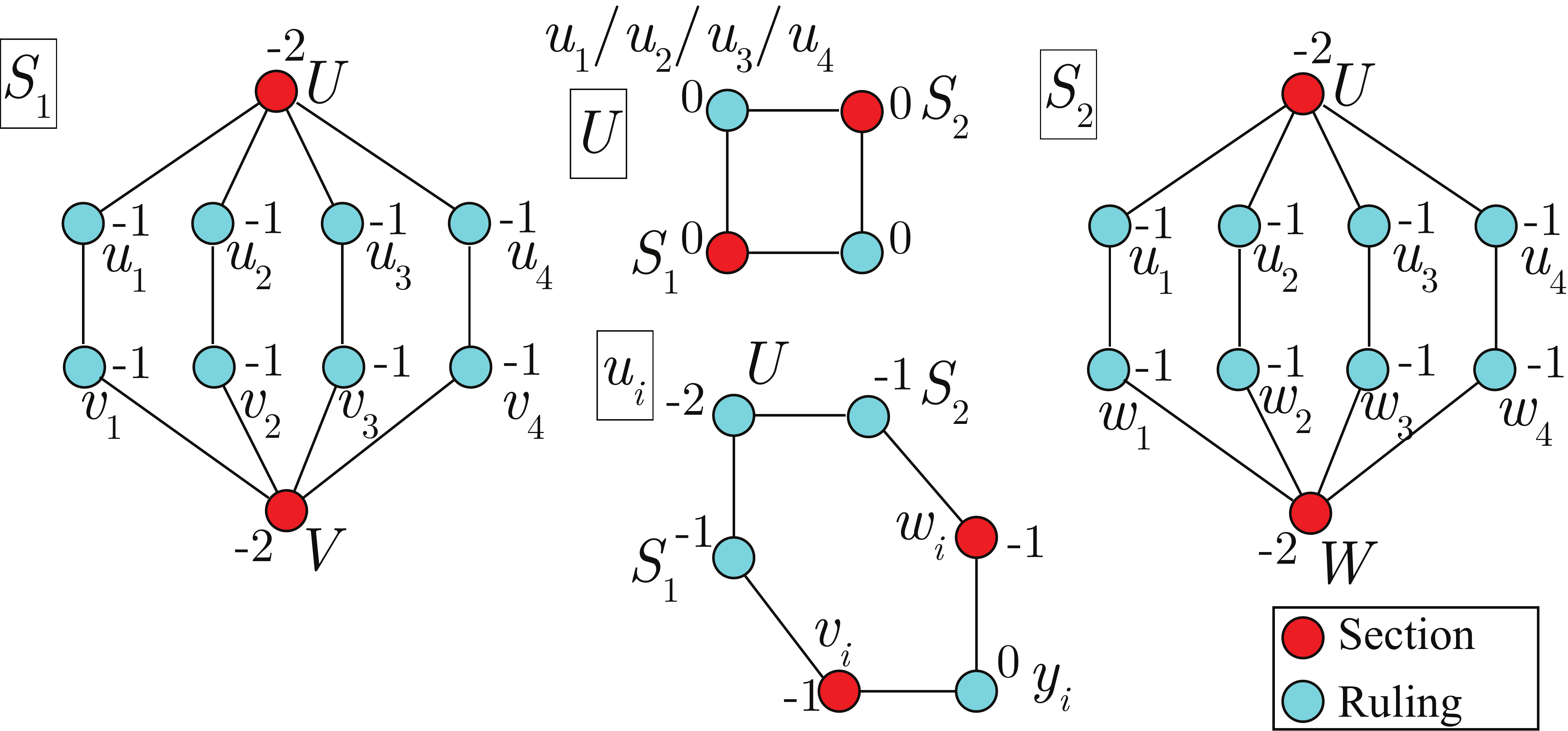}\,.
\ee
The assignment of section/rulings is shown in the figure explicitly. We can hence read off the following quiver description, where each letter in the bracket denotes the Cartan node of the gauge group
\be
\label{D4D4N=2:quiver1}
\begin{array}{ccccc}
& & SU(2)\ (u_2)& &\\
& & \vert & &\\
SU(2)\ (u_1)& - & SU(4)\ (S_1,U,S_2)& - & SU(2)\ (u_3)\,.
\end{array}
\ee
Finally, we can generalize this story to higher $N$, with more $(-1)$ and $(-4)$ curves in the tensor branch:
\be
[SO(8)]-1-\overset{\mathfrak{so}(8)}{4}-1-...-\overset{\mathfrak{so}(8)}{4}-1-[SO(8)]\,.
\ee
The resolution geometry then would become
\be
\label{D4D4N-topresol}
\includegraphics[width=15cm]{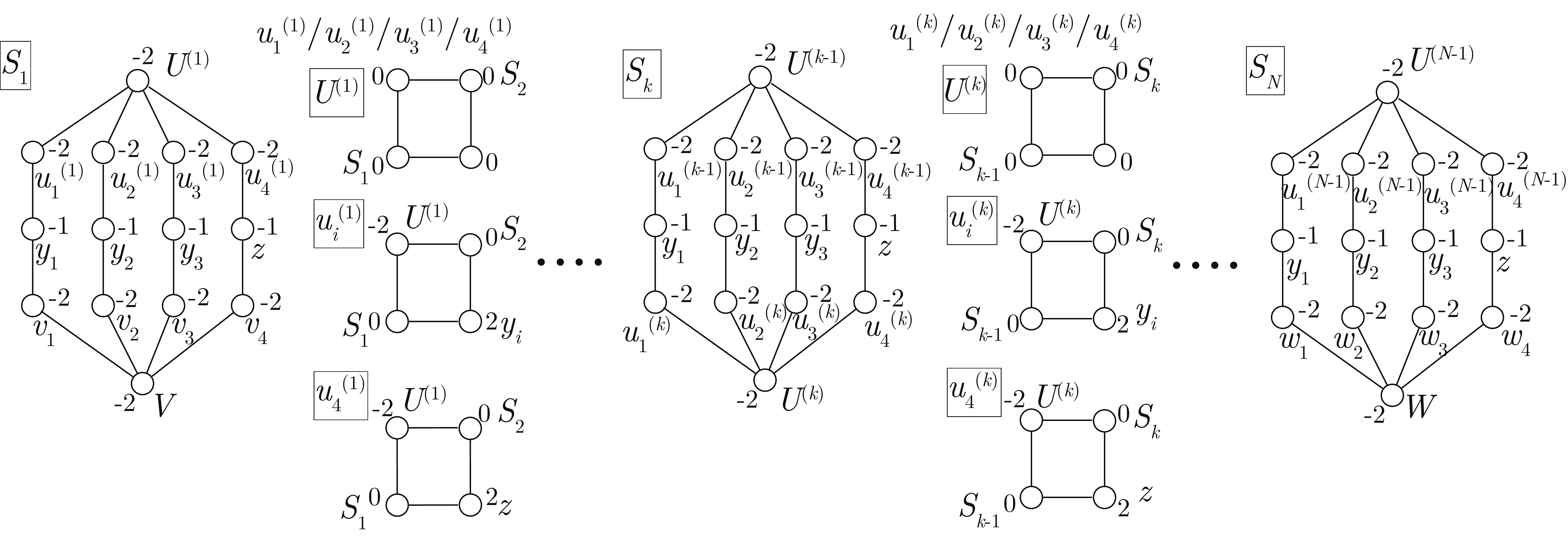}\,,
\ee
where $S_1,\dots,S_N$ denotes the $N$ rational elliptic surfaces over the $(-1)$-curves, and $U^{(k)}$, $u^{(k)}_1$, $u^{(k)}_2$, $u^{(k)}_3$, $u^{(k)}_4$ $(k=1,\dots,N-1)$ are the compact surfaces corresponding to the $k$-th affine $SO(8)$. In this case, the curves $z\cdot S_i$ and $y_j\cdot S_i$ for $i=2,\dots,N-1$, $j=1,\dots,3$ have non-trivial multiplicity factors $\xi_{S_i,z}=\xi_{S_i,y_j}=2$, because they are $\mc{O}(-1)\oplus\mc{O}(-1)$ curves which intersect two other compact surfaces. Hence the correct $n(z)$ and $n(y_j)$ are computed as:
\be
\ba
n(z)=&z^2\cdot (S_1+S_N+2\sum_{i=2}^{N-1} S_i+\sum_{i=1}^{N-1} u^{(i)}_4)=0\cr
n(y_j)=&y_k^2\cdot (S_1+S_N+2\sum_{i=2}^{N-1} S_i+\sum_{i=1}^{N-1} u^{(i)}_j)=0\quad (j=1,\dots,3)\,.
\ea
\ee

The resulting marginal CFD is exact the same as the $N=2$ case (\ref{CFD:D4-D4N=2marginal}), which 
has no descendant. In the flop and decoupling process to get a 5d SCFT, we we make all the surfaces $u_4^{(k)}$ $(k=1,\dots,N-1)$ non-compact, such that the extra $SU(N)$ vector multiplet is decoupled. Moreover, we need to shrink all the $z\cdot S_i$ $(i=1,\dots,N)$ in (\ref{D4D4N=2-subtopresol}), which results in the following geometry:
\be
\label{D4D4N-subtopresol}
\includegraphics[width=15cm]{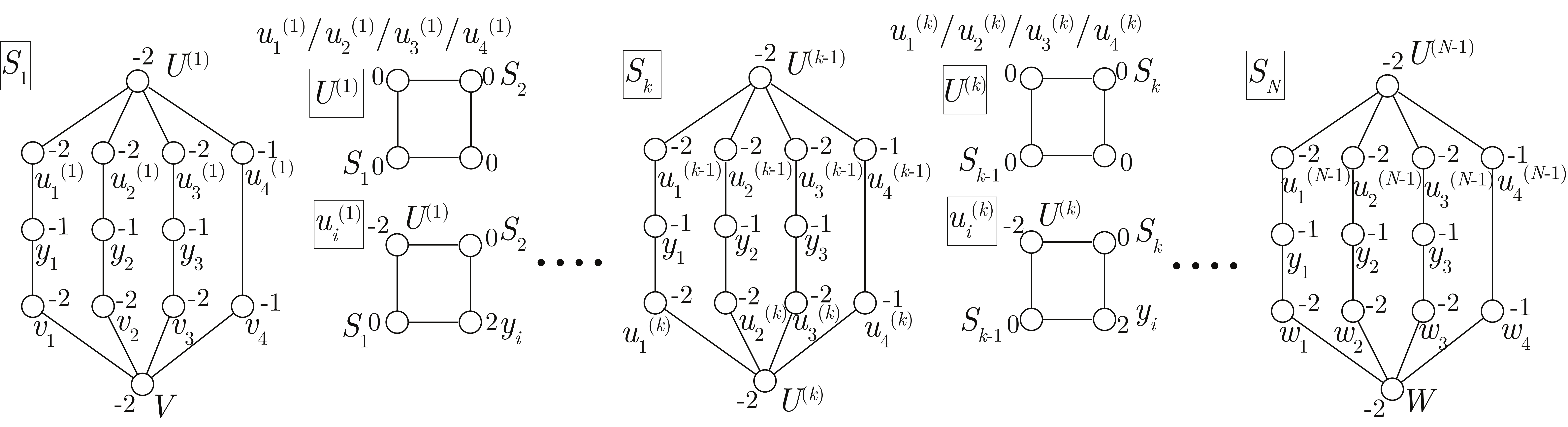}
\ee

\begin{table}
\begin{tabular}{|c|c|c|}\hline
$(G,G)$ & CFD before decoupling & CFD after decoupling \cr \hline\hline
&& \cr 
$(A_{n-1}, A_{n-1})$ &  \includegraphics[width=7cm]{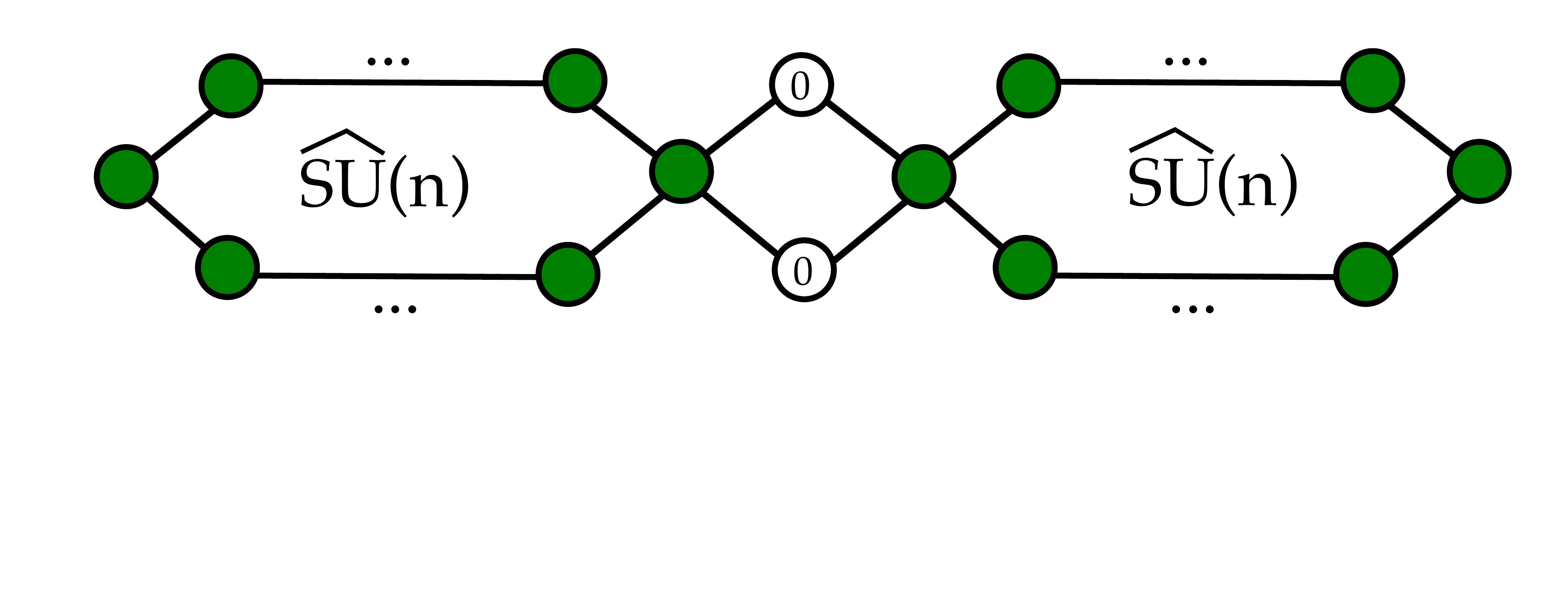}  &\includegraphics[width=4cm]{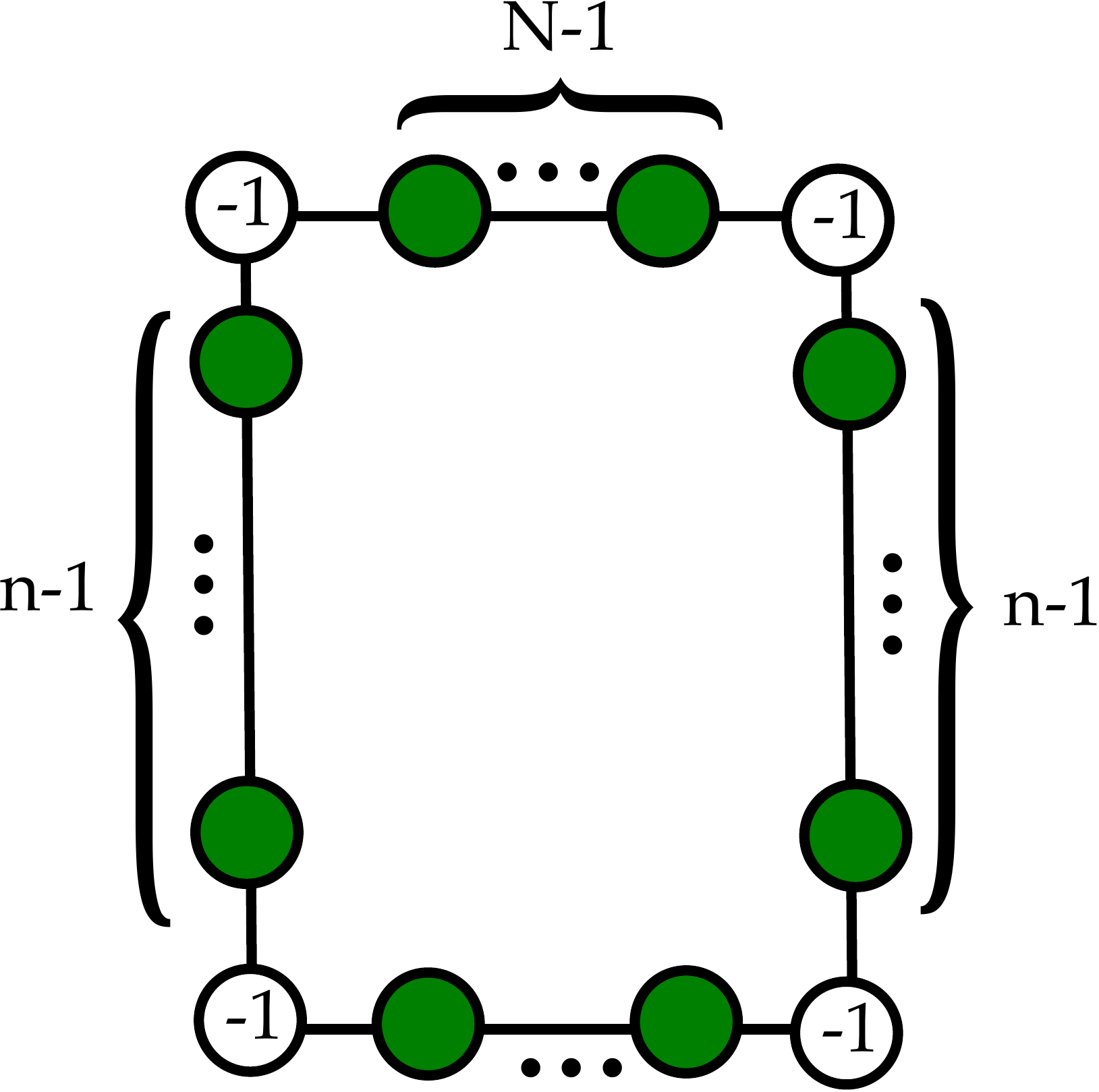} \cr 
 &&\cr  \hline 
 && \cr 
$(D_{4}, D_{4})$ & \includegraphics[width=4cm]{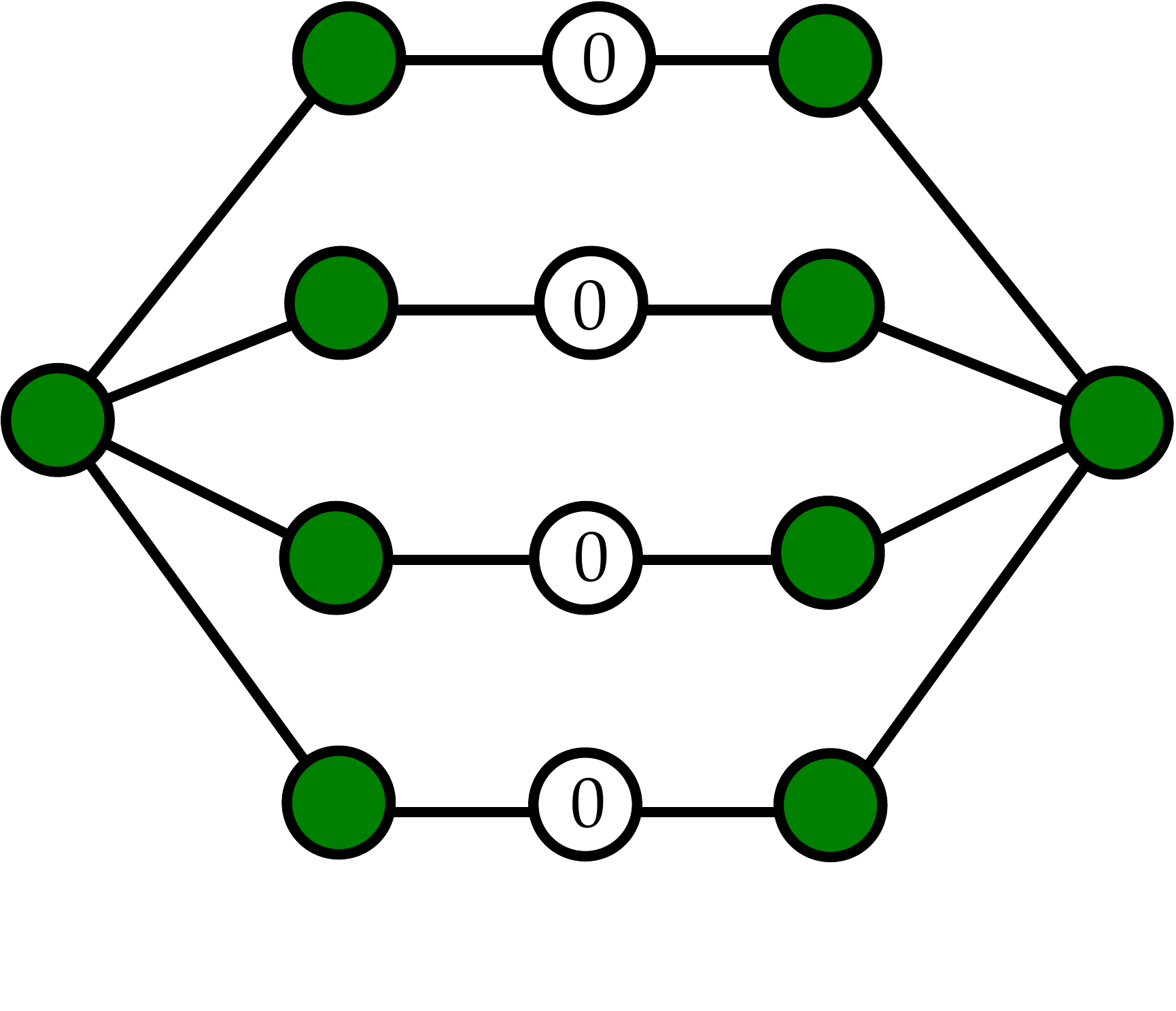}  & \includegraphics[width=4cm]{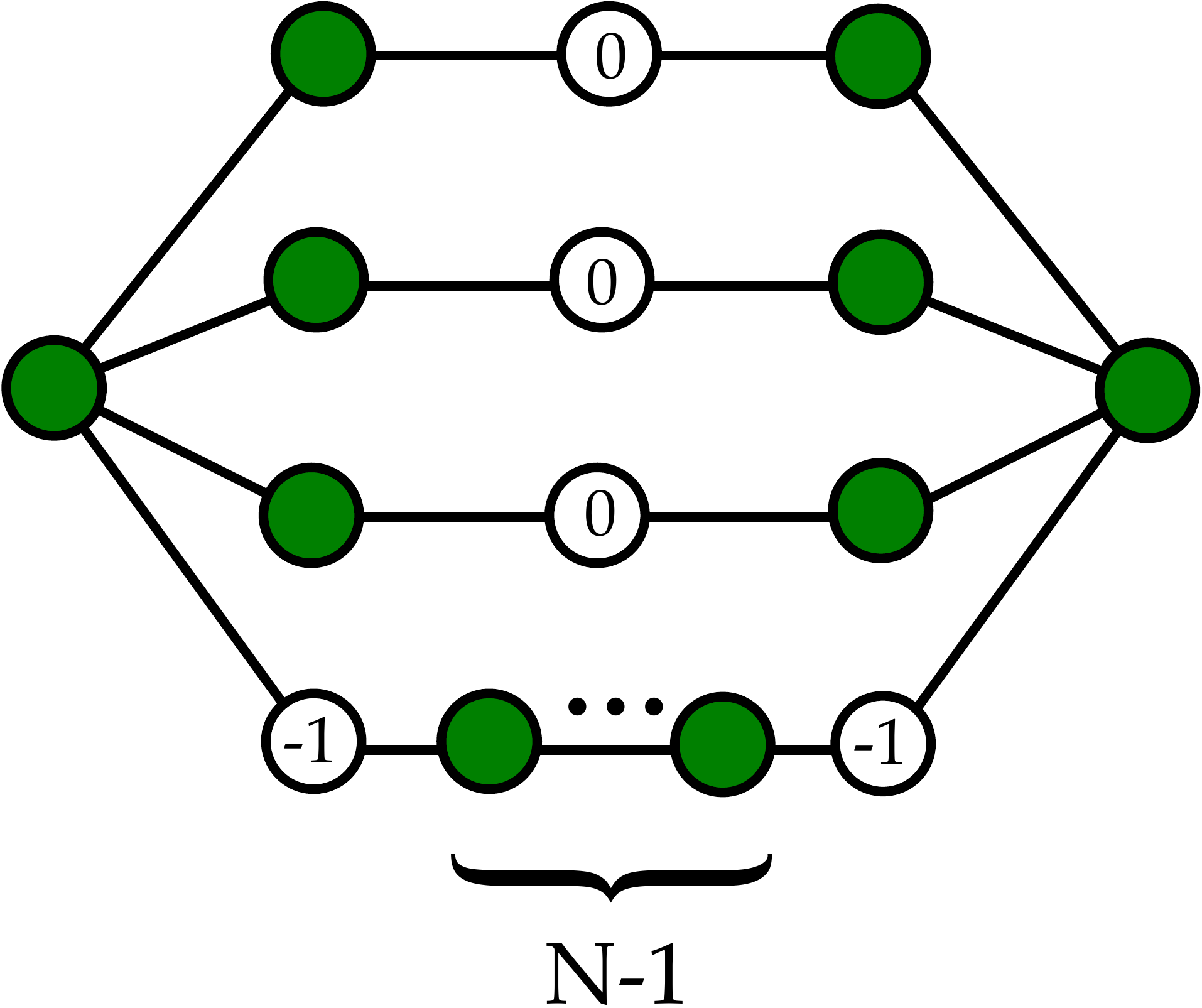}  \cr 
&& \cr \hline 
&& \cr 
$(D_{k}, D_{k})$ & \includegraphics[width=6cm]{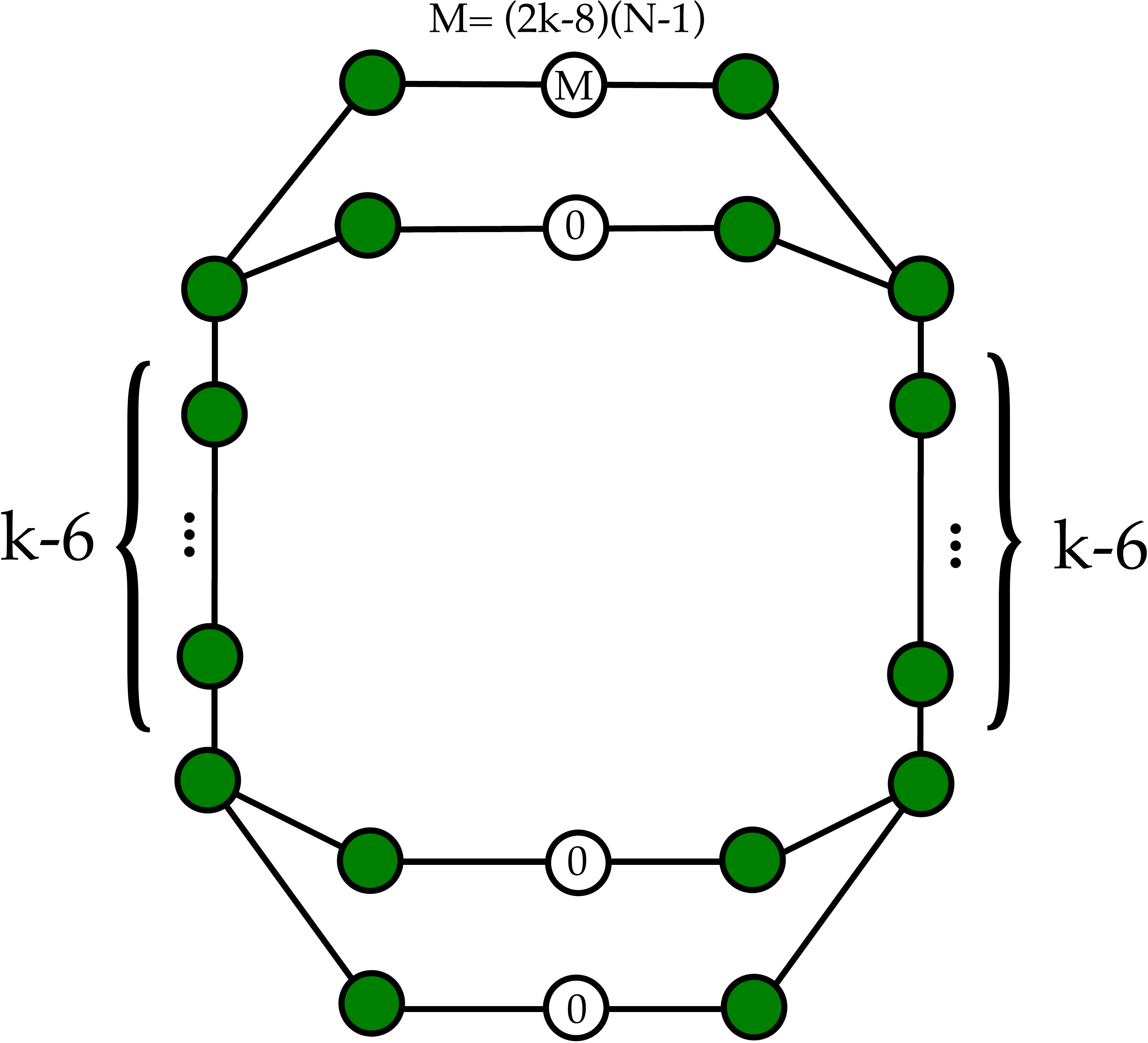} & \includegraphics[width=6cm]{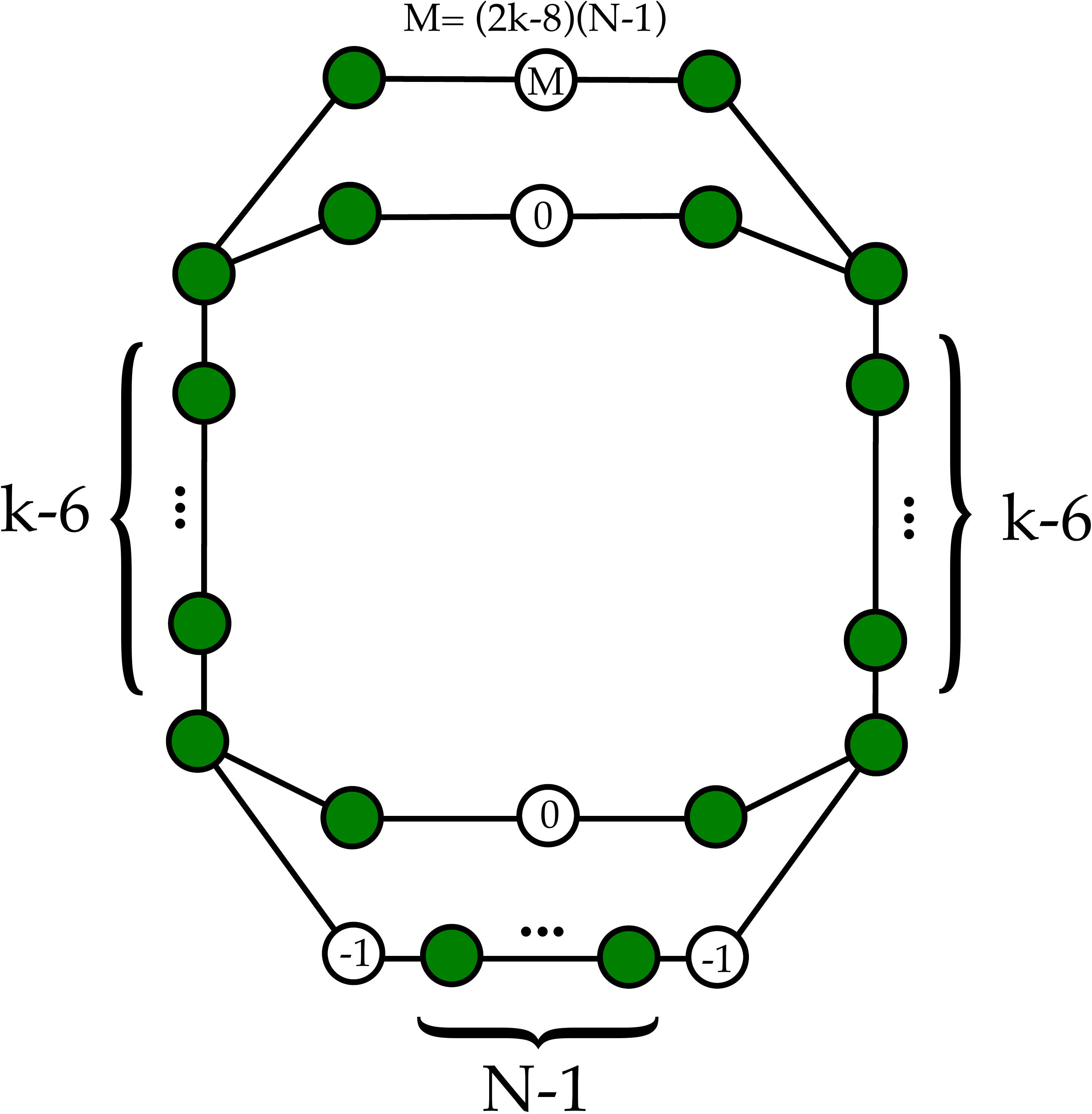} \cr
  \hline 
&& \cr 
$(E_6, E_6)$ &\includegraphics[width=4.5cm]{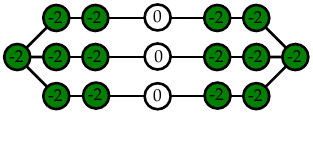} &
 \includegraphics[width=4.5cm]{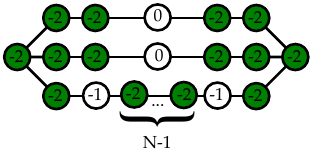} \cr 
 && \cr \hline 

\end{tabular}
\caption{CFDs for non-minimal $N$ $(G,G)$ conformal matter. The left hand picture shows the CFD before decoupling, the right hand one after. }\label{tab:nm-CM-CFD}
\end{table}


In the final CFD, the non-compact surfaces $u_4^{k}$ $(k=1,\dots,N-1)$ give rise to a chain of $N-1$ flavor nodes with $(n,g)=(-2,0)$, which give rise to an extra $SU(N)$ flavor symmetry. The CFD is exactly given by the $(D_4,D_4)$ row in table \ref{tab:nm-CM-CFD}.
The superconformal flavor symmetry $G_F=SO(8)\times SO(8)\times SU(N)$, which is consistent with \cite{Ohmori:2015pia}.

However, in this geometry, there is no consistent assignment of rulings that give rise to a weakly coupled quiver gauge theory
\be
3\bm{F}-SU(2)-SO(8)-SU(2)\dots -SU(2)-SO(8)-SU(2)-3\bm{F}\,.
\ee
The reason is that on the middle surfaces $S_k$ $(k=2,\dots,N-1)$, we need to assign the following linear combination of curves as the ruling
\be
\ba
f(S_k)&=(U^{(k-1)}+u_1^{(k-1)}+u_2^{(k-1)}+u_3^{(k-1)}+u_4^{(k-1)}+u_4^{(k)})\cdot S_k\cr
f(S_k)'&=(U^{(k)}+u_1^{(k)}+u_2^{(k)}+u_3^{(k)}+u_4^{(k-1)}+u_4^{(k)})\cdot S_k\,.
\ea
\ee
Although they are both curves with self-intersection number zero and genus zero, they mutually intersect at two points. Hence they cannot both be the ruling curve of a $\mb{P}^1$ fibration structure. This point was already discussed in section \ref{sec:NMCMGauge}.
Nonetheless, the theory will have a strongly coupled quiver description with $SO(6)\times SO(6)$ classical flavor symmetry, which will be discussed in section~\ref{sec:CM-gauge}.

\subsection{CFDs for Non-Minimal Conformal Matter}
\label{sec:CM-CFD}

In this section, we summarize the CFDs of the non-minimal $(G,G)$ conformal matter with order $N$ in tables \ref{tab:nm-CM-CFD} and \ref{tab:nm-E-CM-CFD}, before and after the decoupling of the extra $SU(N)$ vector multiplet. The figures on the left correspond to KK reduction of the 6d non-minimal $(G,G)$ conformal matter, which is not a 5d SCFT and has no unfactorized descendants (i.e. all descendants would arise from decoupling bifundamentals, and thus factorizing the theory). The figures on the right are the CFDs associated to 5d SCFTs with superconformal flavor symmetry $G_F=G\times G\times SU(N)$. In the $(A_{n-1},A_{n-1})$ CFD after decoupling, there are four $(n,g)=(-1,0)$ nodes that generate CFD transitions. For the other cases, there are two nodes with $(n,g)=(-1,0)$ that generate CFD transitions.

The CFDs of the $(D_4,D_4)$ case have already been derived in the previous section, and we will present the geometric derivation of $(D_n,D_n)$ and $(E_6,E_6)$ cases in appendix~\ref{app:CMN}. The $(A_{n-1},A_{n-1})$ type case follows from the geometry of the single node $SU(n)$ on a $(-2)$ gauge theory that we derive in appendix \ref{sec:BB}. The CFD after decoupling will be derived in \cite{Eckhard:2020jyr} using toric methods. 
 For the cases $(E_7, E_7)$ and $(E_8, E_8)$, we also derive the $(n,g)=(-2,0)$ and $(-1,0)$ vertices from the geometry in appendix~\ref{app:CMN}. 

Given these non-minimal conformal matter CFDs, as well as the quiver structure of the 6d parent theory, it is natural to wonder, whether there is a gluing construction for CFDs. We will return to this in section \ref{sec:gluing}, where we propose building blocks for CFDs and gluing rules. In this context we will re-derive the CFDs for the  $(D_n,D_n)$ and $(E_6,E_6)$  non-minimal conformal matter theories  lower rank theories in section~\ref{sec:CMN}. In this context we also give evidence for the CFDs of $(E_7, E_7)$ and $(E_8, E_8)$.

\begin{table}\centering
\begin{tabular}{|c|c|c|}\hline
$(G,G)$ & CFD before decoupling & CFD after decoupling \cr \hline
&& \cr 
$(E_7, E_7)$ & \includegraphics[height=3cm]{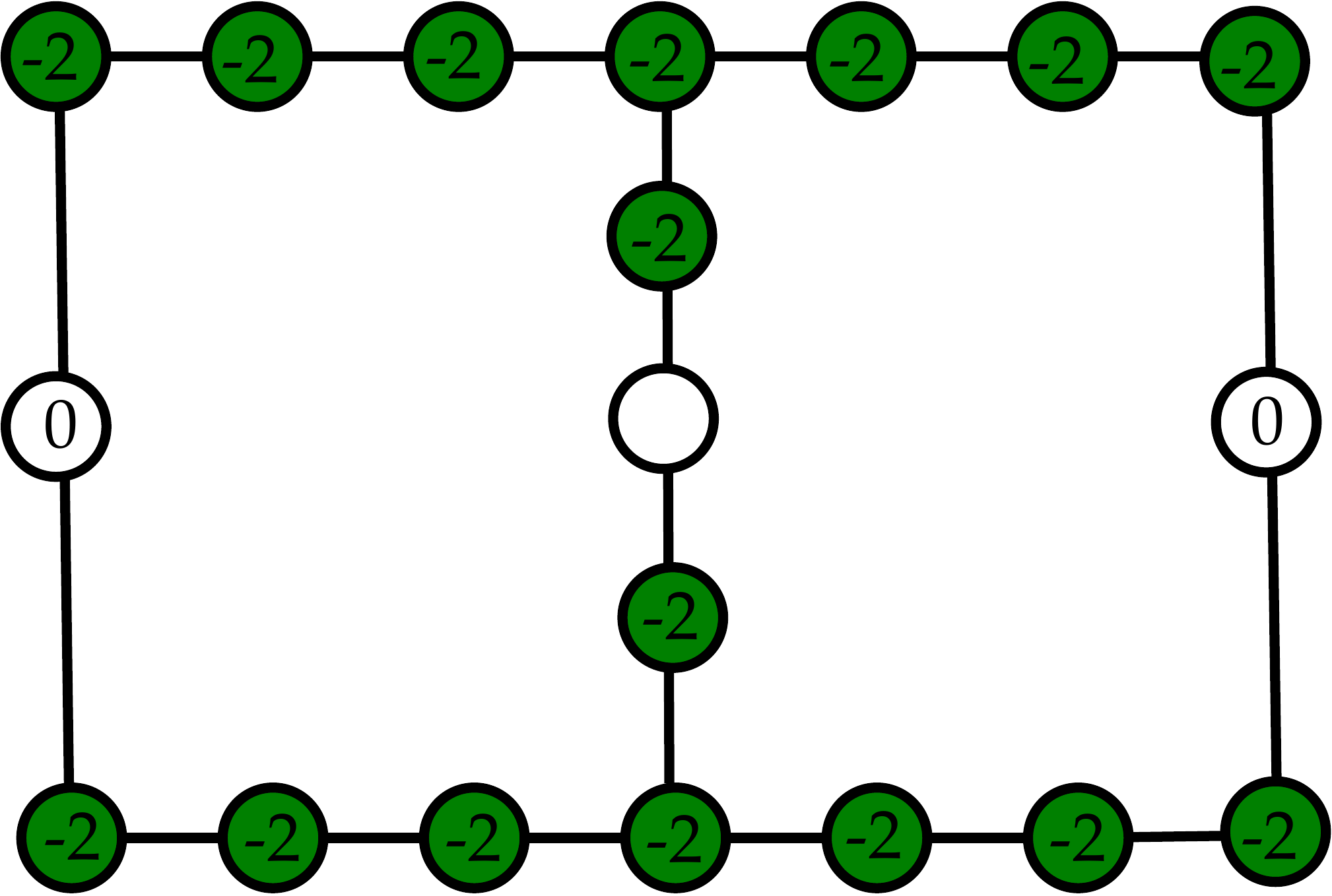} & 
\includegraphics[height=3.5cm]{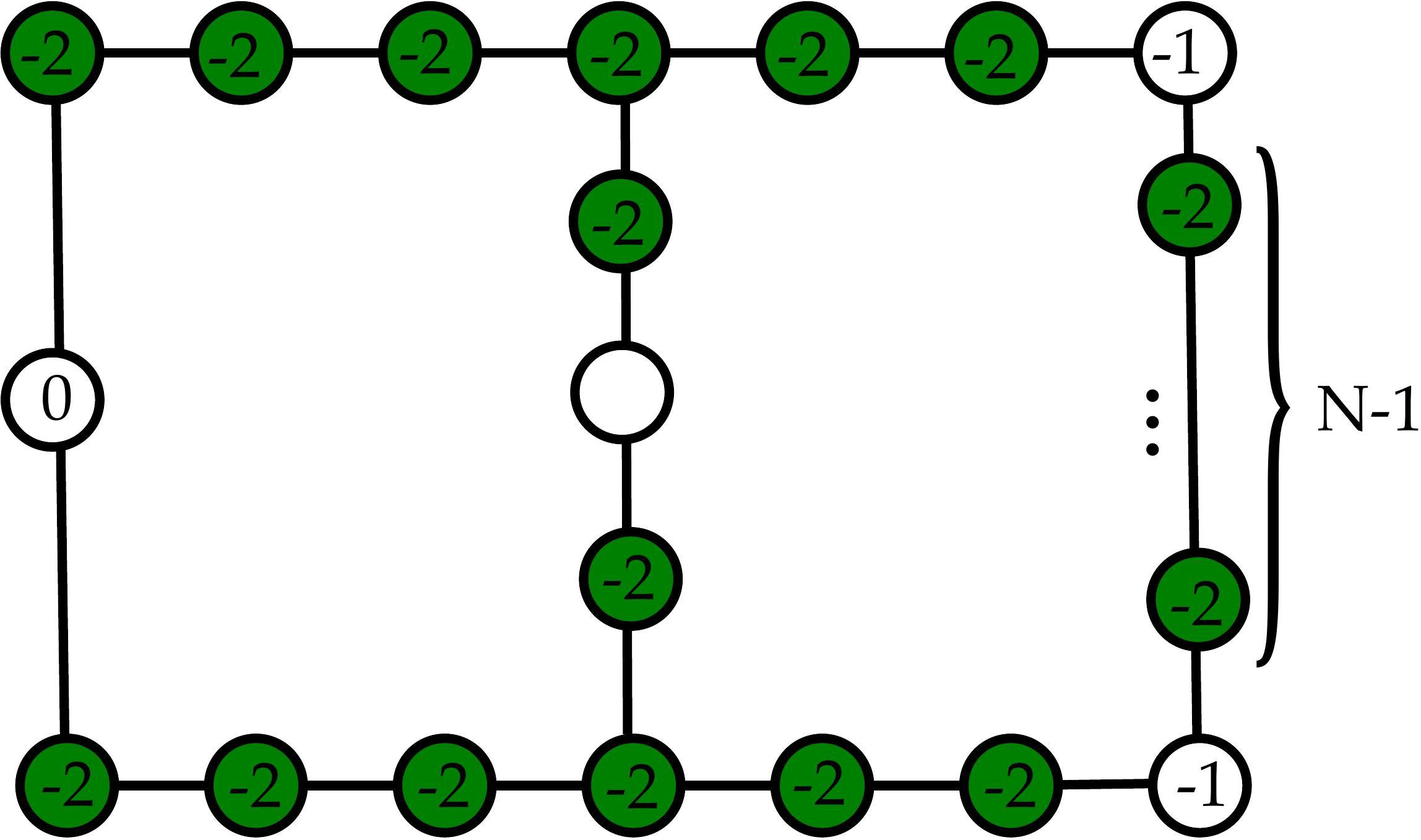}  \cr && \cr \hline
&& \cr 
$(E_8, E_8)$ & \includegraphics[height=3cm]{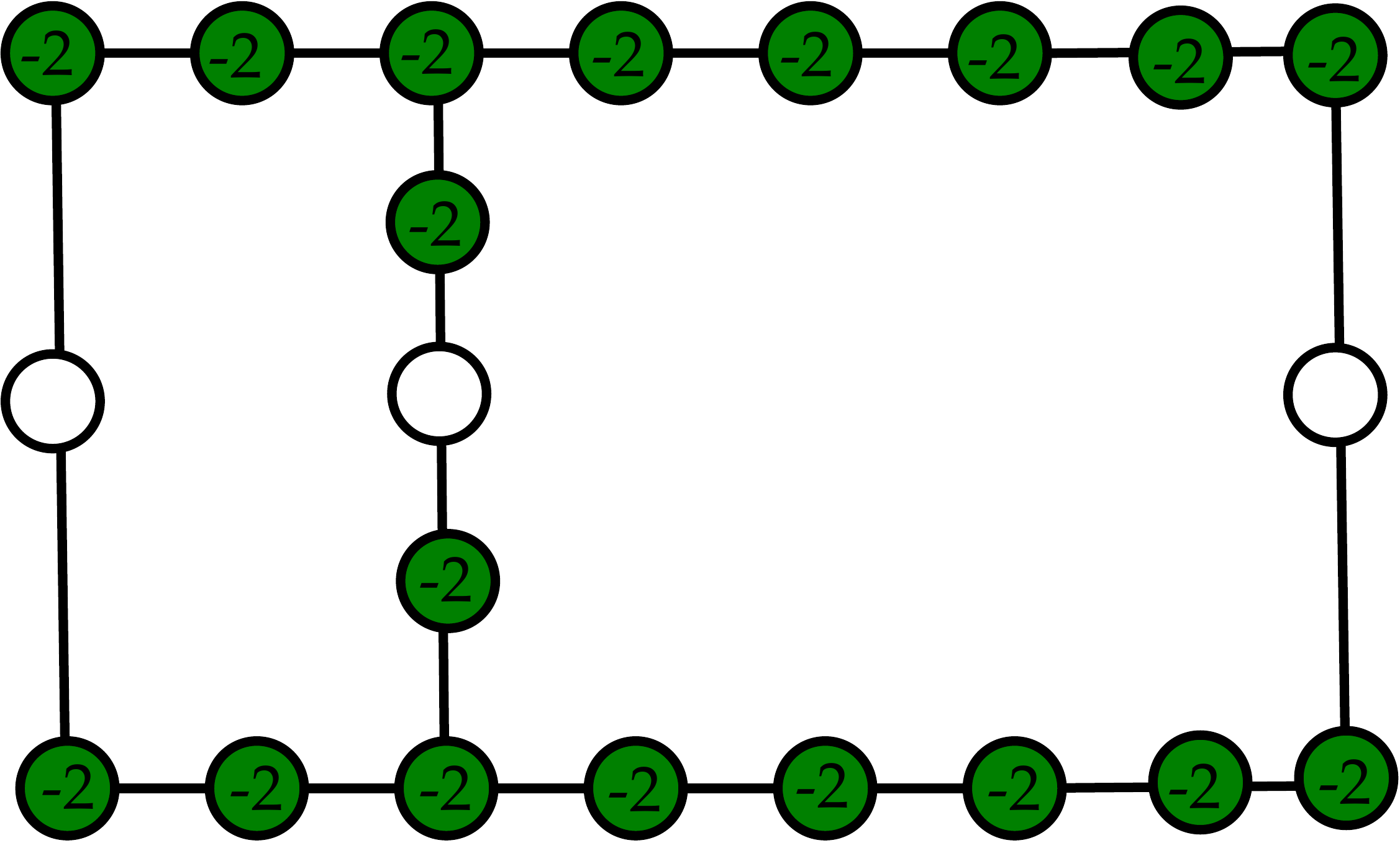} & 
\includegraphics[height=3.5cm]{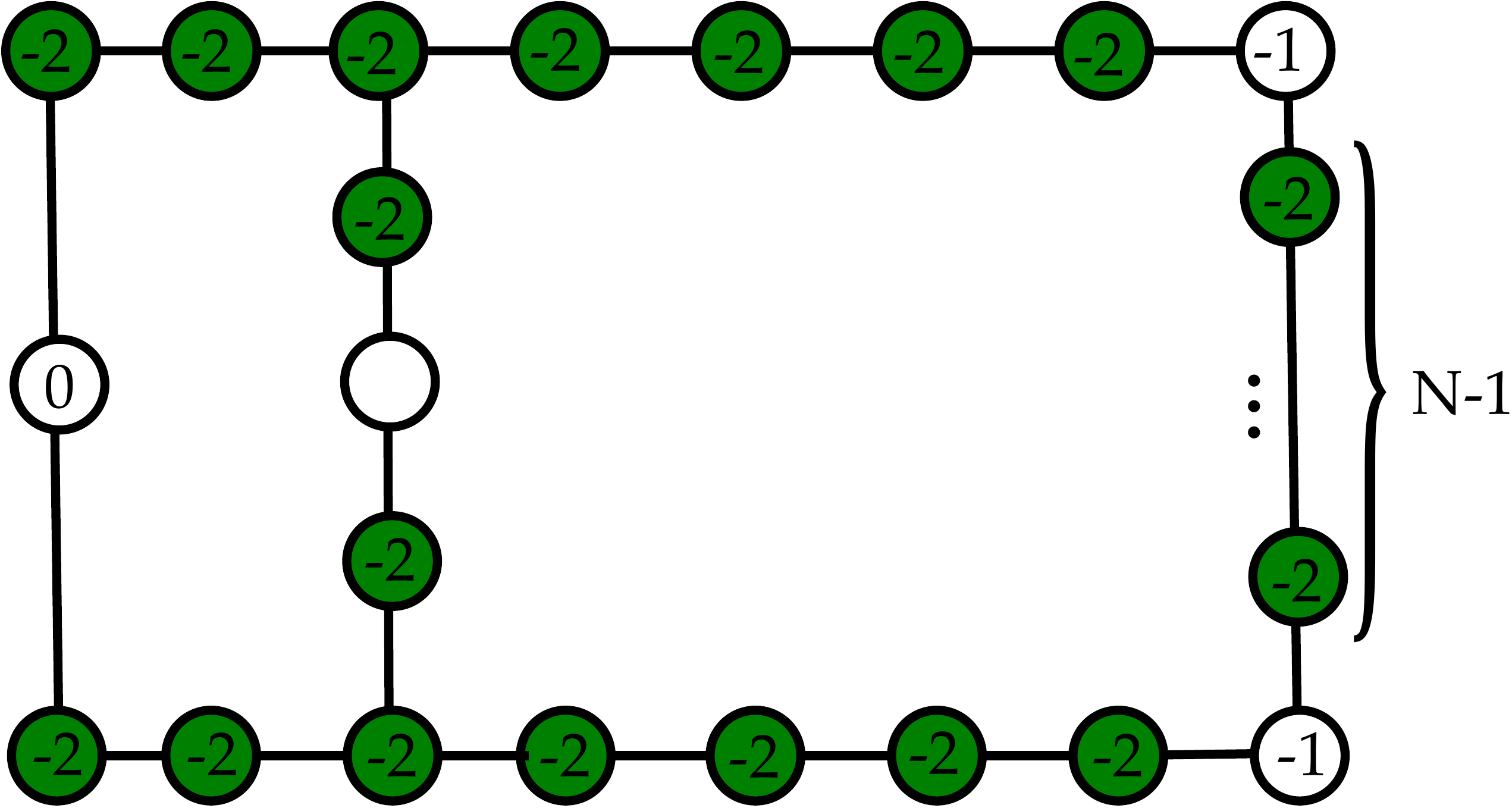}  \cr && \cr \hline
\end{tabular}
\caption{CFDs for non-minimal $N$ $(E_7,E_7)$ and $(E_8, E_8)$ conformal matter. The left hand picture shows the CFD before decoupling, the right hand one after. }\label{tab:nm-E-CM-CFD}
\end{table}



%
%
%
%
%
%
%

\subsection{Low-Energy Descriptions and Dualities}
We will now describe the possible low-energy effective descriptions of the 5d SCFTs and corresponding geometries discussed in this section. We also remind that not all the effective theories will be weakly coupled. For instance, it will sometimes be necessary to introduce strongly coupled matter, e.g. the 5d analog of conformal matter. In fact, it can happen that the non-perturbative part of the flavor symmetry is gauged. For example a subgroup, $H$, of the superconformal flavor symmetry, $G_F^{\rm 5d}$, has to be gauged and, in particular, it contains {some of the $U(1)_T$ symmetries} associated to the gauge vectors, 
\begin{equation}
{ \prod U(1)_T} \subset H \subset G_F^{\rm 5d}\, . 
\end{equation}
Geometrically this corresponds to two surface components $S_1, S_2$  intersecting along $C_{12}$, which is a section for the ruling of $S_1$ and a fiber for the ruling of $S_2$. 

A straightforward set of examples is given by the 5d theories originating from the circle reduction of 6d theories, which are single curve with a gauge groups in the tensor branch. Upon decompactification, or decoupling, we get exactly the 5d analog of the 6d gauge theory in the tensor branch. For instance, the geometry corresponding to the NHC
\begin{equation}
\overset{\mathfrak{su}_3}{3}
\end{equation}
consists of three $\mathbb F_1$ intersecting along $(-1)$ curves. Decompactifying one of the surfaces leads to two $\mathbb F_1$ intersecting along the $(-1)$ curve, and this geometry exactly corresponds to the $SU(3)_0$ theory, as we already seen from the CFD prospective in section \ref{sec:NHCs}. This procedure applies also to the other single $(-n)$-curve theories with $n>1$. 

We now list some of the possible low-energy descriptions of 5d SCFTs coming from decompactification of the geometries corresponding to the 6d non-minimal conformal matter, which are determined by embedding the BG-CFDs into the CFDs. In order to construct these dual IR theories of the same UV SCFT, we will sometimes need to locally dualize gauge nodes of known quiver theory description, and in particular we will use the following duality,
\begin{equation} \label{eq:locduality}
SU(N)_0 - {2N\bm{F}}\quad  \longleftrightarrow \quad  {2\bm{F}}- \underset{N-1}{\underbrace{SU(2)-...-SU(2)-...-SU(2)}}-{2\bm{F}},
\end{equation}
which descend from higher rank $(D_n,D_n)$ conformal matter theories \cite{Apruzzi:2019vpe}.

In addition, we will obtain some description with maximum amount of flavor matter. The descendant 5d SCFTs are obtained from matter mass deformation, which consists of decoupling the flavor hypermultiplets in the IR gauge theory descriptions. Their superconformal flavor symmetries can be straightforwardly read off from the CFD transition, i.e shrinking {$(n,g)=(-1,0)$} vertices.

\begin{figure}
\begin{centering}
\subfloat[]
{
\includegraphics[width=0.8\textwidth]{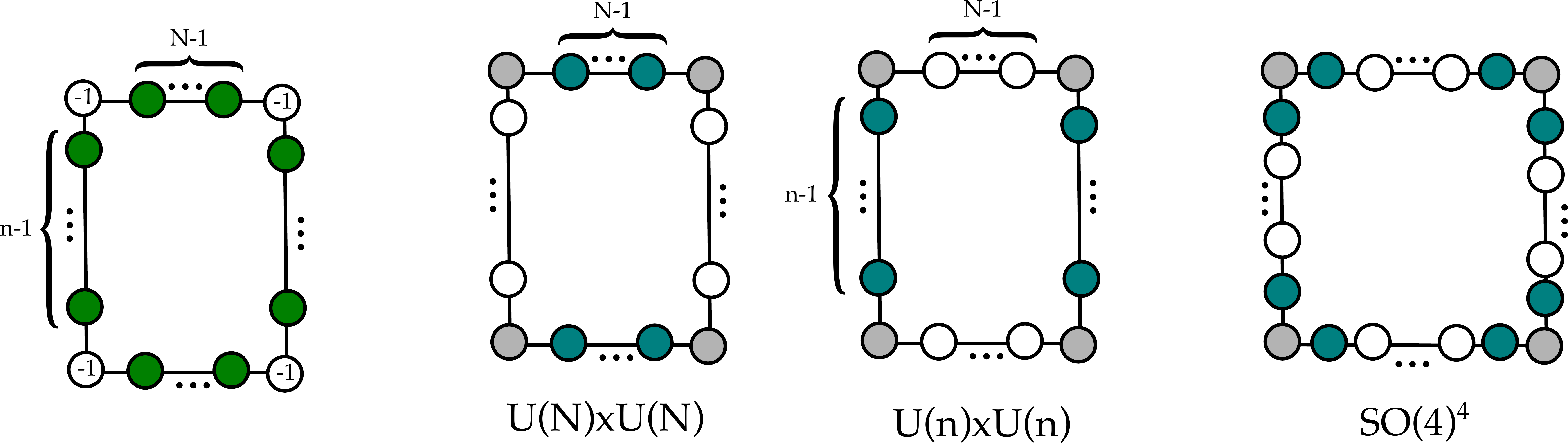}
\label{fig:BGCFDSUn2}
}

\bigskip

\subfloat[]
{
\includegraphics[width=0.9\textwidth]{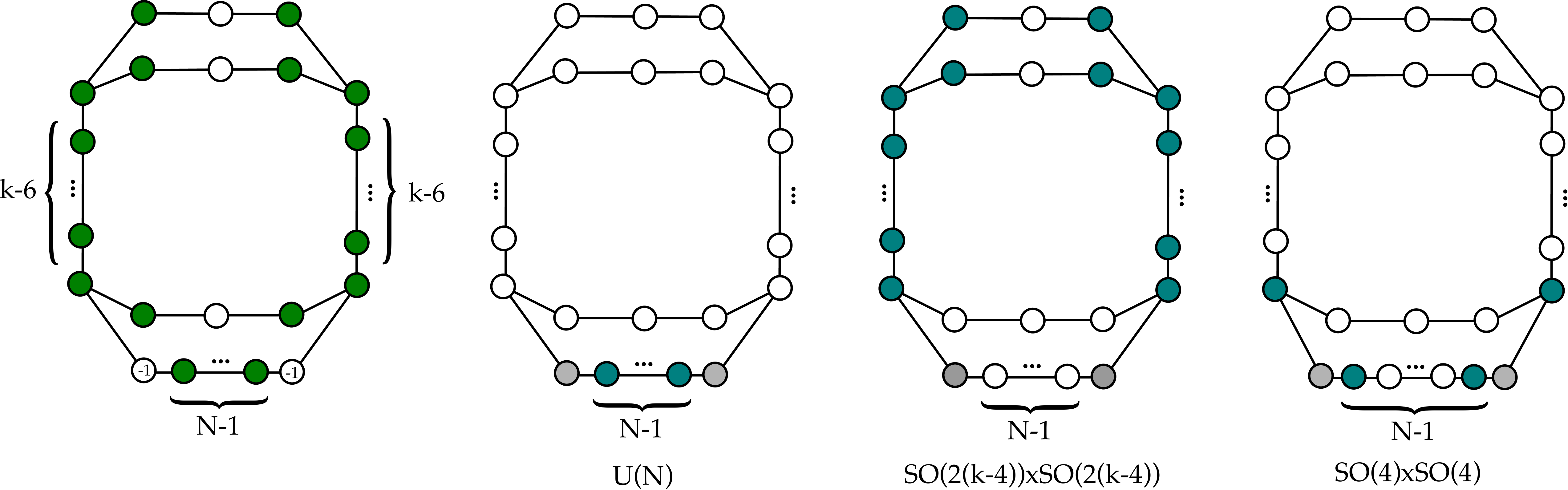}
\label{fig:BGCFDSO2n2}
}

\bigskip

\subfloat[]
{
\includegraphics[width=\textwidth]{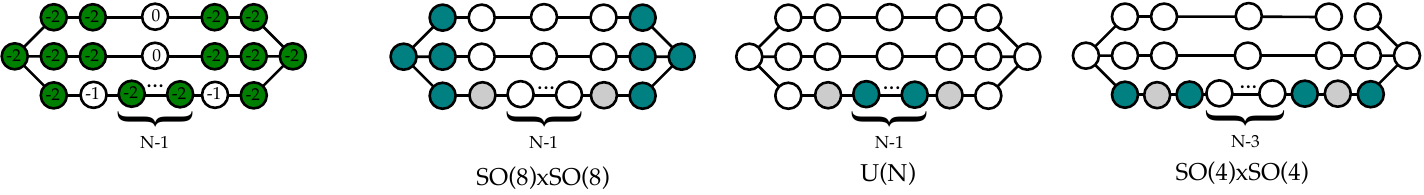}
\label{BG-CFD-E6E6-N-CM}
}

\bigskip

\subfloat[]
{
\includegraphics[width=\textwidth]{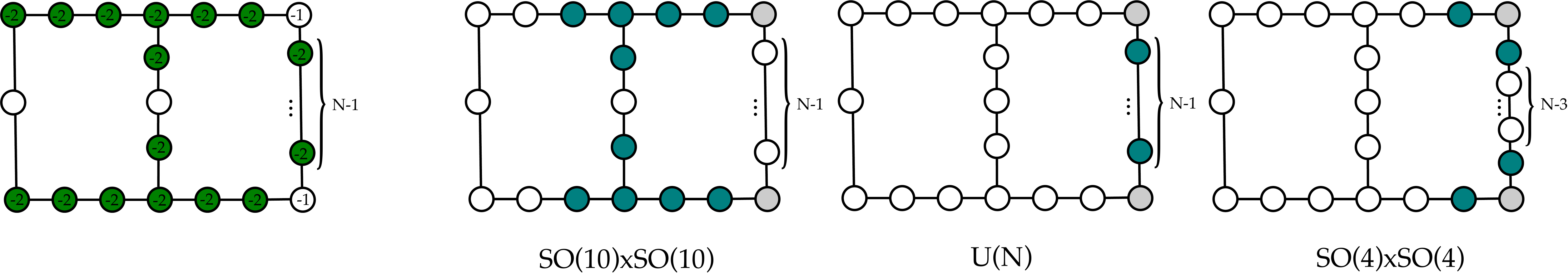}
\label{BG-CFD-E7E7-N-CM}
}

\bigskip

\subfloat[]
{
\includegraphics[width=\textwidth]{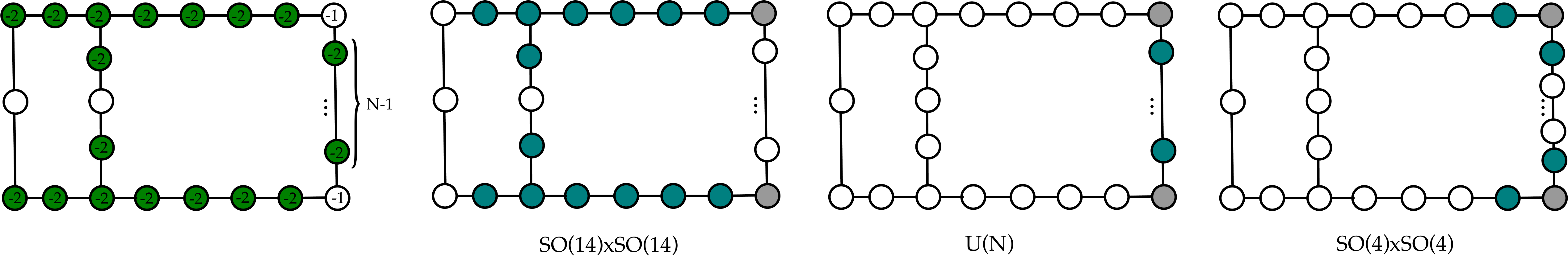}
\label{BGCFDE8}
}

\end{centering}

\caption{CFDs (on the LHS) of tables \ref{tab:nm-CM-CFD} and \ref{tab:nm-E-CM-CFD} and the embeddable BG-CFDs. Below the BG-CFDS we note the classical flavor symmetry.}
\end{figure}
\subsubsection*{$A$-Type non-minimal conformal matter\label{sec:CM-gauge}} 
As already explained a 5d SCFT can be obtained from $(SU(n), SU(n))$ non-minimal ($(N>1$) conformal matter upon decoupling of the extra gauge theory. We can deform the SCFT and study the theory in the IR, which can be a quiver gauge theory. The embedding of the classical flavor symmetries are shown in figure \ref{fig:BGCFDSUn2}. Two weakly coupled descriptions are pretty manifest, and the 5d SCFT in the UV after mass deformation leads in the IR to \cite{Ohmori:2015pia}:
\be \label{eq:Agaugeth}
\ba
U(N)^2: \qquad & {N\bm{F} - \underset{n-1}{\underbrace{SU(N) -\cdots - SU(N)}} - N\bm{F}} \cr 
U(n)^2:\qquad & {n\bm{F}-\underset{N-1}{\underbrace{SU(n) -\cdots -  SU(n)}} - n\bm{F}} \cr 
\ea
\ee
Both of these description have been already anticipated in section \ref{sec:strategy}. The first one is simply obtained from decoupling the $SU(N)$ from the affine circular quiver. The second is the 5d analog of the tensor branch gauge theory in 6d. 

There is a third embedding in figure \ref{fig:BGCFDSUn2} with $SO(4)^4$ classical flavor symmetry, which happens generically for $N,n>3$. This exactly comes from applying the local duality, \eqref{eq:locduality}, at the two tails of the quivers \eqref{eq:Agaugeth}, where we also need to gauge an $SU(N)$ subgroup of the superconformal flavor symmetry of this $SU(2)$ quiver at strong coupling. From the point of view of the $SU(2)$ quiver on the right hand side of \eqref{eq:locduality}, the gauging of $SU(N)$ implies that we are gauging part of the non-perturbative flavor symmetry. The low-energy effective description is, 
\begin{equation}
\includegraphics[scale=0.25]{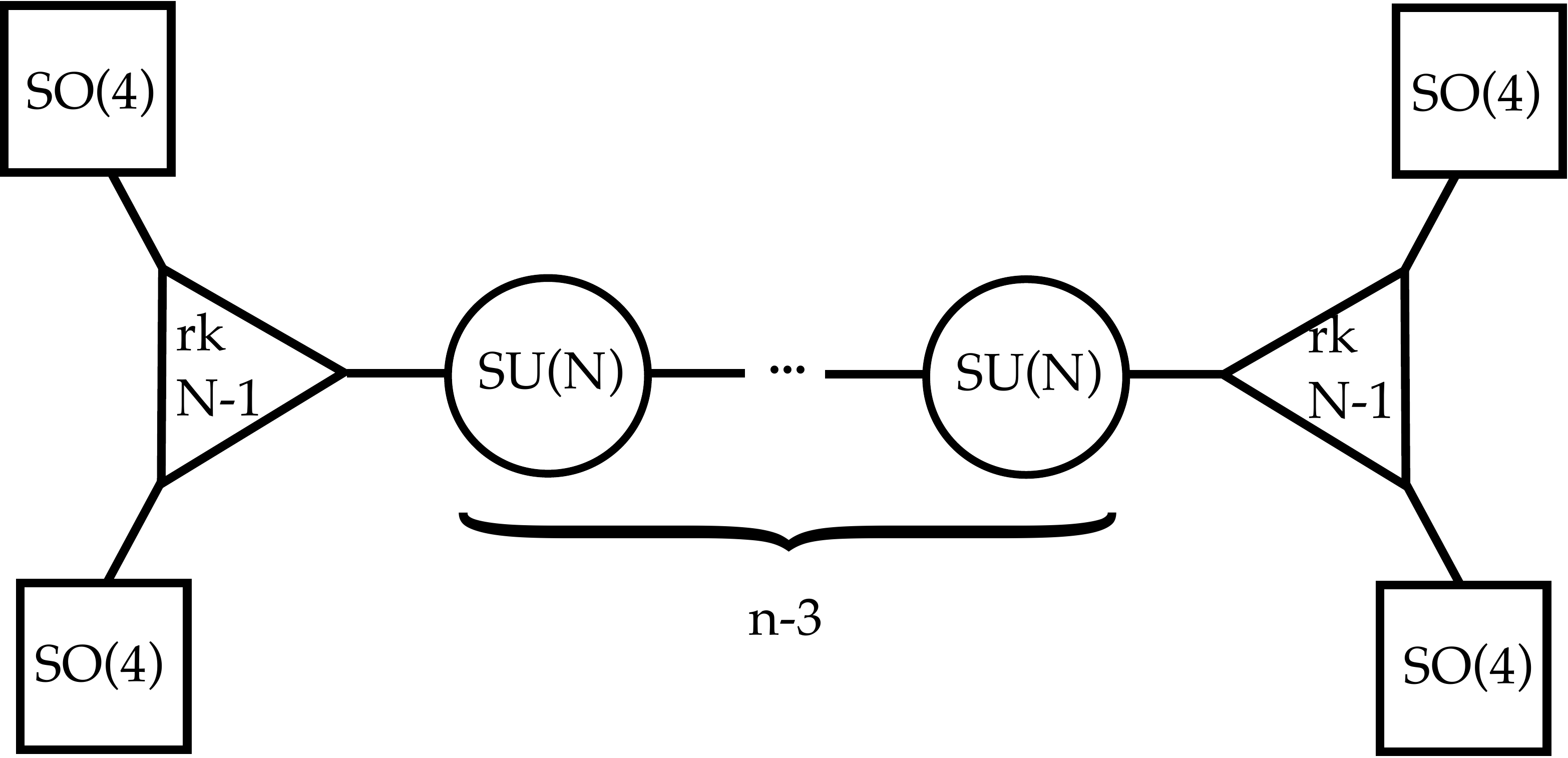}
\end{equation}
where at the two ends we have rank {$(N-1)$} strongly coupled trivalent matter, which only in the Coulomb branch of the neighbor gauge $SU(N)$ is described by
\begin{equation}
\includegraphics[scale=0.25]{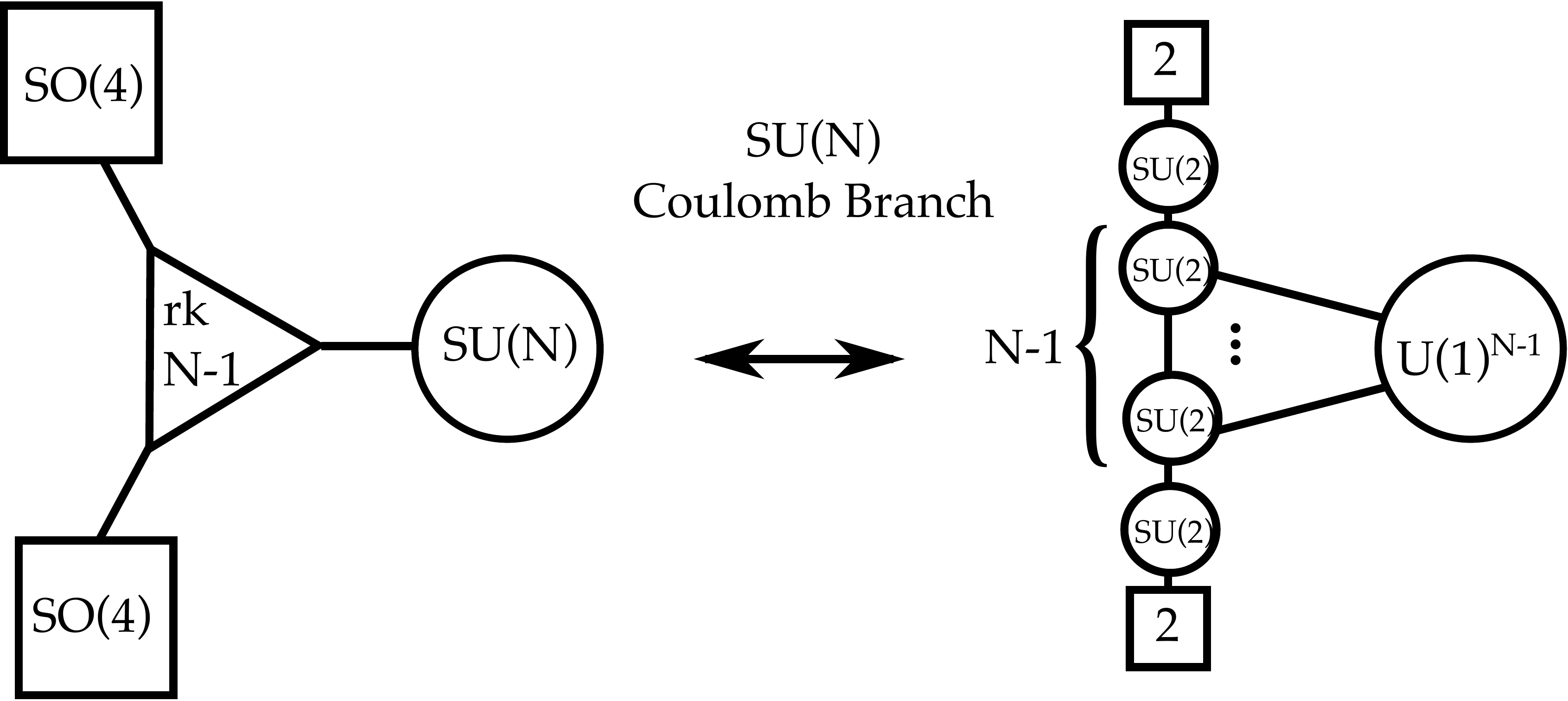}
\end{equation}
The $U(1)^{N-1}$ Coulomb branch scalars can couple to the $SU(2)$ gauge groups kinetic terms and their $J_T(SU(2))$ topological current. They can also couple to the flavor currents corresponding to the baryonic symmetries rotating the fundamental hypers of $SU(2)\times SU(2)$ in the quiver. 
{This is geometrically realized when two surfaces are glued along the section and a fiber of a consistent ruling. 
To provide some more evidence for this, one can construct a local description of the surfaces that realize the $SU(N)_0 + 2N \bm{F}$, e.g. using a toric description \cite{Closset:2018bjz, Eckhard:2020jyr, Xie:2017pfl}, and reinterpret the diagram in terms of an $SU(2)^{N-1}$-quiver. In this case there are $(N-1)$ $U(1)_T$ as well as $U(1)_B$ baryonic symmetry currents. $(N-1)$ independent linear combinations of these are gauged and they correspond to the Coulomb branch $U(1)^{N-1}$ of the neighbor $SU(N)$ in the quiver. The precise linear combination depends on the triangulation, i.e. Coulomb branch phase, of the geometry in question. 

This observation should generalize to the D and E-types we will consider next, by constructing the corresponding rulings. This would be interesting to develop further.

}

\subsubsection*{$D$-Type non-minimal conformal matter}
The 5d SCFTs resulting resulting from non-minimal $(SO(2n), SO(2n))$ conformal matter after decoupling the extra vector multiplet has the two following IR effective descriptions:
\be \label{eq:Dgaugeth}
\ba
U(N): \qquad & 
[SU(N)]-\underset{n-3}{\underbrace{\overset{%
\begin{array}
[c]{c}%
SU(N)\\
|
\end{array}
}{SU(2N)}-SU(2N)-...-SU(2N)-\overset{%
\begin{array}
[c]{c}%
SU(N)\\
|
\end{array}
}{SU(2N)}}}-SU(N),\cr 
SO(2n)^2:\qquad & [SO(2n)] \overset{\rm cm}{-} \underset{N-1}{\underbrace{SO(2n)\overset{\rm cm}{-}\dots \overset{\rm cm}{-} SO(2n)}}\overset{\rm cm}{-} [SO(2n)]  \cr 
\ea
\ee
where the first one has been already discussed in section \ref{sec:strategy} as decoupling of the affine  $SU(N)$ quiver node, and matches the first BG-CFD embedding in figure \ref{fig:BGCFDSO2n2}, whereas the second one corresponds to the 5d copy of the partial tensor branch quiver. The links $\overset{\rm cm}{-}$ are the first descendant of $(SO(2n), SO(2n))$ conformal matter KK-theory \cite{Apruzzi:2019vpe}. We notice that this 5d low-energy effective description has already some strongly coupled sectors. At the interior of the quiver the matter cannot have a direct weakly coupled description. This is due to the fact that the full superconformal flavor is gauged by $SO(2n)\times SO(2n)$. More precisely also the non-perturbative topological symmetry of a putative gauge theory description is also gauged. On the other hand, at the two tails there is still an global $[SO(2n)]$, and in fact the IR theory is also described by the quiver:
\begin{equation}
 [SO(2n-2)] -Sp(n-3) - \underset{N-1}{\underbrace{SO(2n)\overset{\rm cm}{-}\dots \overset{\rm cm}{-} SO(2n)}} -Sp(n-3) - [SO(2n-2)]
\end{equation}
The classical flavor at the two quiver ends matches the second BG-CFD embedding in figure \ref{fig:BGCFDSO2n2}.

The third BG-CFD in figure \ref{fig:BGCFDSO2n2} comes from locally dualizing the first {$SU(2N)$} from the left in the decompactified affine IR description in \eqref{eq:Dgaugeth} for $N>3$. Applying \eqref{eq:locduality}, we get,
\begin{equation}
\includegraphics[scale=0.25]{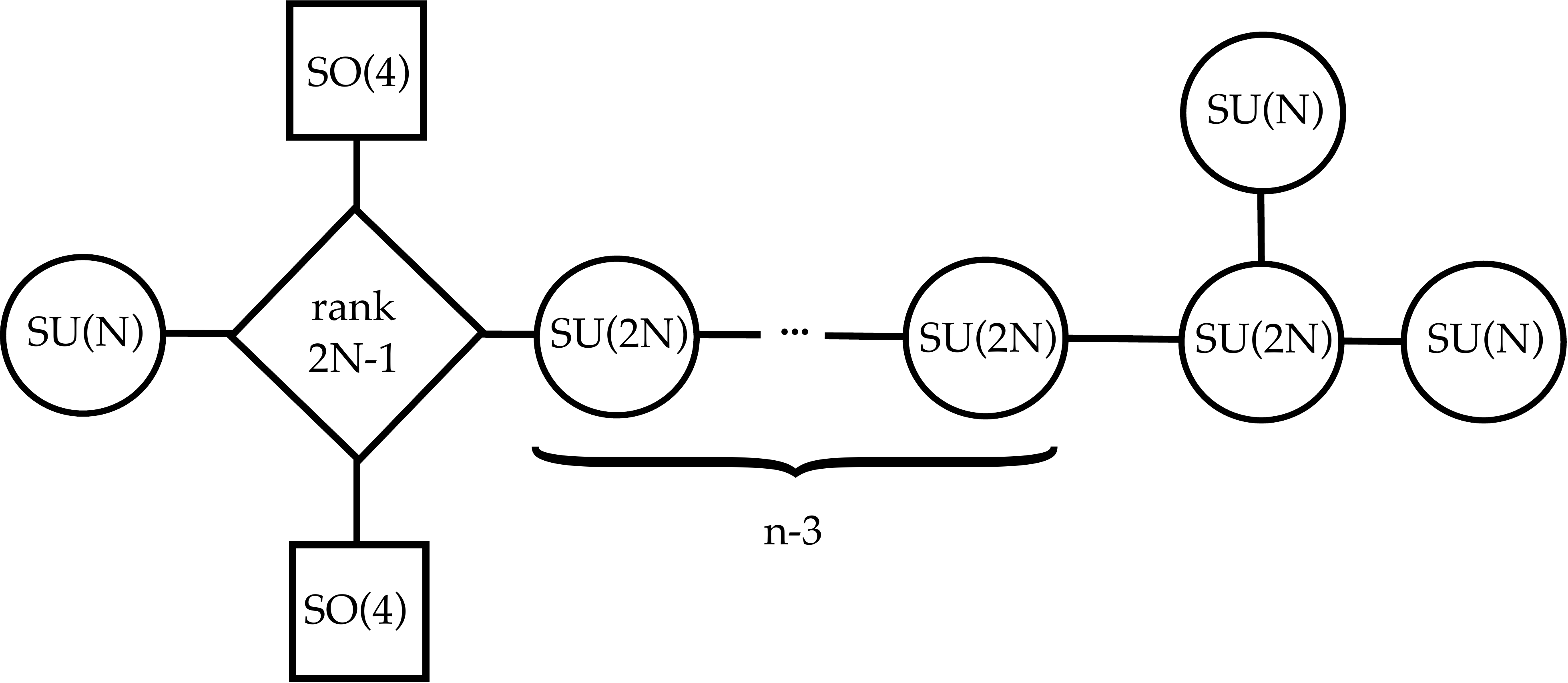}
\end{equation}
In the Coulomb branch of the neighbors {$SU(N)$} and {$SU(2N)$}, the strongly coupled matter theories are given by
\begin{equation}
\includegraphics[scale=0.25]{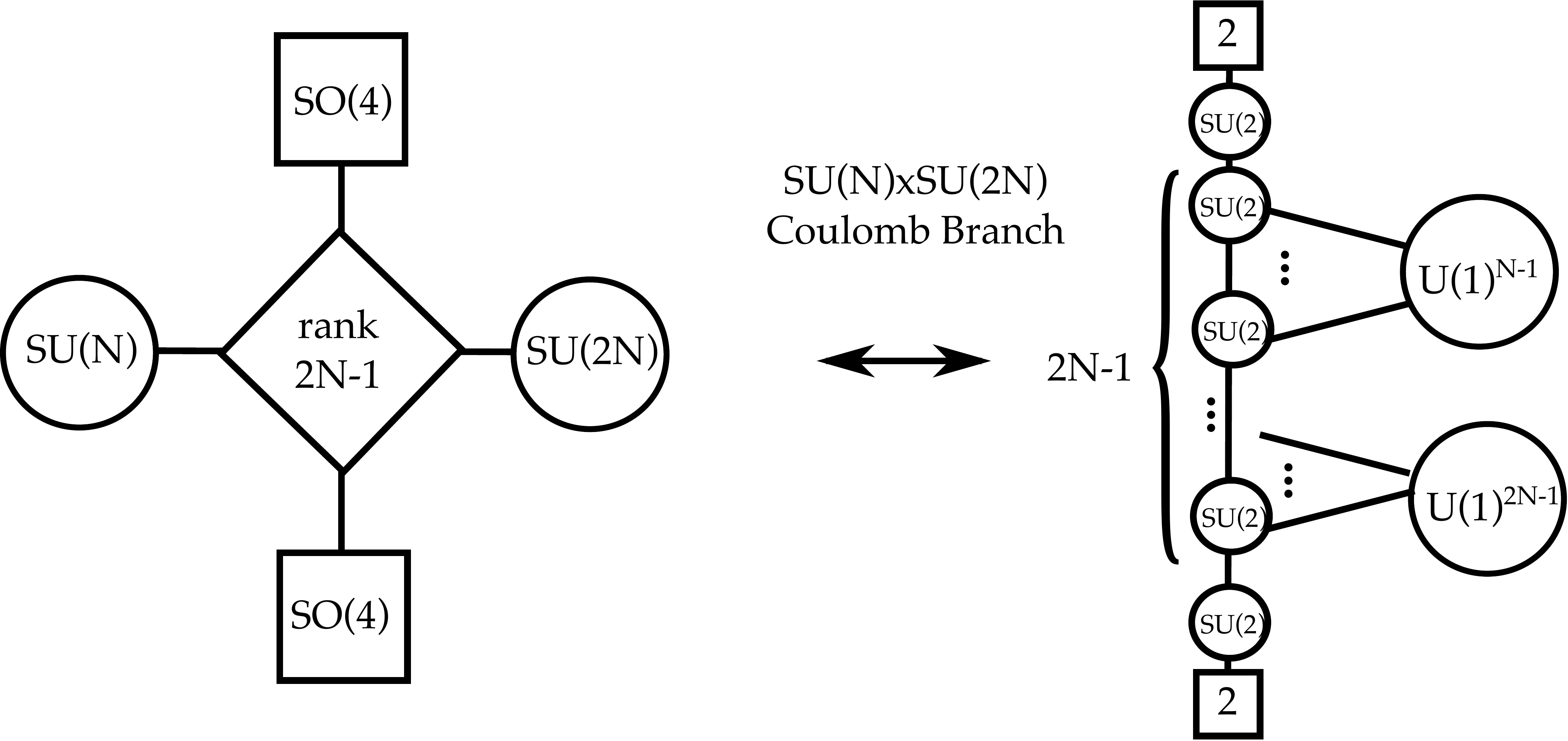}\,,
\end{equation}
where the $U(1)^{N-1}$ couple to $(N-1)$ independent linear combinations of $J_T(SU(2))$ and currents for the baryonic symmetries charging the bifundamental $SU(2) \times SU(2)$ hypermultiplets. Similarly,  $U(1)^{2N-1}$  couple again to $(2N-1)$ independent linear combinations of $J_T(SU(2))$ and baryonic symmetries. At least one $SU(2)$, which does not couple to any of the $U(1)$, separates the two set of couplings for $U(1)^{N-1}$ and $U(1)^{2N-1}$.

For $N=2$ we do actually have an effective Lagrangian description in terms of the following weakly coupled theory,
\begin{equation}
 [SO(2n-2)] -Sp(n-3) - SO(2n) -Sp(n-3) - [SO(2n-2)]
\end{equation}
which matches with the ruling of the geometric resolution. 

\subsubsection*{$E_6$-Type non-minimal conformal matter}
The 5d SCFT from non-minimal $(E_6,E_6)$ conformal matter has the two dual low-energy descriptions:
\be \label{eq:E6gaugeth}
\ba
U(N): \qquad & 
[SU(N)]-SU(2N)-\overset{%
\begin{array}
[c]{c}%
SU(N)\\
|\\
SU(2N)\\
|
\end{array}
}{SU(3N)}-SU(2N)-SU(N),\cr 
E_6^2:\qquad & [E_6] \overset{\rm cm}{-} \underset{N-1}{\underbrace{E_6\overset{\rm cm}{-}\dots \overset{\rm cm}{-} E_6}}\overset{\rm cm}{-} [E_6]  \cr 
\ea
\ee
The first one is again given by the decoupling of the affine gauge node vector multiplet, and matches the second BG-CFD embedding in figure \ref{BG-CFD-E6E6-N-CM}. The second description is the the 5d analog of the 6d tensor branch after decompactification, where the links are given by the first mass deformation of the KK-theory coming from straight circle compactification of $N=1$ $(E_6,E_6)$ conformal matter \cite{Apruzzi:2019opn}. In the interior the link do not have a direct weakly coupled description in terms of gauge theory, since the full superconformal flavor symmetry is gauged. In this case, also for the tails of the quiver we cannot have a complete description in terms of a weakly coupled Lagrangian theory, because gauging the $E_6$ also implies the gauging of a non-perturbative symmetry in the putative weakly coupled description. 
On the other hand, at the two ends of the quiver there might exist a description of this strongly coupled sector, where some gauge theory with flavor matter can be extracted, but is still coupled to a residual strongly coupled part. Applying the strategy of \cite{Apruzzi:2019enx}, we propose a quiver which is compatible with the embedding of the classical flavor symmetry, see figure \ref{BG-CFD-E6E6-N-CM}. That is 
\begin{equation}
{4\bm{F}}-Sp(n_1) \overset{\rm *}{-} E_6 \overset{\rm cm}{-} \ldots \overset{\rm cm}{-} E_6  \overset{\rm *}{-} Sp(n_2)-{4\bm{F}}, \qquad n_1,n_2=1,2,3,4
\end{equation}
for some strongly coupled matter $\overset{\rm *}{-}$ transforming in $Sp(n_1) \times E_6$ or $Sp(n_2) \times E_6$. As we can see this gives the first embedding in  \ref{BG-CFD-E6E6-N-CM}.

Finally, for $N>3$ the last BG-CFD embedding in figure \ref{BG-CFD-E6E6-N-CM} comes again by locally dualizing the left most gauge quiver node in the first case of \eqref{eq:E6gaugeth}. The result is given by 
\begin{equation}
\includegraphics[scale=0.25]{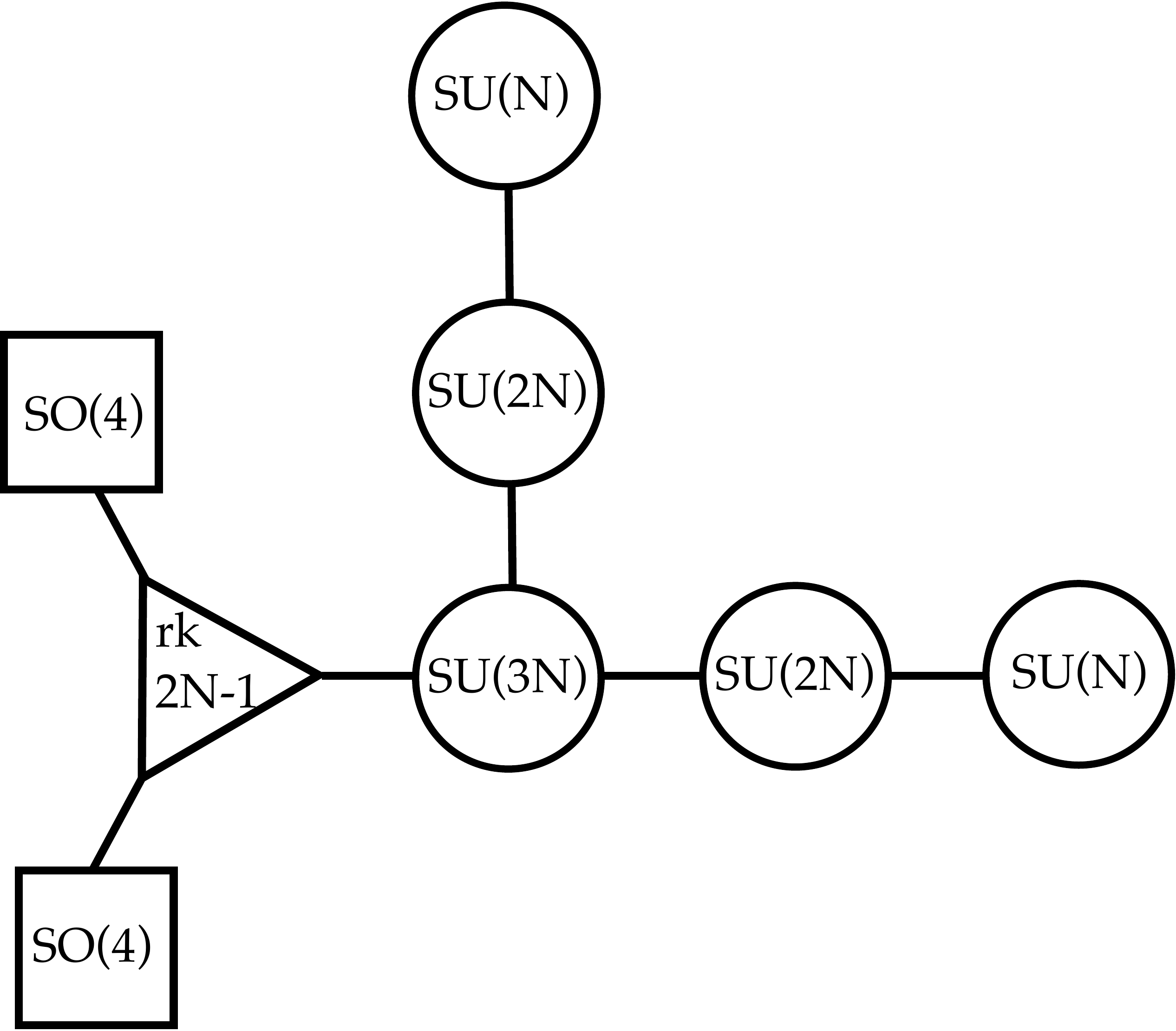}
\end{equation}
where the strongly coupled trivalent node resolves in the coulomb branch of the neighbor $SU(3N)$ gauge theory as
\begin{equation} \label{eq:quiv6}
\includegraphics[scale=0.25]{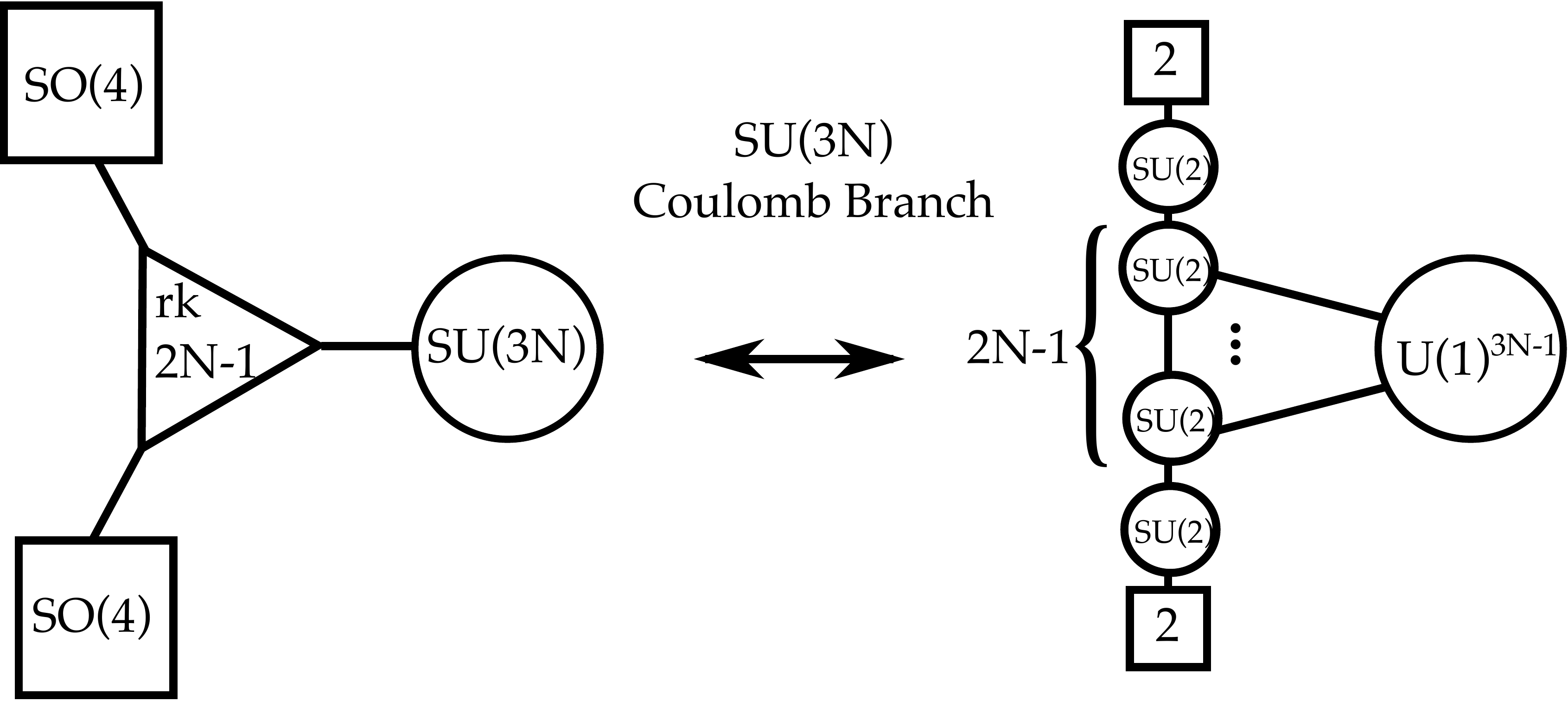}
\end{equation}
The $U(1)^{3N-1}$ couple to  $(3N-1)$ independent linear combinations of $J_T(SU(2))$ and baryonic symmetry currents of the $SU(2)$ quiver.

\subsubsection*{$E_7$-Type non-minimal conformal matter}
Non-minimal $(E_7,E_7)$ conformal matter on a circle and after decoupling of the extra gauge theory and mass deformation has the following dual low-energy descriptions:
\be \label{eq:E7gaugeth}
\ba
U(N):\  &{{[SU(N)]-SU(2N)-SU(3N)-SU(4N)-SU(5N)-\overset{%
\begin{array}
[c]{c}%
SU(3N)\\
|
\end{array}
}{SU(6N)}-SU(4N)-SU(2N)}}\cr 
E_7^2: \  & [E_7] \overset{\rm cm}{-} \underset{N-1}{\underbrace{E_7\overset{\rm cm}{-}\dots \overset{\rm cm}{-} E_7}}\overset{\rm cm}{-} [E_7]  \cr 
\ea
\ee
They correspond to the decoupling of the affine gauge node of the affine quiver in section \ref{sec:NMCMGauge} and to the analog of the 6d tensor branch respectively. The first one is compatible with the second classical flavor symmetry embedding in figure \ref{BG-CFD-E7E7-N-CM}. In addition, we can observe that this last one does not have a complete weakly coupled description, because of the gauging of non-perturbative symmetries of a putative gauge theory describing the conformal matter link.  The links are given by the first descendant of the KK-theory coming from circle reduction of $N=1$ $(E_8, E_8)$ 6d conformal matter.

In the spirit of \cite{Apruzzi:2019enx}, we propose a description of the 5d conformal matter at the two tails, which has a weakly coupled part compatible with the classical flavor symmetry embedding into the CFD in figure \ref{BG-CFD-E7E7-N-CM}. That is given by a gauge theory with some matter hypermultiplets, which is also coupled to a residual strongly coupled theory, $\overset{\rm *}{-}$ transforming in $Sp(n_1) \times E_7$ or $Sp(n_2) \times E_7$. That is
\begin{equation}
{5\bm{F}}-Sp(n_1) \overset{\rm *}{-} E_7 \overset{\rm cm}{-} \ldots \overset{\rm cm}{-} E_7  \overset{ *}{-} Sp(n_2)-{5\bm{F}}, \qquad n_1,n_2=1,\ldots, 9\, .
\end{equation}
As we can see this gives the first embedding in  the CFD in figure \ref{BG-CFD-E7E7-N-CM}.

For $N>3$ the $SO(4)^2$ BG-CFD embedding in  figure \ref{BG-CFD-E7E7-N-CM} is derived by locally dualizing the left most gauge quiver node in the first case of \eqref{eq:E7gaugeth}. The result is given by 
\begin{equation}
\includegraphics[scale=0.25]{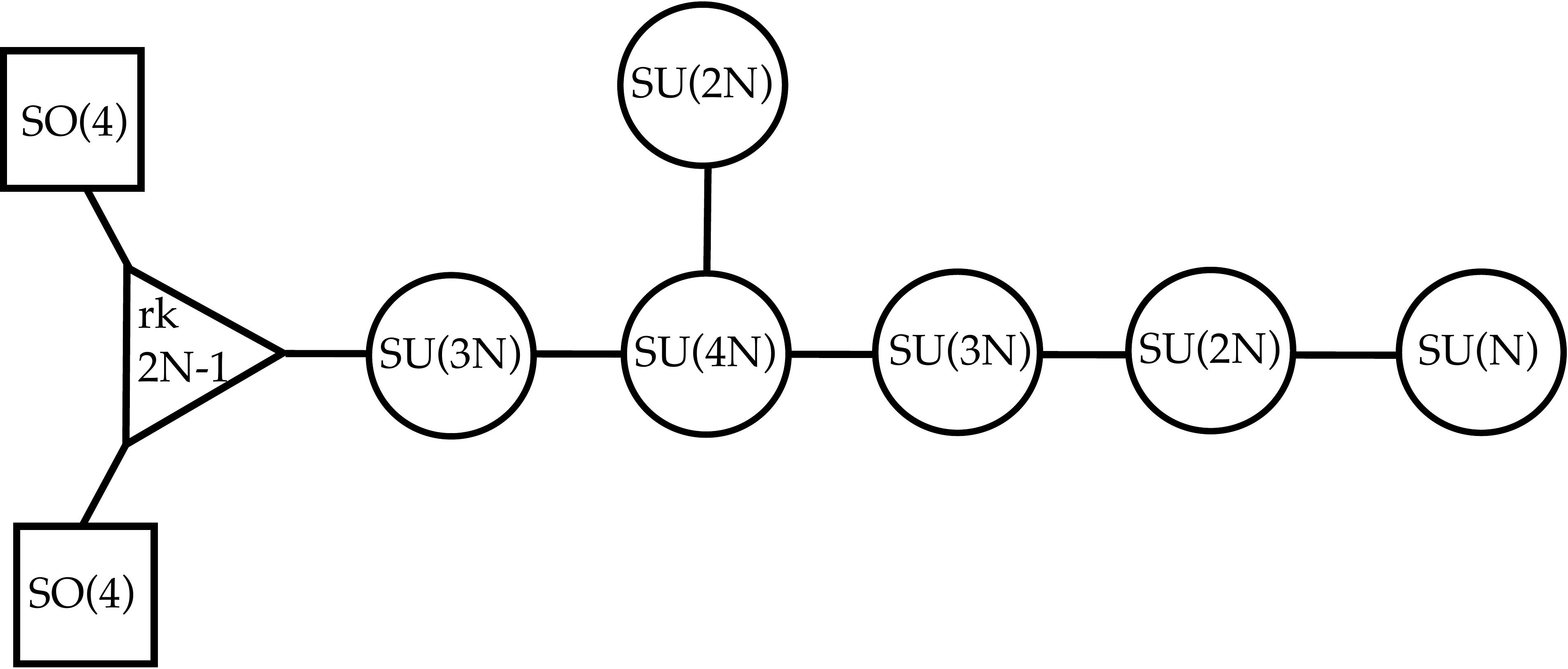}
\end{equation}
where the strongly coupled theory resolves in the coulomb branch of the neighbor $SU(3N)$ gauge theory like in \eqref{eq:quiv6}.

\subsubsection*{$E_8$-Type non-minimal conformal matter}
The 5d SCFT descending from non-minimal $(E_8,E_8)$  conformal matter can be described in the IR by the following dual theories:
\be \label{eq:E8gaugeth}
\ba
U(N): \  & {
[SU(N)]-SU(2N)-SU(3N)-SU(4N)-SU(5N)-\overset{%
\begin{array}
[c]{c}%
SU(3N)\\
|
\end{array}
}{SU(6N)}-SU(4N)-SU(2N)}  \cr 
E_8^2:\  & [E_8] \overset{\rm cm}{-} \underset{N-1}{\underbrace{E_8\overset{\rm cm}{-}\dots \overset{\rm cm}{-} E_8}}\overset{\rm cm}{-} [E_8]  \cr 
\ea
\ee
The first theory comes from decoupling the gauge vector of the affine node of the affine quiver, and the classical flavor symmetry $U(N)$ shows that is corresponds to the first BG-CFD embedding in \ref{BGCFDE8}. The second one is the 5d analog of the partial tensor branch quiver where we have strongly coupled conformal matter transforming under $E_8\times E_8$. This matter is the first descendant of the KK-theory coming from circle reduction of $N=1$ $(E_8, E_8)$ 6d conformal matter.

Similarly to  \cite{Apruzzi:2019enx}, since only a single $E_8$ has been gauged, we propose a partial weakly coupled description which is compatible with the CFD in figure \ref{BGCFDE8}. That is 
\begin{equation}
{6\bm{F}}-Sp(n_1) \overset{\rm *}{-} E_8  \overset{\rm cm}{-}  \ldots  \overset{\rm cm}{-}  E_8  \overset{\rm *}{-} Sp(n_1)-{6\bm{F}}, \qquad n_1, n_2=1,\ldots, 20.
\end{equation}
where $\overset{\rm *}{-}$ is a strongly coupled matter link transforming in $Sp(n_1) \times E_8$ or $Sp(n_2) \times E_8$. We can notice that this corresponds to the first embedding in figure \ref{BGCFDE8}.

At last, the $SO(4)^2$ BG-CFD embedding in figure \ref{BGCFDE8} comes from applying \eqref{eq:locduality} to the left most gauge quiver node in the first quiver of \eqref{eq:E8gaugeth}. The result is given by 
\begin{equation}
\includegraphics[scale=0.25]{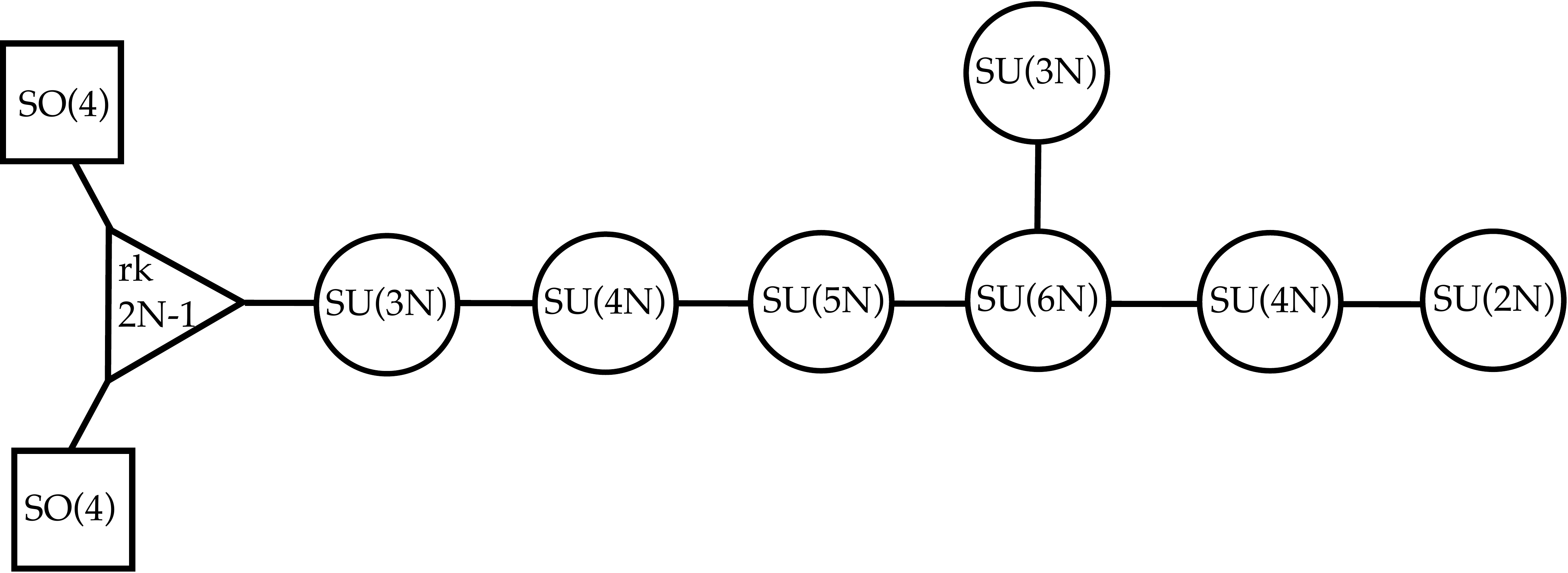}
\end{equation}
where the strongly coupled theory in the coulomb branch of the neighbor $SU(3N)$ gauge theory is described by \eqref{eq:quiv6}.

\section{Gluing CFDs from Building Blocks \label{sec:gluing}}

\subsection{Building Blocks and Gluing}

Any 6d SCFT, in particular NVH theories, in its partial tensor branch can be seen as a generalized quiver \cite{Heckman:2015bfa}, where the nodes are given by
\begin{equation}
\mathcal T^{\text{6d}} (G,G^{\text{6d}}_F)\equiv \overset{G}{n}  \, [G^{\text{6d}}_{F}] \,,
\end{equation}
where $n=\Sigma^2\leq -1$ is the self-intersection number of a compact rational curve $\Sigma$. Over $\Sigma$, the elliptic fiber can be singular, which is associated to the gauge group $G$ in 6d. There can be matter hypermultiplet transforming under the flavor symmetry $G^{\text{6d}}_{F}$. The matter can be either given by standard (half) hypermultiplet, or by VH 6d SCFTs with
\begin{equation}
\mathcal H^{\text{6d}}_{ij}(G_i,G_j) \equiv [G^{\text{6d}}_{F_i}] \overset{\ast}{-} [G^{\text{6d}}_{F_j}] \,,
\end{equation}
where $\overset{\ast}{-}$ the notation means that the link is non-conventional matter and it has a $G^{\text{6d}}_{F_i} \times G^{\text{6d}}_{F_j}$ manifest flavor symmetry. An important class of examples of this type is the minimal $(G^{\text{6d}}_{F_i},G^{\text{6d}}_{F_j})$ conformal matter theory. A link $\mathcal H^{\text{6d}}_{ij}(G^{\text{6d}}_{F_i},G^{\text{6d}}_{F_j})$ is connected to a node $\mathcal T^{\text{6d}} (G,G^{\text{6d}}_F)$ by gauging the flavor symmetry $G^{\text{6d}}_{F_i}$, which should be exactly identical to $G$ of the $\mathcal T^{\text{6d}} (G,G^{\text{6d}}_F)$. Repeating this procedure leads to the generalized quivers of \cite{Heckman:2015bfa}. In this way we can construct general 6d tensor branches, whose origin corresponds to a 6d SCFT.

We implement a similar strategy in 5d based on the M-theory geometry.  The building blocks are defined by the resolution geometries associated to the tensor branch building blocks in 6d:
\begin{enumerate}

\item $\mathcal{S}^{\text{5d}} (G,G^{\text{6d}}_{F})$, which is constructed from $\mathcal{T}^{\text{6d}} (G,G^{\text{6d}}_F)$ by $S^1$ reduction and decoupling. If the self-intersection number of $\Sigma$ in $\mathcal{T}^{\text{6d}} (G,G^{\text{6d}}_F)$ is $n=-1$, then $\mathcal{S}^{\text{5d}} (G,G^{\text{6d}}_{F})$ is simply the KK reduction of $\mathcal{T}^{\text{6d}} (G,G^{\text{6d}}_F)$, which in fact corresponds to matter. If $n<-1$, then we need to decompactify one compact surface in the M-theory geometry, in order to decouple the extra $SU(2)$ gauge theory.

\item 
$\mathcal{H}^{\text{5d}}_{ij}(G^{\text{6d}}_{F_i},G^{\text{6d}}_{F_j})$, which is similarly constructed from $\mathcal H^{\text{6d}}_{ij}(G^{\text{6d}}_{F_i},G^{\text{6d}}_{F_j})$. 
When  $\mathcal{H}^{\text{5d}}_{ij}(G^{\text{6d}}_{F_i},G^{\text{6d}}_{F_j})$ is glued to a building block $\mathcal{S}^{\text{5d}} (G,G^{\text{6d}}_{F})$, where a decoupling occurs, we first need to mass deform $\mathcal{H}^{\text{5d}}_{ij}(G_i,G_j)$ before the gluing.
 In the corresponding M-theory geometry, we flop a curve out of the reducible surface. 
This geometric transition is usually necessary to  decouple the $U(1)_T$ of the extra gauge theory when we start from the 6d tensor branch, since  otherwise, $\mathcal{H}^{\text{5d}}_{ij}(G^{\text{6d}}_{F_i},G^{\text{6d}}_{F_j})$ is simply a $S^1$ reduction of $\mathcal H^{\text{6d}}_{ij}(G^{\text{6d}}_{F_i},G^{\text{6d}}_{F_j})$.
\end{enumerate} 

The bottom-up construction of 6d SCFTs is guided by the definition of a consistent tensor branch with a superconformal fixed point at its origin. Inspired by the tensor branch geometries, we propose a set of rules which allow us to glue the geometries associated to 
$\mathcal{S}^{\text{5d}} (G,G^{\text{6d}}_{F})$ and $\mathcal{H}^{\text{5d}}_{ij}(G^{\text{6d}}_{F_i},G^{\text{6d}}_{F_j})$.

Furthermore, we propose a gluing rule for CFDs, which then allows determining the 5d superconformal flavor symmetries through a gluing. 
The input for this construction are the geometries/theories/CFDs for building blocks that are descendants of simple constituent of the 6d tensor branch.
A class of  these building blocks are the circle-reduction of single curve tensor branches, as listed in \cite{Bhardwaj:2018yhy, Bhardwaj:2019fzv}.  In appendix~\ref{sec:BB}, determine and summarize these constituents and their CFDs, including a single $(-1)$-curve, gauge group on single curves (including NHCs) and minimal conformal matter. This is not a comprehensive list of building blocks, e.g. we do not consider those 6d tensor branch geometries where the flavor symmetry is not manifest. The building blocks that we computed in appendix \ref{sec:BB} are single gauge node components, where the flavor symmetry is manifest as well as minimal conformal matter. 
These will then be used to propose a gluing construction.

\subsection{CFDs from Gluing}

We now propose a gluing rule on CFDs, which proceeds in two steps: first we gauge a common flavor symmetry, and then define how to combine the CFDs.
Suppose that we have already constructed the CFDs for $S^{\text{5d}} (G,G^{\text{6d}}_{F})$ and $H^{\text{5d}}_{ij}(G^{\text{6d}}_{F_i},G^{\text{6d}}_{F_j})_{S^1}$, which are $\text{CFD}^{(1)}$ and $\text{CFD}^{(2)}$. Denote their vertices by $C_\alpha^{(1)}$ and $C_\beta^{(2)}$, respectively. Then the gluing consists of the following two steps:

\begin{enumerate}
\item{\underline{Gauge:}

Geometrically, the gauge part essentially corresponds to identifying complex curves in each building block, and the two set of complex surfaces are glued together. In the CFD language, this corresponds to identifying linear combinations of vertices in each CFD:
\be
v_i= \sum_\alpha a_\alpha^{(i)} C_\alpha^{(i)}\,,\qquad i=1,2\,,
\ee
and then remove all the vertices $C_\alpha^{(i)}$ from both CFDs.

The $(n,g)$ of such linear combinations needs to satisfy the following ``gauge conditions'':
\be
\label{gauge-condition}
\ba
n(v_1)+n(v_2)&=2g(v_1)-2\cr
g(v_1)&=g(v_2)\,.
\ea
\ee
The reasoning is that each $C_\alpha^{(i)}$ can be considered as a linear combination of curves/vertices
\be
C_{\alpha}^{(i)}=\sum_j\xi_{j,\alpha}D_\alpha^{(i)}\cdot S_j^{(i)}\,,\qquad i=1,2\,.
\ee
Then $v_i$ can be written as the following curve in the Calabi-Yau threefold
\be
v_i= \sum_\alpha \sum_j a_\alpha^{(i)}\xi_{j,\alpha}D_\alpha^{(i)}\cdot S_j^{(i)}\,,\qquad i=1,2\,.
\ee
Now assume that all the weight factors $\xi_{j,\alpha}=1$ identically\footnote{Of course there is a similar analysis when the weight factors are not equal to one. For simplicity of notation, we discuss here the simpler case.}, then this is a well-defined complete intersection curve
\be
v_i=\left(\sum_\alpha a_\alpha^{(i)}D_\alpha^{(i)}\right)\cdot \left(\sum_j S_j^{(i)}\right)\,.
\ee
Thus we can identify the two curves $v_1$ and $v_2$ with the following gluing condition
\be
\sum_\alpha a_\alpha^{(1)}D_\alpha^{(1)}\equiv\sum_j S_j^{(2)}\,,\qquad \quad 
 \sum_\alpha a_\alpha^{(2)}D_\alpha^{(2)}\equiv\sum_j S_j^{(1)}\,.
\ee
In this way, we have made $D_\alpha^{(1)}$ and $D_\alpha^{(2)}$ compact. Then one can check that the equalities (\ref{gauge-condition}) indeed hold, with the formula (\ref{CFD-node-n}), (\ref{CFD-node-g}). 

In general, if the building block $S^{\text{5d}} (G,G^{\text{6d}}_{F})$ has non-Abelian flavor symmetry $G^{\text{6d}}_{F}$, then we need to choose two sets of $v_1$ and $v_2$ to gauge both $G^{\text{6d}}_{F}$ and $G^{\text{6d}}_{F_i}$ of  $H^{\text{5d}}_{ij}(G^{\text{6d}}_{F_i},G^{\text{6d}}_{F_j})$.

}

\item{\underline{Combine:}

After we have removed the vertices $C_\alpha^{(i)}$, we need to connect the remaining parts of the two CFDs. We define the set of vertices connected to $C_\alpha^{(i)}$ in the two building block CFDs by $\mc{S}_i=\{C_\beta^{(i)}\}$. For a well defined gluing process, the number of vertices in $\mc{S}_1$ and $\mc{S}_2$ should be the same, such that they can be combined pair-wise. The combined vertex in the glued CFD, $C_\beta^{\rm glued}$ should satisfy
\be
\label{glue-node-combine}
n(C_\beta^{\rm glued})=\sum_{i=1}^2\mu_\beta^{(i)}n(C_\beta^{(i)})\,,
\ee
where $\mu_\beta^{(i)}$ is a ``weight factor'' or multiplicity of vertices in the CFD, which appears in certain gluing processes, for which we will give an explicit formula in (\ref{CFD-BB-mu}). This is an analogy of the weight factor $\xi_{i,\alpha}$ of curves in each surface component. We will give concrete examples of this in the following.
Finally, after  the vertices in $\mc{S}_1$ and $\mc{S}_2$ are combined pair-wise, the other parts of the two CFDs connecting to $\mc{S}_1$ and $\mc{S}_2$ remain the same.

}
\end{enumerate}
This gluing is motivated by the geometric structure that we observe in higher rank theories. We will now exemplify it with gluing of NHCs and E-strings, as well as higher rank conformal matter theories, and show that it provides a consistent framework.

\begin{table}
\centering
\begin{tabular}{|c|c|c|}\hline 
6d Quiver & 6d Tensor Branch & CFD \cr \hline \hline
$(-1)-(-3)$ & $[E_6]-1-\overset{\mathfrak{su}(3)}{3}$ & \includegraphics[height=2cm]{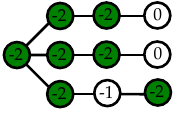}\cr  \hline 
$(-1)-(-4)$ & $[SO(8)]-1-\overset{\mathfrak{so}(8)}{4}$ &\includegraphics[height=3cm]{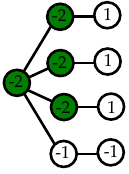}\cr \hline 
$(-1)-(-6)$ & $[SU(3)]-1-\overset{\mathfrak{e}_6}{6}$& \includegraphics[height=3.9cm]{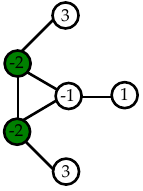}\cr \hline 
$(-1)-(-8)$ & $[SU(2)]-1-\overset{\mathfrak{e}_7}{8}$ &\includegraphics[width=5.3cm]{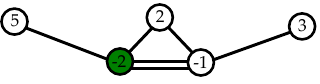}\cr \hline 
$(-1)-(-12)$ & $1-\overset{\mathfrak{e}_8}{12}$ &\includegraphics[width=3.56cm]{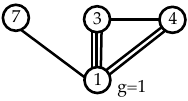}\cr \hline 
\end{tabular}
\caption{CFDs for  5d theories obtained by reduction from 6d quivers with two nodes $(-1)- (-n)$, where $(-n)$ corresponds to an NHC. The tensor branch geometry is shown in the middle and the CFDs in the right-most column.\label{tab:-1-n}}
\end{table}

\subsection{Example: $(-1)$-NHC Quivers}

In this section, we present the simplest example of the gluing philosophy, which is a single $(-1)$-curve glued to a single curve with non-Higgsable cluster (NHC) gauge group $G$, i.e. we consider quivers 
\be
(-1)(-n) \,,\qquad n= -3, -4, -5, -6, -8, -12 \,.
\ee
 In 6d, the theory will have flavor symmetry group $H$ that is the maximal commutant of $G\subset E_8$. This is because the rank-1 E-string theory over a $(-1)$-curve has flavor symmetry group $E_8$, and after the gluing, the subgroup $G\subset E_8$ is gauged, while the maximal commutant $H$ still remains as the flavor symmetry. The tensor branches of these theories are listed in table~\ref{tab:-1-n}.

In these cases, after blowing down the $(-1)$-curve on the tensor branch, we always end up with a curve with self-intersection $-n+1\leq -2$. Hence the 5d KK theory has an $SU(2)$ vector multiplet associated to it, which we need to be decoupled. 
Geometrically, we need to decompactify the surface associated to this, and the CFDs can be derived from directly in appendix \ref{app:Geo-1-n}. We summarize the results in table \ref{tab:-1-n}.

From the perspective of CFD gluing, it is useful to pick a convenient representation of the rank-one E-string marginal CFD in table~\ref{tab:Spn1CFDs}, such that the apparent flavor symmetry is $G\times H$. In other words, the rank-one E-string theory can be thought as a rank-one $(G,H)$ conformal matter, which acts as the link theory $H^{\text{5d}}_{ij}(G,H)$. After the decoupling process, the marginal CFD needs to be flopped once, as we have discussed before. The actual building blocks should be this ``sub-marginal'' CFD with $G_F=G\times H$ 5d flavor symmetry and the CFD of the NHC after decoupling in table~\ref{t:CFD-NHC}.

\begin{sidewaysfigure}
\centering
\includegraphics*[width=.8\textwidth]{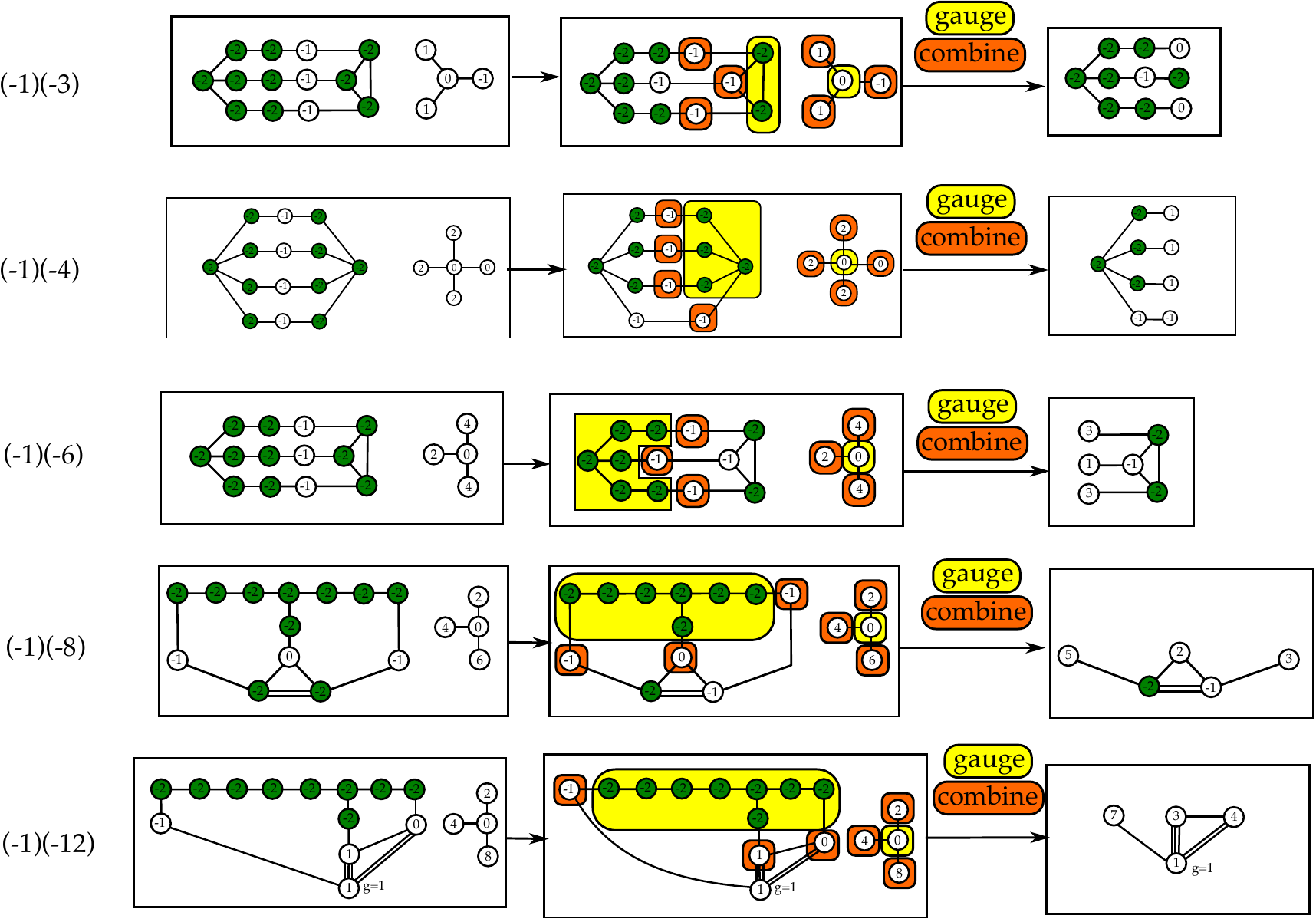}
\caption{Gluing of $(-1)$ and $(-n)$ curves. The E-string is realized in various way that allow gauging the NHC gauge groups. 
The first step corresponds to blowing down one $(-1)$ curves in the E-string CFD. The second step is gauging (i.e. removing from the CFD) and gluing the remaining curves. 
\label{fig:gluegauge}}
\end{sidewaysfigure}

We summarize the gluing process of these two building blocks for $n=3,4,6,8,12$ in figure \ref{fig:gluegauge}. 
In order to satisfy the condition for decoupling in section \ref{sec:nvht}, we first need to flop one of the $(-1)$ curves in the E-string CFD. This is the first step figure \ref{fig:gluegauge}. Then we identify the curves that we use to gauge a flavor symmetry.
The combinations of curves involved in the gauging part are encircled in yellow, and the vertices that get combined are colored orange in the gluing process. As one can see, the orange vertices are matched pair-wise, and they are never flavor vertices in the building block CFDs. The details of matching the orange vertices should be read off from the geometry in appendix~\ref{app:Geo-1-n}. 
However, in many cases we observe that the discrete symmetries of the CFDs select which curves (orange) need to be combined. It would be interesting to understand better the role of these discrete symmetry and their interplay with the 6d and 5d flavors. Note that the multiplicity factors $\mu$ in (\ref{glue-node-combine}) are always trivially one in these cases.

\subsection{Non-Minimal Conformal Matter from Gluing}
\label{sec:CMN}

Another class of theories that can be studied also from the gluing, are the higher rank conformal matter theories. 
The gluing of the $(A_{n-1}, A_{n-1})$ will be discussed in \cite{Eckhard:2020jyr}, where using a toric description it will be even simpler. 
The first interesting non-trivial case to consider is $(D_4, D_4)$ non-minimal conformal matter. We already discussed the geometry of the tensor branches as well as the CFDs from the geometry of the tensor branch plus decoupling in section \ref{sec:NMCM}. 

We can also get $(D_4, D_4)$ $N=2$ non-minimal conformal matter by gluing two rank-one E-string the $\Sigma^2=(-4)$ NHC (see table \ref{t:CFD-NHC}). 
The CFD for rank-one E-string is taken to be the one with explicit $G_i\times G_j=SO(8)\times SO(8)$. We again first take the descendants of the conformal matter theories and then gauge the $SO(8)$ flavor symmetries with the $\mathfrak{so}(8)$ of the NHC (shown in yellow) and then gluing the remaining curves (shown in orange), which results in the following $N=2$ $(D_4, D_4)$ non-minimal conformal matter CFD
\be
\includegraphics[width=\textwidth]{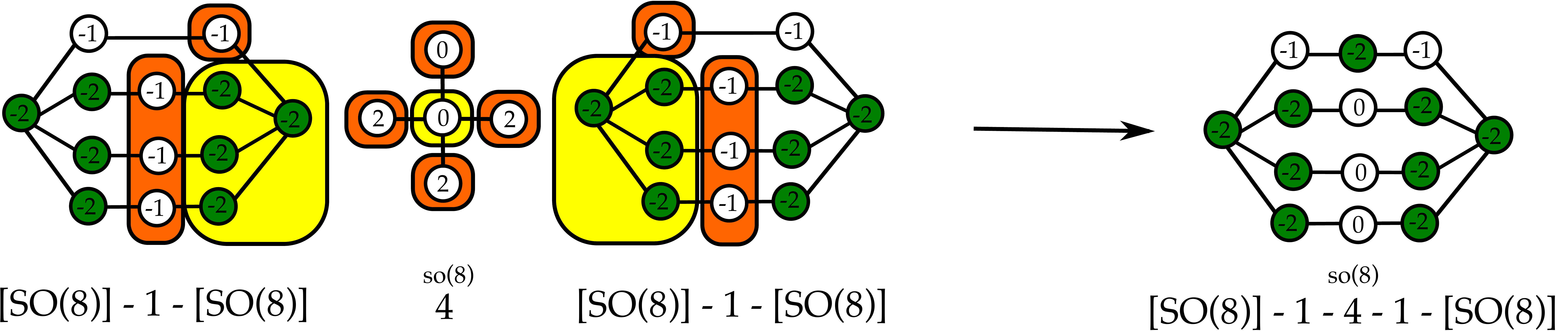} \,.
\ee
Below the graphs we shown the 6d tensor branch quiver building blocks. The resulting CFD is  in agreement with the one derived directly from the geometry in section \ref{sec:NMCM}. 

To obtain higher $N$, we iterate this process  as follows 
\be \label{eq:glue=2SO8CM}
\includegraphics[width=\textwidth]{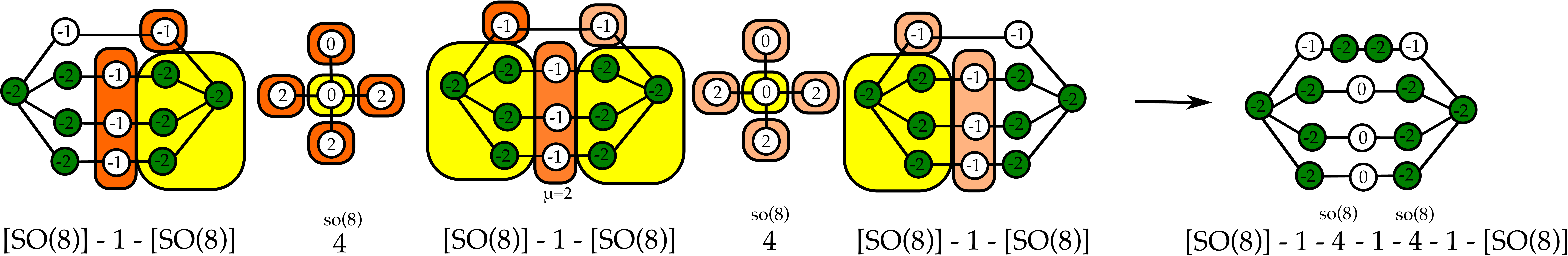} \,.
\ee
Note that the $(-1)$ vertices in the middle building block have $\mu=2$ multiplicity. 
The gluing for the general $(D_k, D_k)$ non-minimal conformal matter theory works along the same logic. 

We should now comment on the matter of the multiplicities that are key in the gluing: 
one might think that determining this requires considering the full resolved geometry as in section \ref{sec:CFD-geo}. 
However, we will be able to extract a relatively simple rule, from considerations of the tensor branch structure.  
The main point is that the multiplicity $\mu$  of the curves (\ref{glue-node-combine}) used to combine the CFDs, the orange-colored curves, 
has contributions from both the multiplicities $\xi$ in the building block as well as from the surfaces that are getting compactified in the gauging. For the building blocks in this paper, only the minimal conformal matter building blocks in appendix~\ref{app:BB-MCM} have non-zero intrinsic $\xi$s. For the other simpler building blocks, we have $\xi=0$.

Let us consider a set of CFD vertices $\{ C_\beta, \beta \in \Phi_{\text{gauge}}\}$, where $\Phi_\text{gauge}$ is the set of roots that we gauge. They correspond to the set of non-compact surfaces that get compactified in the gluing process, i.e. these are associated with the Cartans of the flavor symmetry that is getting gauged. The multiplicity of the orange curves gets modified, as they intersect these surfaces,  according to the \eqref{multiplicity-flop}. In general, the new multiplicity $\mu$ is computed with the following formula:
\be
\label{CFD-BB-mu}
\mu_\beta=\mathrm{max}\left[\left(\sum_{\gamma\in\Phi_{\text{gauge}}}m_{\beta\gamma}\right)\cdot(\xi(C_\beta)+1),1\right],
\ee
where $\xi(C_\beta)$ is the multiplicity factor intrinsic to a $(n,g)=(-1,0)$ vertex $C_\beta$ of the building block CFD (e.g. for $(E_6, E_6)$ there are multiplicity $\xi=1$ curves, see appendix~\ref{app:BB-MCM}). Note that $m_{\beta\gamma}$ is the number of edges between the vertices $C_\beta$ and $C_\gamma$ in the CFD. 

For instance in \eqref{eq:glue=2SO8CM}, we observe that the gluing procedure requires to identify the middle orange $(-1)$ vertices, as well as the yellow $SO(8)$ curves on the left, on the right and the $(0)$ in the middle, which are then removed by gauging. In the case of $(D_k,D_k)$ conformal matter, there is no intrinsic multiplicity factor $\xi$. Then the multiplicity (\ref{CFD-BB-mu}) of a $(-1)$ in the gluing procedure is simply given by the number of its adjacent vertices that are gauged (marked yellow). 

Let us apply this to the $N=2$ $(E_6,E_6)$  non-minimal conformal matter theory. This is glued from two minimal conformal matter theories  along an NHC with $\Sigma^2=(-6)$, which results in
\be
\includegraphics[width=15cm]{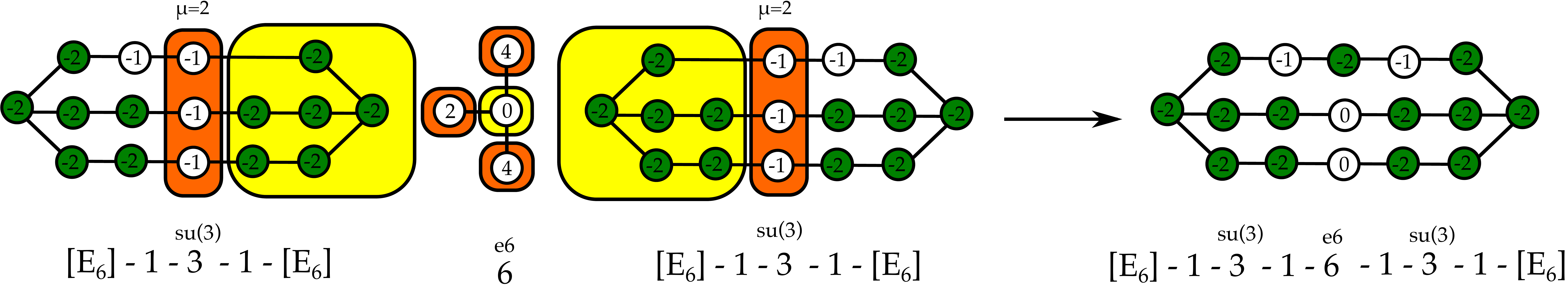} \,.
\ee
Here the $(-1)$ gluing nodes have multiplicity $\mu=2$, following the general rule stated above. 
Again the CFD is the one we obtained from a direct computation in the geometry in {appendix \ref{app:E6E6}}. 
We can iterate this and obtain the $N=3$ $(E_6, E_6)$ from gluing as follows -- note the additive nature of the multiplicity $\mu$:
\be
 \includegraphics[width=14.5cm]{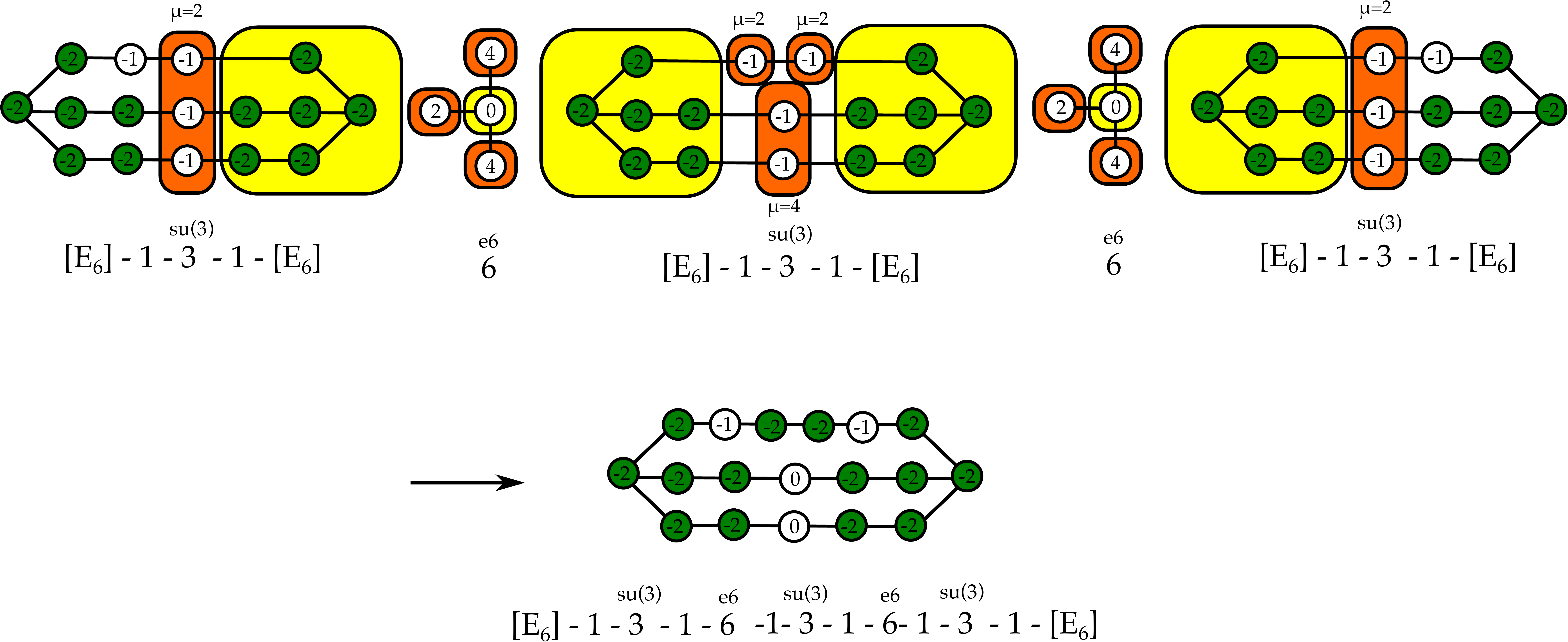} \,.
\ee
Finally, we consider the case of the $(E_7, E_7)$ and $(E_8, E_8)$ for which we earlier conjectured the CFDs. Although we will require a detailed knowledge of the geometry to compute the multiplicity factors, which are key to deriving the labels of the unmarked vertices with $n>-1$, we can determine the part of the CFD, that encodes the superconformal flavor symmetry (marked vertices) as well as the mass deformations (i.e. the $(-1)$ vertices).  
For $(E_7, E_7)$ theories with $N=2$ the tensor branch suggests the following gluing of two minimal conformal matter theories of type $(E_7, E_7)$ with the NHC with $\Sigma^2 =-8$ in table \ref{t:CFD-NHC}
\be
 \includegraphics[width=14.5cm]{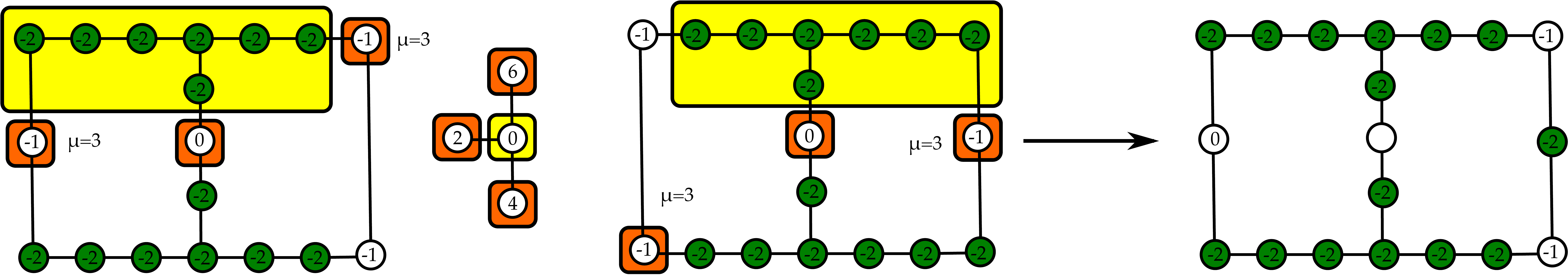} 
\ee
The multiplicity factor is $\mu=3$. 
Likewise for the $(E_8, E_8)$ theory with $N=2$, we glue two minimal conformal matter theories with the NHC $\Sigma^2= -12$ to obtain
\be
\includegraphics[width=15cm]{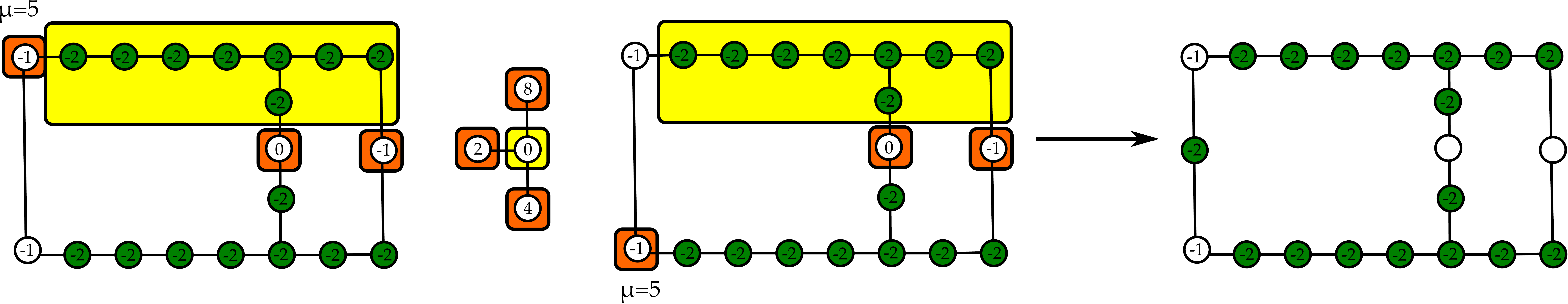} \,.
\ee
The multiplicity factor here is $\mu=5$. Note that these multiplicity factors can be exactly seen from the resolution geometry after the flop and decompactification, see Appendix~\ref{app:E7E7} and Appendix~\ref{app:E8E8}.


\section{Conclusions and Outlook}
\label{sec:conclusions}

In this paper we investigate the possible ways of getting 5d superconformal field theories (SCFTs) coming from 6d on a circle. In general, the circle reduction of a 6d SCFTs leads to a KK-theory, which in the UV completes back into the original 6d theory. 
To obtain a genuine 5d SCFT we need to consider mass deformations of the KK-theory, or equivalently, holonomies in the flavor symmetry. 
There are two type of possible mass deformation which we studied, which lead to 5d SCFTs:
\begin{enumerate}
\item The first corresponds to in the gauge theory description in the IR to the decoupling of matter hypermultiplets. In the M-theory Calabi-Yau geometry this corresponds to flopping the associated $(-1)$-curves out of the compact surfaces of the geometry. 
\item The second one is more drastic and require the decoupling an entire sector of the theory, like a gauge vector multiplet. In geometry this corresponds to the decompactification of some surfaces. 
\end{enumerate}
In \cite{Apruzzi:2019enx, Apruzzi:2019opn, Apruzzi:2019vpe}, we mainly focused on the so-called very Higgsable theories, where the natural mass deformations are those of the first type: giving masses to the hypermultiplets. We determined the starting points for such 5d RG-flows, and encoded these in the CFDs, which enabled tracking the complete tree of descendants and their superconformal flavor symmetries. 

In this work we focused on the exploration of the second possibility, in particular it turns out that for many not very Higgsable theories the first possibility is not an option if we want to get a single unfactorized 5d SCFT, and the second approach is unavoidable. We prove that the decoupling of an entire sector can be necessary for instance if the starting point is a 6d SCFT single curve with gauge group theory in the tensor branch, as is the case for non-Higgsable clusters (NHCs). In addition, we studied the circle-reduction of non-minimal conformal matter theories, i.e. the 6d theory of $N>1$ M5-branes probing an ADE singularity. We show that by decoupling an $SU(N)$ gauge theory, which geometrically correspond to decompactifying resolution surfaces, we obtain  5d SCFTs of arbitrary rank. In particular, the  geometries describing the Coulomb branched of these theories present very interesting features. We characterize these theories in terms of CFDs, which again encode flavor symmetries, mass deformations, descendant structure, and BPS states. 
We did not study in detail the descendants for these theories, but they are easily accessible by applying the descendant rules. 

Finally, inspired by the 6d classification, we propose a gluing procedure in order to get higher rank 5d SCFTs from lower rank building blocks. We first define the building blocks, which are the single node tensor branch theories (which are not very Higgsable), and the minimal conformal matter theories. We then use the tensor branch resolution of the 6d SCFT, to motivate the gluing rules, and cross-check these against direct computations for simple quivers and non-minimal conformal matter theories. 


It would be interesting to generalize this gluing procedure further, in order to capture the vast landscape of 5d SCFTs which originate from 6d on a circle. In particular we did not consider the building blocks in 6d  \cite{Bertolini:2015bwa}, which do not have the manifest 6d superconformal flavor symmetry realized geometrically. It would be interesting to compute the CFDs for such models, and consider the BPS states to determine the 5d CFDs and flavor symmetry enhancements in those cases as well.


\subsection*{Acknowledgements}

We thank C.~Closset, J.~Eckhard, T.~Rudelius, and G.~Zafrir for discussions. 
This work is supported by the  ERC Consolidator Grant number 682608 ``Higgs bundles: Supersymmetric Gauge Theories and Geometry (HIGGSBNDL)''. 



\appendix

\section{Building Blocks}
\label{sec:BB}

\subsection{Rank 1 E-string Building Blocks}
\label{sec:1-curve}

\begin{table}
\centering
\begin{tabular}{|c|c|}\hline
$(\mathfrak{g}_1, \mathfrak{g}_2)$ &CFD \cr \hline
$(E_8, \emptyset)$& \includegraphics[height=3cm]{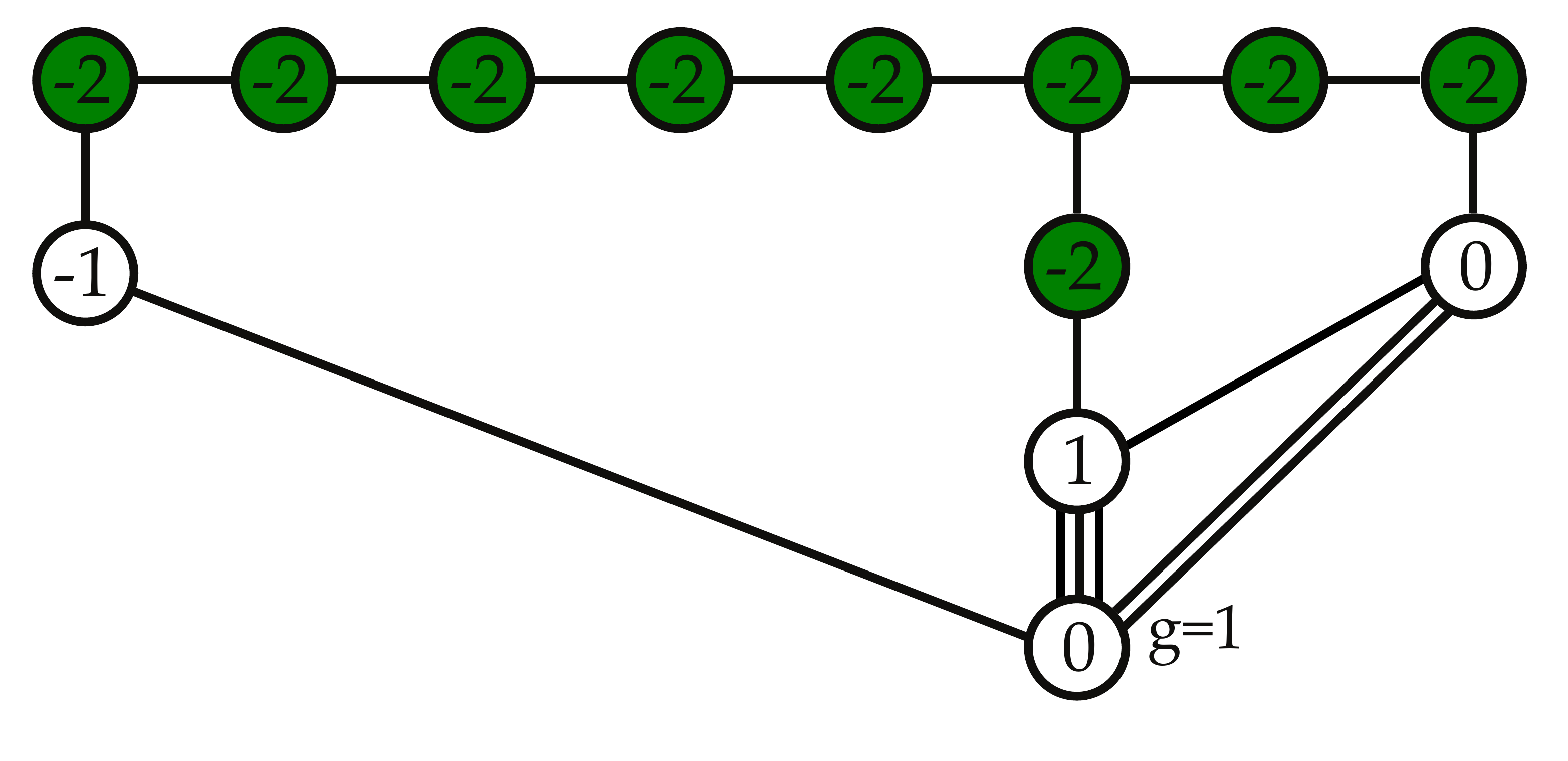} \cr \hline
$(E_7, SU(2))$ & \includegraphics[height=3cm]{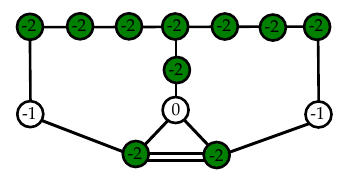} \cr \hline
$(E_6, SU(3))$ & \includegraphics[height=2cm]{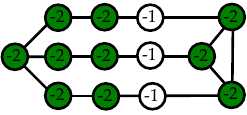} \cr \hline
$(SO(8),SO(8))$ & \includegraphics[height=3.5cm]{1-SO_8_-SO_8_.pdf} \cr \hline
$(SO(7), SO(9))$ & \includegraphics[height=3.9cm]{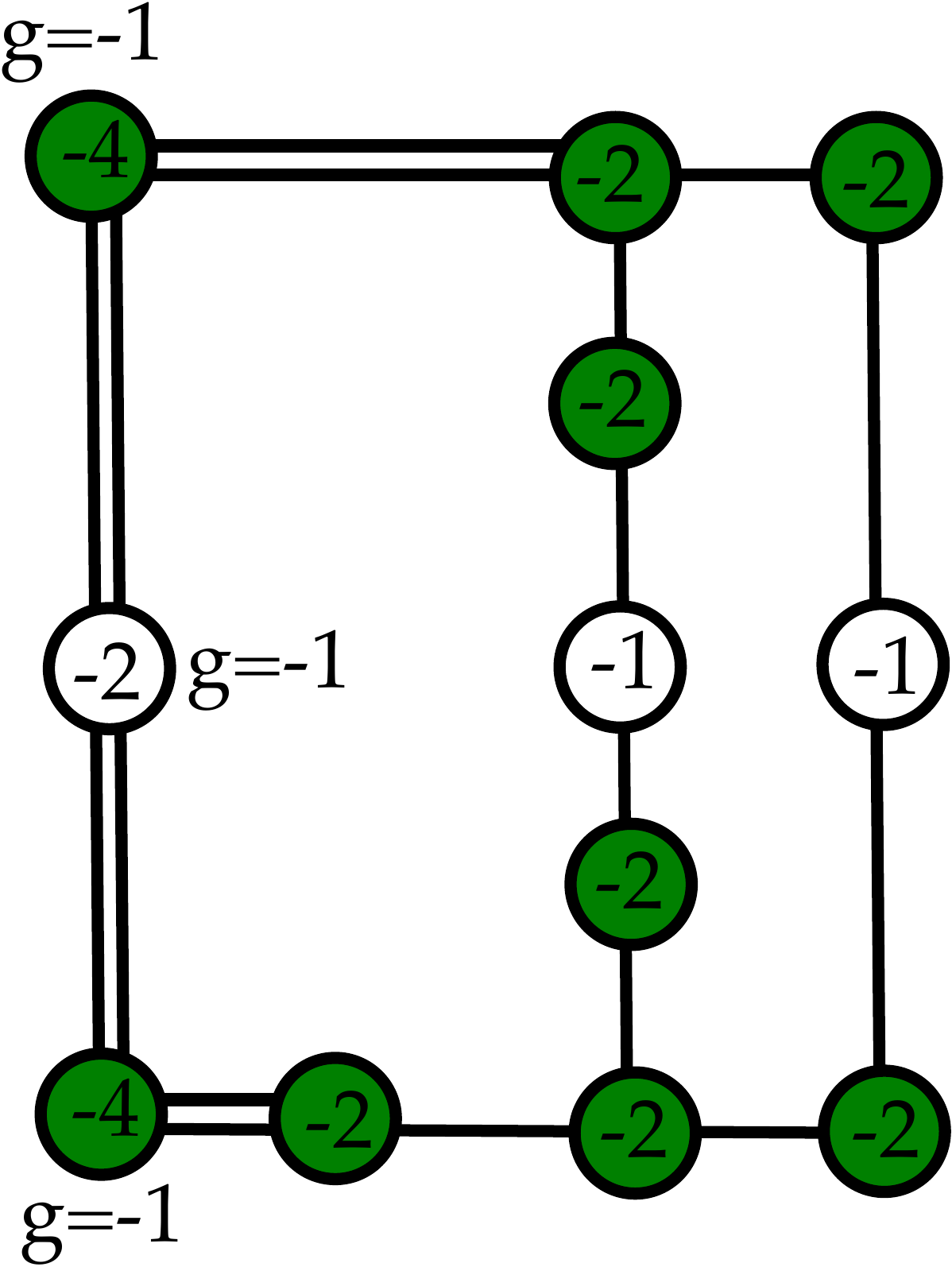} \cr \hline
\end{tabular}
\caption{CFDs for the marginal rank 1 E-string obtained by different collisions of $G_1$ and $G_2$ singularities. The maximal manifest flavor symmetry is realized only in the $(E_8, \emptyset)$ model. For all other the manifest flavor symmetry is $\widehat{G}_1\times \widehat{G}_2$, and enhances to $\hat{E}_8$ by including the additional BPS states. However these different realizations are useful for the gluing process. \label{tab:Rank1CFDs}}
\end{table}

In the 6d tensor branch descriptions, a single $(-1)$-curve by itself corresponds to the rank-1 E-string theory with flavor symmetry $G_F=E_8$. In the tensor branch resolution geometry, the compact surface $S$ will be a rational elliptic surface (generalized dP$_9$) over the base $(-1)$-curve. Nonetheless, if there are two curves with simple gauge groups $\mathfrak{g}_1$ and $\mathfrak{g}_2$ connected to it:
\be
\overset{\mathfrak{g}_1}{m}-1-\overset{\mathfrak{g}_2}{n},
\ee
then the surface $S$ serves as a connection surface between the Cartan divisors of $\mathfrak{g}_1$ and $\mathfrak{g}_2$. On the rational elliptic surface $S$, there are degenerate elliptic fibers corresponding to affine Lie algebra $\hat{\mathfrak{g}}_1$ and $\hat{\mathfrak{g}}_2$. Namely, the $(-2)$-curves on $S$ form the affine Dynkin diagram of $\hat{\mathfrak{g}}_1\times\hat{\mathfrak{g}}_2$, and there are $(-1)$-curves connected between them.

In principle, for any $\mathfrak{g}_1\oplus\mathfrak{g}_2\subset E_8$, such a rational elliptic surface exists. However, only for a subset of $(\mathfrak{g}_1,\mathfrak{g}_2)$, the number of $(-1)$-curve on the rational elliptic surface is finite. These surfaces are called ``extremal'' and has been classified in \cite{miranda1986extremal,miranda1990persson}. In this paper,  we only list the ones that will be used in the later gluing discussions, which happens to satisfy the extremal criterion. They can be generated by blowing up the generalized del Pezzo surfaces in \cite{derenthal2014singular} or putting together set of $(-2)$-curves. These are summarized in table \ref{tab:Rank1CFDs}

In some cases where decoupling happens, the rational elliptic surface needs to be blown down to gdP$_8$. The set of curves are transformed according to the usual rule of shrinking $(-1)$-curves.
\subsection{Single Curves with Gauge Group}

In this section, we discuss the building block of a single curve with a (tuned) non-Abelian gauge group on it, which is not an NHC. The flavor symmetry and F-theory realization of such building blocks are discussed in \cite{Bertolini:2015bwa}. We focus on the cases where the flavor symmetry is identical in Table 2 and Table 3 of \cite{Bertolini:2015bwa}, as it is easier to construct the maximal global symmetry from geometry. While the CFDs are presented in this section, the detailed resolution geometries are put in the appendix ~\ref{app:geo-single}. Because of the presence of non-Abelian flavor symmetry, there are typically multiple equivalent CFD building blocks for the same theory, similar to the rank-one E-string case.

\subsubsection{$Sp(n)$ on a $(-1)$-curve}

In this case, the 6d global symmetry is $SO(4n+16)$. It can be realized by the following tensor branch:
\be
[SO(2m)]-\overset{\mathfrak{sp}(n)}{1}-[SO(4n+16-2m)]\,,
\ee
where $0\leq m\leq n+4$. When $m=n+4$, it is realized by $(D_{n+4},D_{n+4})$ minimal conformal matter theory. The CFDs are shown in table \ref{tab:Spn1CFDs}. 

\begin{table}
\centering
\subfloat[][]{\begin{tabular}{|c|c|}\hline
$m$ &$[D_m]-\overset{\mathfrak{sp}(n)}{1}-[D_{2n+8-m}]$ \cr \hline\hline
$m=0$ & \includegraphics[height=1.4cm]{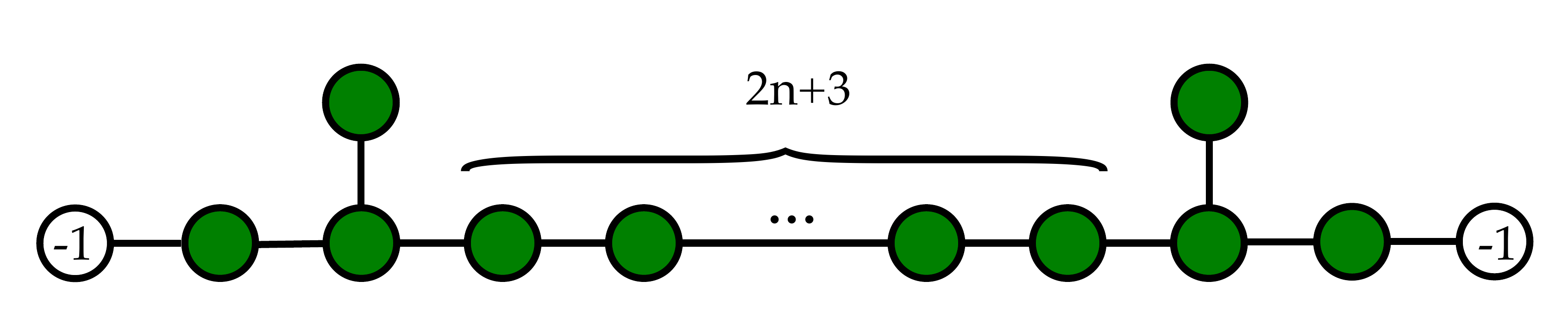} \cr \hline
$m=2$ & \includegraphics[height=5cm]{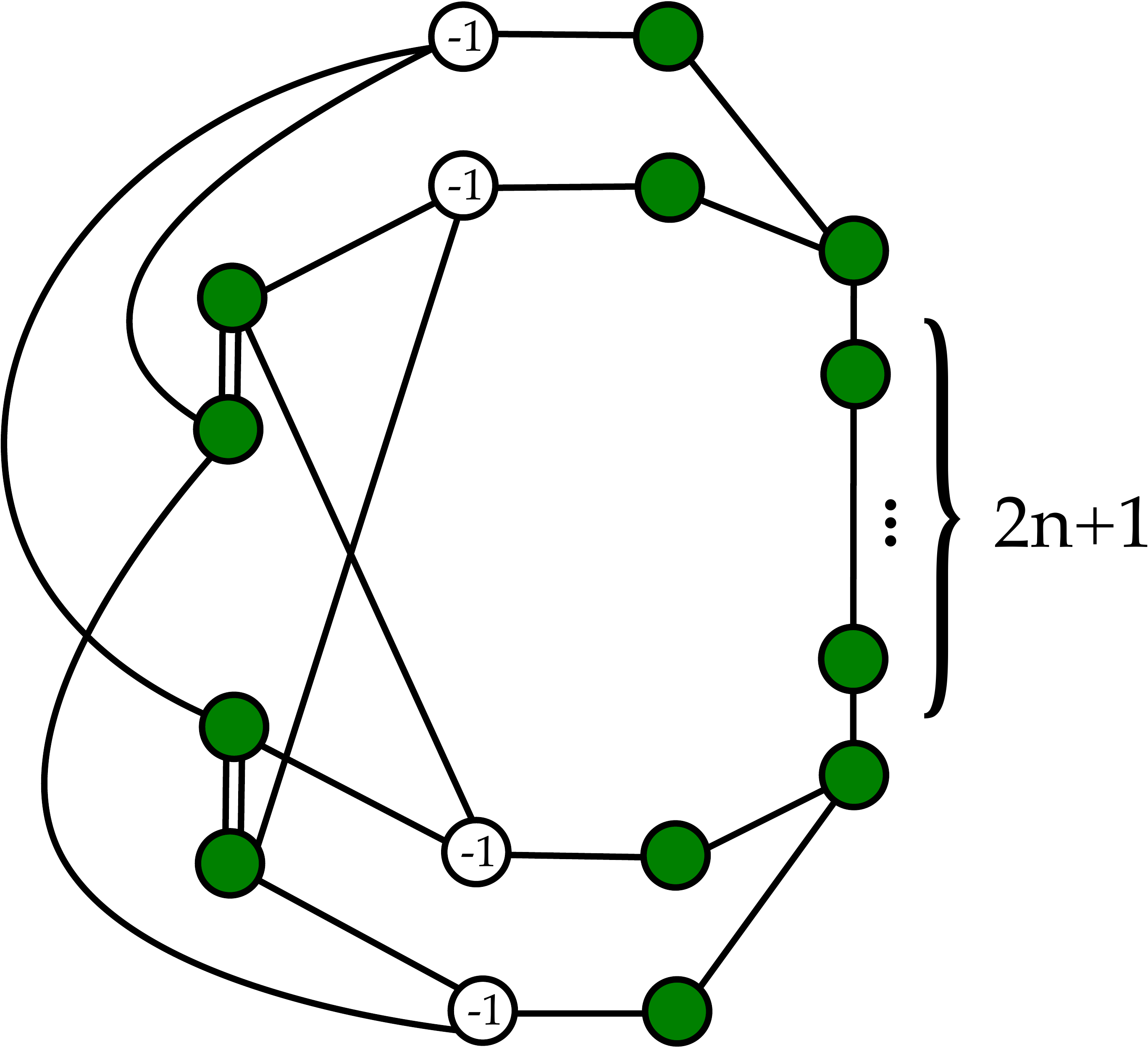} \cr \hline
 $m=3$ & \includegraphics[height=5cm]{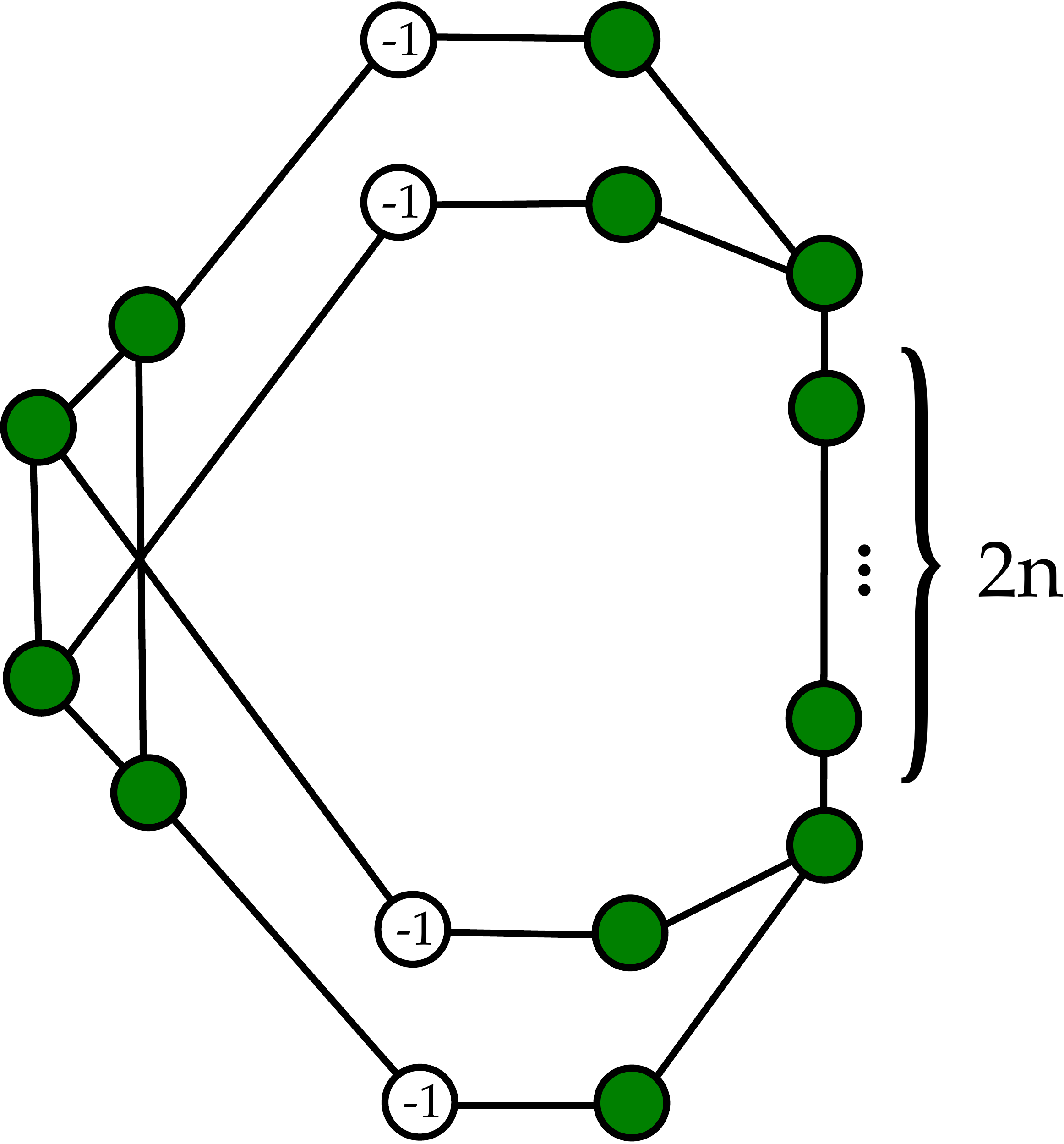}\cr \hline
$m\geq 4$ & \includegraphics[height=5cm]{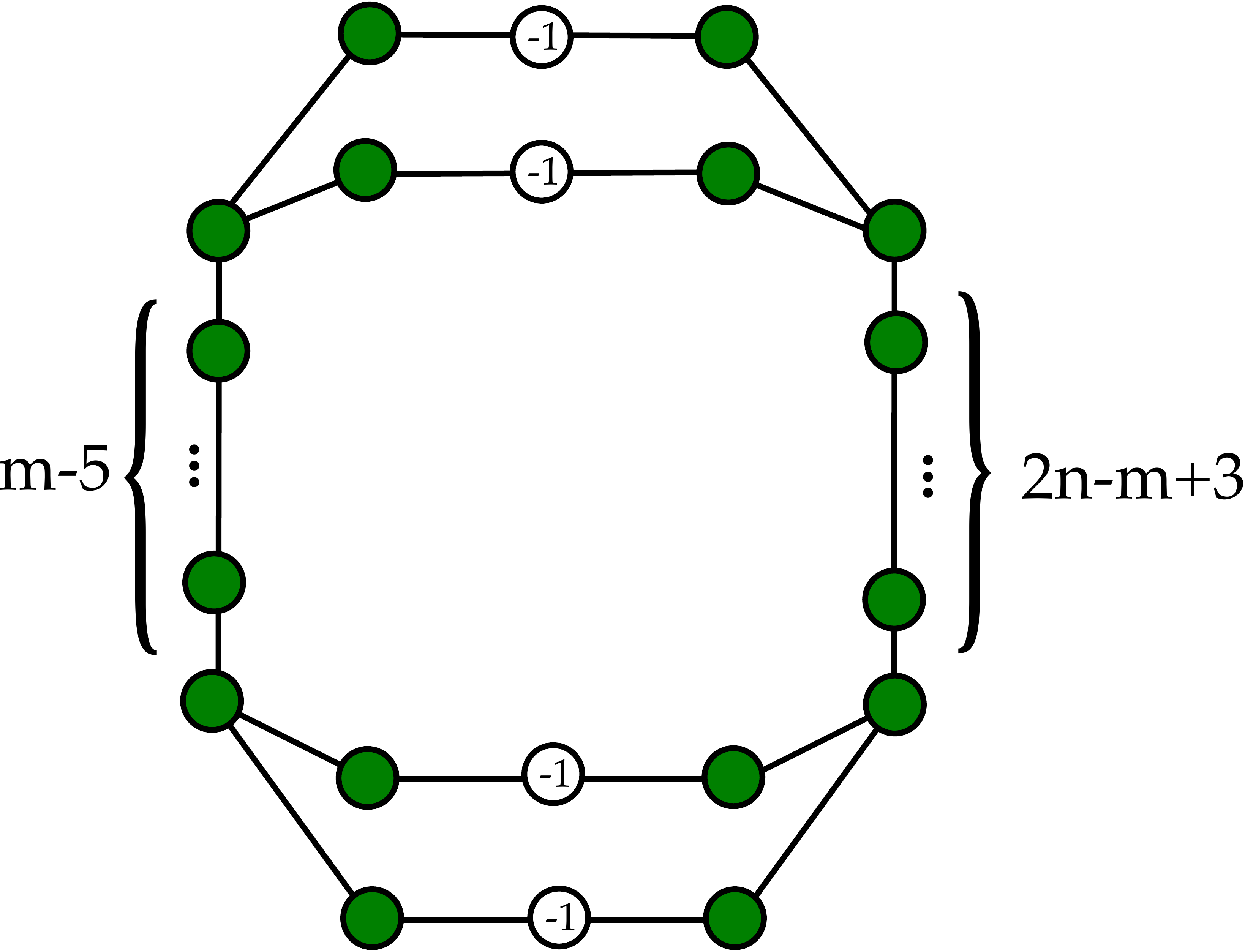}  \cr \hline
\end{tabular}}
\subfloat[][]{
\begin{tabular}{|c|c|}\hline
$m$ &$[B_m]-\overset{\mathfrak{sp}(n)}{1}-[B_{2n+7-m}]$ \cr \hline\hline
$m>2$ & \includegraphics[height=5cm]{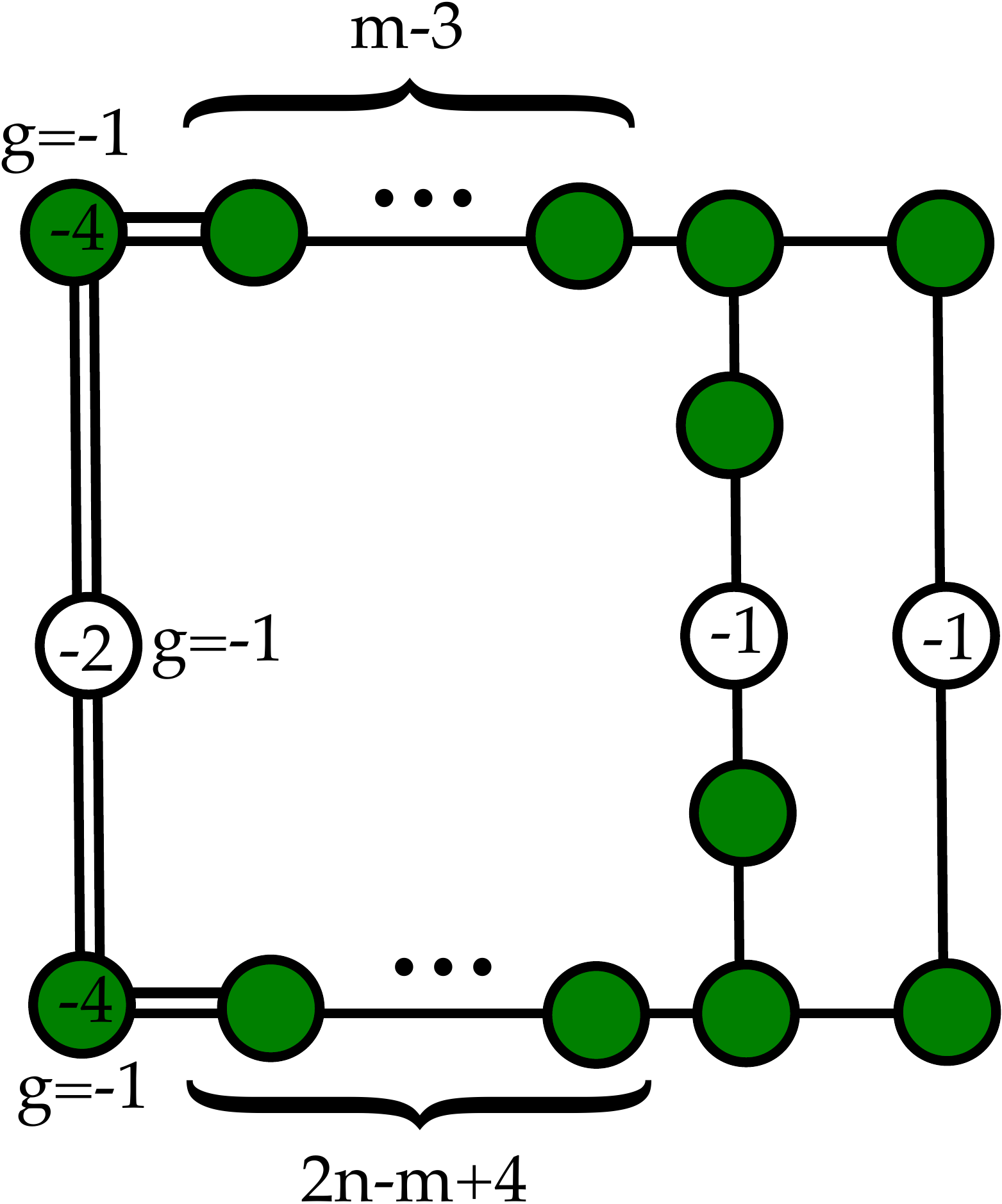}\cr \hline
\end{tabular}}
\caption{CFDs for $\mathfrak{sp}(n)$ on a $(-1)$-curve in the description on the tensor branch in terms of $[SO(2m)]-\overset{\mathfrak{sp}(n)}{1}-[SO(4n+16-2m)]$ (a) and for $[SO(2m+1)]-\overset{\mathfrak{sp}(n)}{1}-[SO(4n+15-2m)]$ (b). 
\label{tab:Spn1CFDs}}
\end{table}

%
%
%
%
%

\subsubsection{$SU(n)$ on $(-2)$-curve}

The 6d global symmetry is $SU(2n)$, and the tensor branch can be chosen as:
\be
[SU(m)]-\overset{\mathfrak{su}(n)}{2}-[SU(2n-m)]\,,
\ee
where $0\leq m\leq n$. After the decompling process, the 5d gauge theory description is $SU(n)_0+2n\bm{F}$, which is a descendant of $(D_{n+2},D_{n+2})$ conformal matter.
For $m=0$, the CFD is a descendant of the $(D_{2n+4},D_{2n+4})$ CFD tree in \cite{Apruzzi:2019vpe}.
The CFDs before and after decoupling are shown in table \ref{tab:Sun2CFDs}. 

\begin{table}
\centering
\begin{tabular}{|c|c|c|}\hline
$m$ & CFD before decoupling & CFD after decoupling\cr  \hline\hline
$m=0$ & \includegraphics[width=5cm]{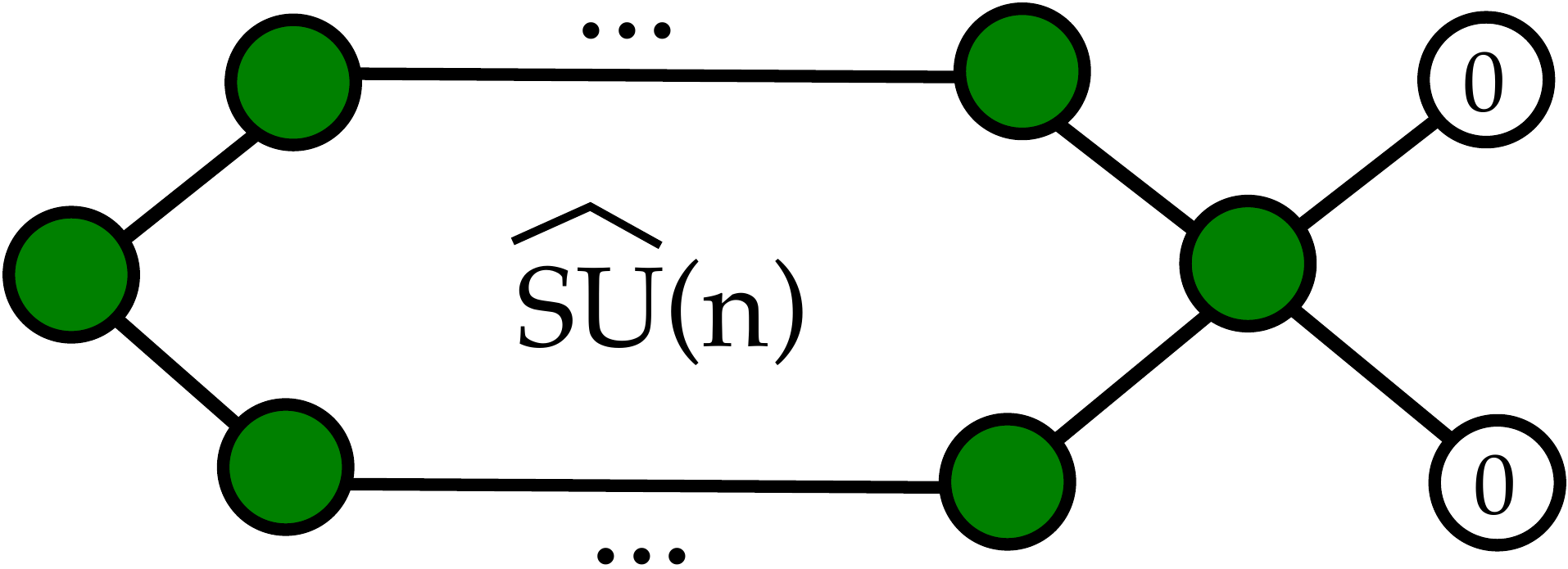} & \includegraphics[width=5cm]{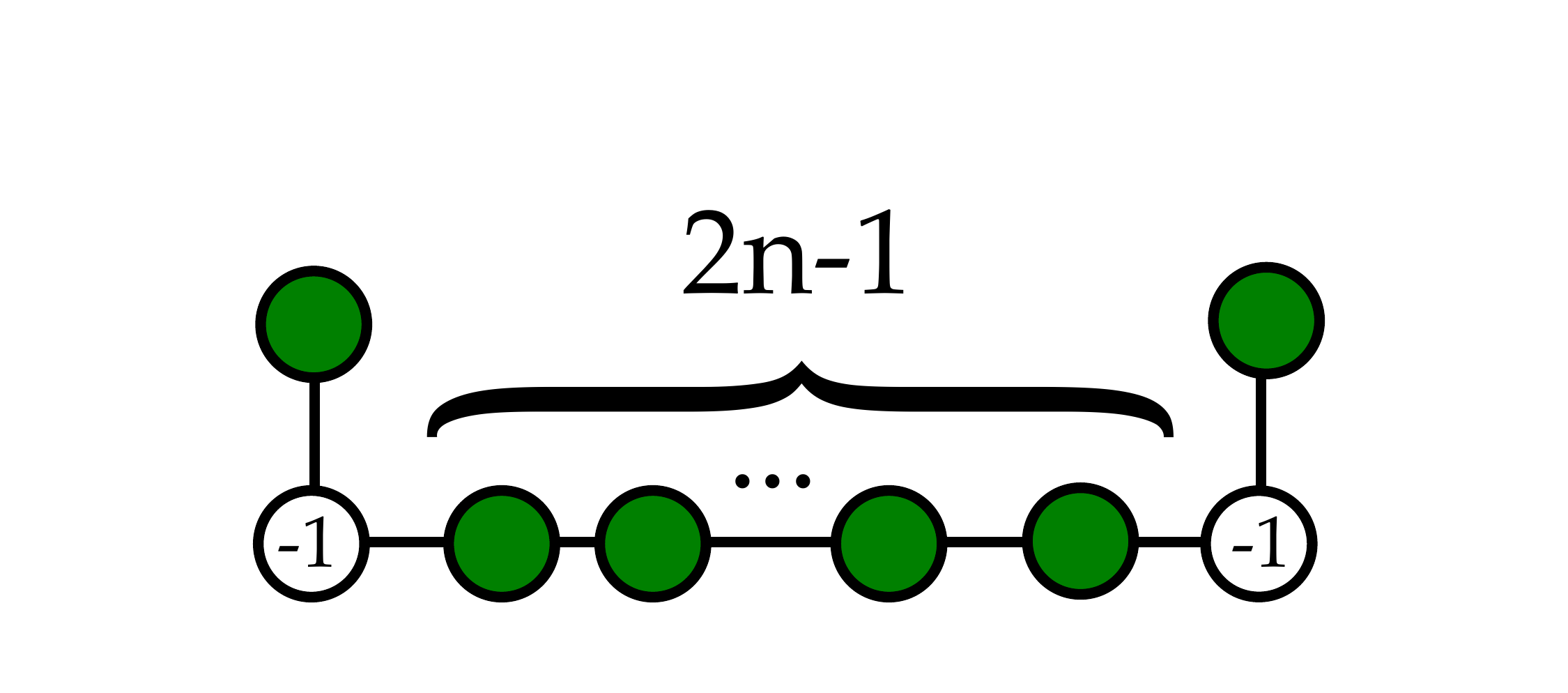} \cr \hline
$m>0$ & \includegraphics[width=8cm]{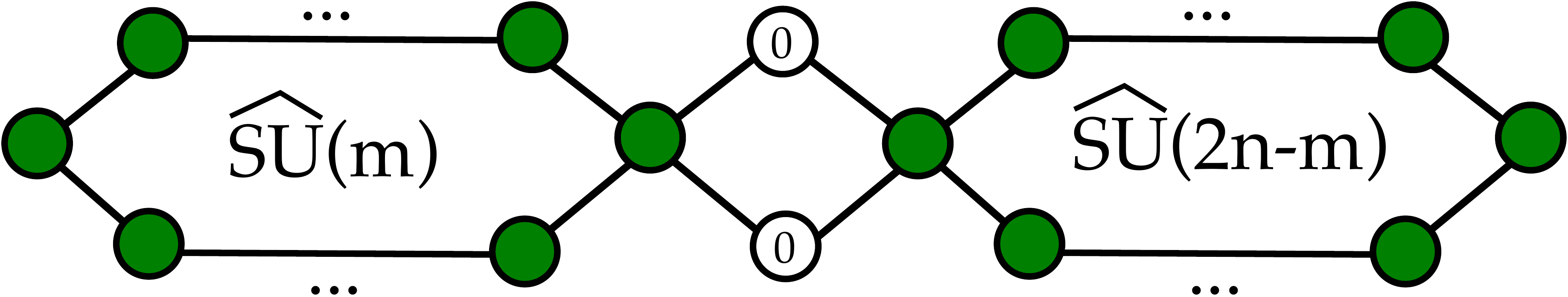} & \includegraphics[width=5cm]{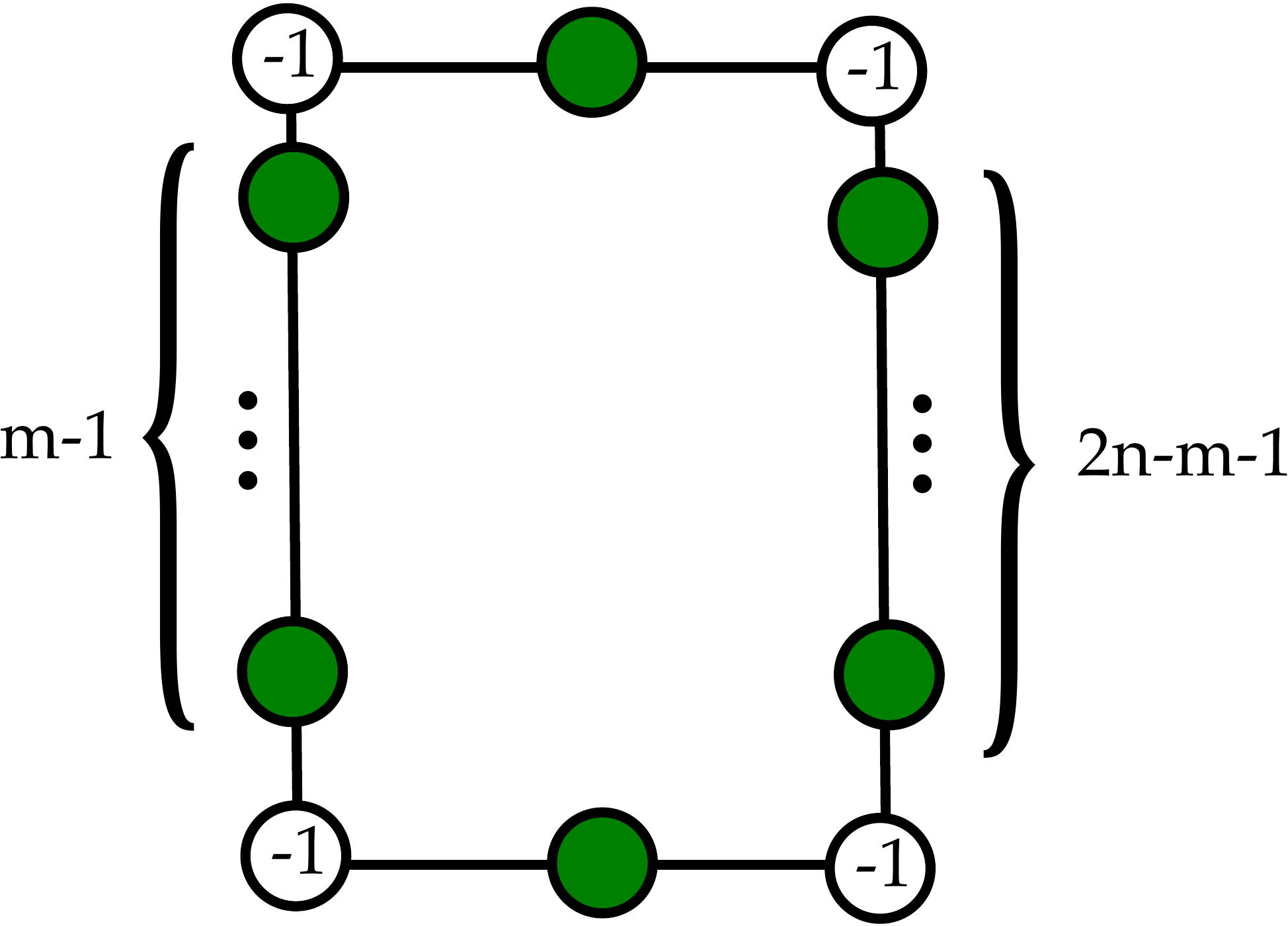} \cr \hline
 \end{tabular}
\caption{CFDs for  for $\mathfrak{su}(n)$ on a $(-2)$-curve  in the description on the tensor branch in terms of $[SU(m)]-\overset{\mathfrak{su}(n)}{2}-[SU(2n-m)]$. 
\label{tab:Sun2CFDs}}
\end{table}

\begin{table}
\centering
\begin{tabular}{|c|c|}\hline
$m$ & CFD \cr  \hline\hline
$m=0$ & \includegraphics[width=7cm]{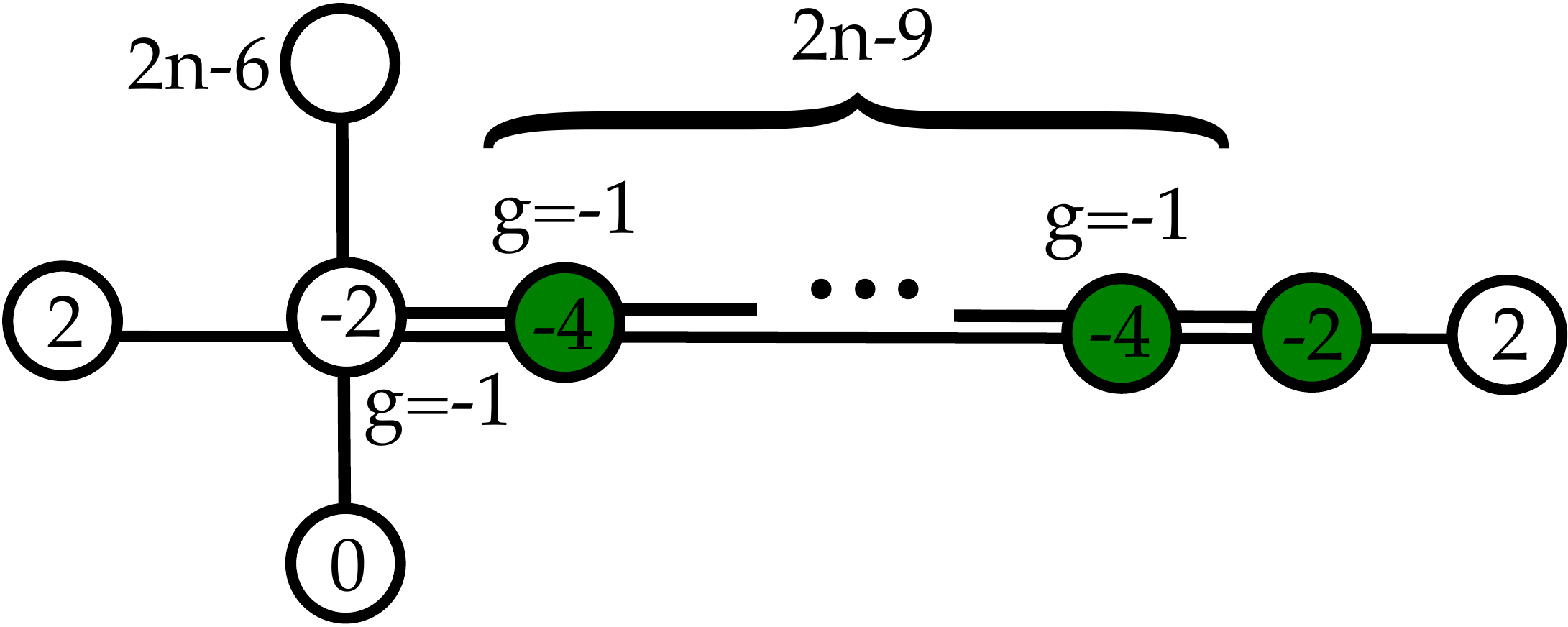} \cr \hline
$m>0$ & \includegraphics[width=6cm]{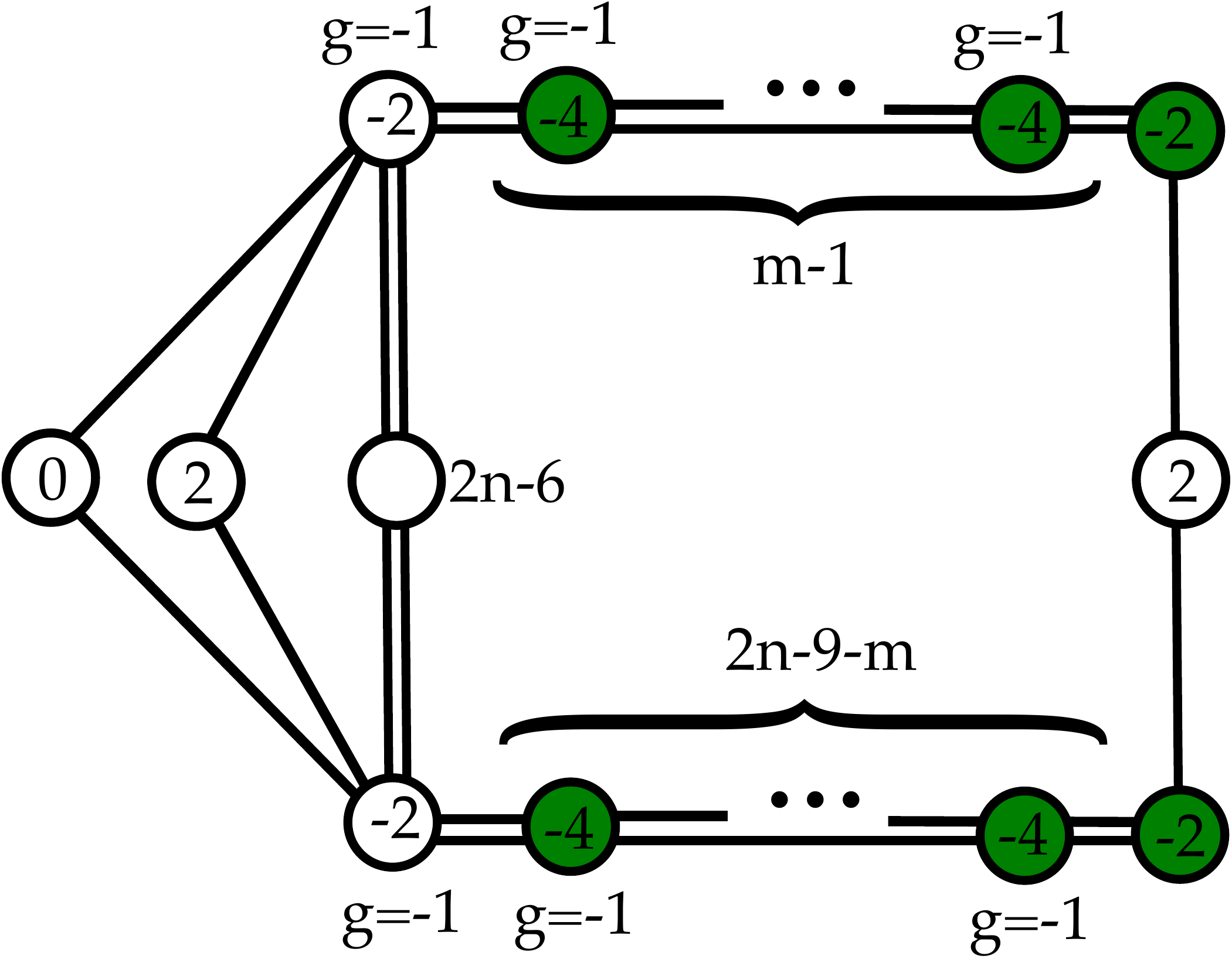} \cr \hline
 \end{tabular}
\caption{CFDs for  for $\mathfrak{so}(2n)$ on a $(-4)$-curve  in the description on the tensor branch in terms of $[Sp(m)]-\overset{\mathfrak{so}(n)}{4}-[Sp(2n-8-m)]$.\label{tab:So2n4CFDs}}
\end{table}

\subsubsection{$SO(2n)$ on $(-4)$-curve}

Since the non-Higgsable gauge group on a $(-4)$-curve is already $SO(8)$, we require that $n>4$. The 6d global symmetry is $Sp(2n-8)$, and the tensor branch can be chosen as:
\be
[Sp(m)]-\overset{\mathfrak{so}(n)}{4}-[Sp(2n-8-m)],
\ee
where $0\leq m\leq n-4$. After the decoupling process, the 5d gauge theory description is $SO(2n)+(2n-8)\bm{V}$.

When $m=0$, the CFD can be read off from the geometry in appendix \ref{app:SO-4}. 
The CFDs are summarized in table \ref{tab:So2n4CFDs}.

\subsection{Minimal Conformal Matter}
\label{app:BB-MCM}

Another useful class of building blocks is the minimal conformal matter where the 6d tensor branch have rank higher than one. For many of these theories, their marginal CFDs have been constructed in \cite{Apruzzi:2019vpe,Apruzzi:2019opn,Apruzzi:2019enx}. Here we summarize their marginal CFDs in table \ref{tab:MinCM}. We are going to shortly discuss the subtleties associated to non-trivial intrinsic multiplicities involved in the gluing section~\ref{sec:CMN} and non-simply laced Lie algebra.

\begin{table}
\centering
\begin{tabular}{|c|c|}\hline
$(\mathfrak{g}_1, \mathfrak{g}_2)$ &CFD \cr \hline
$(E_6, E_6)$ & \includegraphics[height=1.7cm]{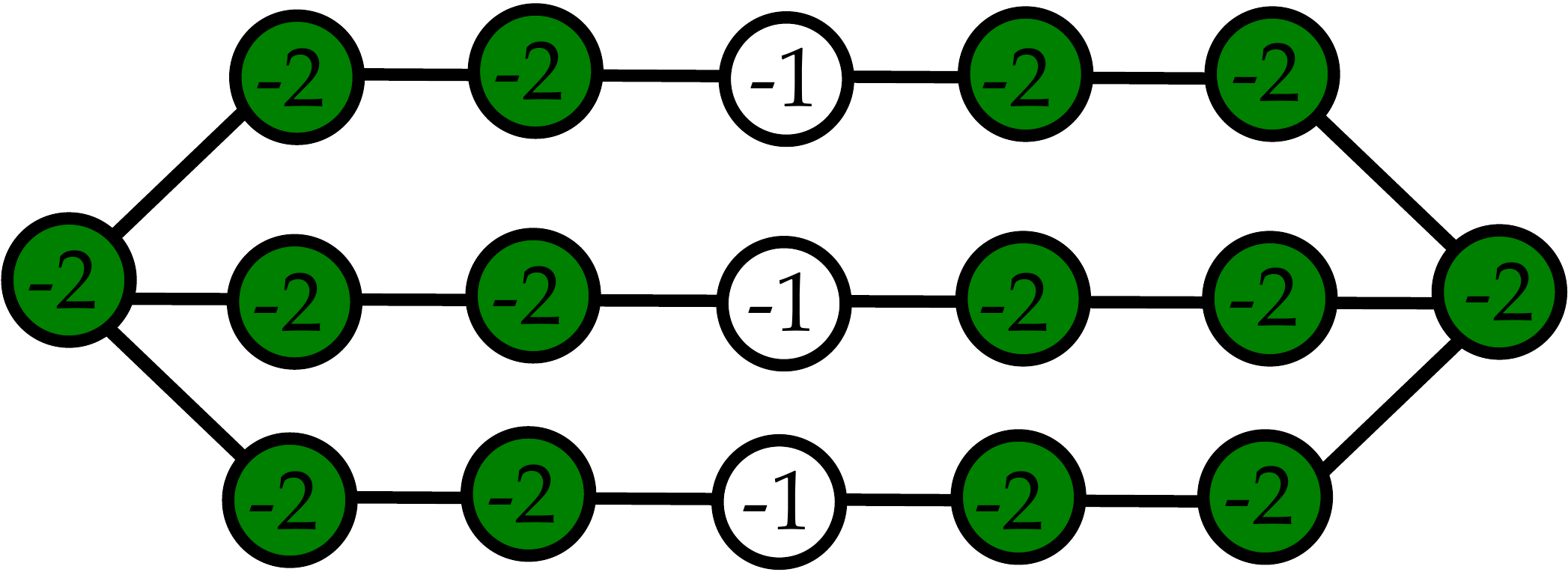} \cr \hline
$(E_7, E_7)$ & \includegraphics[height=2.5cm]{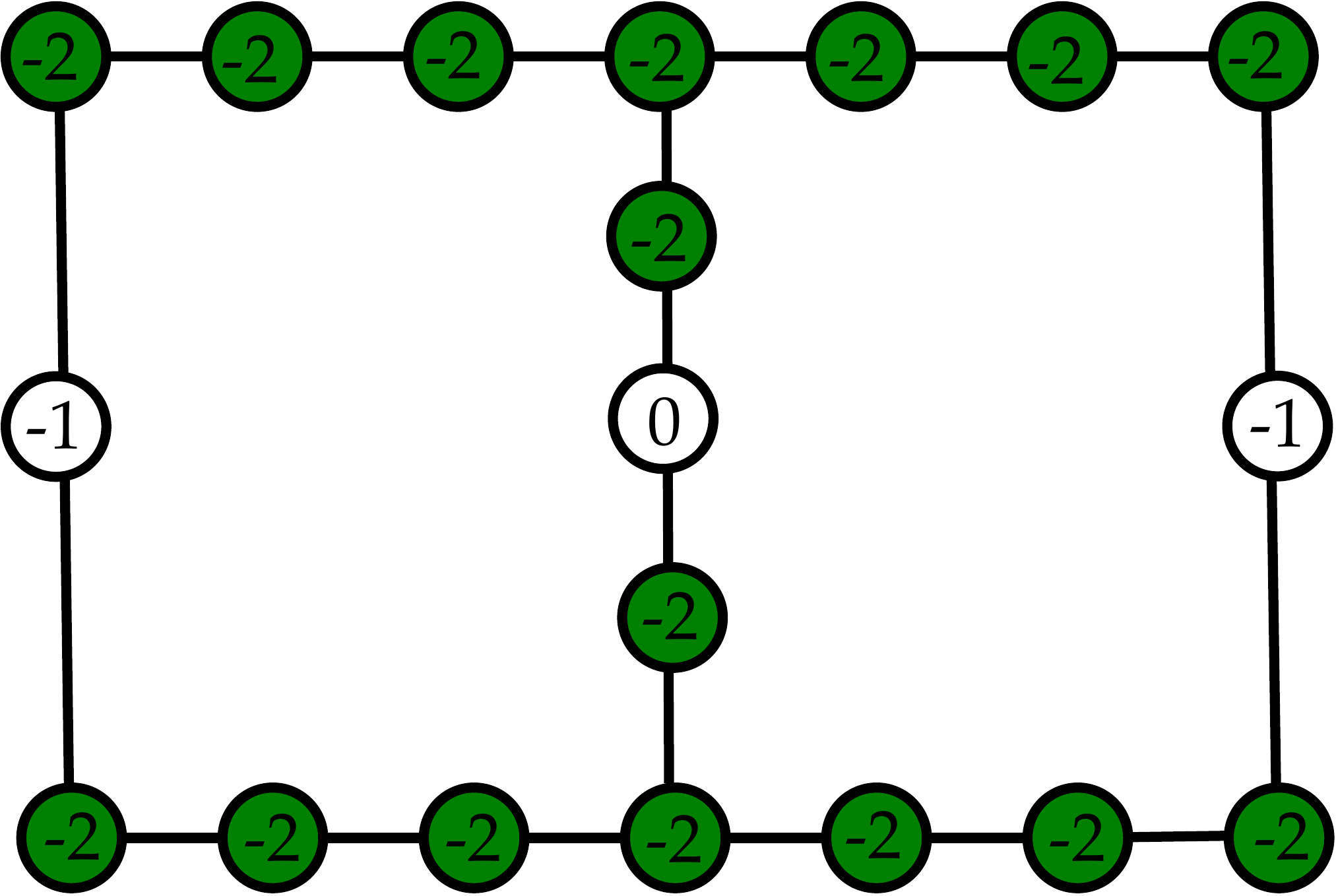} \cr \hline
$(E_8, E_8)$ & \includegraphics[height=2.5cm]{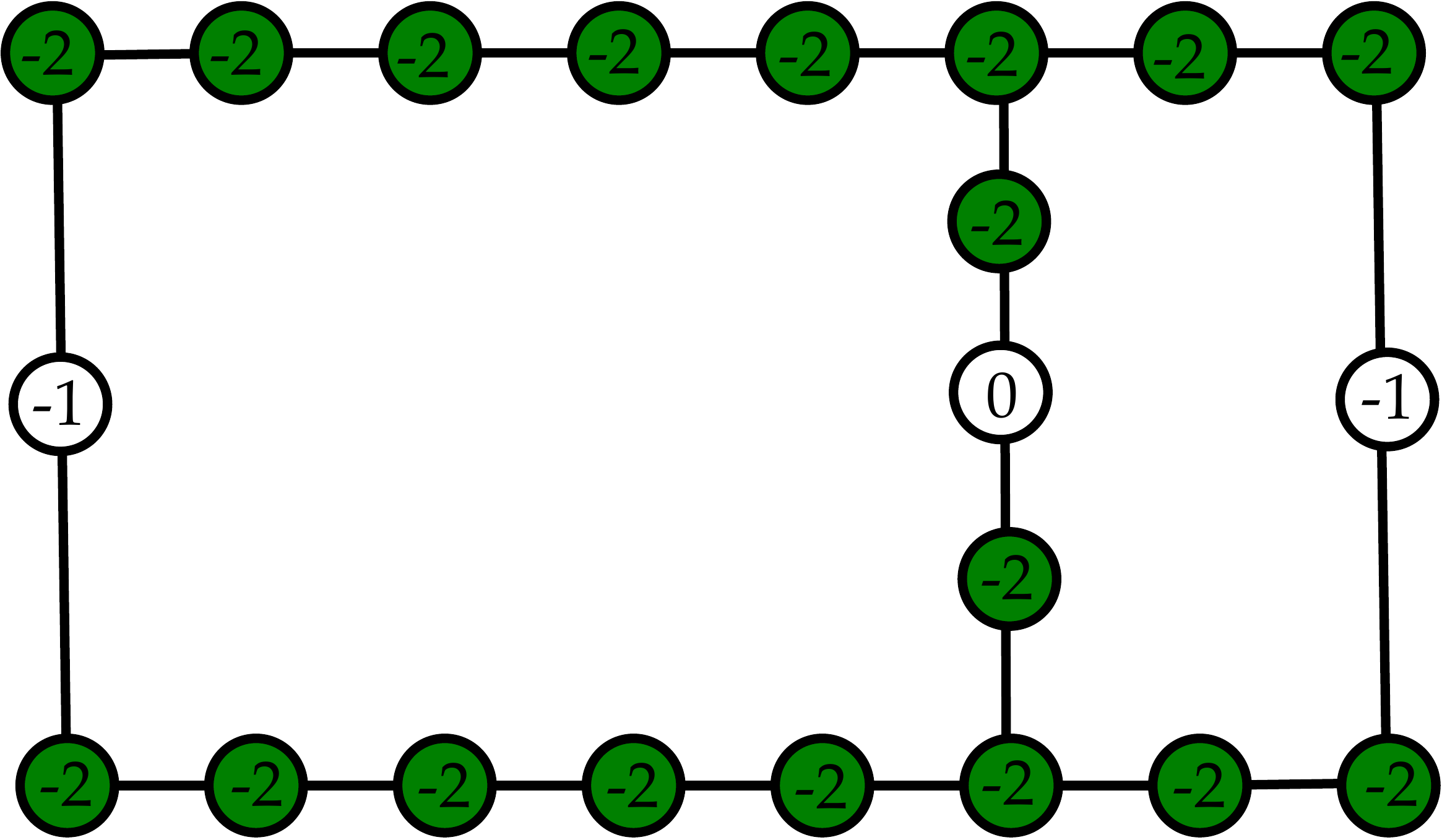} \cr \hline
$(E_8,SU(2k+1))$ & \includegraphics[height=2.5cm]{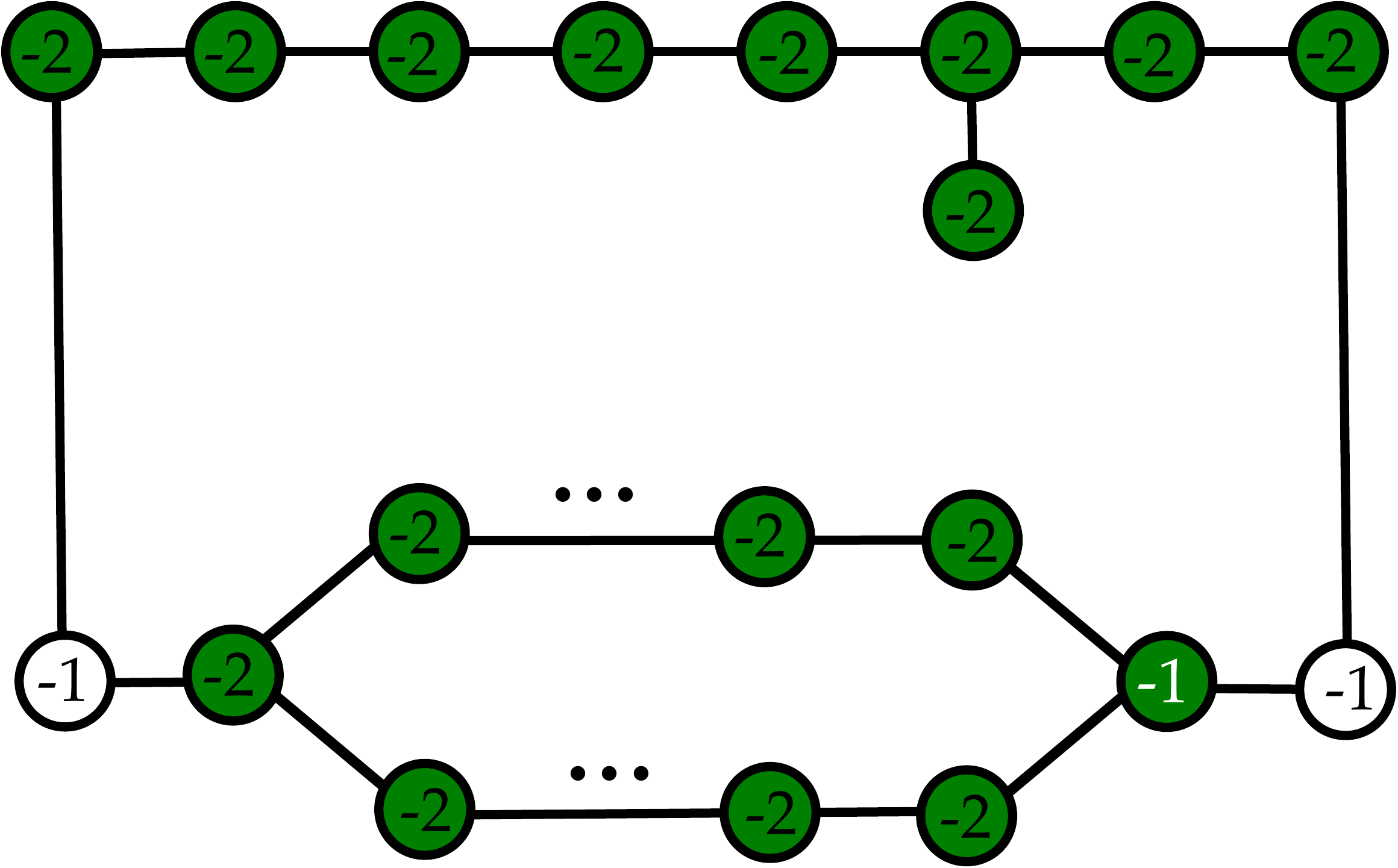} \cr \hline
$(E_8,SU(2k))$ & \includegraphics[height=2.5cm]{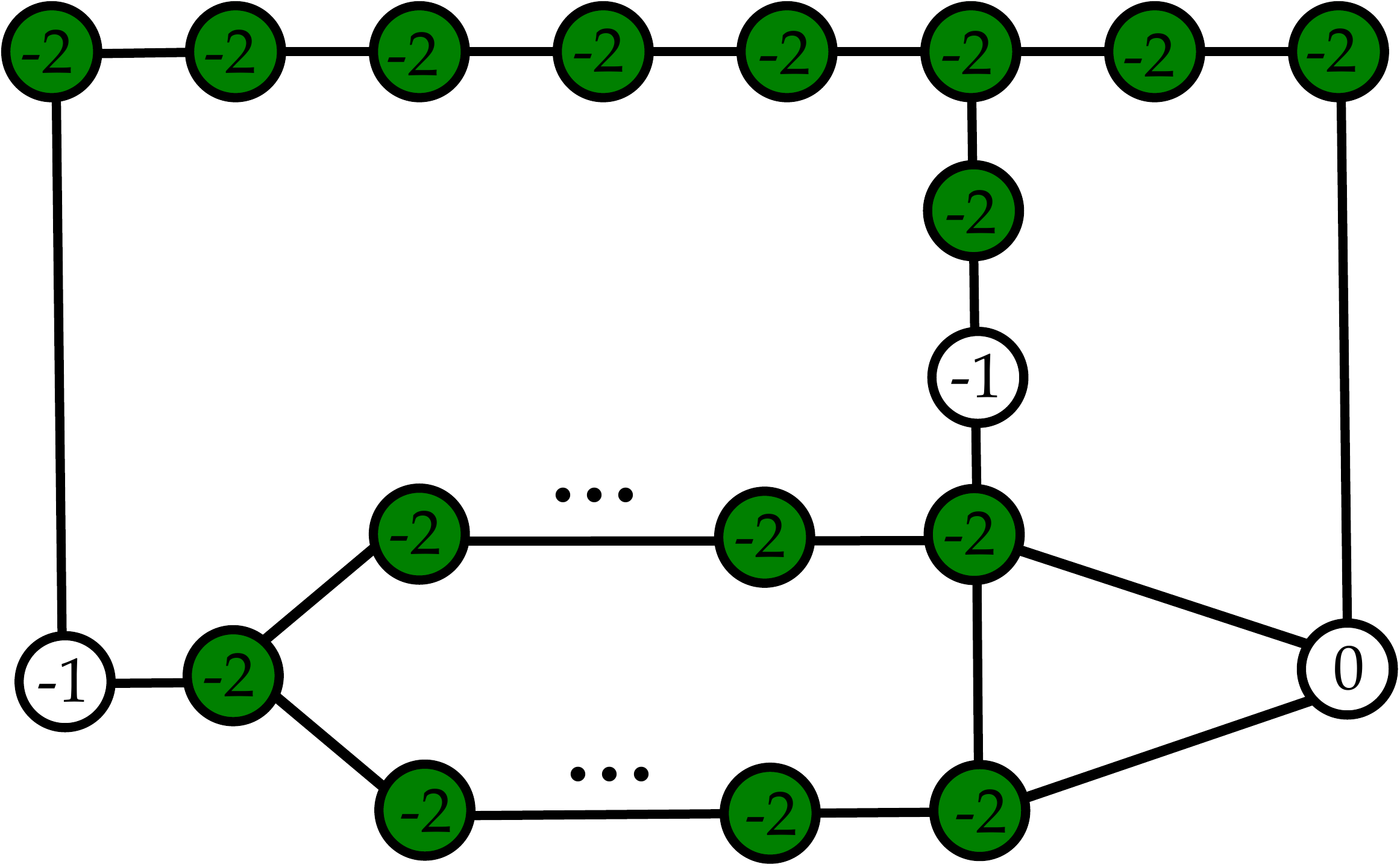}\cr \hline
$(E_7,SO(7))$ &\includegraphics[height=2.5cm]{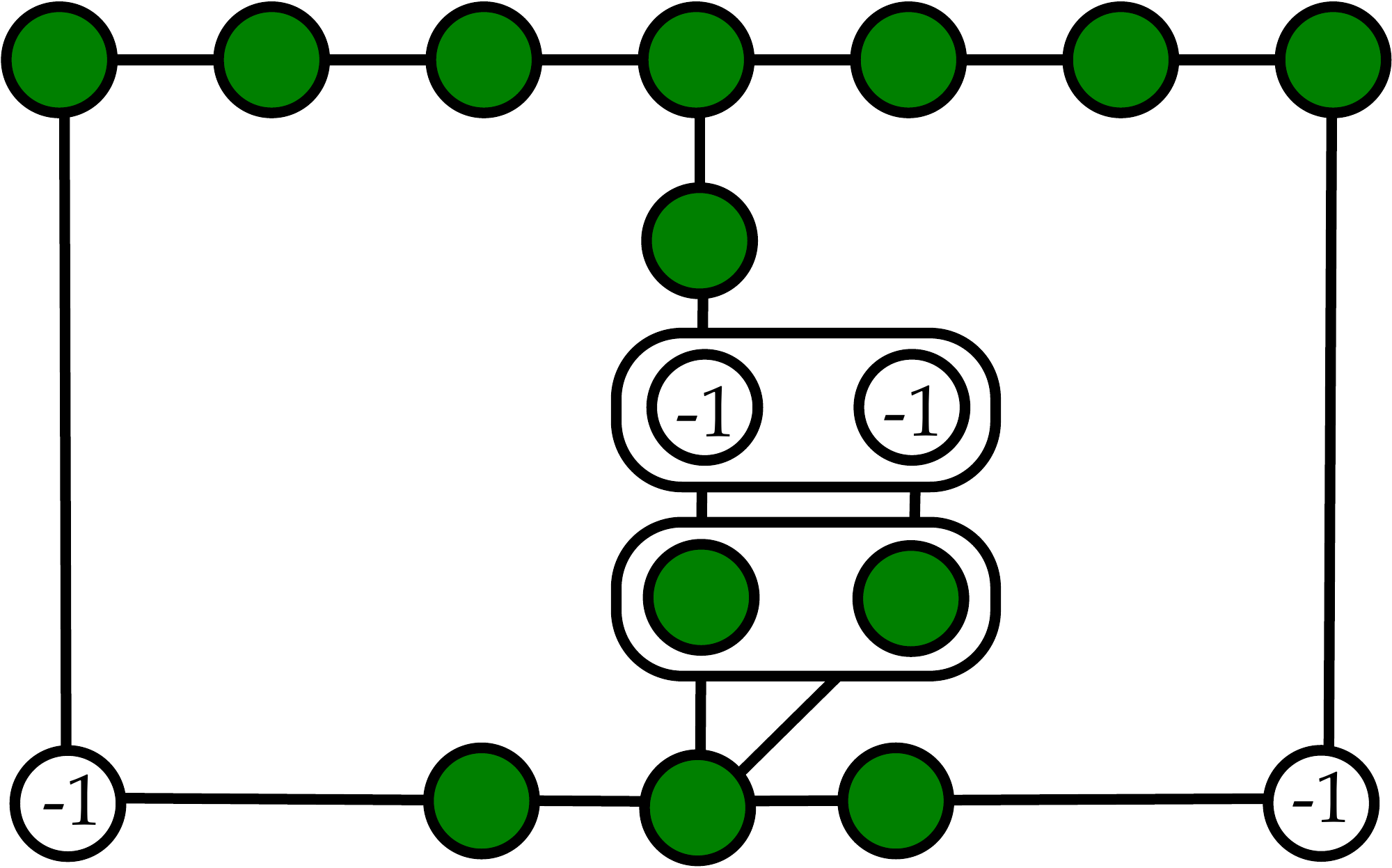} \cr \hline
$(E_8,G_2)$ & \includegraphics[height=2.5cm]{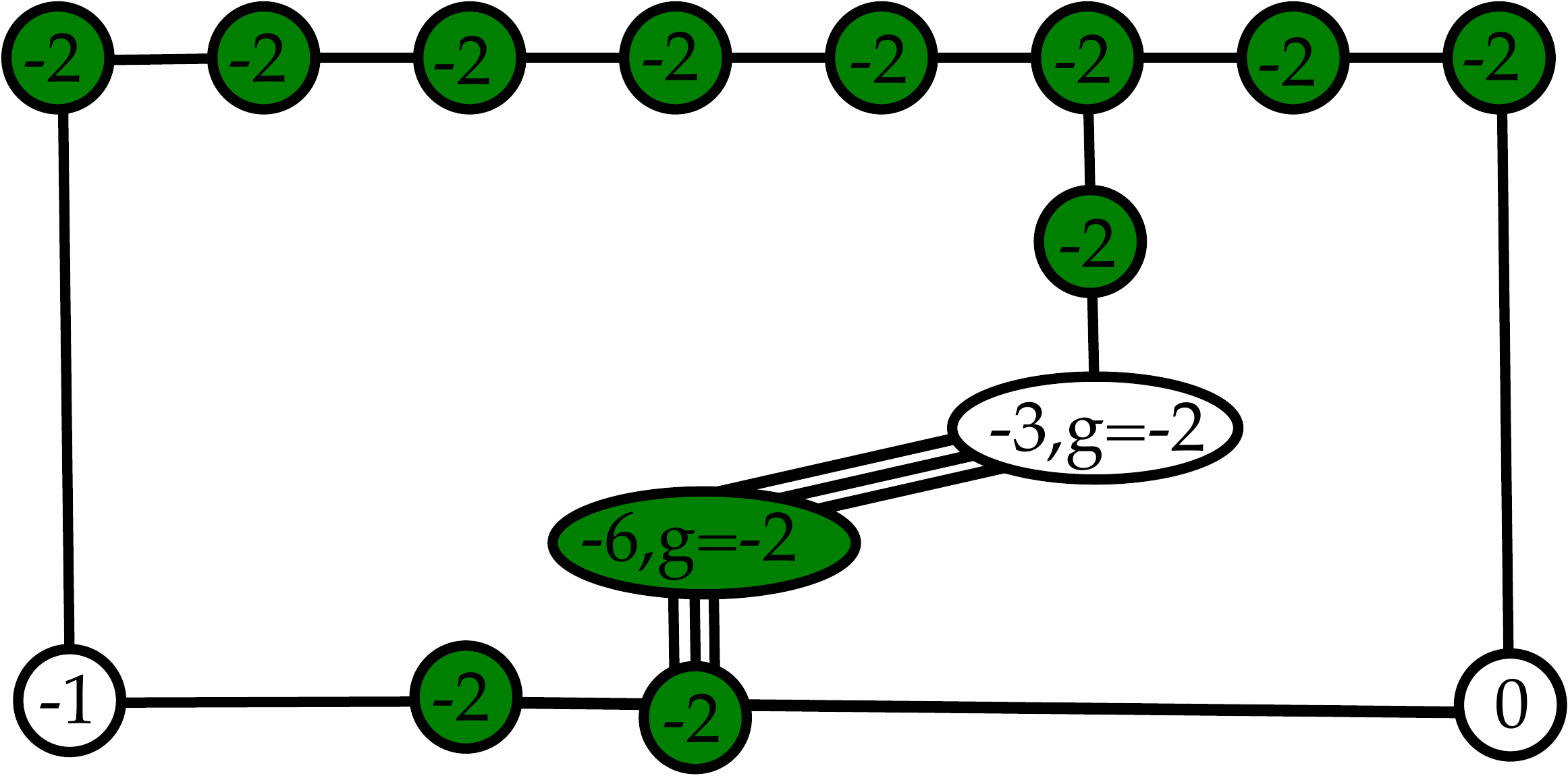}\cr \hline
\end{tabular}
\caption{CFDs for minimal conformal matter building blocks. \label{tab:MinCM}}
\end{table}

\begin{enumerate}
\item $(E_6,E_6)$:\\
In this case, there is a non-trivial intrinsic multiplicity factor $\xi=1$ for the three $(n,g)=(-1,0)$ vertices. In the sub-marginal CFD generated after one CFD transition, the two new $(-1,0)$ vertices both have $\xi=1$ too.

\item $(E_7,E_7)$:\\
In this case, there is a non-trivial intrinsic multiplicity factor $\xi=2$ for the two $(n,g)=(-1,0)$ vertices. In the sub-marginal CFD generated after one CFD transition, the two new $(-1,0)$ vertices both have $\xi=2$ as well.

\item $(E_8,E_8)$:\\
In this case, there is a non-trivial intrinsic multiplicity factor $\xi=4$ for the $(n,g)=(-1,0)$ vertex that connects the two affine node of $\hat{E}_8$. In the sub-marginal CFD generated after one CFD transition, the two new $(-1,0)$ vertices both have $\xi=4$ as well.

\item $(E_8,SU(2k))$:\\
Similar to the rank-two E-string case in section~\ref{sec:CFD-geo}, one can choose to draw a ``green $(-1)$-node'' that still contributes to the non-Abelian flavor symmetry. Alternatively, one can draw a $(n,g)=(-2,0)$ node instead, but the $(n,g)=(-1,0)$ node will become an ``interpolating node'' that connects to the $E_8$ node at one edge but connects to the $SU(2)$ node at two edges.

\item $(E_7,SO(7))$: \\
Note that there is a green node with $(n,g)=(-4,-1)$, which is a linear combination of two disjoint $(n,g)=(-2,0)$ nodes. In the Calabi-Yau threefold geometry, they correspond to two homologous $\mc{O}\oplus\mc{O}(-2)$ curves. It can also be interpreted as a $(-2)$-curve with multiplicity two. 

Above this node, there is an interpolating $(n,g)=(-2,-1)$ node with a similar property. It connects to the $E_7$ node with one edge but connects to the $SO(7)$ node with two edges. After the CFD transition where it is removed, the $E_7$ node above it will become an $(n,g)=(-1,0)$ node, while the $SO(7)$ node below it becomes an $(n,g)=(-2,-1)$ node.

\item $(E_8,G_2)$:\\
Similar to the $(E_7,SO(7))$ case, there is a green node with $(n,g)=(-6,-2)$, which is a linear combination of three disjoint $(n,g)=(-2,0)$ nodes, or a $(-2)$-curve with multiplicity three. Above that node, there is an interpolating $(n,g)=(-3,-2)$ node that is three copies of a $(n,g)=(-1,0)$ nodes. After the CFD transition where it is removed, the $E_8$ node above it will become an $(n,g)=(-1,0)$ node, while the $G_2$ node below it becomes an $(n,g)=(-3,-2)$ node.

\end{enumerate}

\section{Geometry for NHCs}
\label{app:NHCGeo}

\subsection{$(-4)$ with $SO(8)$}

For a single $(-4)$ curve, the 6d non-Higgsable gauge group is $SO(8)$. In the marginal geometry, the five surface components are arranged as:
\be
\begin{array}{ccccc}
& & \mb{F}_2(u_2) & &\\
& & \vert & &\\
\mb{F}_2(u_1) & - & \mb{F}_0(U) & - & \mb{F}_2(u_3)\,,\\
& & \vert & &\\
& & \mb{F}_2(u_4) & &\\
\end{array}
\ee

Here $u_4$ denotes the affine node.

We plot the curve configurations on the surfaces here:

\be
\label{4-topresol}
\includegraphics[width=10cm]{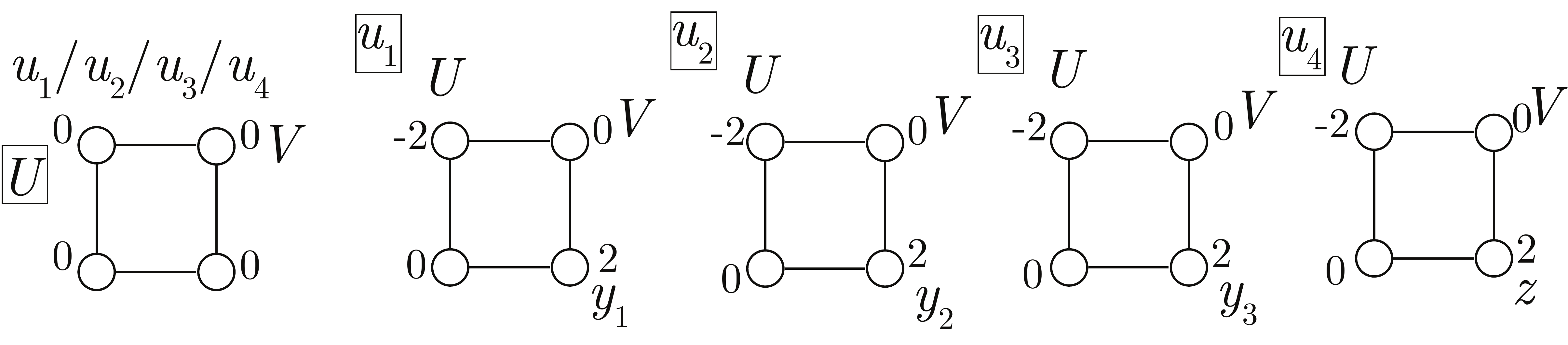}.
\ee

$V$, $z$ and $y_i$ $(i=1,\dots,3)$ are non-compact surfaces. Before the decoupling, the CFD is read off to be:

\be
\includegraphics[height=2.2cm]{NHC-CFD-4-pre.pdf},
\ee
where the middle node with genus $g=1$ corresponds to the non-compact surface $V$ and the four $(+2)$-nodes correspond to $z$ and $y_i$.

To get a 5d SCFT, we decompactify the surface component $u_4$. Then the remaining compact surfaces are three $\mb{F}_2$ with one $\mb{F}_1$, with the following triple intersection numbers:
\be
U^3=u_1^3=u_2^3=u_3^3=8\ ,\ u_1^2 U=u_2^2 U=u_3^2 U=0\ ,\ U^2 u_1=U^2 u_2=U^2 u_3=-2.
\ee

The compact surface components are connected via the sections of the $\mb{P}^1$ fibration on the Hirzebruch surface components, and there are no $\mc{O}(-1)\oplus\mc{O}(-1)$ curves in the geometry. From this, we conclude that the 5d gauge theory description is a pure $SO(8)$ gauge theory. We can read off the following CFD from the geometry, where the three $(+2)$-curves correspond to $y_1$, $y_2$, $y_3$ and the 0-curve corresponds to $V$:

\be
\includegraphics[height=2.2cm]{NHC-CFD-4.pdf}.
\ee

\subsection{$(-5)$ with $F_4$}

For a single $(-5)$ curve, the 6d non-Higgsable gauge group is $F_4$. In the marginal geometry, the five surface components are arranged as:
\be
\begin{array}{ccccccccc}
\mb{F}_3(U) & - & \mb{F}_1(u_1) & - & \mb{F}_1(u_2) & = & \mb{F}_6(u_3) & - & \mb{F}_8(u_4)
\end{array}
\ee
$U$ is the affine Cartan divisor that will be decompactified.

The curve configuration is
\be
\label{5-topresol}
\includegraphics[width=11cm]{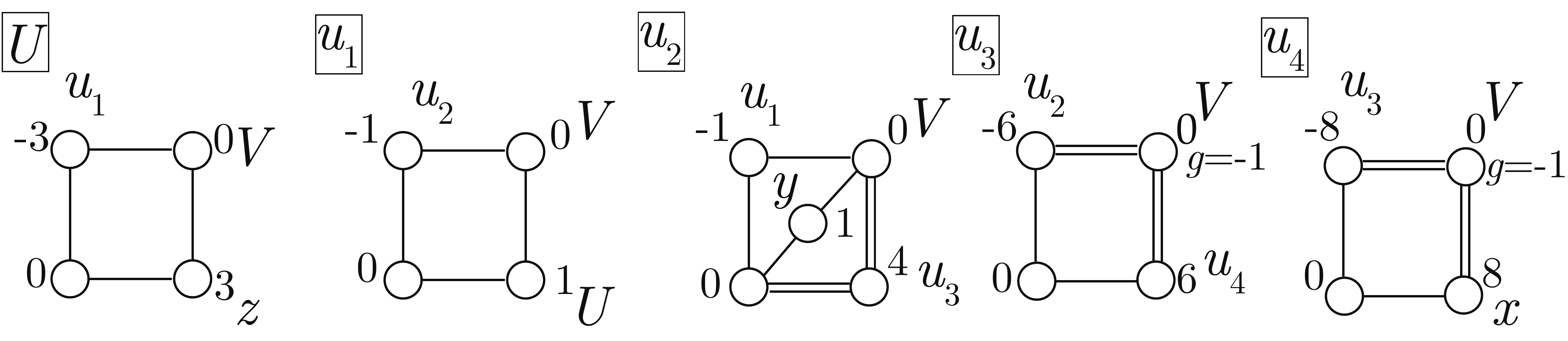}.
\ee
Note that on $u_3$ and $u_4$, the intersection curve with non-compact divisor $V$ is a reducible 0-curve with two identical fiber components. $V$, $x$, $y$ and $z$ are non-compact surfaces. Before the decoupling, the CFD is given by:

\be
\includegraphics[height=1.5cm]{NHC-CFD-5-pre.pdf},
\ee

where the $(+3)$-node corresponds to $z$, the middle $(n,g)=(0,1)$ node corresponds to $V$ and the $(+8)$-node corresponds to $y$. From the geometry of compact surface $u_4$ in (\ref{5-topresol}), we can clearly see that the $y$-node should connect to the $V$ node via a double line.

After we decompactify $U$, the remaining compact surfaces have the following triple intersection numbers:
\be
u_1^3=u_2^3=u_3^3=u_4^3=8\ ,\ u_1^2 u_2=u_2^2 u_1=-1\ ,\ u_2^2 u_3=-6\ ,\ u_3^2 u_2=4\ ,\ u_3^2 u_4=-8\ ,\ u_4^2 u_3=6.
\ee

After the ruling curves on $u_i$ $(i=1,\dots,4)$ are shrunk to zero size, the intersection matrix between the ruling curves and surface components on each surface component are:
\be
F_i\cdot S_j=\bp -2 & 1 & 0 & 0\\ 1 & -2 & 2 & 0 \\ 0 & 1 & -2 & 1\\ 0 & 0 & 1 & -2\ep,
\ee
which is the $(-\mc{C})_{ij}$ of Lie algebra $F_4$. Hence the 5d gauge theory description of this geometry is a pure $F_4$ gauge theory. The CFD can be read off as follows, where the $(+1)$-curve corresponds to $U$, the 0-curve corresponds to $V$ and the 8-curve corresponds to $y$:
\be
\includegraphics[height=1.5cm]{NHC-CFD-5.pdf}.
\ee

\subsection{$(-6)$ with $E_6$}

For a single $(-6)$ curve, the 6d non-Higgsable gauge group is $E_6$. In the marginal geometry, the seven surface components are arranged as:
\be
\begin{array}{ccccccccc}
& & & & \mb{F}_4(u_6) & & & & \\
& & & & \vert & & & & \\
& & & & \mb{F}_2(u_5) & & & & \\
& & & & \vert & & & & \\
\mb{F}_4(U) & - & \mb{F}_2(u_1) & - & \mb{F}_0(u_2) & - & \mb{F}_2(u_3) & - & \mb{F}_4(u_4)
\end{array}
\ee
$U$ is the affine Cartan divisor that will be decompactified.

The curve configuration is
\be
\label{6-topresol}
\includegraphics[width=10cm]{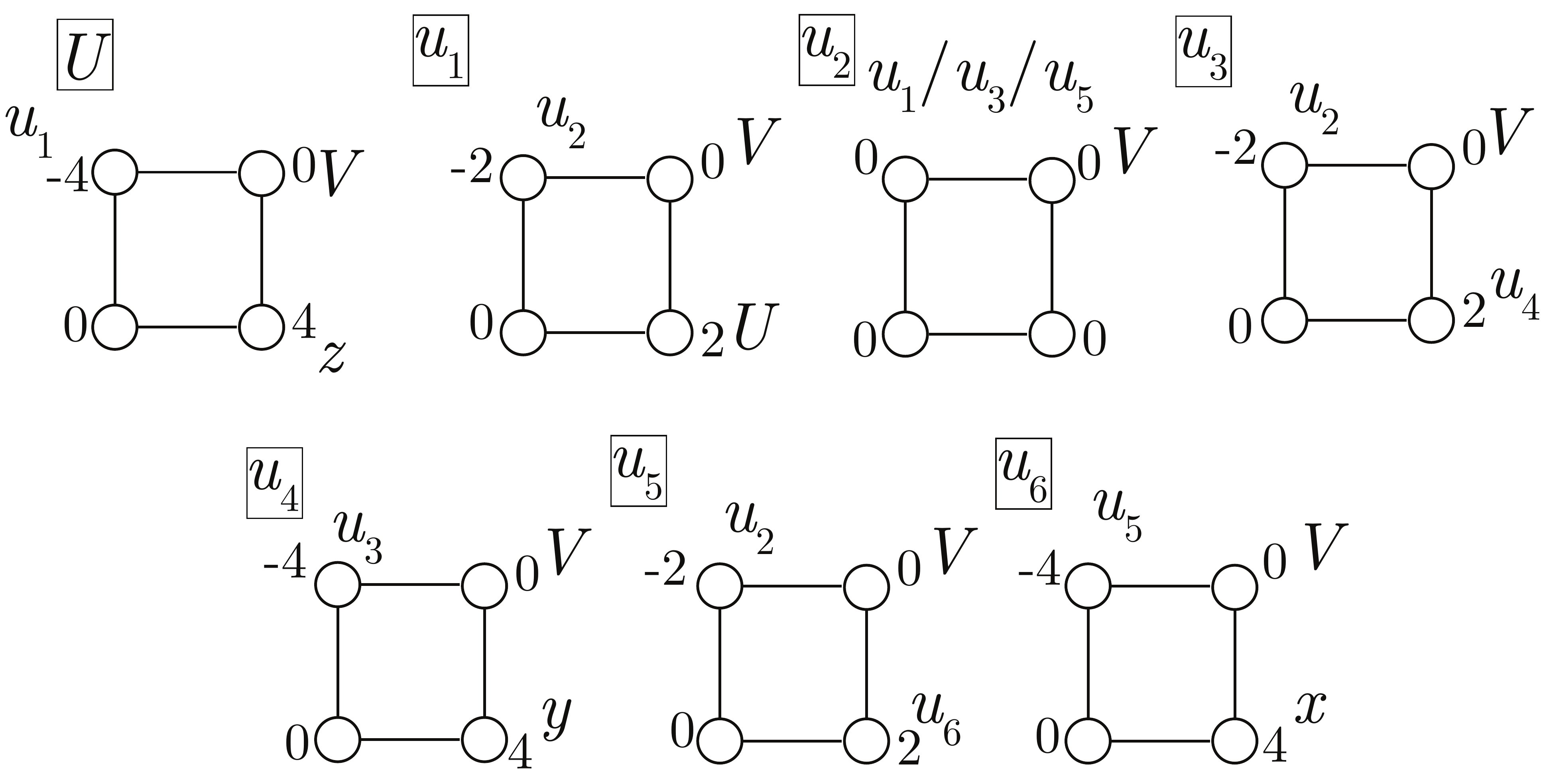}.
\ee
Before the decoupling, the CFD is read off as:
\be
\includegraphics[height=2cm]{NHC-CFD-6-pre.pdf},
\ee
where the middle node corresponds to the non-compact surface $V$ and the $(+4)$-nodes correspond to non-compact surfaces $x$, $y$, $z$.

After we decompactify $U$, the remaining compact surfaces have the following triple intersection numbers:
\be
\ba
&u_1^3=u_2^3=u_3^3=u_4^3=u_5^3=u_6^3=8\ ,\ u_1^2 u_2=u_3^2 u_2=u_5^2 u_2=0\ ,\ u_2^2 u_1=u_2^2 u_3=u_2^2 u_5=-2\ ,\ \cr
&u_6^2 u_5=u_4^2 u_3=2\ ,\ u_5^2 u_6=u_3^2 u_4=-4.
\ea
\ee

After the ruling curves on $u_i$ $(i=1,\dots,4)$ are shrunk to zero size, the 5d gauge theory description is a pure $E_6$ gauge theory since there is no $\mc{O}(-1)\oplus\mc{O}(-1)$ curve. The CFD can be read off as follows, where the $(+2)$-curve corresponds to $U$, the 0-curve corresponds to $V$ and the $(+4)$-curves correspond to $x$, $y$:
\be
\includegraphics[height=2cm]{NHC-CFD-6.pdf}.
\ee

\subsection{$(-7)$ with $E_7+\frac{1}{2}\mbf{56}$}

For a single $(-7)$ curve, the 6d non-Higgsable gauge group is $E_7$ with a half-hypermultiplet in the fundamental representation $\mbf{56}$ of $E_7$. This case is different from others in the sense that the surface components have more connections than the affine $E_7$ Dynkin diagram.

The curve configurations on the eight surface components $U,u_i$ $(i=1,\dots,7)$ are
\be
\label{7-topresol}
\includegraphics[width=10cm]{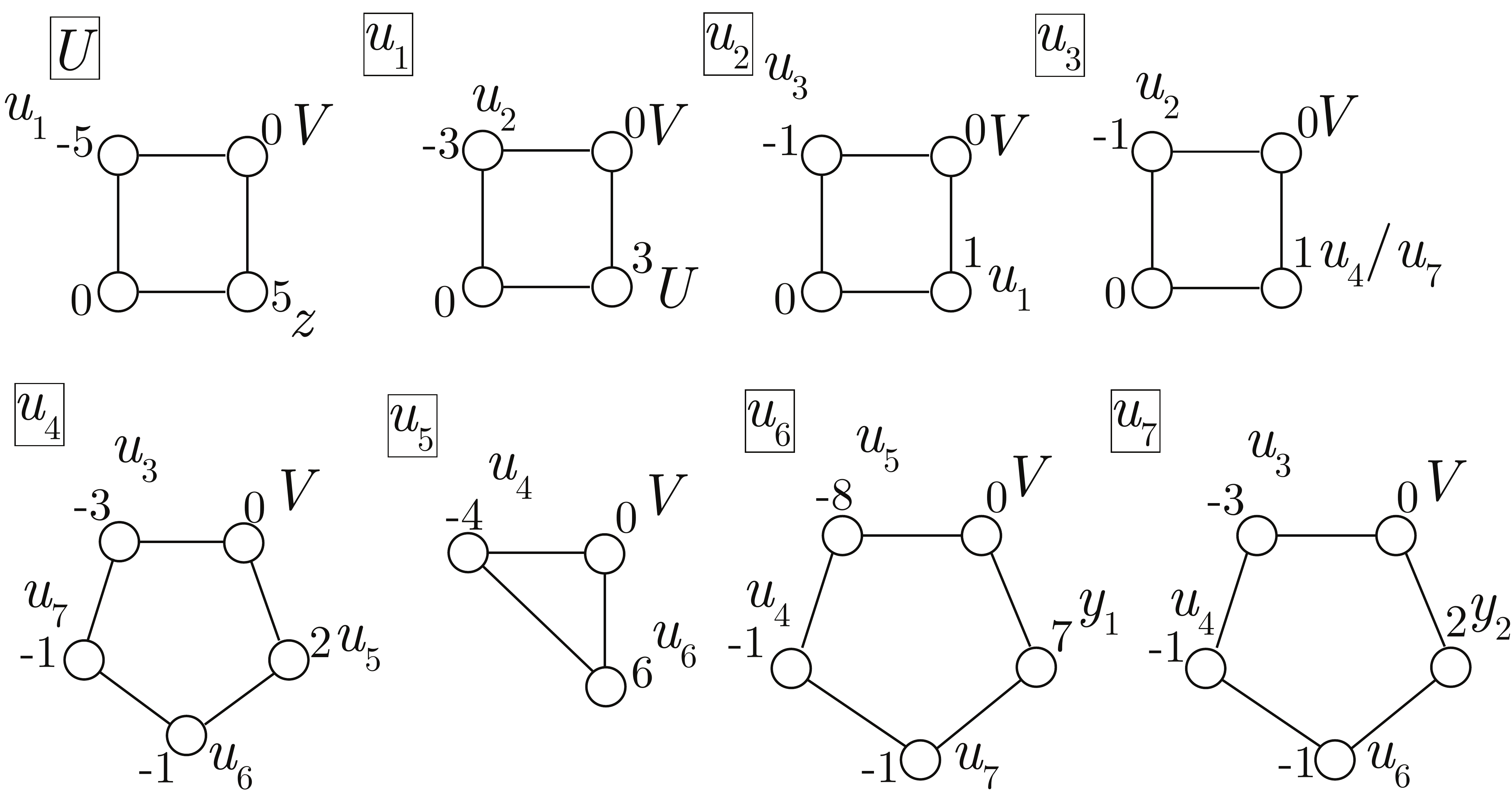}.
\ee
As one can see, $U$, $u_1$, $u_2$, $u_3$ and $u_5$ are $\mb{F}_5$, $\mb{F}_3$, $\mb{F}_1$, $\mb{F}_1$ and $\mb{F}_4$ respectively, while $u_4$, $u_6$ and $u_7$ are blow ups of Hirzebruch surfaces. They are arranged into the affine $E_7$ Dynkin diagram as:

\be
\begin{array}{ccccccccccccc}
& & & & & & Blp_1\mb{F}_3(u_7) & & & & & &\\
& & & & & & \vert & & & & & &\\
\mb{F}_5(U)&-&\mb{F}_3(u_1)&-&\mb{F}_1(u_2)&-&\mb{F}_1(u_3) &- &Blp_1\mb{F}_3(u_4)&-&\mb{F}_4(u_5)&-&Blp_1\mb{F}_8(u_6)
\end{array}
\ee

where $U$ is the affine node. Before the decoupling, the CFD is read off as:
\be
\includegraphics[height=2cm]{NHC-CFD-7-pre.pdf},
\ee
where the middle node corresponds to $V$, the $(+5)$-node corresponds $z$, the $(+7)$-node corresponds to $y_1$ and the $(+2)$-node corresponses to $y_2$.

After we decompactify $U$, the remaining compact surfaces have the following triple intersection numbers:
\be
\ba
&u_1^3=u_2^3=u_3^3=u_5^3=u_6^3=8\ ,\ u_4^3=u_6^3=u_7^3=7\ ,\ u_2^2 u_1=-3\ ,\ u_3^2 u_2=-1\ ,\ u_1^2 u_2=1\ ,\ \cr
&u_2^2 u_3=-1\ ,\ u_4^2 u_3=u_7^2 u_3=1\ ,\ u_3^2 u_4=-3\ ,\ u_6^2 u_4=u_7^2 u_4=-1\ ,\ u_4^2 u_5=-4\ ,\ u_6^2 u_5=6\ ,\ \cr
&u_5^2 u_6=-8\ ,\ u_4^2 u_6=u_7^2 u_6=-1\ ,\ u_5^2 u_7=-3\ ,\ u_4^2 u_7=u_6^2 u_7=-1
\ea
\ee
The ruling structures of the surface components are
\be
\ba
f(u_1)&=f(u_2)=f(u_3)=f(u_5)=V\cr
f(u_4)&=V=u_6+u_7\cr
f(u_6)&=V=u_4+u_7\cr
f(u_7)&=V=u_4+u_6\,.
\ea
\ee
The 5d gauge group is $E_7$ after the ruling curves are shrunk to zero size. In this case, there are three $\mc{O}(-1)\oplus\mc{O}(-1)$ curves in the ruling: 
\be
\ba
C_1&=u_4\cdot u_7\cr
C_2&=u_4\cdot u_6\cr
C_3&=u_6\cdot u_7.
\ea
\ee 
Their charge $C_i\cdot u_j$ under the Cartans of $E_7$ are weight vectors in the $\mbf{56}$ representation of $E_6$. However, we cannot flop them out of the compact surface, since the surface components $u_4$, $u_6$ and $u_7$ cannot be blown down twice. From these information, we conclude that the 5d gauge theory description should be $E_7+\frac{1}{2}\mbf{56}$.

The CFD can be read off as:
\be
\includegraphics[height=2cm]{NHC-CFD-7.pdf}.
\ee

\subsection{$(-8)$ with $E_7$}

For a single $(-8)$ curve, the 6d non-Higgsable gauge group is $E_7$ and there is no matter field. In the marginal geometry, the eight surface components are arranged as:
\be
\begin{array}{ccccccccccccc}
& & & & & & \mb{F}_2(u_7) & & & & & &\\
& & & & & & \vert & & & & & &\\
\mb{F}_6(U)&-&\mb{F}_4(u_1)&-&\mb{F}_2(u_2)&-&\mb{F}_0(u_3) &- &\mb{F}_2(u_4)&-&\mb{F}_4(u_5)&-&\mb{F}_6(u_6)
\end{array}
\ee

The configuration of curves are:
\be
\label{8-topresol}
\includegraphics[width=10cm]{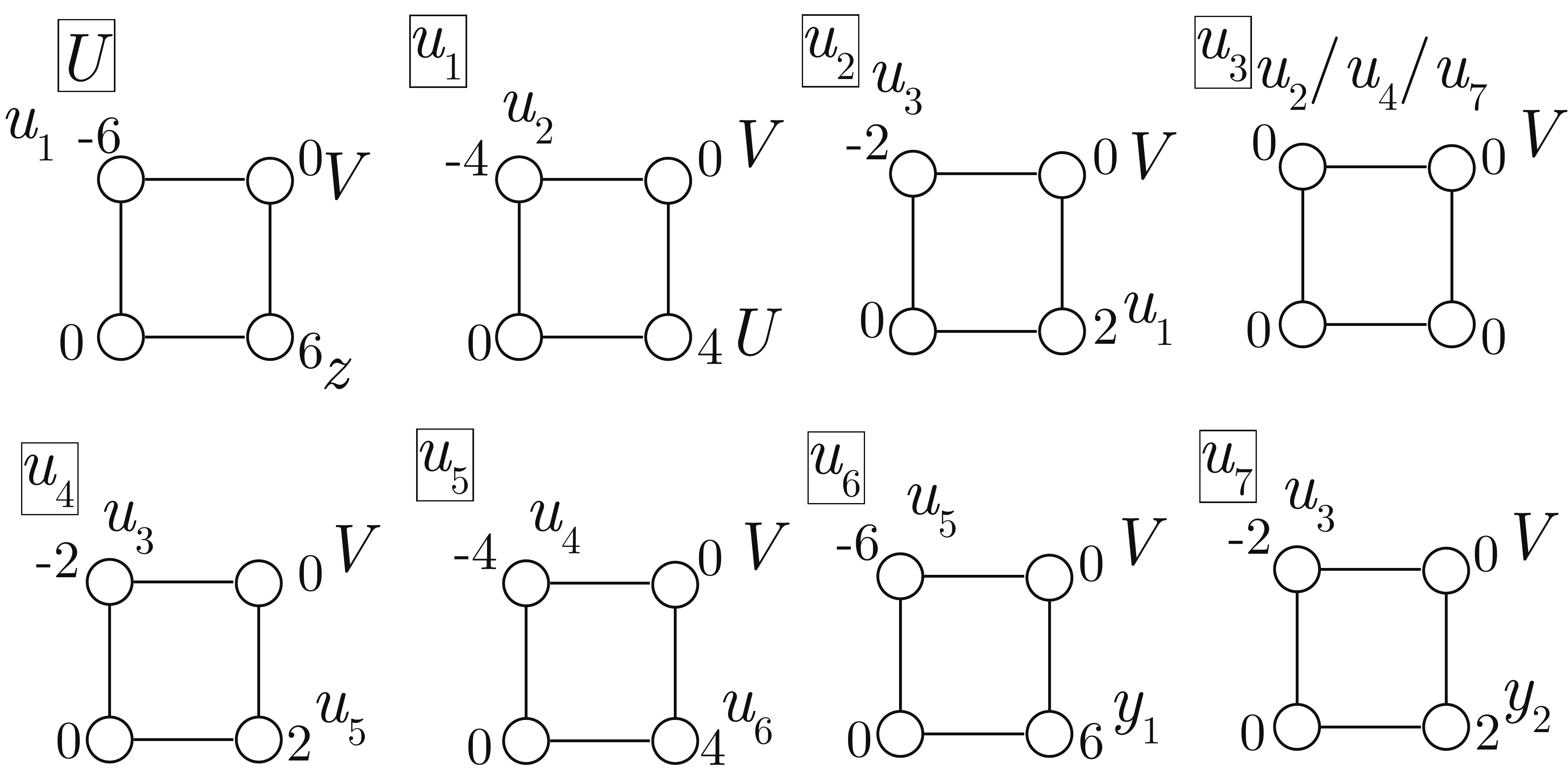}.
\ee

Before the decoupling, the CFD is read off as:
\be
\includegraphics[height=2cm]{NHC-CFD-8-pre.pdf},
\ee
where the middle node corresponds to $V$, the $(+2)$-node corresponds to $y_2$ and the $(+6)$-nodes correspond to $z$ and $y_1$.

After we decompactify $U$, the remaining compact surfaces have the following triple intersection numbers:
\be
\ba
&u_1^3=u_2^3=u_3^3=u_4^3=u_5^3=u_6^3=u_7^3=8\ ,\ u_2^2 u_1=-4\ ,\ u_3^2 u_2=-2\ ,\ u_1^2 u_2=2\ ,\ \cr
&u_2^2 u_3=u_4^2 u_3=u_7^2 u_3=0\ ,\ u_3^2 u_4=-2\ ,\ u_5^2 u_4=2\ ,\ u_4^2 u_5=-4\ ,\ u_6^2 u_5=4\ ,\ u_5^2 u_6=-6\ ,\ \cr
&u_3^2 u_7=-2
\ea
\ee

The assignment of sections on each surface component is the same as the $(-7)$ case. Along with the fact that there is no $\mc{O}(-1)\oplus\mc{O}(-1)$ curve in the geometry, the 5d gauge theory description is a pure $E_7$ gauge theory. The CFD is read off as:

\be
\includegraphics[height=2cm]{NHC-CFD-8.pdf}.
\ee

\subsection{$(-12)$ with $E_8$}

For a single $(-12)$ curve, the 6d non-Higgsable gauge group is $E_8$ and there is no matter field. In the marginal geometry, the nine surface components are arranged as:
\be
\begin{array}{ccccccccccccccc}
& & & & & & & & & & \mb{F}_2(u_8) & & & &\\
& & & & & & & & & & \vert & & & &\\
\mb{F}_{10}(U)&-&\mb{F}_8(u_1)&-&\mb{F}_6(u_2)&-&\mb{F}_4(u_3) &- &\mb{F}_2(u_4)&-&\mb{F}_0(u_5)&-&\mb{F}_2(u_6)&-&\mb{F}_4(u_7)
\end{array}
\ee

The configuration of curves are:
\be
\label{12-topresol}
\includegraphics[width=11cm]{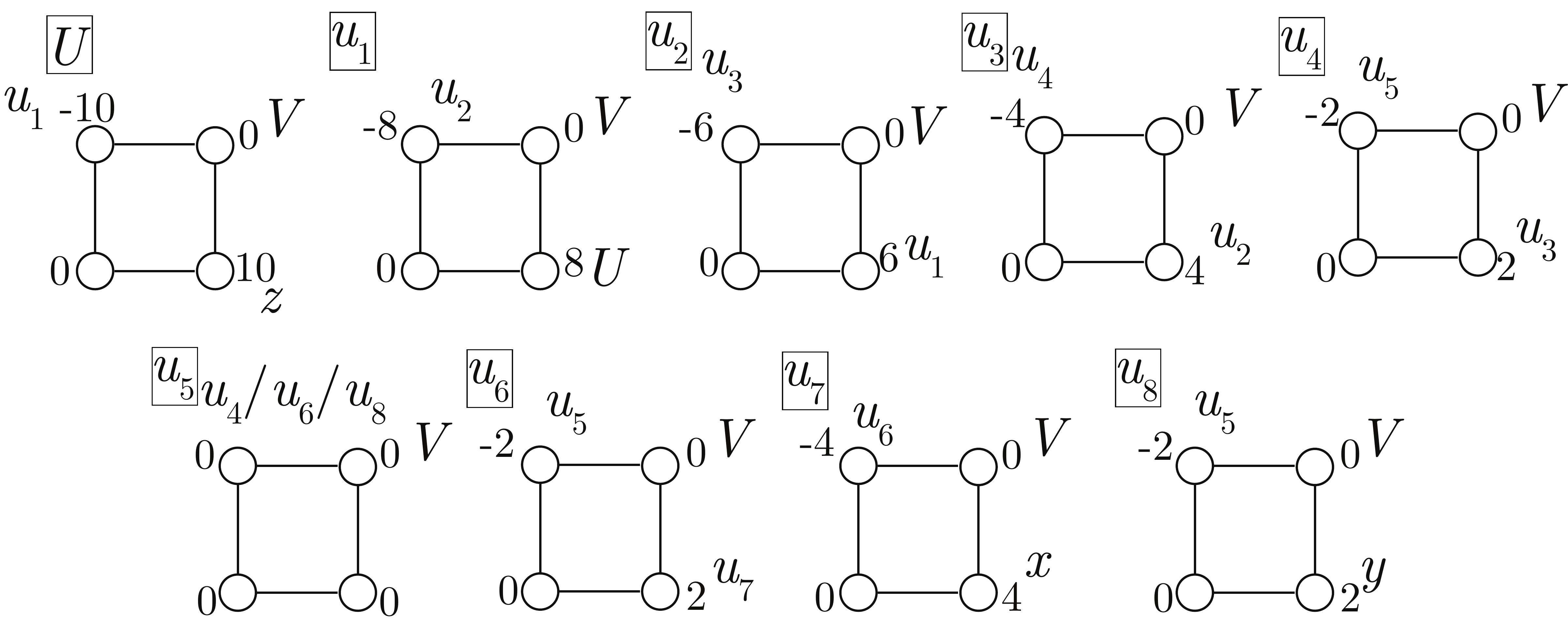}.
\ee
Before the decoupling, the CFD is read off as:
\be
\includegraphics[height=2cm]{NHC-CFD-12-pre.pdf},
\ee
where the middle node corresponds to $V$, the $(+10)$-node corresponds to $z$, the $(+2)$-node corresponds to $y$ and the $(+4)$-node corresponds to $x$.

After we decompactify $U$, the remaining compact surfaces have the following triple intersection numbers:
\be
\ba
&u_1^3=u_2^3=u_3^3=u_4^3=u_5^3=u_6^3=u_7^3=u_8^3=8\ ,\ u_2^2 u_1=-8\ ,\ u_3^2 u_2=-6\ ,\ u_1^2 u_2=6\ ,\ \cr
&u_4^2 u_3=-4\ ,\ u_2^2 u_3=4\ ,\ u_5^2 u_4=-2\ ,\ u_3^2 u_4=2\ ,\ u_4^2 u_5=u_6^2 u_5=u_8^2 u_5=0\ ,\ u_5^2 u_6=-2\ ,\ \cr
&u_7^2 u_6=2\ ,\ u_6^2 u_7=-4\ ,\ u_5^2 u_8=-2
\ea
\ee

The 5d gauge theory is a pure $E_8$ gauge theory since there is no $\mc{O}(-1)\oplus\mc{O}(-1)$ curve in the geometry. The CFD is read off as
\be
\includegraphics[height=2cm]{NHC-CFD-12.pdf}\,.
\ee

\section{Resolution Geometries for the Single Curve Building Blocks}
\label{app:geo-single}

\subsection{$SU(3)$ on $(-2)$-curve}
\label{app:SU3-2}

We first consider the following tensor branch
\be
[SU(6)]-\overset{\mathfrak{su}(3)}{2} \,.
\ee
In the resolution geometry, we denote the non-compact Cartan divisors of the $SU(6)$ by $V,v_1,\dots,v_5$ and the Cartan divisors of the $SU(3)$ by $U,U_1,U_2$. The configuration of curves on $U,u_1,u_2$ are
\be
\includegraphics[width=8cm]{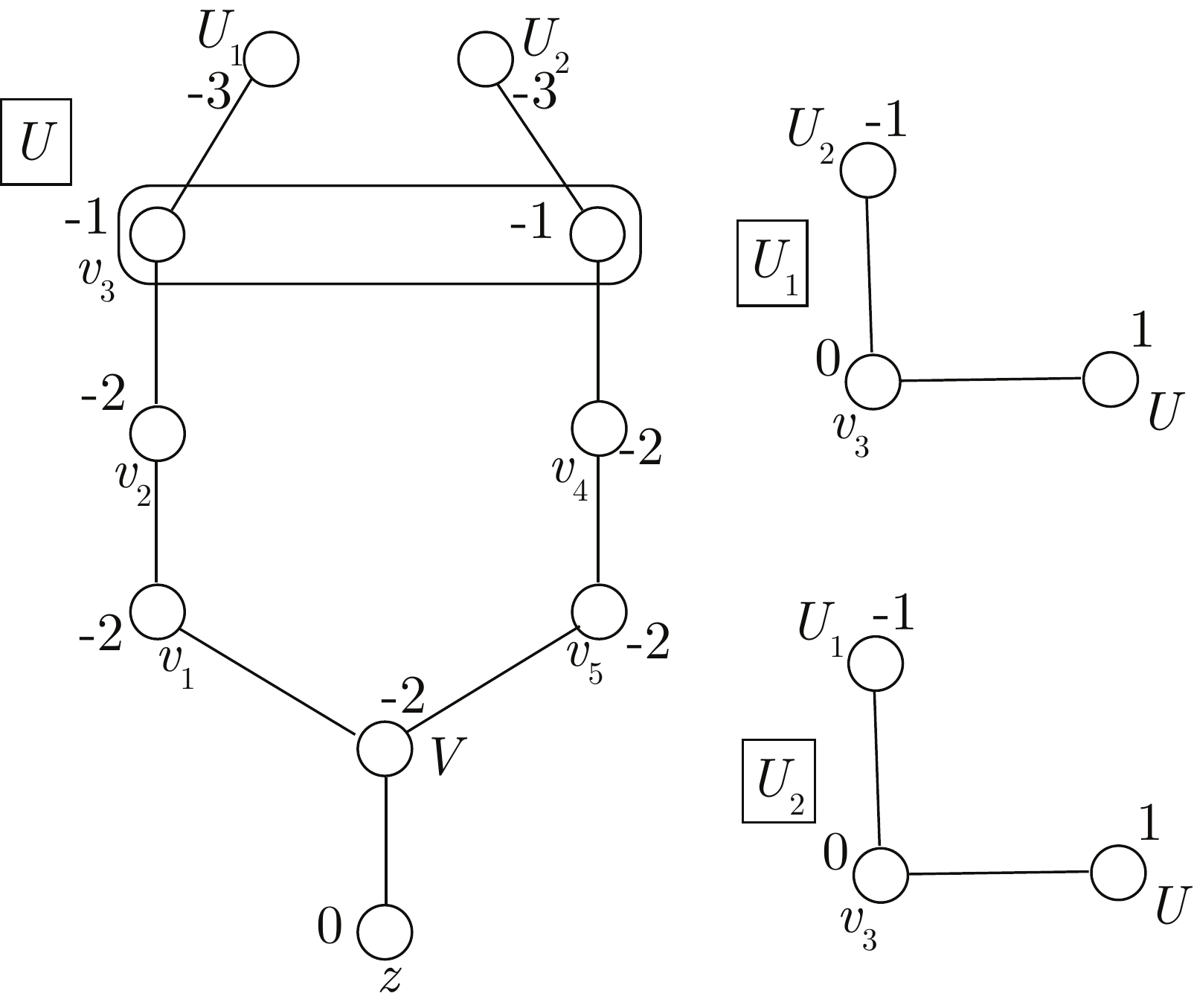} \,.
\ee
Note that the intersection curve $U\cdot v_3$ is reducible, with two $(-1)$-curve components. Note that this collection of  surfaces does not satisfy the shrinkability condition~\cite{Jefferson:2018irk}, because there exists a genus-one fibration structure where the singular fiber is a ring of three $\mb{P}^1$s and the sections are the intersection curves $U\cdot U_1$, $U\cdot U_2$, $U_1\cdot U_2$. 

To get a 5d SCFT geometry, we need to either flop curves out of these compact surfaces or decompactify a surface component. The former choice is only possible if we shrink the curve $U_1\cdot U_2$ and results in the same surface geometry $U$ with two $\mb{P}^2$s. However, this geometry has no gauge theory description either, since the surface $\mb{P}^2$ does not have a ruling structure.

For the latter choice, we can flop curves out of $U$ and then decompactify $U$, and we consequently get a theory with more descendants. We first shrink the two $(-1)$-curves that consist of $U\cdot v_3$, and then shrink $U\cdot v_2$, $U\cdot v_4$, $U\cdot v_1$ and $U\cdot v_5$ consequently.

In this process, $U$ is blown down six times, while $U_1$ and $U_2$ are blown up three times for each. The final surface geometry after the process is:

\be
\includegraphics[width=10cm]{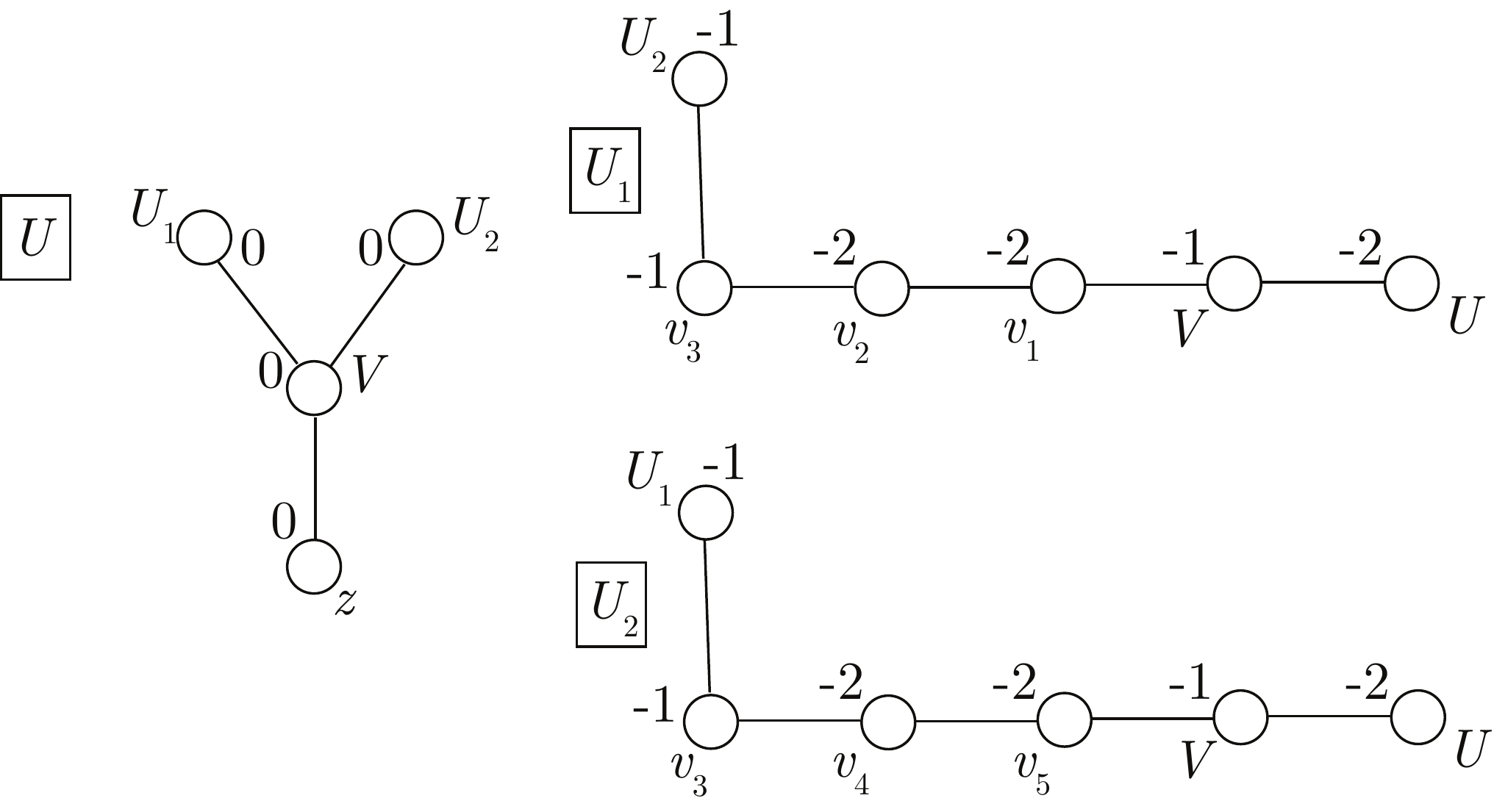}.
\ee

In this geometry, the surface $U$ is $\mb{F}_0$ and $U_1,U_2$ are two identical gdP$_4$s. After $U$ is decompactified, the $(-2)$ curves $U\cdot u_1$ and $U\cdot u_2$ are actually unrelated. Similarly, the $(-1)$-curves $V\cdot u_1$ and $V\cdot u_2$ become independent, since we can shrink one of these $(-1)$-curves without changing the geometry of the other surface component. Hence in this case, these curves should not be combined, in contrary to the usual rule of extracting CFD from the geometry. The only combined curves are $v_3\cdot u_1$ and $v_3\cdot u_2$, and the CFD is read off as:
\be
\includegraphics[width=4cm]{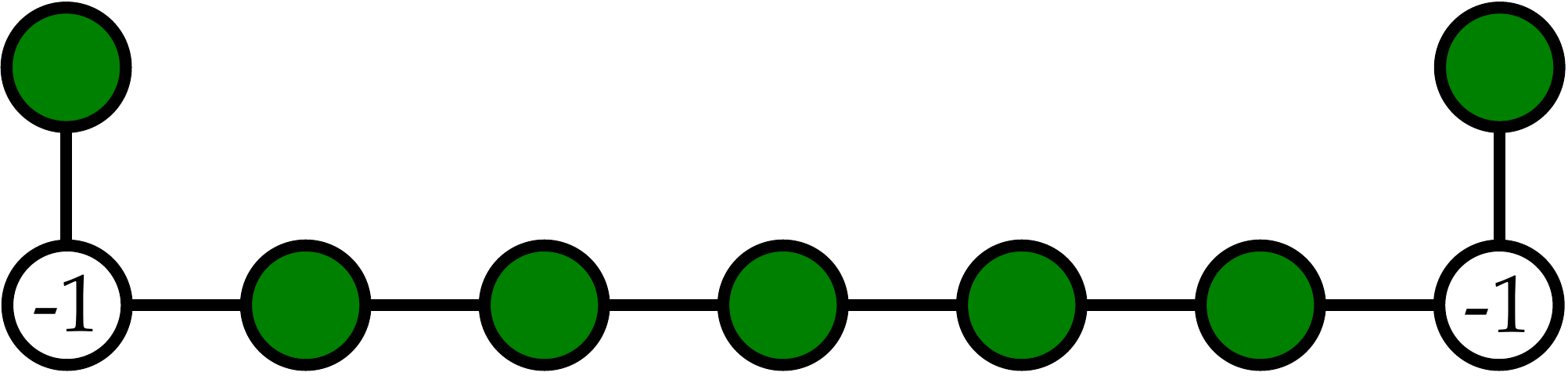},
\ee
which is exactly the CFD of the 5d rank-2 gauge theory $SU(3)_0+6\bm{F}$\cite{Apruzzi:2019opn}.

Similarly, we can study the resolution geometry for the equivalent tensor branch
\be
[SU(m)]-\overset{\mathfrak{su}(3)}{2}-[SU(6-m)]
\ee
for other $m$ as well, which gives rise to the CFDs in table~\ref{tab:Sun2CFDs}.

\subsection{$SU(N)$ on $(-2)$-curve}
\label{app:SUN-2}

We can generalize the $SU(3)$ discussions to arbitrary $SU(N)$ as well, with the tensor branch geometry 
\be
[SU(2N)]-\overset{\mathfrak{su}(N)}{2}.
\ee

In the resolution geometry, the non-compact Cartan divisors of $SU(2N)$ are $V$, $v_1,\dots,v_{2N-1}$, and the Cartan divisors of $SU(N)$ are $U$, $U_1,\dots,U_{N-1}$. The configuration of curves on $U$, $u_i$ are:

\be
\includegraphics[width=8cm]{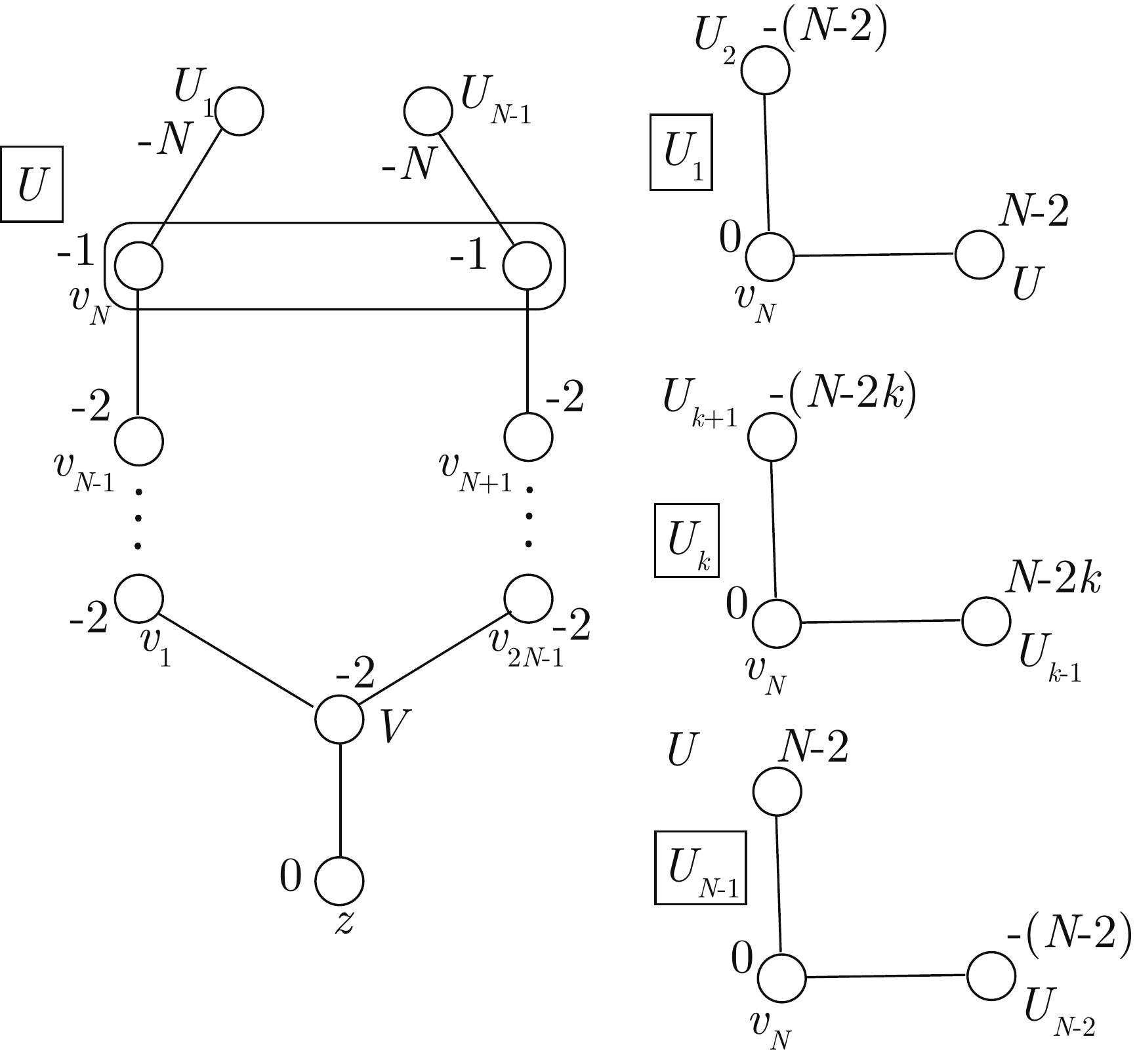}.
\ee

The intersection $U\cdot v_N$ is reducible. We can flop curves out of the affine Cartan divisor $U$ $2N$ times by shrinking $U\cdot v_N$, $U\cdot v_{N-1},\dots U\cdot v_1$, $U\cdot v_{N+1}\dots U\cdot v_{2N-1}$ consequently. As a result, the surface components $U_1$ and $U_{N-1}$ are blown up $N$ times for each, and the resulting surface geometry is 

\be
\includegraphics[width=10cm]{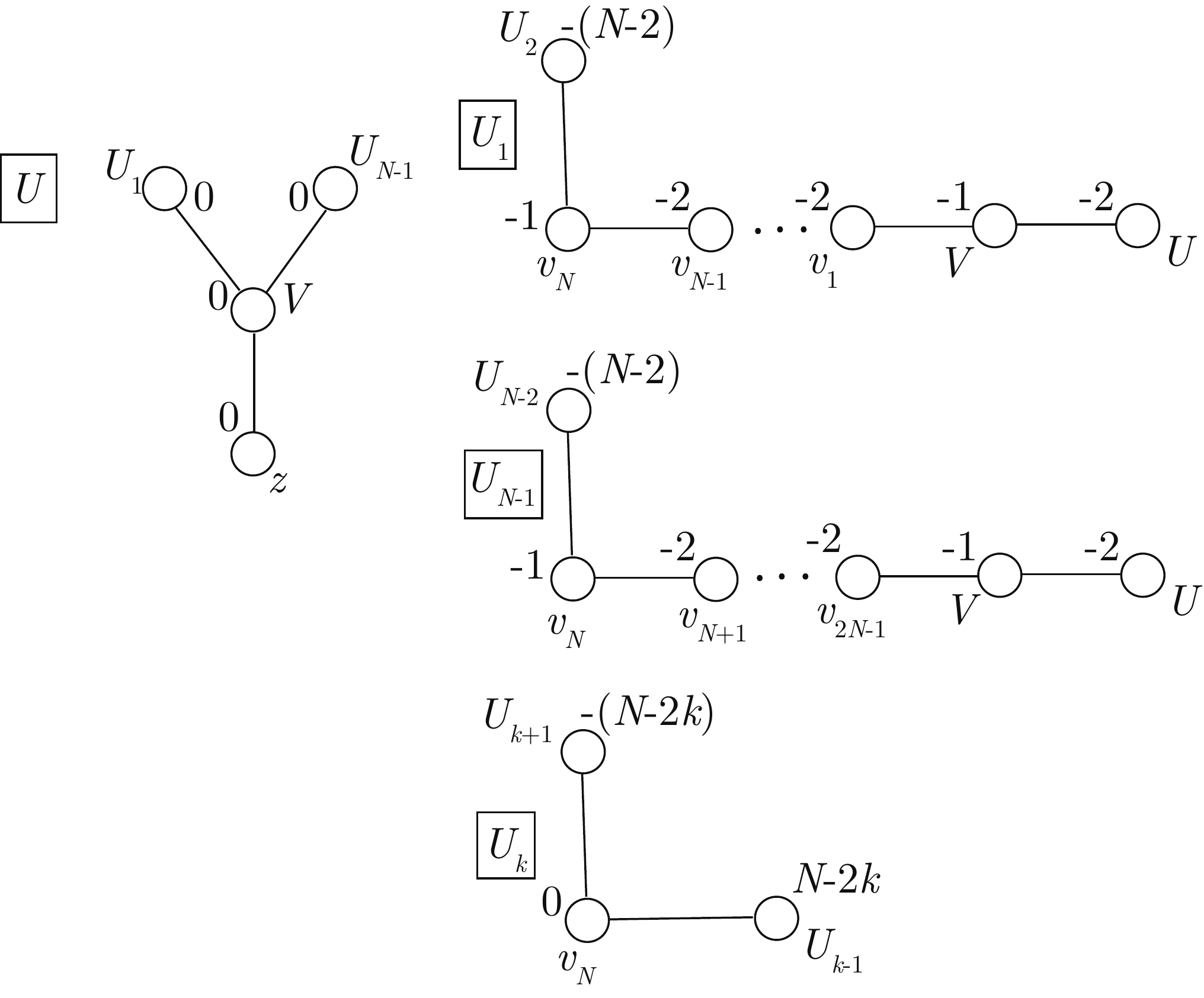}.
\ee

After $U$ is decompactified, the curves $V\cdot U_1$ and $V\cdot U_2$ can be independently shrinked. Similarly, $U\cdot U_1$ and $U\cdot U_2$ becomes independent curves. The CFD is then 
\be
\includegraphics[width=4cm]{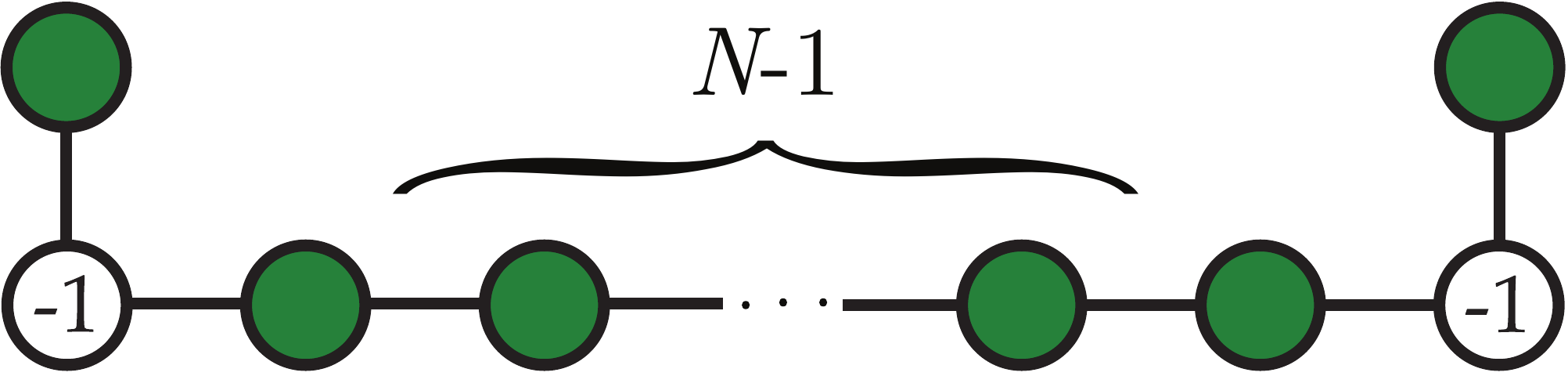},
\ee
which is consistent with the CFD for the gauge theory $SU(N)_0+2N\bm{F}$ as a descendant of the marginal $(D_{N+2},D_{N+2})$ CFD\cite{Apruzzi:2019vpe}.

For other equivalent tensor branch
\be
[SU(m)]-\overset{\mathfrak{su}(N)}{2}-[SU(2N-m)]\,,
\ee
with other $m$, a similar resolution gives rise to the CFDs in table~\ref{tab:Sun2CFDs}.

\subsection{$SO(2n)$ gauge group on $(-4)$-curve}
\label{app:SO-4}

We study the resolution geometry of the tensor branch:
\be
[Sp(2n-8)]-\overset{\mathfrak{so}(2n)}{4}.
\ee
We use the resolution sequence of $SO(2n)$ in \cite{Lawrie:2012gg}, and the resolution sequence for $Sp(2n-8)$ is given by:
\be
(x,y,V,v_1)\ ,\ (x,y,v_1,v_2)\ ,\dots, (x,y,v_{2n-9},v_{2n-8})
\ee

Denote the Cartan divisors of $SO(2n)$ by $U,U_1,U_2,\dots,U_n$, the configuration of curves on each surface component  are:
\be
\includegraphics[width=13cm]{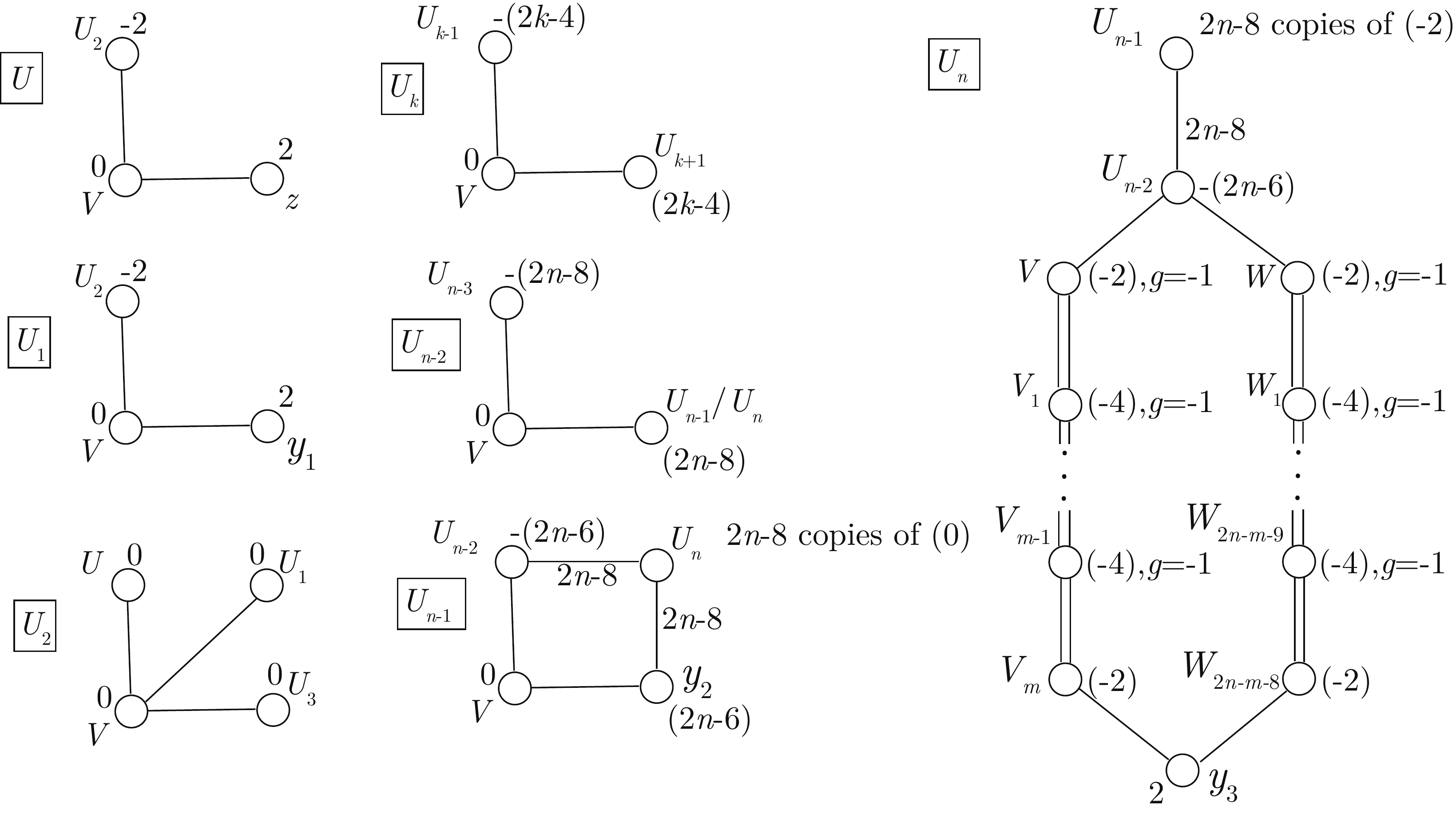}.
\ee
Note that the curves with $g=-1$ are a double copy of a rational curve on the surface. Moreover, the intersection curve $U_{n-1}\cdot U_n$ consists of $(2n-8)$ copies of rational curve. The intersection relations among surface components are:
\be
\includegraphics[width=6cm]{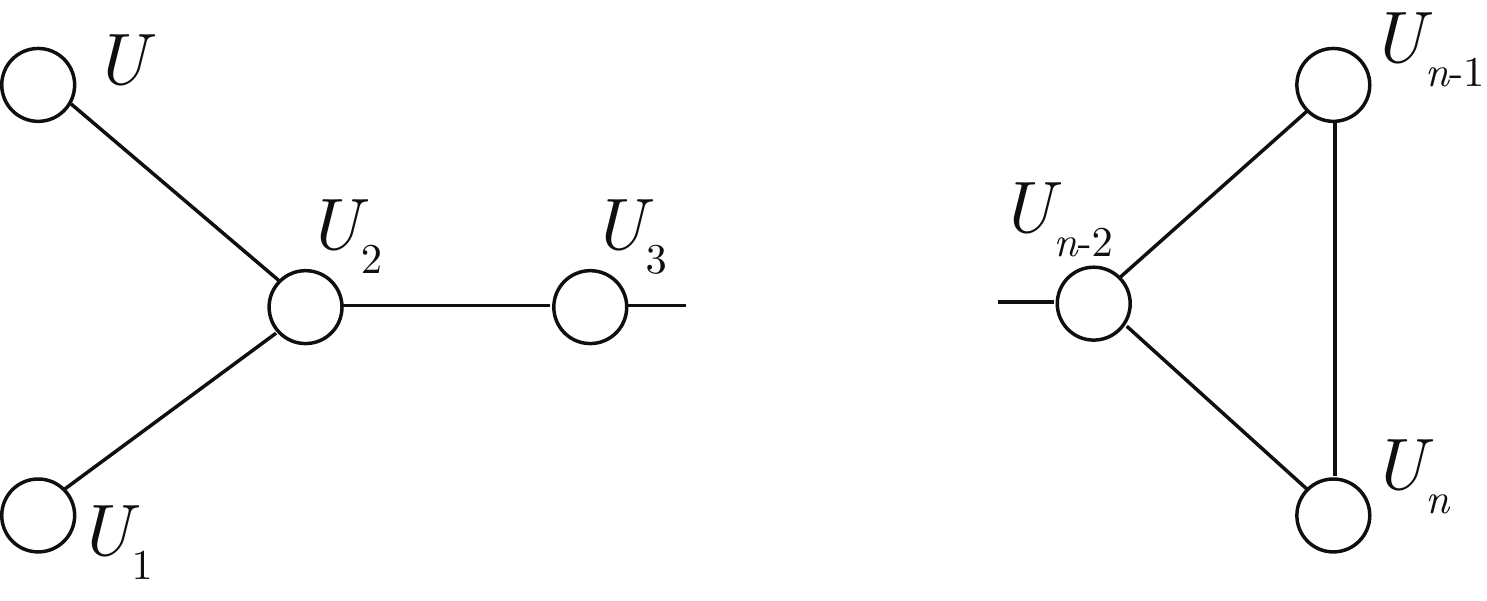},\label{f:4-SO(2n)surfaces}
\ee
which is consistent with the geometric picture in \cite{Bhardwaj:2018vuu}. 

To get a 5D SCFT, we need to decompactify the surface $U$, and the CFD can be read off as the one in  table~\ref{tab:So2n4CFDs}. Note that the vertices $V,V_1,\dots,V_{2n-8}$ form the BG-CFD of $Sp(2n-8)$, see table~\ref{tab:BGCFDs}. The nodes with $n>0$ are given by $U$, $y_1$, $y_2$ and $y_3$.

We can similarly work out the resolution geometry for
\be
[Sp(m)]-\overset{\mathfrak{so}(2n)}{4}-[Sp(2n-m-8)]
\ee
as well.


\section{$(-1)$-NHC Gluing From Geometry}
\label{app:Geo-1-n}

In this appendix we derive the CFDs from the tensor branch geometry. 

\subsection*{$(-1)(-3)$: $G=SU(3)$}

In the tensor branch resolution, we denote the non-compact Cartan divisors of $E_6$ by $V,v_1,v_2,\dots,v_6$, the compact vertical divisor over $(-1)$-curve by $S$ and the compact Cartan divisors of $SU(3)$ by $U,u_1,u_2$. The curve configurations on the compact surfaces are:
\be
\label{f:1-3-gluing-topresol}
\includegraphics[width=7.5cm]{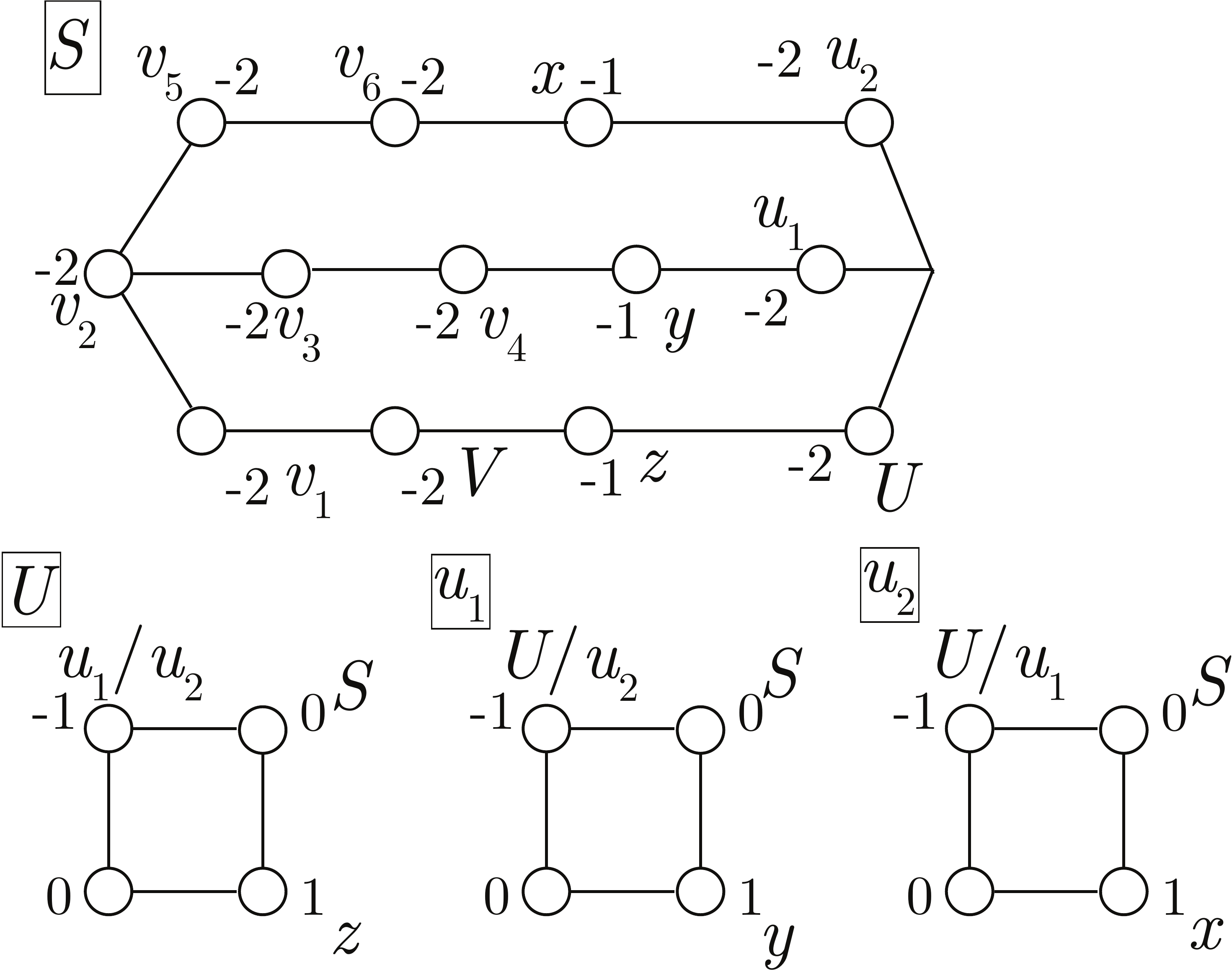}\,.
\ee

In this case, to circumvent the multiplicity factor subtlety, we do a flop on (\ref{f:1-3-gluing-topresol}) by shrinking the single intersection curve among the surfaces $U,u_1$ and $u_2$. Consequently, the surface $S$ is blown up at the point where three curves $U\cdot S$, $u_1\cdot S$ and $u_2\cdot S$ intersect. The curve configurations after this flop are:
\be
\label{f:1-3-gluing-topflop}
\includegraphics[width=6.5cm]{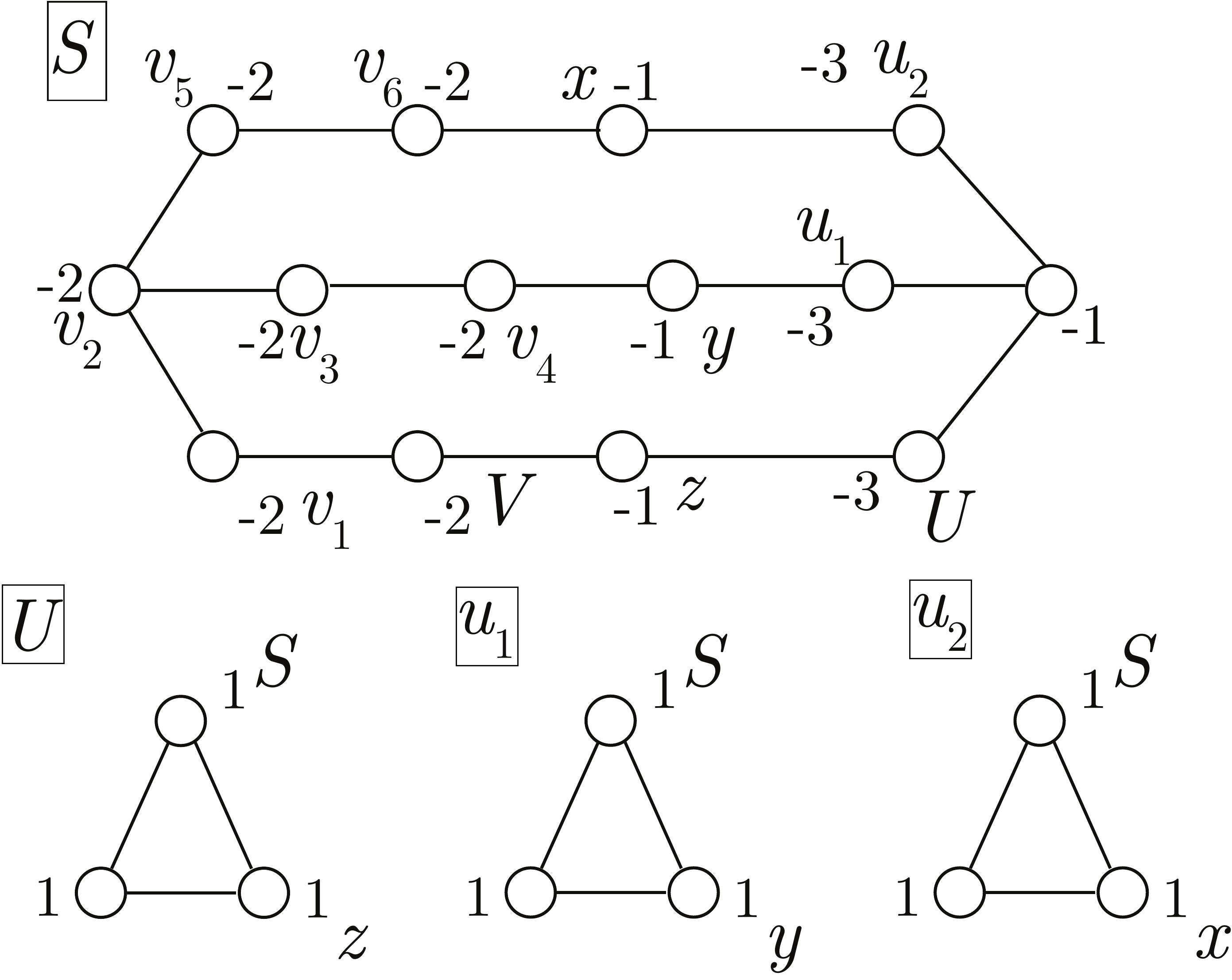}\,.
\ee

To get a valid 5d theory, we need to flop the curve $z\cdot S$ on $S$ into $U$, and then decompactify $U$. This leads to the following curve configurations on the remaining compact surfaces:

\be
\label{f:1-3-gluing-subtopflop}
\includegraphics[width=7cm]{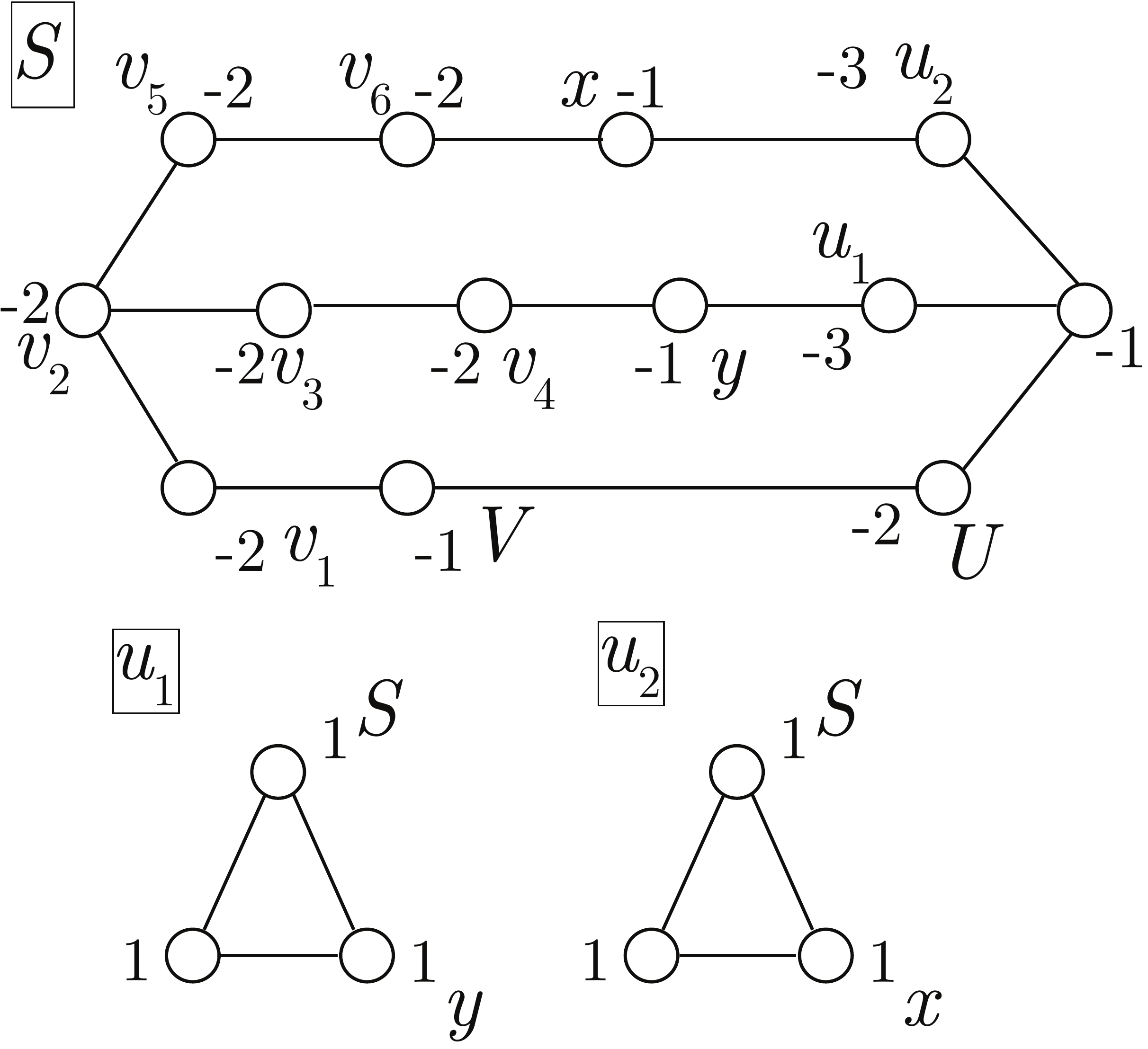}\,.
\ee
Hence we can read off the following CFD:
\be
\label{f:1-3-gluing-CFD}
\includegraphics[width=4.5cm]{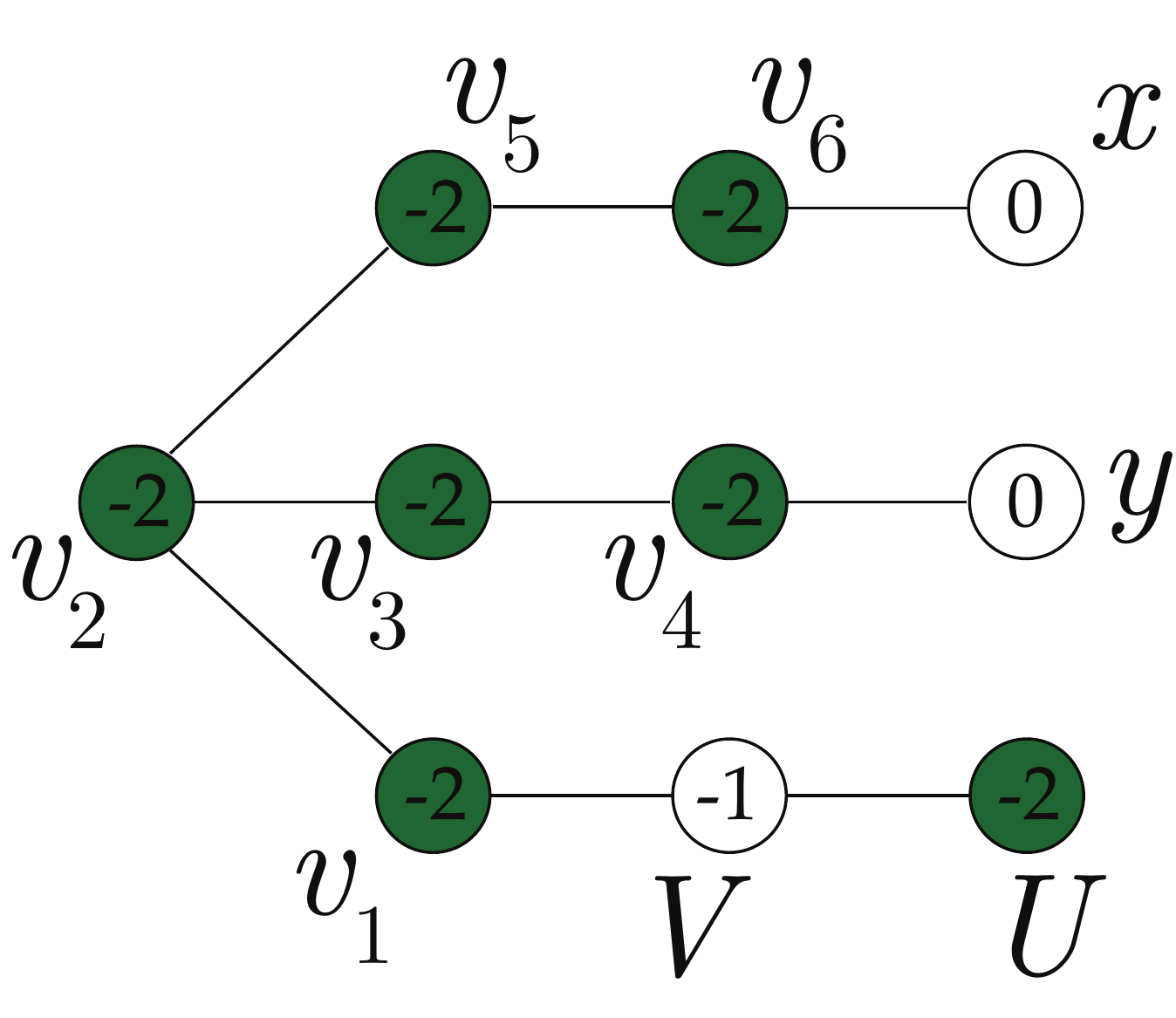}\,.
\ee
The rank-3 5d theory has $G_F=E_6\times SU(2)$ superconformal flavor symmetry and an IR quiver gauge theory description of:
\be
4\bm{F}-SU(2)-SU(3)\,.
\ee
The BG-CFD of $SU(2)+4\bm{F}$ can be embedded in (\ref{f:1-3-gluing-CFD}).

\subsection*{$(-1)(-4)$: $G=SO(8)$}

The curve configurations in the tensor branch resolution are 
\be
\label{f:1-4-gluing-topresol}
\includegraphics[width=7cm]{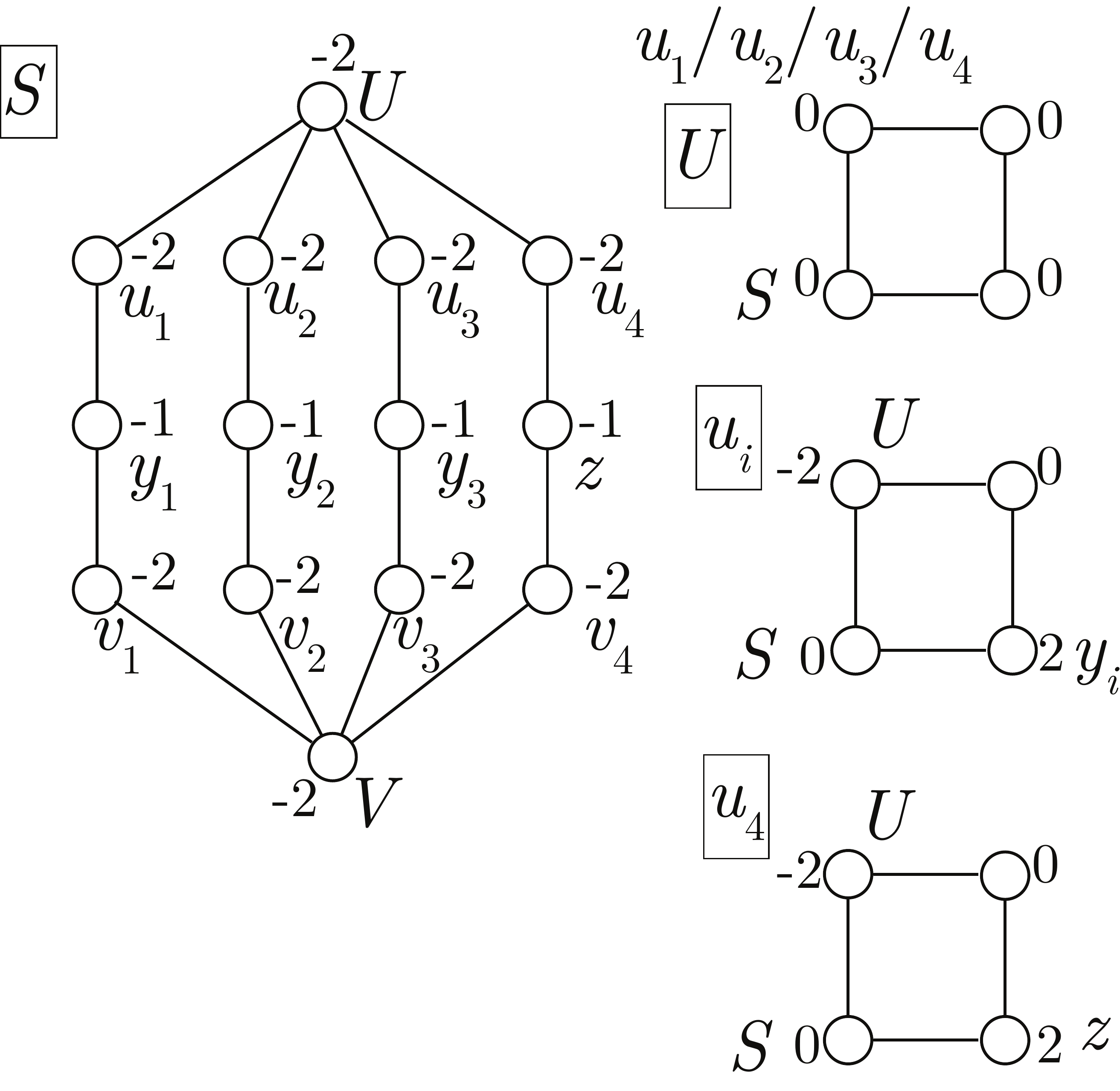}.
\ee
$V$, $v_1,\dots,v_4$ are the non-compact Cartan divisors of the $SO(8)$ in the tensor branch
\be
[SO(8)]-1-\overset{\mathfrak{so}(8)}{4},
\ee
and $U$, $u_1,\dots,u_4$ are the compact Cartan divisors of the $\mathfrak{so}(8)$ on the $(-4)$-curve. The affine nodes are $u_4$ and $v_4$. $S$ is the vertical divisor over the $(-1)$-curve, which is a gdP$_9$. We flop the curve $z\cdot S$ on $S$ into $U$ and then decompactify $U$, which result in the following curve configurations
\be
\label{f:1-4-gluing-subtopresol}
\includegraphics[width=7cm]{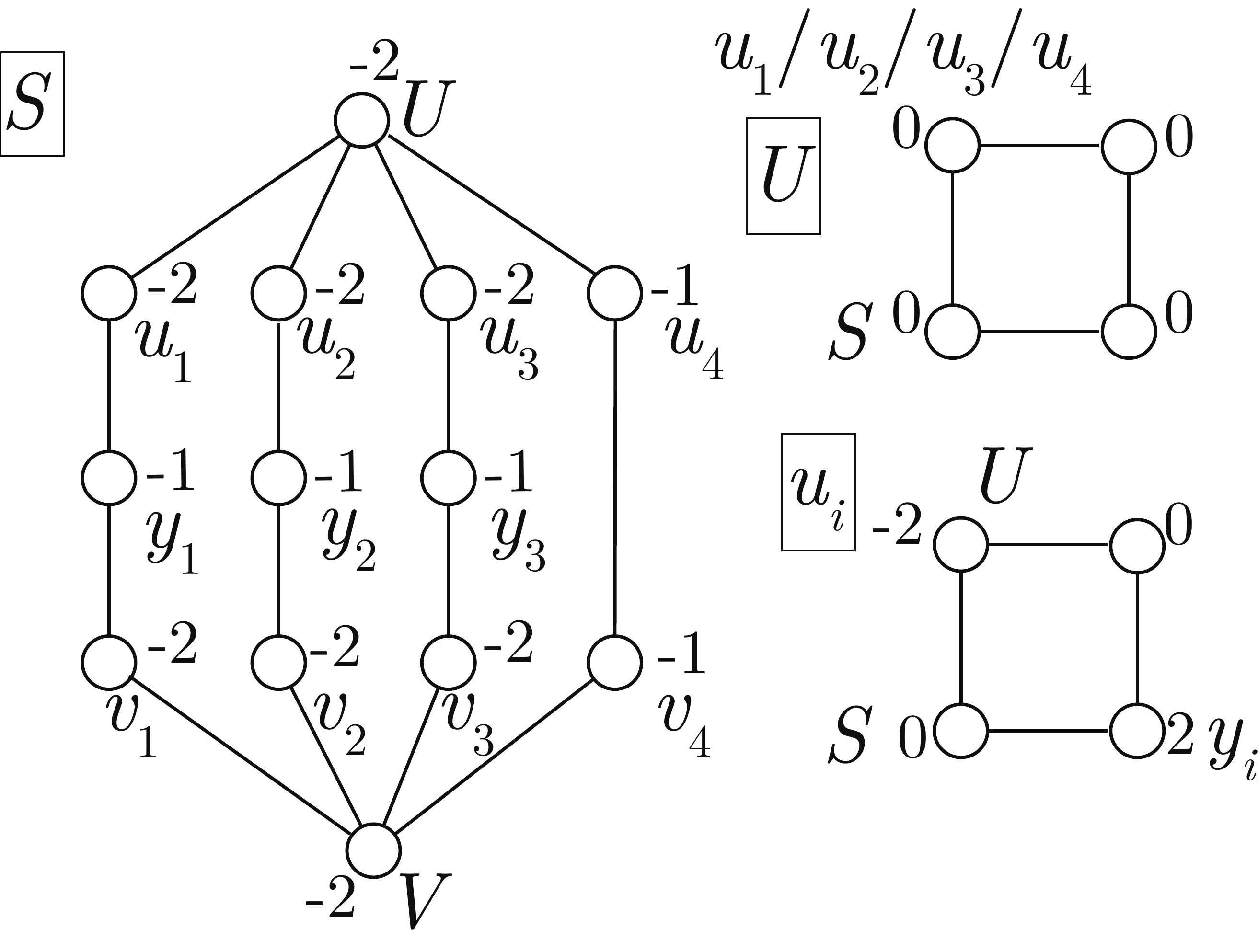}\,.
\ee
The CFD can be read off as
\be
\label{f:1-4-gluing-CFD}
\includegraphics[width=3.5cm]{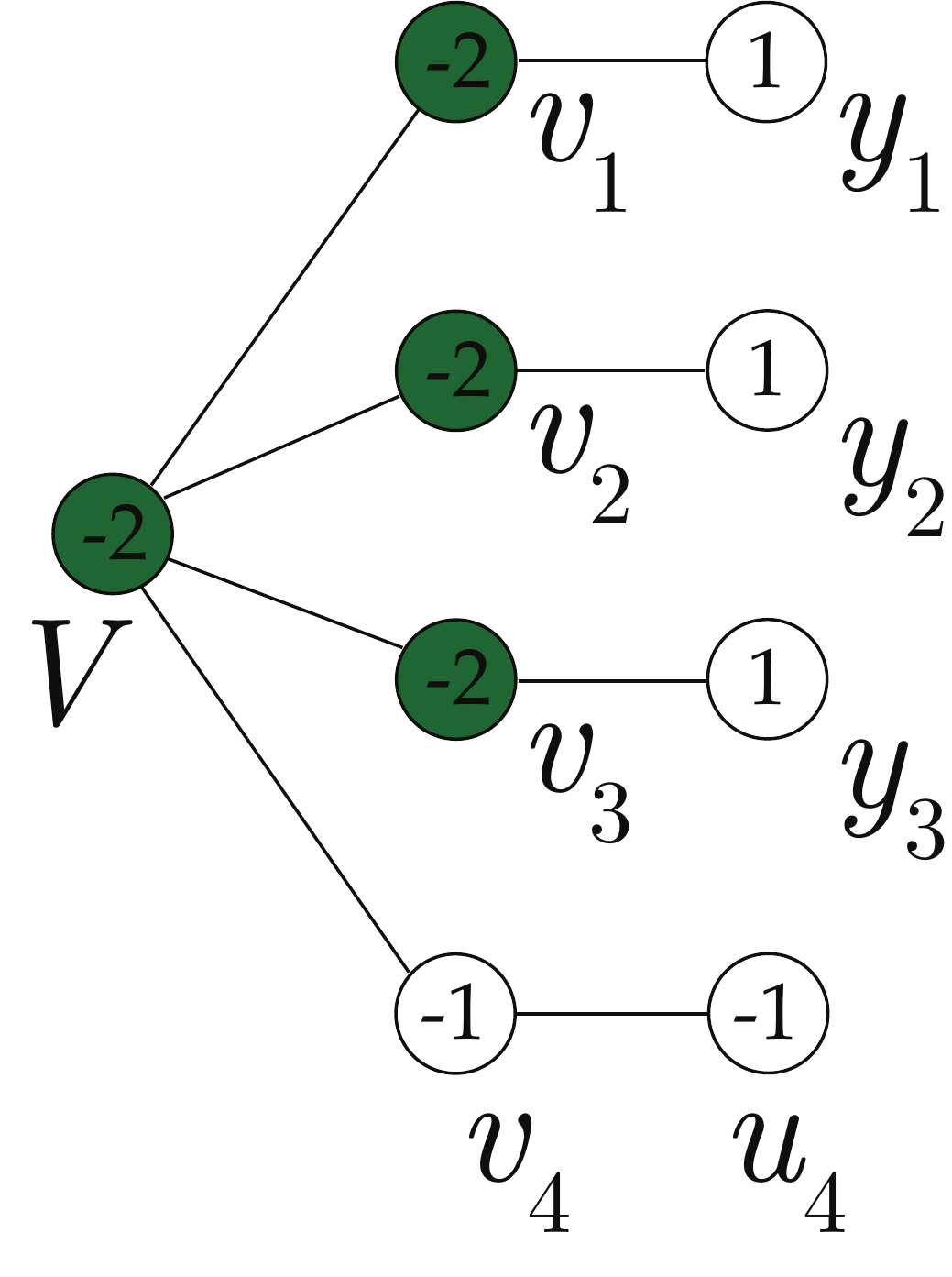}\,.
\ee

\subsection*{$(-1)(-6)$: $G=E_6$}

The resolution geometry is similar to the $G=SU(3)$ case. Here we denote the compact Cartan divisors of $E_6$ by $U,u_1,u_2,\dots,u_6$, the compact vertical divisor over $(-1)$-curve by $S$ and the non-compact Cartan divisors of $SU(3)$ by $V,v_1,v_2$. The curve configurations on the compact surfaces are:
\be
\label{f:1-6-gluing-topresol}
\includegraphics[width=10cm]{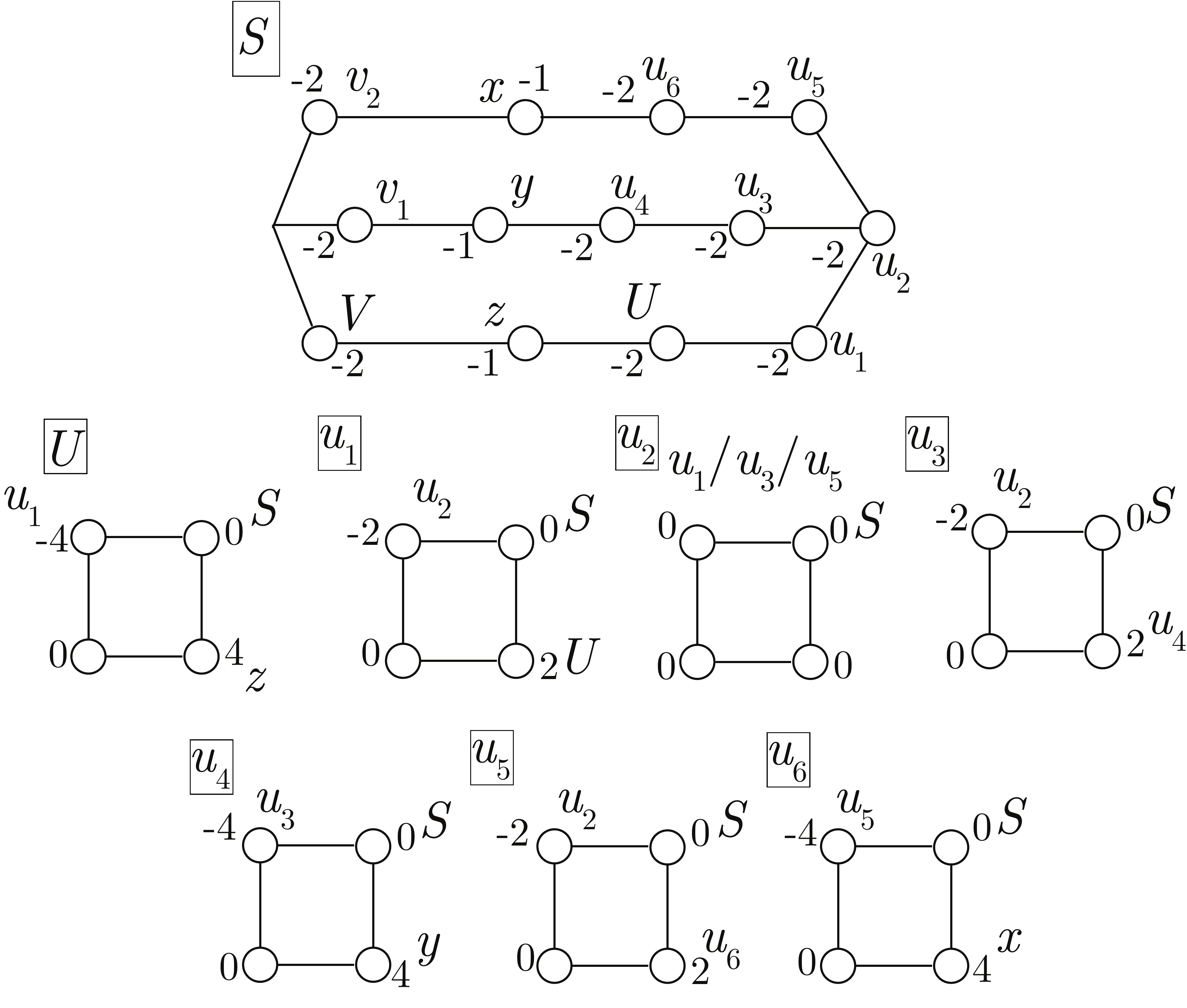}.
\ee
Similar to the previous cases, we shrink the curve $z\cdot S$ on $S$ and decompactify $U$. The resulting CFD is going to be:
\be
\label{f:1-6-gluing-CFD}
\includegraphics[width=4cm]{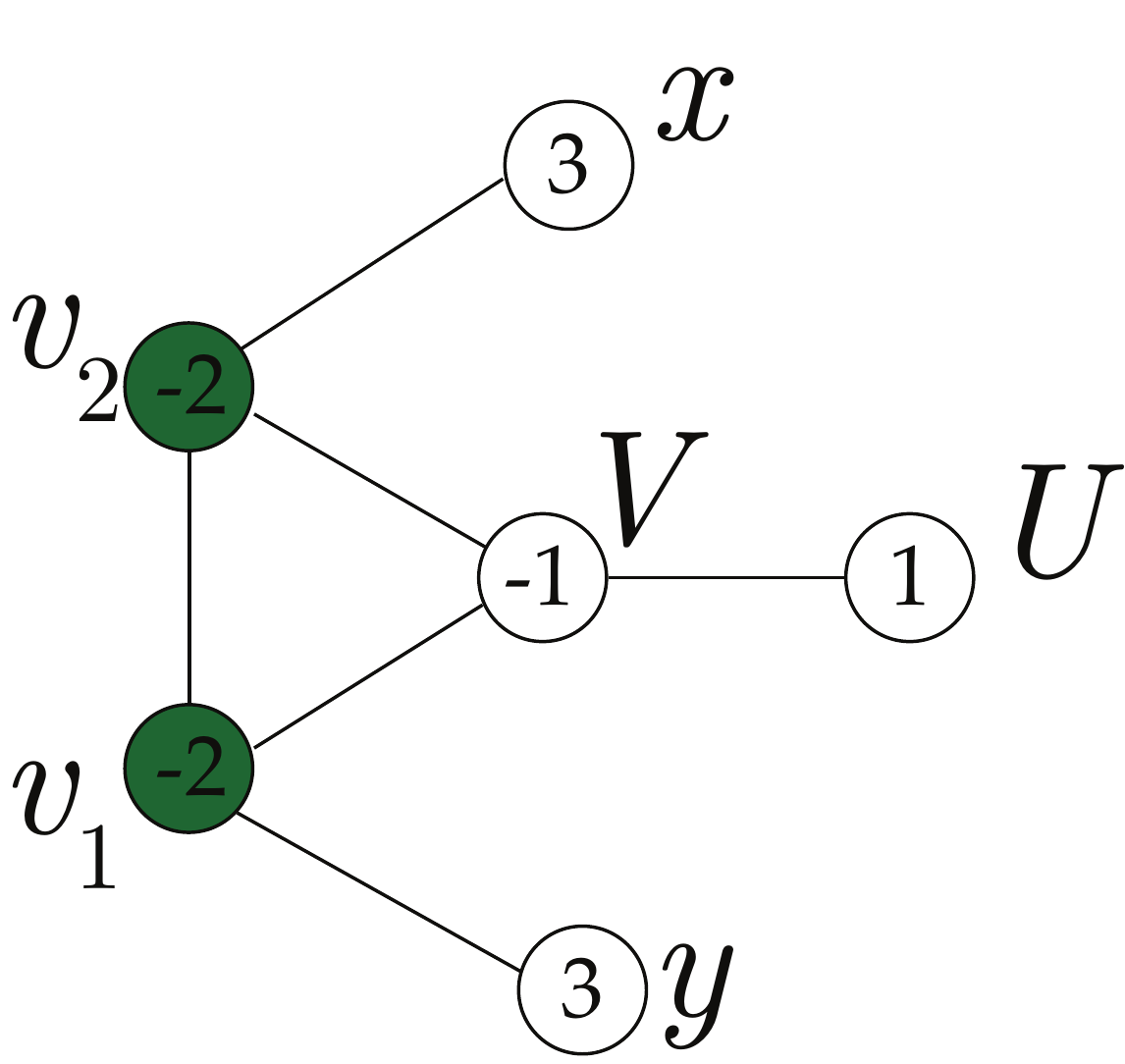}.
\ee

\subsection*{$(-1)(-8)$: $G=E_7$}

In the resolution geometry, we denote the compact Cartan divisors of $E_7$ by $U,u_1,\dots,u_7$, the compact vertical divisor over $(-1)$-curve by $S$ and the non-compact Cartan divisors of $SU(2)$ by $V,v_1$. The curve configurations on the compact surfaces are:
\be
\label{f:1-8-gluing-topresol}
\includegraphics[width=9cm]{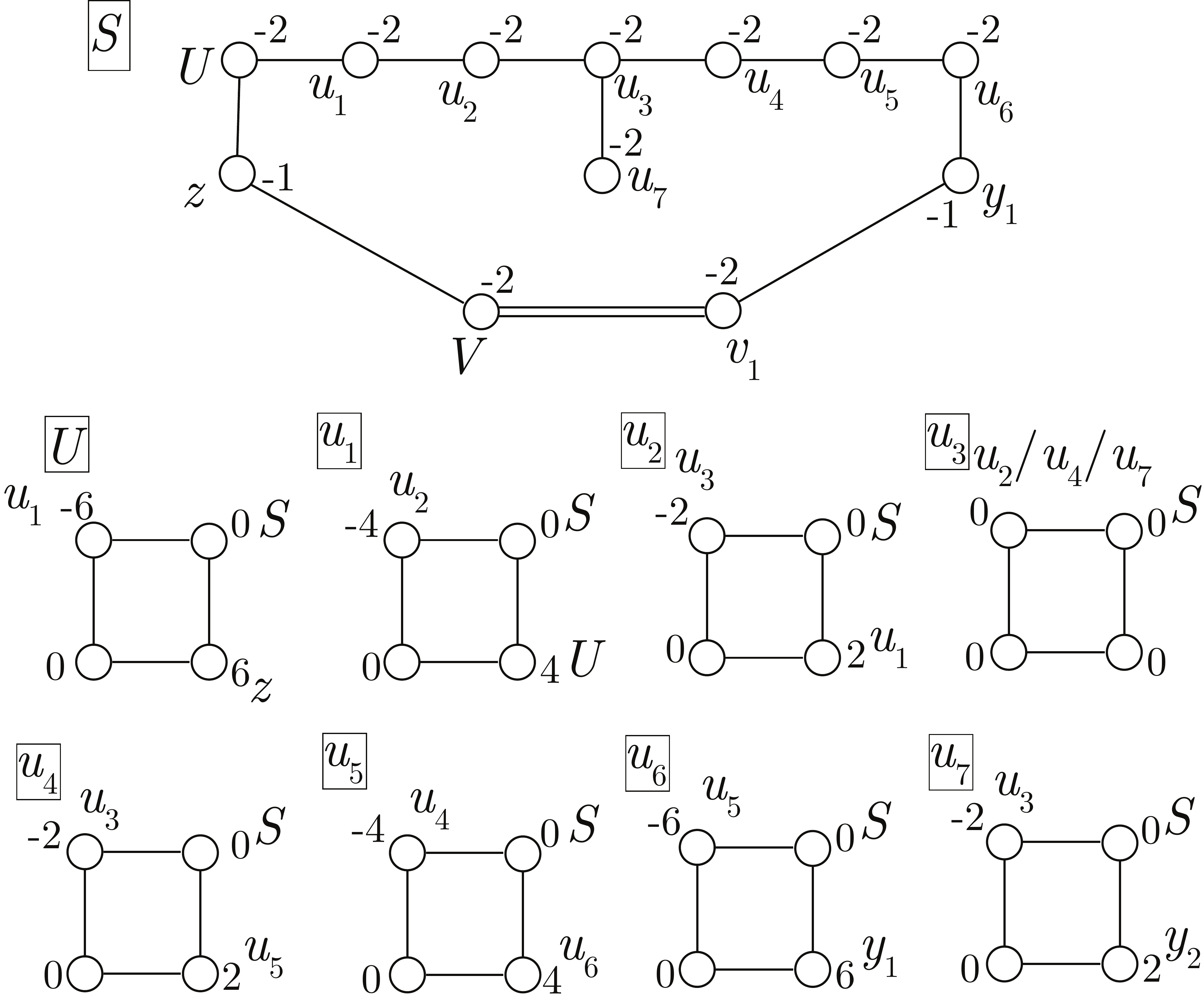}.
\ee
After we shrink the curve $z\cdot S$ on $S$ and decompactify $U$, the resulting CFD is:
\be
\label{f:1-8-gluing-CFD}
\includegraphics[width=5cm]{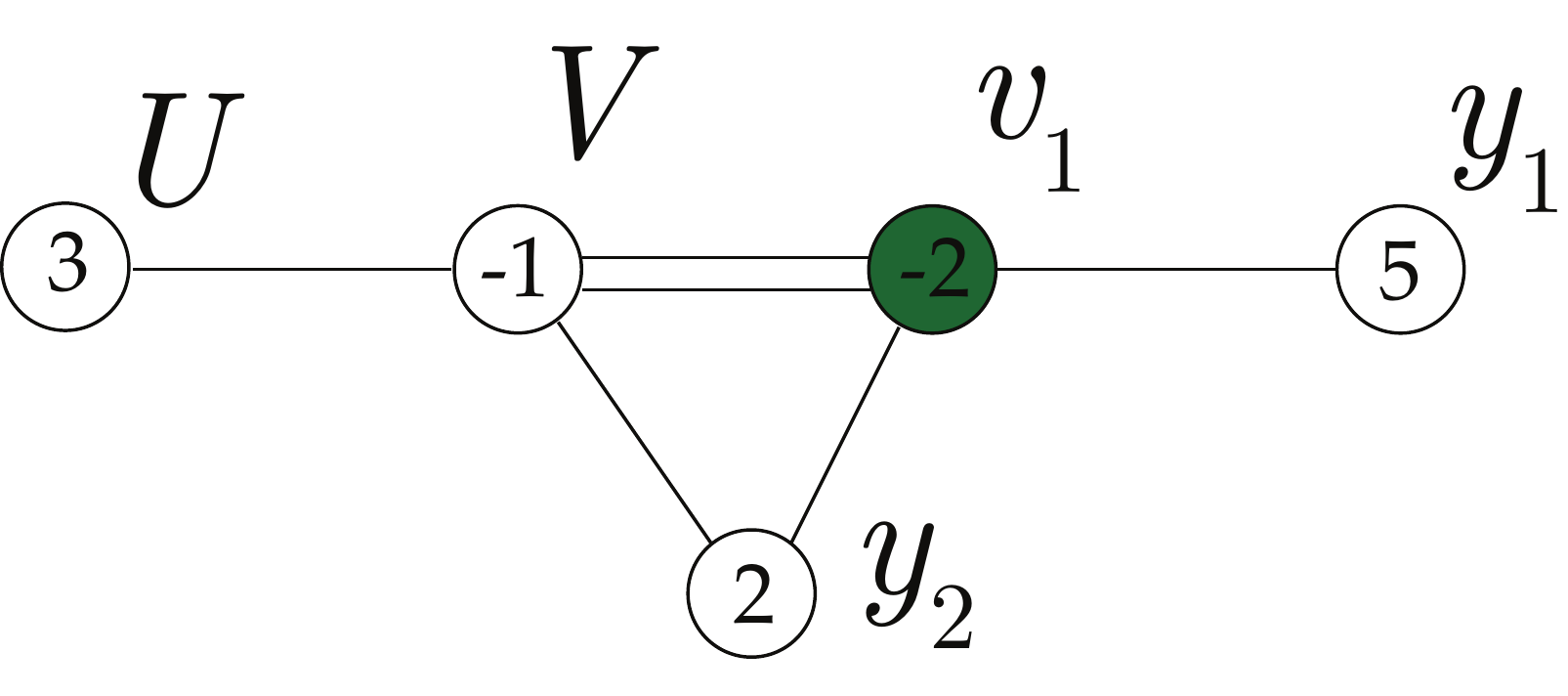}.
\ee

\subsection*{$(-1)(-12)$: $G=E_8$}

In this case, there will not be any non-Abelian flavor symmetry on the non-compact curve. We just have a resolution geometry of a gdP$_9$ glued with the exceptional divisors of a $II^*$ Kodaira singularity. The curve configurations are:
\be
\label{f:1-12-gluing-topresol}
\includegraphics[width=13cm]{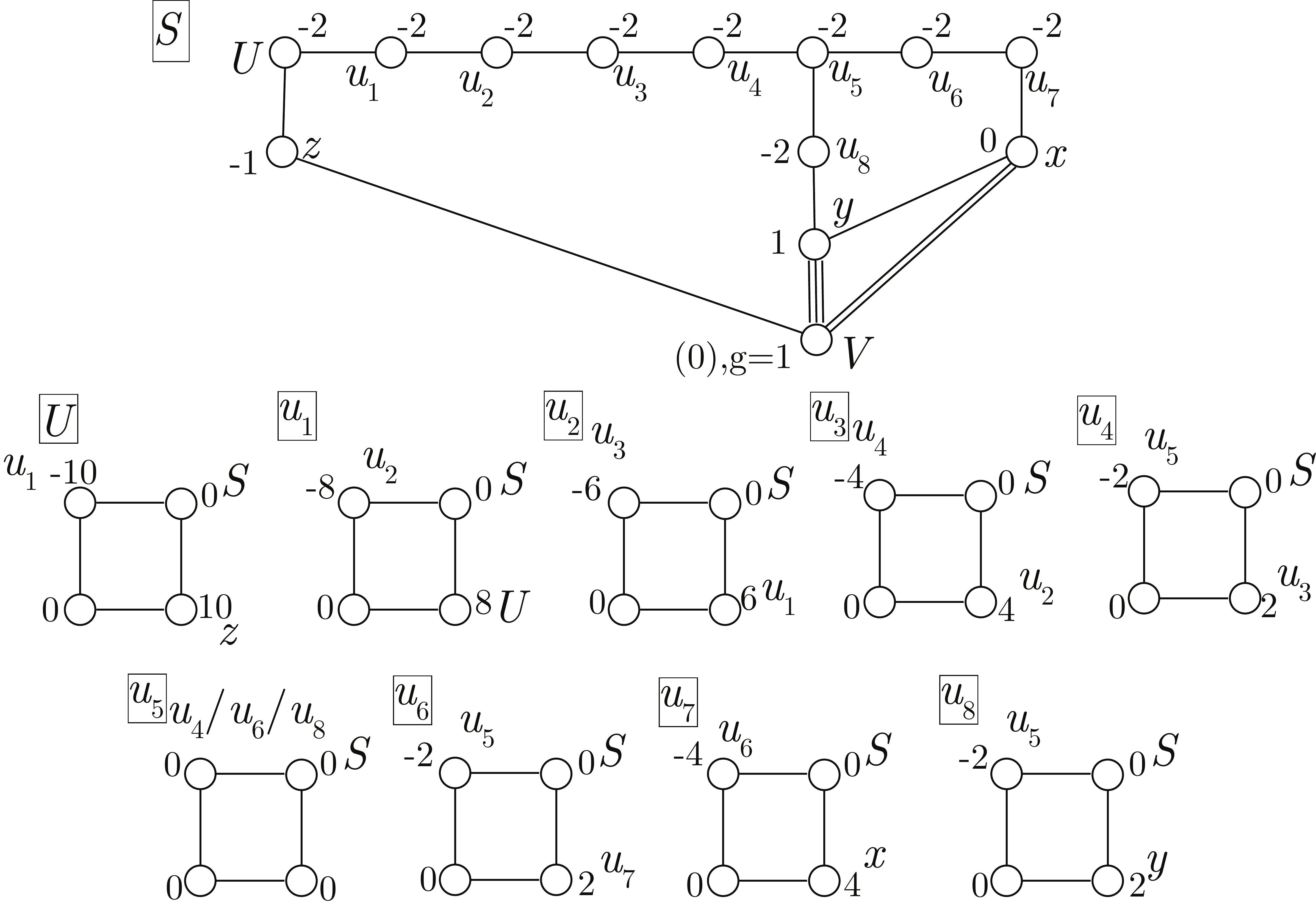}.
\ee
After the shrinking of curve $z\cdot S$ on $S$ and decompactification of $U$, we just get a CFD with no $(-1)$ or lower node:
\be
\label{f:1-12-gluing-CFD}
\includegraphics[width=4cm]{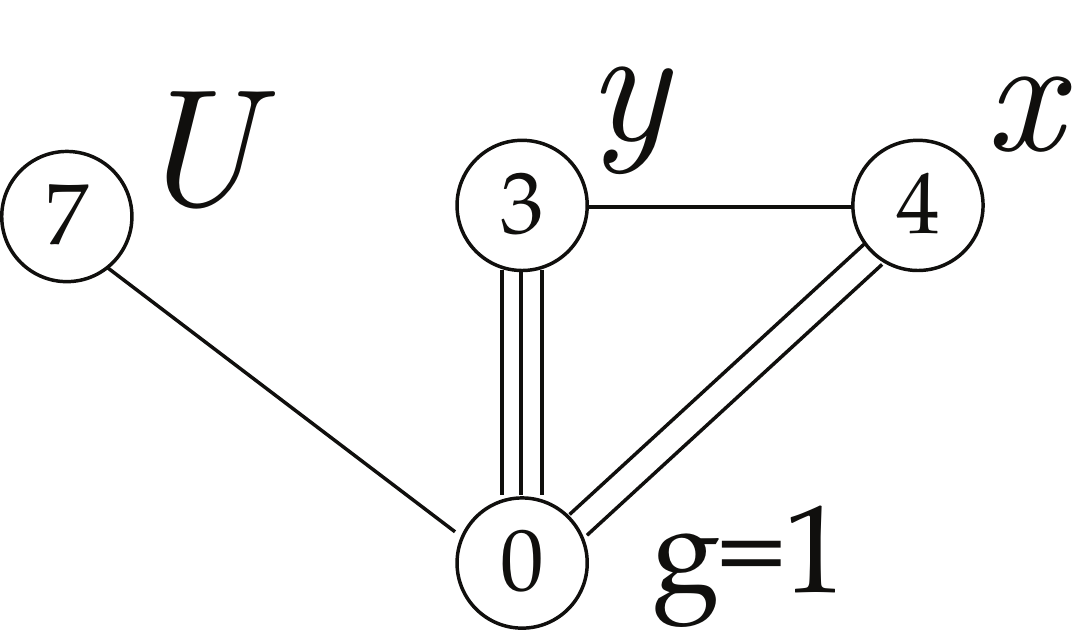}.
\ee

\section{Geometry of Non-Minimal Conformal Matter}
\label{app:CMN}

\subsection{Non-minimal $(D_n, D_n)$ Conformal Matter}
\label{app:DkDk}

The tensor branch of non-minimal $(D_n,D_n)$ $(n>4)$ conformal matter is given by:
\be
[SO(2n)]-\overset{\mathfrak{sp}(n-4)}{1}-\overset{\mathfrak{so}(2n)}{4}-\cdots-\overset{\mathfrak{so}(2n)}{4}--\overset{\mathfrak{sp}(n-4)}{1}-[SO(2n)]\,.
\ee

When $N=1$, the theory is the minimal $(D_n,D_n)$ conformal matter, with the equivalent descriptions of marginal CFD in table~\ref{tab:Spn1CFDs}.

\begin{figure}
\centering
\includegraphics[width=14cm]{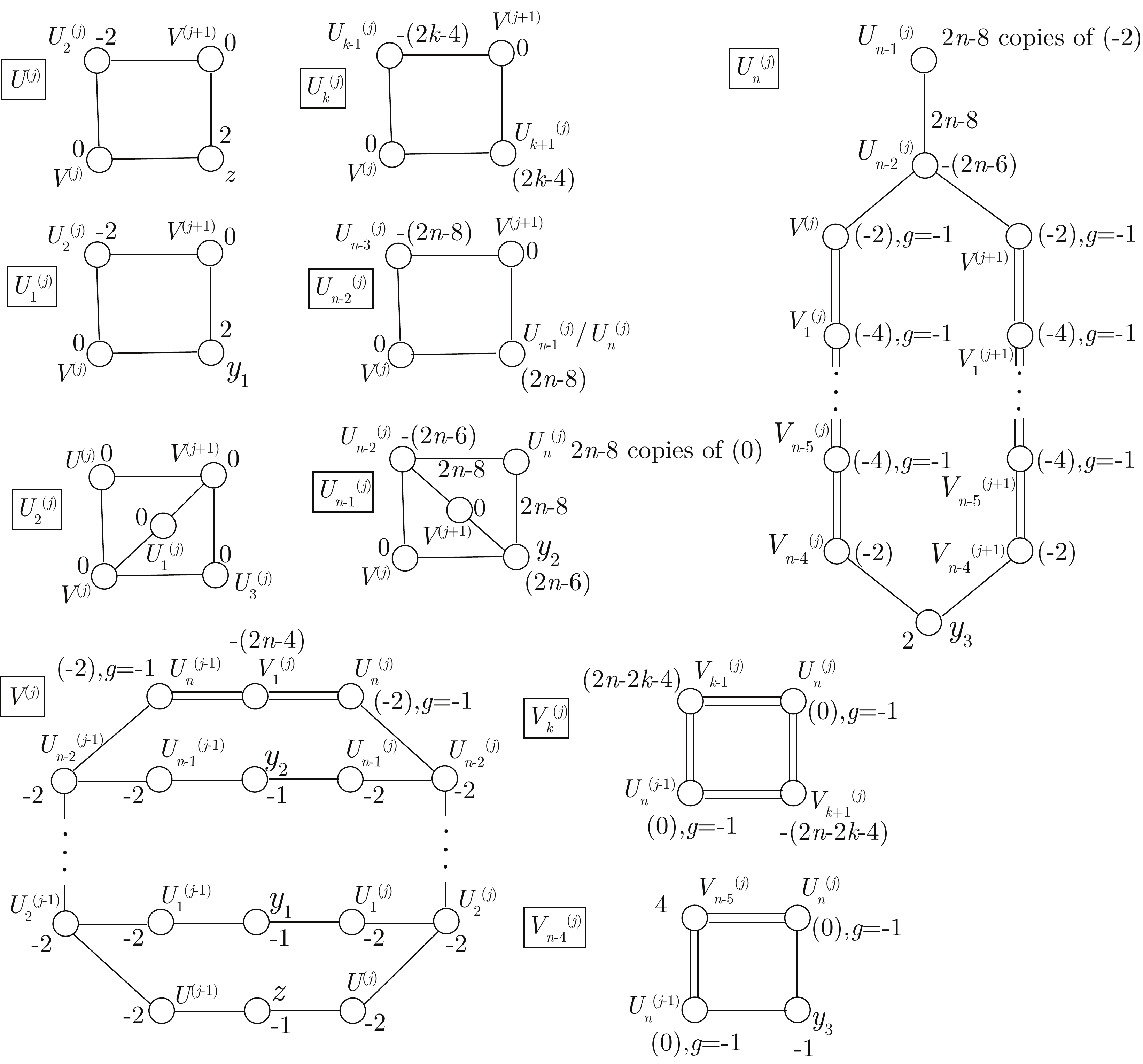}\
\caption{The configuration of curves in the KK resolution geometry of non-minimal $(D_n,D_n)$ conformal matter with order $N$.}
\label{f:DkDktopresol}
\end{figure}

In the tensor branch (KK) resolution geometry, we label the Cartan divisors of each $SO(2n)$ by $U^{(j)}$ and $U_k^{(j)}$, where $j=0,\dots,N$, $k=1,\dots,n$. The ones with $j=1,\dots,N-1$ correspond to the compact surfaces which are fibered over the curves in the middle of tensor branch, while $j=0$ and $j=N$ correspond to the two non-compact $SO(2n)$. For each $Sp(n-4)$, the Cartan divisors are labeled by $V^{(j)}$ and $V_k^{(j)}$, where $j=1,\dots,N$, $k=1,\dots,n-4$. We plot the configuration of curves in figure~\ref{f:DkDktopresol}.

Here $U_k^{(j)}$ denotes the compact surface components with $k=3,\dots,n-3$ and $V_k^{(j)}$ denotes the compact surface components with $k=1,\dots,n-5$. For $V^{(j)}$ with $j=2,\dots,N-1$, the curves $V^{(j)}\cdot z$, $V^{(j)}\cdot y_1$ and $V^{(j)}\cdot y_2$ has non-trivial multiplicity two, and the curves $V_{n-4}^{(j)}\cdot y_3$ on $V_{n-4}^{(j)}$ has multiplicity two as well.

Hence in the CFD from this geometry, the vertices $z$, $y_1$, $y_2$ and $y_3$ have:
\be
\ba
n(z)&=\sum_{j=1}^{N-1}z^2\cdot U^{(j)}+z^2\cdot V^{(1)}+z^2\cdot V^{(N)}+2\sum_{j=2}^{N-1}z^2\cdot V^{(j)}=0\,,\cr
n(y_1)&=\sum_{j=1}^{N-1}y_1^2\cdot U_1^{(j)}+y_1^2\cdot V^{(1)}+y_1^2\cdot V^{(N)}+2\sum_{j=2}^{N-1}y_1^2\cdot V^{(j)}=0\,,\cr
n(y_2)&=\sum_{j=1}^{N-1}y_2^2\cdot U_{n-1}^{(j)}+y_2^2\cdot V^{(1)}+y_2^2\cdot V^{(N)}+2\sum_{j=2}^{N-1}y_2^2\cdot V^{(j)}=(2n-8)(N-1)\,,\cr
n(y_3)&=\sum_{j=1}^{N-1}y_3^2\cdot U_n^{(j)}+y_3^2\cdot V_{n-4}^{(1)}+y_3^2\cdot V_{n-4}^{(N)}+2\sum_{j=2}^{N-1}y_3^2\cdot V_{n-4}^{(j)}=0\,.
\ea
\ee

\begin{figure}
\centering
\includegraphics[width=14cm]{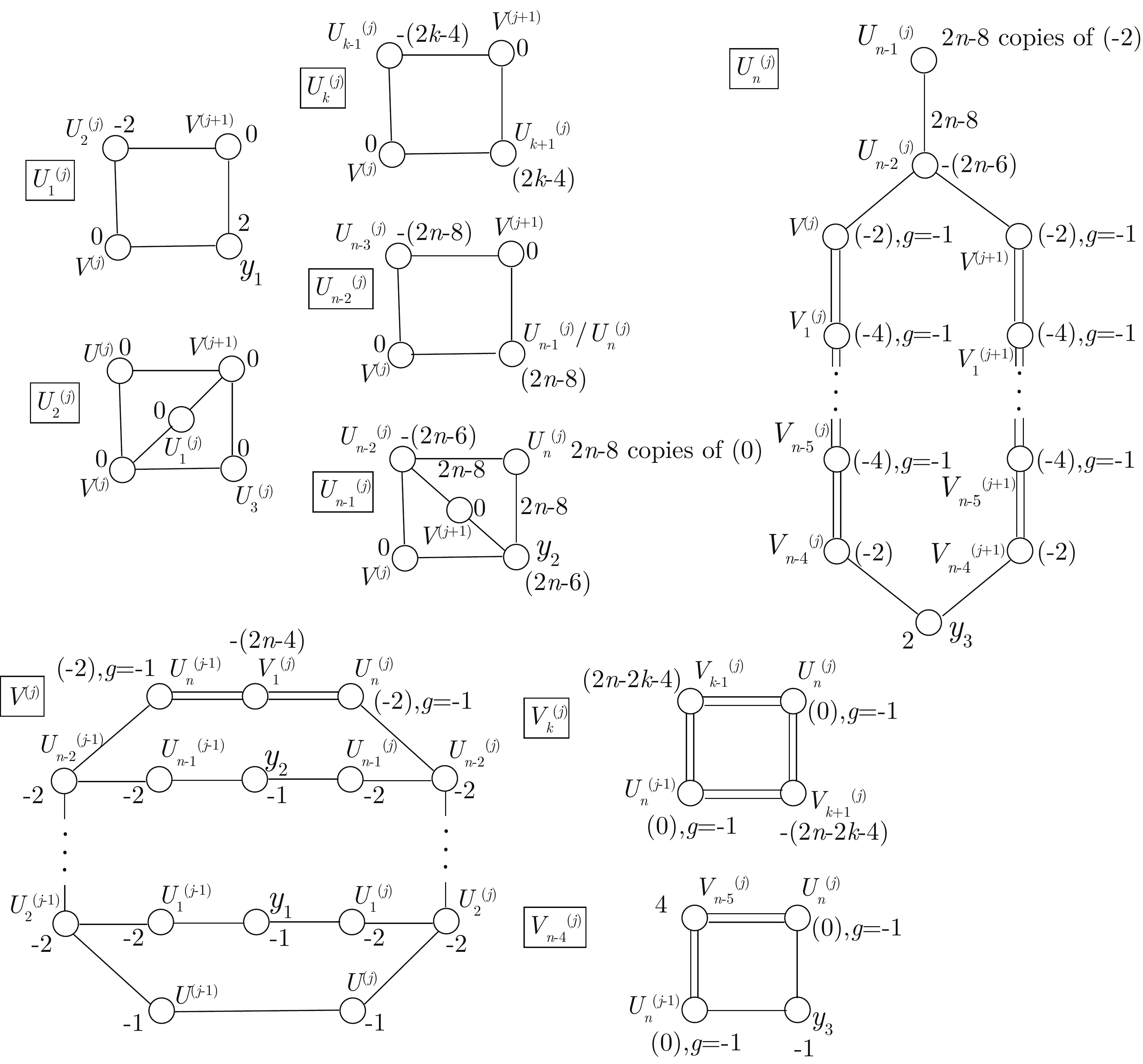}\
\caption{The configuration of curves in the decoupled and flopped geometry of figure~\ref{f:DkDktopresol}. Note that the decompactified surfaces $U^{(j)}$ has been removed.}
\label{f:DkDksubtopresol}
\end{figure}

The CFD, see tables~\ref{tab:nm-CM-CFD} and \ref{tab:nm-E-CM-CFD}, then does not have any possible transitions to a 5d SCFT descendant. To get a 5d SCFT, we shrink the curves $z\cdot V^{(j)}$ for $j=1,\dots,N$ and then decompactify the surfaces $U^{(j)}$ for $j=1,\dots,N-1$. The configuration of curves on the new compact surfaces is plotted in figure~\ref{f:DkDksubtopresol}.

The curves with multiplicity two are still $V^{(j)}\cdot y_1$, $V^{(j)}\cdot y_2$ and $V_{n-4}^{(j)}\cdot y_3$ with $j=2,\dots,N-1$, and we can read off:
\be
\ba
n(U^{(0)})&=(U^{(0)})^2\cdot V^{(1)}=-1\,,\cr
n(U^{(N)})&=(U^{(N)})^2\cdot V^{(N)}=-1\,,\cr
n(U^{(j)})&=(U^{(j)})^2\cdot (V^{(j)}+V^{(j+1)})=-2\quad (j=1,\dots,N-1)\,,\cr
n(y_1)&=\sum_{j=1}^{N-1}y_1^2\cdot U_1^{(j)}+y_1^2\cdot V^{(1)}+y_1^2\cdot V^{(N)}+2\sum_{j=2}^{N-1}y_1^2\cdot V^{(j)}=0\,,\cr
n(y_2)&=\sum_{j=1}^{N-1}y_2^2\cdot U_{n-1}^{(j)}+y_2^2\cdot V^{(1)}+y_2^2\cdot V^{(N)}+2\sum_{j=2}^{N-1}y_2^2\cdot V^{(j)}=(2n-8)(N-1)\,,\cr
n(y_3)&=\sum_{j=1}^{N-1}y_3^2\cdot U_n^{(j)}+y_3^2\cdot V_{n-4}^{(1)}+y_3^2\cdot V_{n-4}^{(N)}+2\sum_{j=2}^{N-1}y_3^2\cdot V_{n-4}^{(j)}=0\,.
\ea
\ee
Along with the edge multiplicities computed from (\ref{CFD-edge-m}), we can exactly read off the CFD in tables \ref{tab:nm-CM-CFD} and \ref{tab:nm-E-CM-CFD}. It has an extra chain of $N$ $(-2,0)$-vertices that correspond to the decompactified surfaces $U^{(j)}$ $j=1,\dots,N-1$. 


\subsection{Non-Minimal $(E_6, E_6)$ Conformal Matter}
\label{app:E6E6}

\begin{figure}
\centering
\includegraphics[width=15cm]{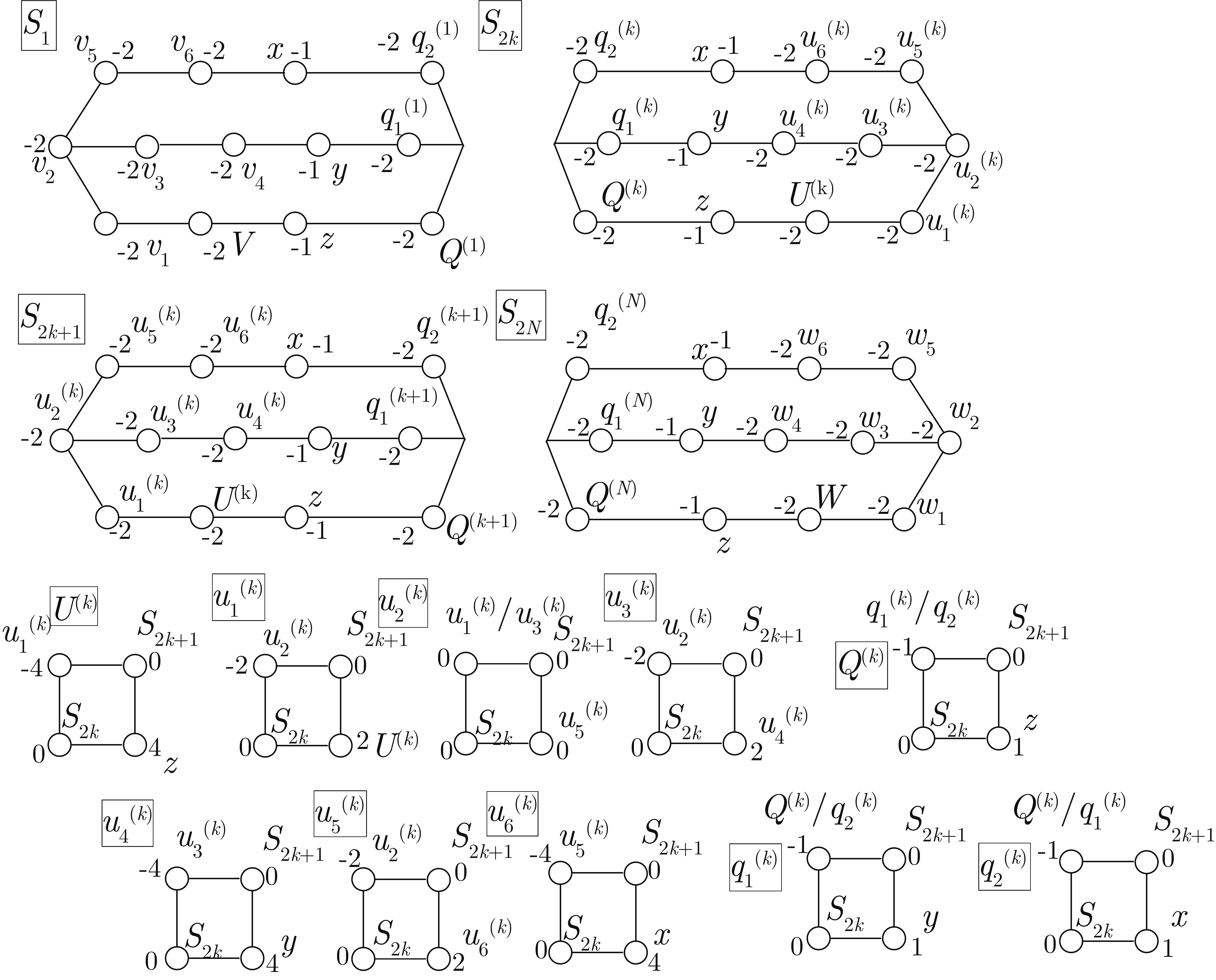}
\caption{The configuration of curves in the KK resolution geometry of non-minimal $(E_6,E_6)$ conformal matter with order $N$.}
\label{f:E6E6topresol}
\end{figure}

In the resolution geometry of the tensor branch, we label the Cartan divisors of the left and right non-compact $E_6$ by $V,v_1,\dots,v_6$ and $W,w_1,\dots,w_6$, including the affine nodes. The Cartan divisors of the $N-1$ compact $E_6$ in the middle are denoted by $U^{(i)},u^{(i)}_1,\dots,u^{(i)}_6$ $(i=1,\dots,N-1)$. The Cartan divisors of the $N$ compact $SU(3)$s are denoted by $Q^{(i)},q^{(i)}_1,q^{(i)}_2$ $(i=1,\dots,N)$. The vertical divisors over the $(-1)$-curves are denoted by $S_i$ $(i=1,\dots,2N)$. Then we plot the curve configurations on each surface components in figure~\ref{f:E6E6topresol}, where $x$, $y$ and $z$ corresponds to non-compact divisors. For the surfaces $Q^{(k)}$, $q_1^{(k)}$, $q_2^{(k)}$ label $k$ goes from 1 to $N$, and we effectively have $U^{(0)}\equiv V$, $U^{(N)}\equiv W$.

As we can see, the surfaces $S_i$ are gdP$_9$s with $\hat{E}_6$ and $\widehat{SU}(3)$ singular fibers, and the configuration of Mori cone generators is exactly given in the table~\ref{tab:Rank1CFDs}. The divisors $Q^{(i)},q^{(i)}_1,q^{(i)}_2$ are Hirzebruch surface $\mb{F}_1$ sharing a common $\mc{O}(-1)\oplus\mc{O}(-1)$ curve. In this geometry, the multiplicity factor of the intersection curves between the compact surfaces with $x$, $y$, $z$ can be computed from the procedure (\ref{fm-shrinking}). For $z$, obviously the intersection curves on the non-compact surface $z$ form the following chain, which is exactly the same as the tensor branch base geometry:
\be
\label{z-tensor-chain}
\underset{S_1}{(-1)}-\underset{Q^{(1)}}{(-3)}-\underset{S_2}{(-1)}-\underset{U^{(1)}}{(-6)}-\underset{S_3}{(-1)}-\underset{Q^{(2)}}{(-3)}-\underset{S_4}{(-1)}-\dots-\underset{Q^{(N-1)}}{(-3)}-\underset{S_{2N-2}}{(-1)}-\underset{U^{(N-1)}}{(-6)}-\underset{S_{2N-1}}{(-1)}-\underset{Q^{(N)}}{(-3)}-\underset{S_{2N}}{(-1)}.
\ee

Then we try to shrink this chain of curves by blow down the $(-1)$-curves in the middle of the chain, and finally we get:
\be
\underset{S_1}{(-1)}-\underset{Q^{(1)}}{(-2)}-\underset{U^{(1)}}{(-3)}-\underset{U^{(2)}}{(-2)}-\dots-\underset{U^{(N-2)}}{(-2)}-\underset{U^{(N-1)}}{(-3)}-\underset{Q^{(N)}}{(-2)}-\underset{S_{2N}}{(-1)}.
\ee
Now we assign weight factor one to all the curves in the chain above, and blow up back to the original chain (\ref{z-tensor-chain}). In the process, the weight factor of a new $(-1)$-curve is given by the sum of its two neighbors. Finally, we get all the weight factors for the chain (\ref{z-tensor-chain}), which is labeled above each curve:
\be
\label{z-tensor-chain-wf}
\overset{1}{\underset{S_1}{(-1)}}-\overset{1}{\underset{Q^{(1)}}{(-3)}}-\overset{2}{\underset{S_2}{(-1)}}-\overset{1}{\underset{U^{(1)}}{(-6)}}-\overset{3}{\underset{S_3}{(-1)}}-\overset{2}{\underset{Q^{(2)}}{(-3)}}-\overset{3}{\underset{S_4}{(-1)}}-\dots-\overset{2}{\underset{Q^{(N-1)}}{(-3)}}-\overset{3}{\underset{S_{2N-2}}{(-1)}}-\overset{1}{\underset{U^{(N-1)}}{(-6)}}-\overset{2}{\underset{S_{2N-1}}{(-1)}}-\overset{1}{\underset{Q^{(N)}}{(-3)}}-\overset{1}{\underset{S_{2N}}{(-1)}}.
\ee

Then for the vertex $z$ in the CFD, we can compute:
\be
\ba
n(z)&=z^2\cdot (S_1+S_{2N}+2S_2+2S_{2N-1}+3\sum_{i=3}^{2N-2}S_i+Q^{(1)}+Q^{(N)}+2\sum_{i=2}^{N-1}Q^{(i)}+\sum_{i=1}^{N-1}U^{(i)})\cr
&=0
\ea
\ee
for any $N$.

Since there is a permutation symmetry among $x$, $y$ and $z$, we can carry over the same analysis to $x$ and $y$, and compute $n(x)=n(y)=0$ as well. The CFD is then given in tables \ref{tab:nm-CM-CFD} and \ref{tab:nm-E-CM-CFD}, with no descendants.

\begin{figure}
\centering
\includegraphics[width=13cm]{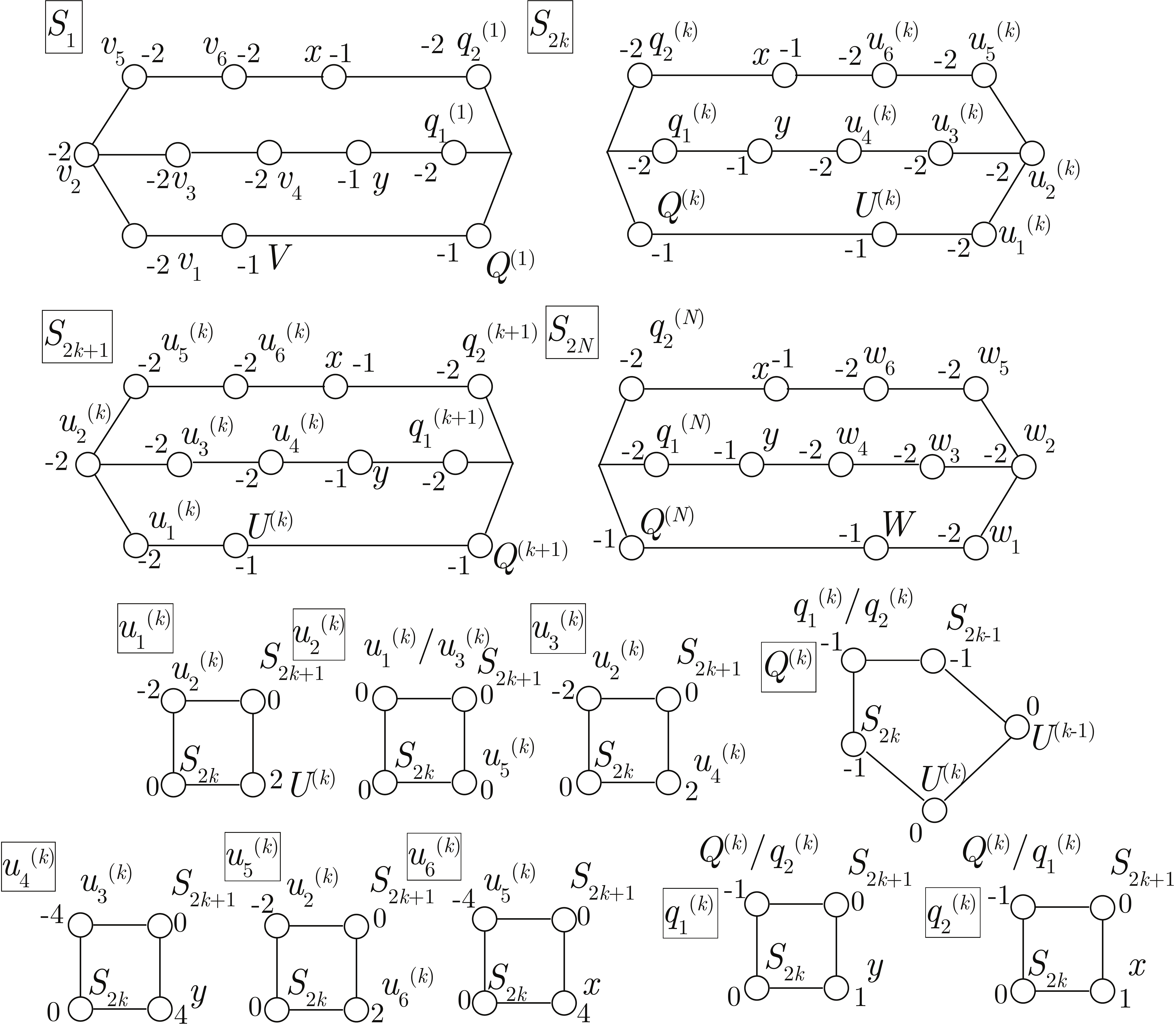}
\caption{The configuration of curves in the decoupled and flopped geometry of figure~\ref{f:E6E6topresol}. Note that the decompactified surfaces $U^{(j)}$ has been removed.}
\label{f:E6E6subtopresol}
\end{figure}

In the decoupling process, we first flop all the $(-1)$-curves $z\cdot S_i$ on each $S_i$ into the surface components $Q^{(k)}$, and then shrink the $(-1)$-curves $z\cdot Q^{(k)}$. After these flops, we decompactify the divisors $U^{(i)}$, $(i=1,\dots,N-1)$. Finally, we get the following configuration of curves on the compact surfaces in figure~\ref{f:E6E6subtopresol}.

The CFD will contain vertices $V$, $v_i$, $W$, $w_i$, $x$, $y$ and $U^{(k)}$ for $k=1,\dots,N-1$. Note that the $\mc{O}(-1)\oplus\mc{O}(-1)$ curves $U^{(k)}\cdot S_{2k}$ and $U^{(k)}\cdot S_{2k+1}$ have multiplicity two as well, which can be derived from the chain of curves on $U^{(k)}$:
\be
\begin{array}{c}
 \overset{1}{\underset{Q^{(k)}}{(-2)}}-\overset{2}{\underset{S_{2k}}{(-1)}}-\overset{1}{\underset{u_1^{(k)}}{(-4)}}-\overset{2}{\underset{S_{2k+1}}{(-1)}}-\overset{1}{\underset{Q^{(k+1)}}{(-2)}}\\
\downarrow\\
 \overset{1}{\underset{Q^{(k)}}{(-1)}}-\overset{1}{\underset{u_1^{(k)}}{(-2)}}-\overset{1}{\underset{Q^{(k+1)}}{(-1)}}
\end{array}
\ee

Hence the vertices $U^{(k)}$ actually have $(n,g)=(-2,0)$ in the CFD:
\be
\ba
n(U^{(k)})&=(U^{(k)})^2\cdot (2S_{2k}+2S_{2k+1}+u_1^{(k)}+Q^{(k)}+Q^{(k+1)})\cr
&=-2\,,\cr
g(U^{(k)})&=1+\frac{1}{2}\left[n(U^{(k)})+U^{(k)}\cdot (2S_{2k}+2S_{2k+1}+u_1^{(k)}+Q^{(k)}+Q^{(k+1)})^2\right]\cr
&=0\,.
\ea
\ee
and we can read off the CFD in table \ref{tab:nm-CM-CFD} and \ref{tab:nm-E-CM-CFD},  with $G_F=E_6\times E_6\times SU(N)$.

\subsection{Non-Minimal $(E_7, E_7)$ Conformal Matter}
\label{app:E7E7}

\begin{figure}
\centering
\includegraphics[width=15cm]{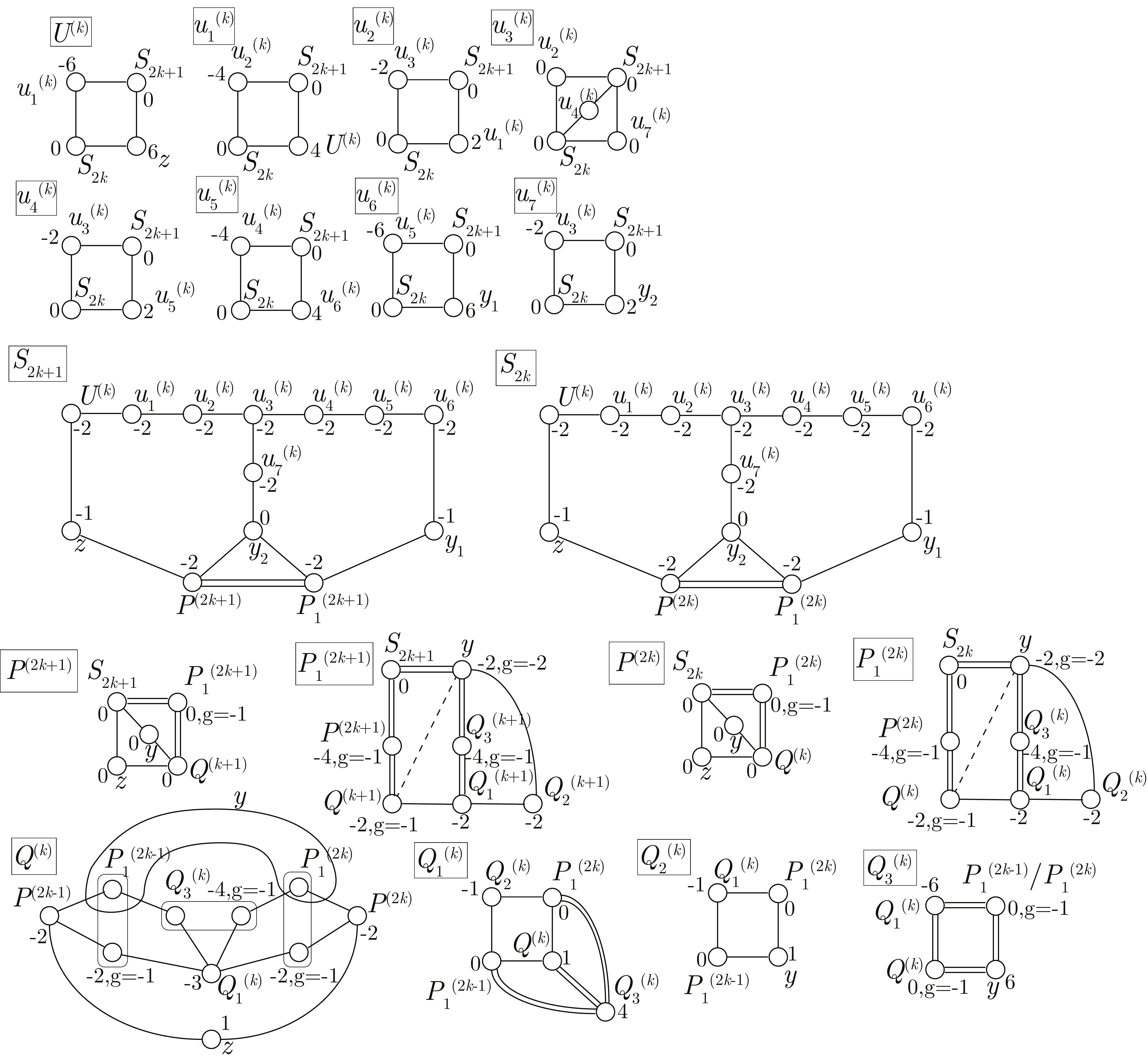}
\caption{The configuration of curves in the KK resolution geometry of non-minimal $(E_7,E_7)$ conformal matter with order $N$. Dotted line means negative intersection number.}
\label{f:E7E7topresol}
\end{figure}

\begin{figure}
\centering
\includegraphics[width=15cm]{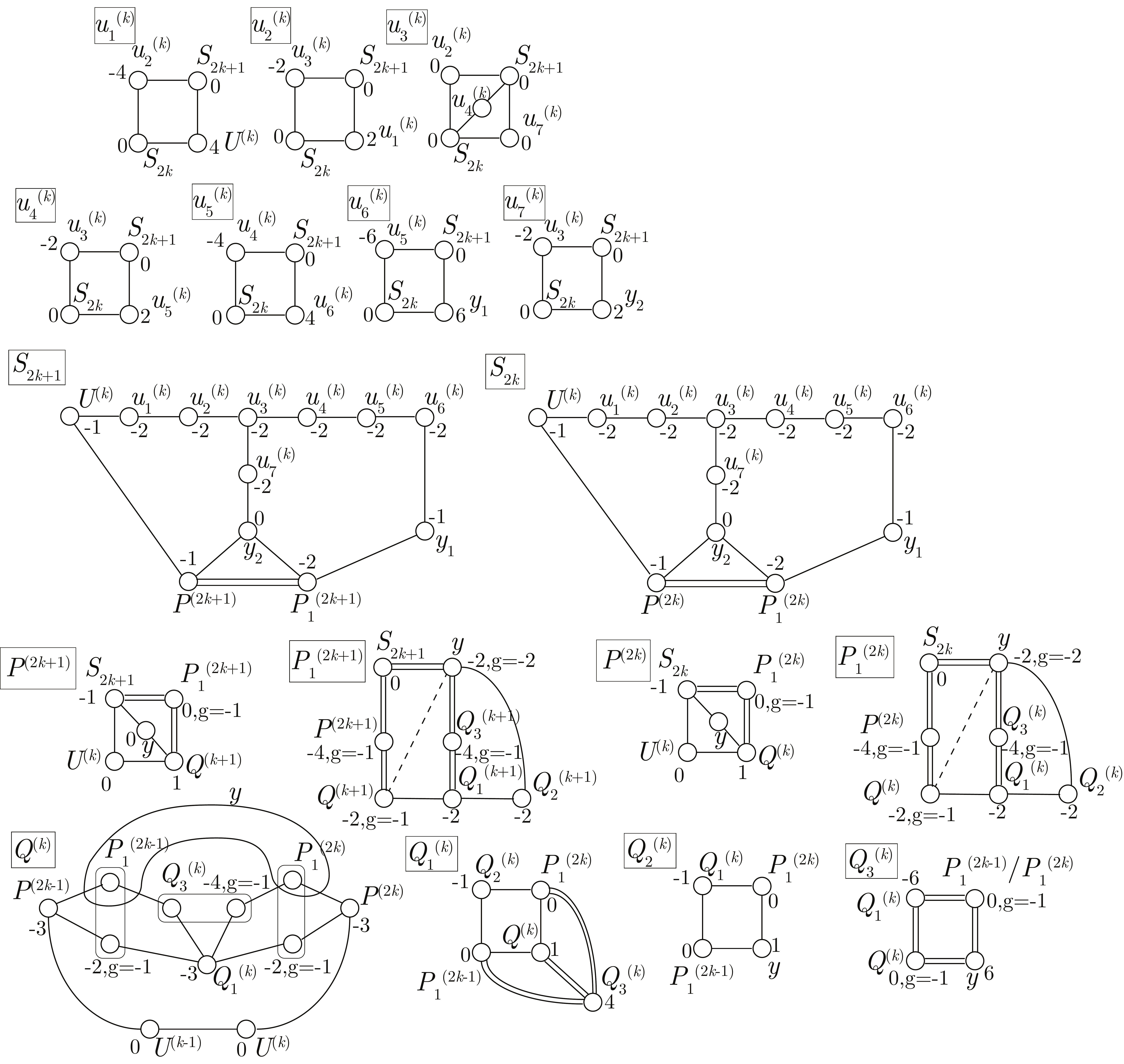}
\caption{The configuration of curves in the flopped geometry of non-minimal $(E_7,E_7)$ conformal matter with order $N$. Dotted line means negative intersection number.}
\label{f:E7E7flopresol}
\end{figure}

In the resolution geometry of the tensor branch, we label the exceptional divisors as follows $(0\leq k\leq N)$:
\be
\dots-\overset{U^{(k)},u_i^{(k)}}{8}-\overset{S_{2k+1}}{1}-\overset{P^{(2k+1)},P_1^{(2k+1)}}{2}-\overset{Q^{(k+1)},Q_i^{(k+1)}}{3}-\overset{P^{(2k+2)},P_1^{(2k+2)}}{2}-\overset{S_{2k+2}}{1}-\overset{U^{(k+1)},u_i^{(k+1)}}{8}-\dots\,.
\ee
The divisors $U^{(0)},u_i^{(0)}$ and $U^{(N)},u_i^{(N)}$ correspond to the non-compact flavor $E_7$. We plot the configuration of curves in figure~\ref{f:E7E7topresol}. Then we do a series of flops on the geometry. First we shrink the $\mc{O}(-1)\oplus\mc{O}(-1)$ curves $z\cdot S_k$ on all $S_k$. After that, we shrink the $\mc{O}(-1)\oplus\mc{O}(-1)$ curves $z\cdot P^{(k)}$ on all $P^{(k)}$. Finally, shrink the $\mc{O}(-1)\oplus\mc{O}(-1)$ curves $z\cdot Q^{(k)}$ on all $Q^{(k)}$. In the end, we get the configuration of curves on the flopped geometry in figure~\ref{f:E7E7flopresol}, where $U^{(k)}$ are decompactified.

To compute the $n(U^{(k)})$ for $k=1,\dots,N-1$, we need to compute the correct multiplicity factors for the curves $U^{(k)}\cdot S_{2k}$ and $U^{(k)}\cdot S_{2k+1}$ in figure~\ref{f:E7E7flopresol}. The chain of curves on $U^{(k)}$ and the multiplicity factors are:
\be
 \overset{1}{\underset{Q^{(k)}}{(-2)}}-\overset{2}{\underset{P^{2k}}{(-2)}}-\overset{3}{\underset{S_{2k}}{(-1)}}-\overset{1}{\underset{u_1^{(k)}}{(-6)}}-\overset{3}{\underset{S_{2k+1}}{(-1)}}-\overset{2}{\underset{P^{2k+1}}{(-2)}}-\overset{1}{\underset{Q^{(k+1)}}{(-2)}}\,.
\ee
The number in the bracket denotes the self-intersection number of that complete intersection curve inside $U^{(k)}$. Hence the multiplicity factors for $U^{(k)}\cdot S_{2k}$ and $U^{(k)}\cdot S_{2k+1}$ ($k=1,\dots,N-1$) are 3, and we can compute that the vertices $U^{(k)}$ ($k=1,\dots,N-1$) in the CFD has $n(U^{(k)})=-2$, $g(U^{(k)})=0$. Then we can read off the CFD in table \ref{tab:nm-CM-CFD} and \ref{tab:nm-E-CM-CFD},  with $G_F=E_7\times E_7\times SU(N)$.

\subsection{Non-Minimal $(E_8, E_8)$ Conformal Matter}
\label{app:E8E8}

In the resolution geometry of the tensor branch, we label the exceptional divisors as follows $(0\leq k\leq N)$:
\be
\ba
&\dots-\overset{U^{(k)},u_i^{(k)}}{12}-\overset{S_{2k+1}}{1}-\overset{P^{(2k+1)}}{2}-\overset{V^{(2k+1)},V_1^{(2k+1)}}{2}-\overset{W^{(2k+1)},W_i^{(2k+1)}}{3}-\overset{T_{2k+1}}{1}-\overset{Q^{(k+1)},Q_i^{(k+1)}}{5}-\cr
&\overset{T_{2k+2}}{1}-\overset{W^{(2k+2)},W_i^{(2k+2)}}{3}-\overset{V^{(2k+2)},V_1^{(2k+2)}}{2}-\overset{P^{(2k+2)}}{2}-\overset{S_{2k+2}}{1}-\overset{U^{(k+1)},u_i^{(k+1)}}{12}-\dots\,.
\ea
\ee

\begin{figure}
\centering
\includegraphics[width=13cm]{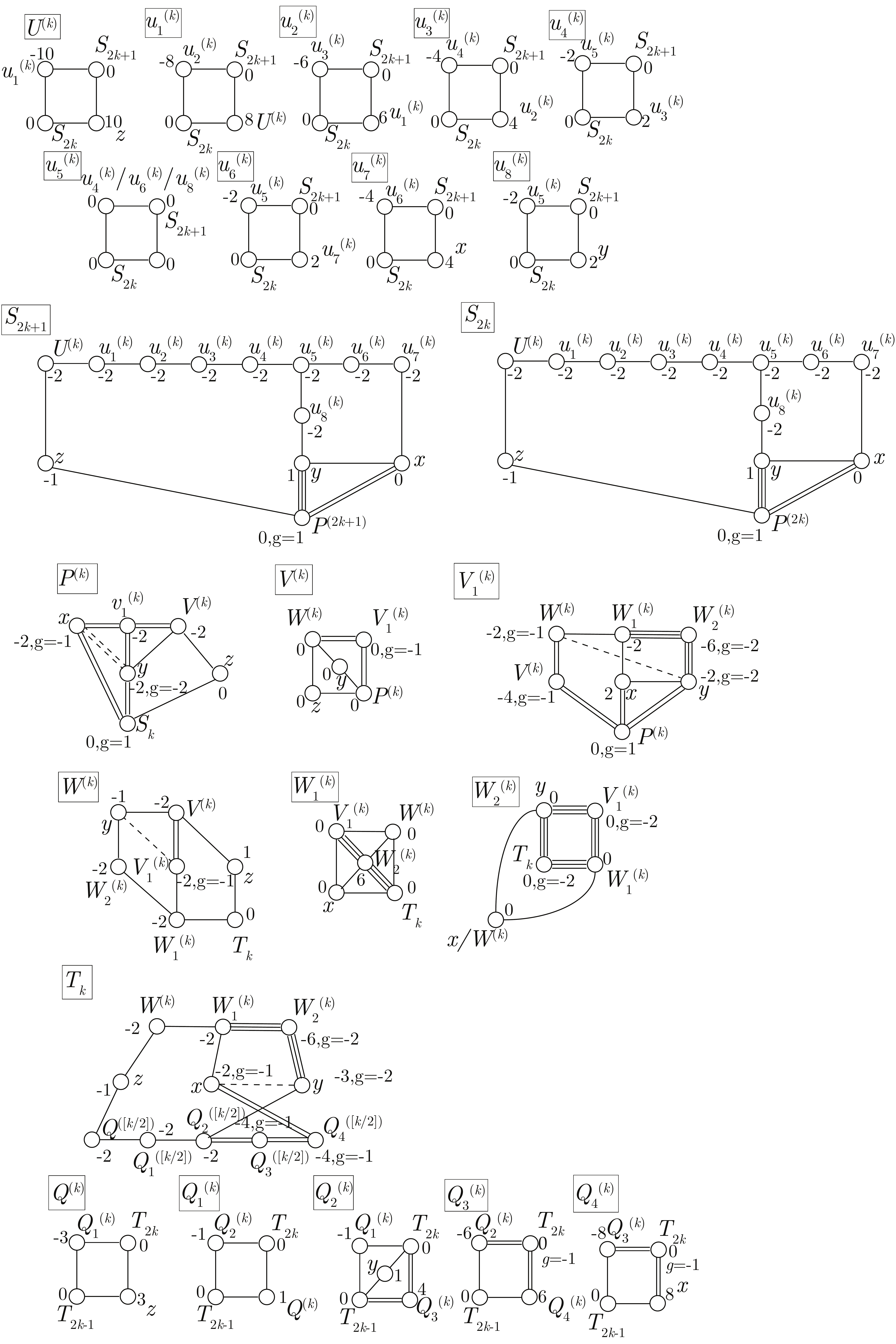}
\caption{The configuration of curves in the KK resolution geometry of non-minimal $(E_8,E_8)$ conformal matter with order $N$. ``[]'' means the rounded up integer value. Dotted line means negative intersection number.}
\label{f:E8E8topresol}
\end{figure}

\begin{figure}
\centering
\includegraphics[width=13cm]{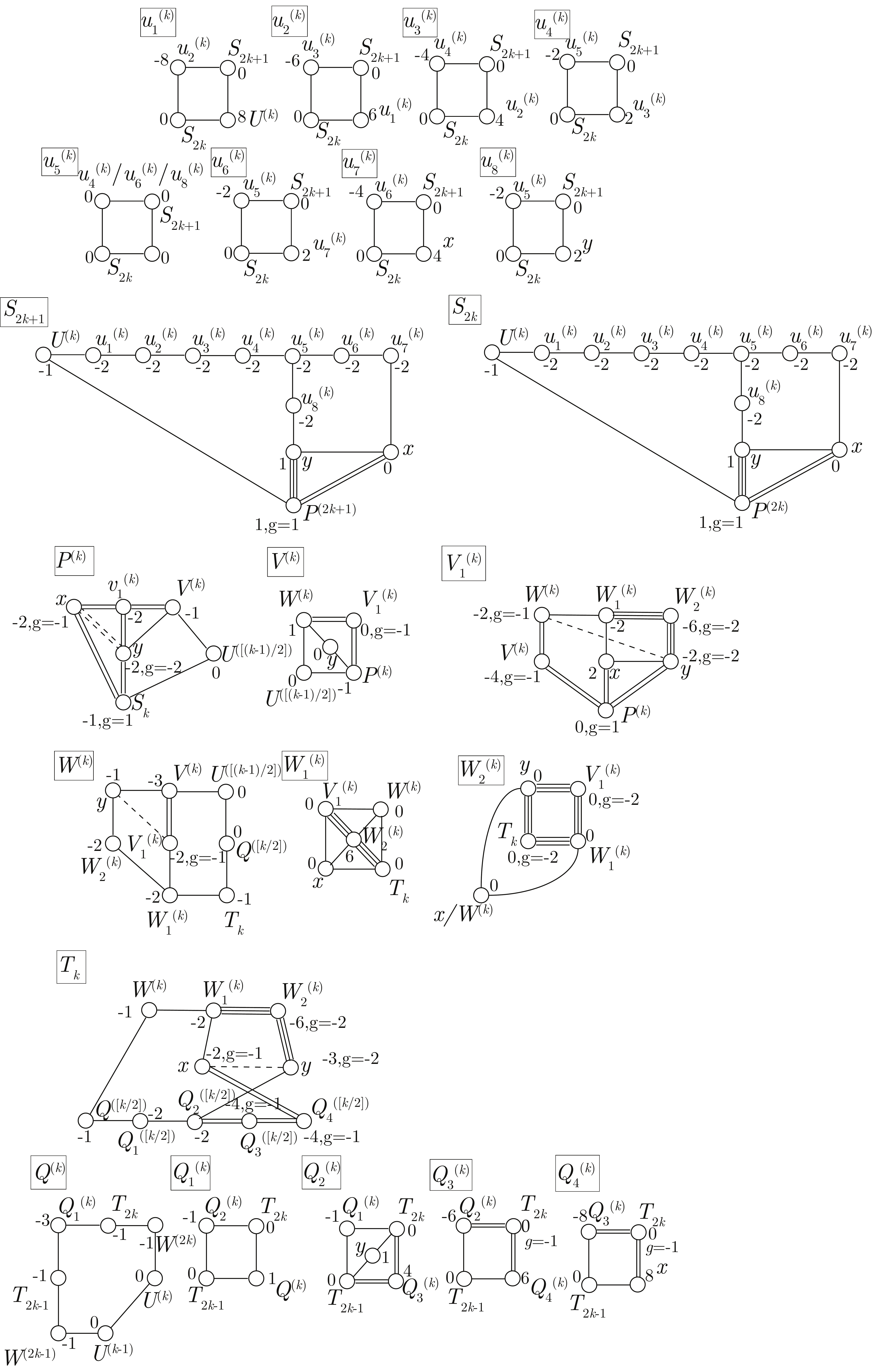}
\caption{The configuration of curves in the flopped geometry of non-minimal $(E_8,E_8)$ conformal matter with order $N$.  ``[]'' means the rounded up integer value. Dotted line means negative intersection number.}
\label{f:E8E8flopresol}
\end{figure}

The divisors $U^{(0)},u_i^{(0)}$ and $U^{(N)},u_i^{(N)}$ correspond to the non-compact flavor $E_8$. We plot the configuration of curves in figure~\ref{f:E8E8topresol}. Then we do a series of flops on the geometry. We first shrink the $\mc{O}(-1)\oplus\mc{O}(-1)$ curves $z\cdot S_k$ on all $S_k$. Then we shrink the $\mc{O}(-1)\oplus\mc{O}(-1)$ curves $z\cdot P^{(k)}$ on all $P^{(k)}$. After that, we shrink the $\mc{O}(-1)\oplus\mc{O}(-1)$ curves $z\cdot V^{(k)}$ on all $V^{(k)}$. After this step, the curves $z\cdot W^{(k)}$ are $0$-curves on $W^{(k)}$, which cannot be shrunk. So we shrink all the curves $z\cdot T_k$ as well, which results in the blow ups of $Q^{(k)}$. Finally, we shrink the $\mc{O}(-1)\oplus\mc{O}(-1)$ curves $z\cdot W^{(k)}$ on all $W^{(k)}$. The final curve configurations after the flop is shown in figure~\ref{f:E8E8flopresol}, where the surfaces $U^{(k)}$ are already decompactified. 

The chain of curves on $U^{(k)}$ and the multiplicity factors are:
\be
 \overset{1}{\underset{Q^{(k)}}{(-2)}}-\overset{2}{\underset{W^{(2k)}}{(-2)}}-\overset{3}{\underset{V^{(2k)}}{(-2)}}-\overset{4}{\underset{P^{(2k)}}{(-2)}}-\overset{5}{\underset{S_{2k}}{(-1)}}-\overset{1}{\underset{u_1^{(k)}}{(-10)}}-\overset{5}{\underset{S_{2k+1}}{(-1)}}-\overset{4}{\underset{P^{(2k+1)}}{(-2)}}-\overset{3}{\underset{V^{(2k+1)}}{(-2)}}-\overset{2}{\underset{W^{(2k+1)}}{(-2)}}-\overset{1}{\underset{Q^{(k+1)}}{(-2)}}\,.
\ee
The number in the bracket denotes the self-intersection number of that complete intersection curve inside $U^{(k)}$. Thus the curves $U^{(k)}\cdot S_{2k}$ and $U^{(k)}\cdot S_{2k+1}$ ($k=1,\dots,N-1$) have multiplicity factors 5, and we can compute that the vertices $U^{(k)}$ ($k=1,\dots,N-1$) in the CFD has $n(U^{(k)})=-2$, $g(U^{(k)})=0$. Then we can read off the CFD in table \ref{tab:nm-CM-CFD} and \ref{tab:nm-E-CM-CFD},  with $G_F=E_8\times E_8\times SU(N)$.

\newpage


\bibliography{FM}
\bibliographystyle{JHEP}

\end{document}